# CROSS-VALIDATION FOR UNSUPERVISED LEARNING

A DISSERTATION
SUBMITTED TO THE DEPARTMENT OF STATISTICS
AND THE COMMITTEE ON GRADUATE STUDIES
OF STANFORD UNIVERSITY
IN PARTIAL FULFILLMENT OF THE REQUIREMENTS
FOR THE DEGREE OF
DOCTOR OF PHILOSOPHY

Patrick O. Perry
September 2009

I certify that I have read this dissertation and that, in my opinion, it is fully adequate in scope and quality as a dissertation for the degree of Doctor of Philosophy.

(Art B. Owen) Principal Adviser

I certify that I have read this dissertation and that, in my opinion, it is fully adequate in scope and quality as a dissertation for the degree of Doctor of Philosophy.

(Iain M. Johnstone)

I certify that I have read this dissertation and that, in my opinion, it is fully adequate in scope and quality as a dissertation for the degree of Doctor of Philosophy.

(Jonathan E. Taylor)

Approved for the University Committee on Graduate Studies.

# Abstract

Cross-validation (CV) is a popular method for model-selection. Unfortunately, it is not immediately obvious how to apply CV to unsupervised or exploratory contexts. This thesis discusses some extensions of cross-validation to unsupervised learning, specifically focusing on the problem of choosing how many principal components to keep. We introduce the latent factor model, define an objective criterion, and show how CV can be used to estimate the intrinsic dimensionality of a data set. Through both simulation and theory, we demonstrate that cross-validation is a valuable tool for unsupervised learning.

# Acknowledgments

This work could not have been done alone. I would like to thank:

- Art Owen for always having an open door and for providing endless advice and encouragement;
- Gunnar Carlsson, Trevor Hastie, Iain Johnstone, and Jonathan Taylor for valuable feedback;
- Mollie Biewald, Ray, Kerry, Clare, and Erin Perry for unwavering support;
- the entire Stanford Statistics Department for community and education.

I would also like to thank Mark Churchland for providing the motor cortex data. This work was supported by a Stanford Graduate Fellowship and by the National Science Foundation under Grant DMS-0652743.

# Contents

# Chapter 1

# Introduction

Cross-validation (CV) is a popular method for model selection. It works by estimating the prediction error of each model under consideration and then choosing the model with the best performance. In unsupervised contexts, though, there is no clear notion as to what exactly "prediction error" is. Therefore, it is difficult to employ cross-validation for model selection in unsupervised or exploratory contexts. In this thesis, we take a look at some extensions of cross-validation to unsupervised learning. We focus specifically on the problem of choosing how many components to keep for principal component analysis (PCA), but many of the concepts we introduce are more broadly applicable.

Before we can do anything, we need a solid theoretical foundation. To this end, Chapter 2 gives a survey of relevant results from multivariate statistics and random matrix theory. Then, Chapter 3 derives the behavior of the singular value decomposition (SVD) for "signal-plus-noise" matrix models. These two chapters are complemented by Appendices A and B, which collect properties of random orthogonal matrices and give some limit theorems for weighted sums of random variables. Collectively, this work provides the groundwork for the rest of the thesis.

In Chapter 4, we introduce the latent factor model. This is a generative model for signal-plus-noise matrix data that expands the setup of PCA to include correlated factor loadings. We motivate loss functions for estimating the signal part, and then show how the SVD performs with respect to these criteria.

The next chapter (Chapter 5) focuses on cross-validation strategies. It covers both Wold-style "speckled" hold-outs as well as Gabriel-style "blocked" hold-outs. We define model error and prediction error for the latent factor model, and present the two cross-validation methods as estimators of prediction error. The chapter includes a comparison of CV methods with parametric model-selection procedures, showing through simulation that CV is much more robust to violations in model assumptions. In situations where parametric assumptions are unreasonable, cross-validation proves to be an attractive method for model selection.

Chapter 6, the final chapter, contains a theoretical analysis of Gabriel-style cross-validation for the SVD, also known as bi-cross-validation (BCV). This chapter shows that BCV is in general a biased estimator of model error, with an explicit expression for the bias. Despite this bias, though, the procedure can still be used successfully for model selection, provided the leave-out sizes are chosen appropriately. The chapter shows how to choose the leave-out sizes and proves a weak form of consistency.

Cross-validation is a valuable and flexible procedure for model selection. Through theory and simulation, this thesis demonstrates the applicability and utility of cross-validation as applied to principal component analysis. Many of the ideas in the following chapters generalize to other unsupervised learning procedures. This thesis shows that cross-validation can successfully be used for model selection in a variety of contexts.

# Chapter 2

# Multivariate statistics background

Multivariate statistics will prove to be a central tool for this thesis. We use this chapter to gather the relevant definitions and results, concentrating mainly on the eigenvalues and eigenvectors from sample covariance matrices. The literature in this area spans over fifty years. We give the basic definitions and properties of the multivariate normal and Wishart distributions in Section 2.1. Then, in Section 2.2, we survey classical results about sample covariance matrices when the number of dimensions, $p$, is fixed and the sample size, $n$, grows to infinity. This material is by now standard and can be found in any good multivariate statistics book (e.g. Muirhead [61]). Lastly, in Section 2.3, we survey modern asymptotics, where $n \to \infty$ and $p$ grows with $n$. Modern multivariate asymptotics is still an active research topic, but today it is possible to give a reasonably-complete description of the objects of interest.

## 2.1 The multivariate normal and Wishart distributions

We start with the definition of the multivariate normal distribution and some basic properties, which can be found, for example, in Muirhead [61][Chapters 1–3].

**Definition 2.1** (Multivariate Normal Distribution). *For mean vector* $\underline{\mu} \in \mathbb{R}^p$ *and*

*positive-semidefinite covariance matrix $\mathbf{\Sigma} \in \mathbb{R}^{p\times p}$, a random vector $\underline{X} \in \mathbb{R}^p$ is said to be distributed from the multivariate normal distribution, denoted $\underline{X} \sim \mathcal{N}\left(\underline{\mu}, \mathbf{\Sigma}\right)$ if for every fixed vector $\underline{a} \in \mathbb{R}^p$, the vector $\underline{a}^{\mathrm{T}}\underline{X}$ has a univariate normal distribution with mean $\underline{a}^{\mathrm{T}}\underline{\mu}$ and variance $\underline{a}^{\mathrm{T}}\mathbf{\Sigma}\underline{a}$.*

The multivariate normal distribution is defined for any positive-semidefinite covariance matrix $\mathbf{\Sigma}$, but it only has a density when $\mathbf{\Sigma}$ is strictly positive-definite.

**Proposition 2.2.** *If $\underline{X} \in \mathbb{R}^p$ follows a multivariate normal distribution with mean $\underline{\mu}$ and positive-definite covariance matrix $\mathbf{\Sigma}$, then its components have density*

$$f(\underline{x}) = (2\pi)^{-p/2}|\mathbf{\Sigma}|^{-1/2}\exp\left(-\frac{1}{2}(\underline{x}-\underline{\mu})^{\mathrm{T}}\mathbf{\Sigma}^{-1}(\underline{x}-\mu)\right). \tag{2.1}$$

A basic fact about the multivariate normal is the following:

**Proposition 2.3.** *Let $\underline{a} \in \mathbb{R}^p$, $\boldsymbol{C} \in \mathbb{R}^{q\times p}$, and $\underline{X} \sim \mathcal{N}\left(\underline{\mu}, \mathbf{\Sigma}\right)$. Define $\underline{Y} = \boldsymbol{C}\underline{X} + \underline{a}$. Then $\underline{Y} \sim \mathcal{N}\left(\boldsymbol{C}\underline{\mu} + \underline{a}, \boldsymbol{C}\mathbf{\Sigma}\boldsymbol{C}^{\mathrm{T}}\right)$.*

Two immediate corollaries are:

**Corollary 2.4.** *Suppose that $\underline{X} \sim \mathcal{N}\left(\underline{0}, \sigma^2\boldsymbol{I}_p\right)$ and that $\boldsymbol{O} \in \mathbb{R}^{p\times p}$ is an orthogonal matrix. Then $\boldsymbol{O}\underline{X} \stackrel{d}{=} \underline{X}$.*

**Corollary 2.5.** *If $\underline{X} \sim \mathcal{N}\left(\underline{0}, \mathbf{\Sigma}\right)$ and $\mathbf{\Sigma} = \boldsymbol{C}\boldsymbol{C}^{\mathrm{T}}$ for a matrix $\boldsymbol{C} \in \mathbb{R}^{p\times p}$, and if $\underline{Z} \sim \mathcal{N}\left(\underline{0}, \boldsymbol{I}_p\right)$, then $\boldsymbol{C}\underline{Z} \stackrel{d}{=} \underline{X}$.*

We are often interested in estimating the underlying parameters from multivariate normal data. The sufficient statistics are the standard estimates.

**Proposition 2.6.** *Say that $\underline{X}_1, \underline{X}_2, \ldots, \underline{X}_n$ are independent draws from a $\mathcal{N}\left(\underline{\mu}, \mathbf{\Sigma}\right)$ distribution. Then the sample mean*

$$\bar{\underline{X}}_n \equiv \frac{1}{n}\sum_{i=1}^{n}\underline{X}_i \tag{2.2}$$

*and the sample covariance*

$$\boldsymbol{S}_n \equiv \frac{1}{n-1}\sum_{i=1}^{n}\left(\underline{X}_i - \bar{\underline{X}}_n\right)\left(\underline{X}_i - \bar{\underline{X}}_n\right)^{\mathrm{T}} \tag{2.3}$$

*are sufficient statistics for* $\underline{\mu}$ *and* $\boldsymbol{\Sigma}$.

To describe the distribution of $\boldsymbol{S}_n$, we need to introduce the Wishart distribution.

**Definition 2.7** (Wishart Distribution)**.** *Let* $\underline{X}_1, \underline{X}_2, \ldots, \underline{X}_n \in \mathbb{R}^p$ *be an iid sequence of random vectors, each distributed as* $\mathcal{N}\left(\underline{0},\, \boldsymbol{\Sigma}\right)$. *Then the matrix*

$$\boldsymbol{A} = \sum_{i=1}^{n}\underline{X}_i\underline{X}_i^{\mathrm{T}}$$

*is said to have the Wishart distribution with* $n$ *degrees of freedom and scale parameter* $\boldsymbol{\Sigma}$. *We denote this by* $\boldsymbol{A} \sim \mathcal{W}_p\left(n,\, \boldsymbol{\Sigma}\right)$.

When $n \geq p$ and $\boldsymbol{\Sigma}$ is positive-definite, the elements of a Wishart matrix have a density.

**Proposition 2.8.** *Suppose that* $\boldsymbol{A} \sim \mathcal{W}_p\left(n,\, \boldsymbol{\Sigma}\right)$. *If* $n \geq p$ *and* $\boldsymbol{\Sigma}$ *is positive-definite, then the elements of* $\boldsymbol{A}$ *have a density over the space of positive-definite matrices, given by*

$$f(\boldsymbol{A}) = \frac{|\boldsymbol{A}|^{\frac{n-p-1}{2}}}{2^{\frac{np}{2}}|\boldsymbol{\Sigma}|^{\frac{n}{2}}\Gamma_p\left(\frac{n}{2}\right)}\exp\left(-\frac{1}{2}\operatorname{tr}\left(\boldsymbol{\Sigma}^{-1}\boldsymbol{A}\right)\right), \tag{2.4}$$

*where* $\Gamma_p\left(\cdot\right)$ *is the multivariate gamma function, computed as*

$$\Gamma_p\left(\frac{n}{2}\right) = \pi^{p(p-1)/4}\prod_{i=1}^{p}\Gamma\left(\frac{n+1-i}{2}\right). \tag{2.5}$$

We can now characterize the distributions of the sufficient statistics of a sequence of iid multivariate normal random vectors.

**Proposition 2.9.** *Let* $\bar{\underline{X}}_n$ *and* $\boldsymbol{S}_n$ *be defined as in Proposition 2.6. Then* $\bar{\underline{X}}_n$ *and* $\boldsymbol{S}_n$ *are independent with* $\bar{\underline{X}}_n \sim \mathcal{N}\left(\underline{\mu},\, \frac{1}{n}\boldsymbol{\Sigma}\right)$ *and* $(n-1)\boldsymbol{S}_n \sim \mathcal{W}_p\left(n-1,\, \boldsymbol{\Sigma}\right)$.

White Wishart matrices—those with scale parameter $\mathbf{\Sigma} = \sigma^2 \boldsymbol{I}_p$ are of particular interest. We can characterize their distribution in terms of eigenvalues and eigenvectors.

**Proposition 2.10.** *Suppose that* $\boldsymbol{A} \sim \mathcal{W}_p(n, \sigma^2 \boldsymbol{I}_p)$ *with* $n \geq p$ *and let* $\boldsymbol{A} = n\boldsymbol{O}\boldsymbol{L}\boldsymbol{O}^{\mathrm{T}}$ *be the spectral decomposition of* $\boldsymbol{A}$*, with* $\boldsymbol{L} = \operatorname{diag}(l_1, l_2, \ldots, l_p)$ *and* $l_1 > l_2 > \cdots > l_p > 0$*. Then* $\boldsymbol{O}$ *and* $\boldsymbol{L}$ *are independent, with* $\boldsymbol{O}$ *Haar-distributed over the group of* $p \times p$ *orthogonal matrices and the elements of* $\boldsymbol{L}$ *having density*

$$\left(\frac{1}{2\sigma^2}\right)^{np/2} \frac{\pi^{p^2/2}}{\Gamma_p\left(\frac{n}{2}\right)\Gamma_p\left(\frac{p}{2}\right)} \prod_{i<j}^{p} |l_i - l_j| \prod_{i=1}^{p} l_i^{(n-p-1)/2} e^{-\frac{l_i}{2\sigma^2}}. \tag{2.6}$$

In the random matrix theory literature, with $\sigma^2 = 1$ the eigenvalue density above is sometimes referred to as the Laguerre Orthogonal Ensemble (LOE).

## 2.2 Classical asymptotics

In this section we present results about sample covariance matrices when the sample size, $n$, tends to infinity, with the number of dimensions, $p$ a fixed constant. A straightforward application of the strong law of large numbers gives us the limits of the sample mean and covariance.

**Proposition 2.11.** *Let* $\underline{X}_1, \underline{X}_2, \ldots, \underline{X}_n$ *be a sequence of iid random vectors in* $\mathbb{R}^p$ *with* $\mathbb{E}[\underline{X}_1] = \underline{\mu}$ *and* $\mathbb{E}\left[\left(\underline{X}_1 - \underline{\mu}\right)\left(\underline{X}_1 - \underline{\mu}\right)^{\mathrm{T}}\right] = \mathbf{\Sigma}$*. Then, as* $n \to \infty$*,*

$$\bar{\underline{X}}_n \equiv \frac{1}{n} \sum_{i=1}^{n} \underline{X}_i \overset{a.s.}{\to} \underline{\mu}$$

*and*

$$\boldsymbol{S}_n \equiv \frac{1}{n-1} \sum_{i=1}^{n} \left(\underline{X}_i - \bar{\underline{X}}_n\right)\left(\underline{X}_i - \bar{\underline{X}}_n\right)^{\mathrm{T}} \overset{a.s.}{\to} \mathbf{\Sigma}.$$

To simplify matters, for the rest of the section we will mostly work in a setting when the variables have been centered. In this case, the sample covariance matrix takes the form $\boldsymbol{S}_n = \frac{1}{n}\sum_{i=1}^n \underline{X}_i \underline{X}_i^{\mathrm{T}}$. To see that centering the variables does not change the theory in any essential way, we provide the following proposition.

**Proposition 2.12.** *Let $\underline{X}_1, \underline{X}_2, \ldots, \underline{X}_n$ be a sequence of random observations in $\mathbb{R}^p$ with mean vector $\underline{\mu}$ and covariance matrix $\boldsymbol{\Sigma}$. Let $\bar{\underline{X}}_n = \frac{1}{n}\sum_{i=1}^n \underline{X}_i$ and $\boldsymbol{S}_n = \frac{1}{n-1}\sum_{i=1}^n \left(\underline{X}_i - \bar{\underline{X}}_n\right)\left(\underline{X}_i - \bar{\underline{X}}_n\right)^{\mathrm{T}}$ be the sample mean and covariance, respectively. Define the centered variables $\tilde{\underline{X}}_i = \underline{X}_i - \underline{\mu}$. Then*

$$\boldsymbol{S}_n = \frac{1}{n}\sum_{i=1}^n \tilde{\underline{X}}_i \tilde{\underline{X}}_i^{\mathrm{T}} + \mathcal{O}_P\left(\frac{1}{n}\right).$$

*In particular, this implies that*

$$\sqrt{n}\left(\boldsymbol{S}_n - \boldsymbol{\Sigma}\right) = \sqrt{n}\left(\frac{1}{n}\sum_{i=1}^n \tilde{\underline{X}}_i \tilde{\underline{X}}_i^{\mathrm{T}} - \boldsymbol{\Sigma}\right) + \mathcal{O}_P\left(\frac{1}{\sqrt{n}}\right).$$

*Proof.* We can write

$$\begin{aligned}
\boldsymbol{S}_n &= \frac{1}{n-1}\sum_{i=1}^n \tilde{\underline{X}}_i \tilde{\underline{X}}_i^{\mathrm{T}} + \frac{n}{n-1}\left(\bar{\underline{X}}_n - \underline{\mu}\right)\left(\bar{\underline{X}}_n - \underline{\mu}\right)^{\mathrm{T}} \\
&= \left(\frac{1}{n} + \frac{1}{n(n-1)}\right)\sum_{i=1}^n \tilde{\underline{X}}_i \tilde{\underline{X}}_i^{\mathrm{T}} + \frac{n}{n-1}\left(\bar{\underline{X}}_n - \underline{\mu}\right)\left(\bar{\underline{X}}_n - \underline{\mu}\right)^{\mathrm{T}}
\end{aligned}$$

The result follows since $\tilde{\underline{X}}_i \tilde{\underline{X}}_i^{\mathrm{T}} = \mathcal{O}_{\mathrm{P}}(1)$ and $\bar{\underline{X}}_n - \underline{\mu} = \mathcal{O}_{\mathrm{P}}\left(\frac{1}{\sqrt{n}}\right)$.

□

The next fact follows directly from the multivariate central limit theorem.

**Proposition 2.13.** *Suppose that $\underline{X}_1, \underline{X}_2, \ldots, \underline{X}_n$ is a sequence of iid random vectors in $\mathbb{R}^p$ with*

$$\mathbb{E}\left[\underline{X}_1 \underline{X}_1^{\mathrm{T}}\right] = \boldsymbol{\Sigma},$$

*and that for all $1 \leq i, j, i', j' \leq p$ there exists finite $\Gamma_{iji'j'}$ with*

$$\mathbb{E}\left[\left(X_{1i}X_{1j} - \Sigma_{ij}\right)\left(X_{1i'}X_{1j'} - \Sigma_{i'j'}\right)\right] = \Gamma_{iji'j'}.$$

*If* $\boldsymbol{S}_n = \frac{1}{n}\sum_{i=1}^{n} \underset{\sim}{X}_i \underset{\sim}{X}_i^{\mathrm{T}}$*, then* $\sqrt{n}\,\mathrm{vec}\,(\boldsymbol{S}_n - \boldsymbol{\Sigma}) \xrightarrow{d} \mathrm{vec}\,(\boldsymbol{G})\,,$ *where* $\boldsymbol{G}$ *is a random* $p \times p$ *symmetric matrix with* $\mathrm{vec}\,(\boldsymbol{G})$ *a mean-zero multivariate normal having covariance* $\mathrm{Cov}\,(G_{ij},\, G_{i'j'}) = \Gamma_{iji'j'}$.

In the previous proposition, vec is an operator that stacks the columns of an $n \times p$ matrix to create an $np$-dimensional vector.

If the elements of $\underset{\sim}{X}_1$ have vanishing first and third moments (for instance if the distribution of $\underset{\sim}{X}_1$ is symmetric about the origin, i.e. $\underset{\sim}{X}_1 \stackrel{d}{=} -\underset{\sim}{X}_1$), and if $\mathbb{E}\left[\underset{\sim}{X}_1 \underset{\sim}{X}_1^{\mathrm{T}}\right] = \mathrm{diag}\,(\lambda_1, \lambda_2, \ldots, \lambda_p)\,,$ then $\Gamma_{iji'j'}$ simplifies to

$$\Gamma_{iiii} = \mathbb{E}\left[X_{1i}^4\right] - \lambda_i^2 \qquad \text{for } 1 \le i \le p, \tag{2.7a}$$

$$\Gamma_{ijij} = \Gamma_{ijji} = \mathbb{E}\left[X_{1i}^2 X_{1j}^2\right] \qquad \text{for } 1 \le i, j \le p,\ i \ne j; \tag{2.7b}$$

all other values of $\Gamma_{iji'j'}$ are 0. In particular, this implies that the elements of $\mathrm{vec}\,(\boldsymbol{G})$ are independent. If we also have that $\underset{\sim}{X}_1$ is multivariate normal, then

$$\Gamma_{iiii} = 2\lambda_i^2 \qquad \text{for } 1 \le i \le p, \text{ and} \tag{2.8a}$$

$$\Gamma_{ijij} = \Gamma_{ijji} = \lambda_i \lambda_j \qquad \text{for } 1 \le i, j \le p,\ i \ne j. \tag{2.8b}$$

It is inconvenient to study the properties of the sample covariance matrix when the population covariance $\boldsymbol{\Sigma}$ is not diagonal. By factorizing $\boldsymbol{\Sigma} = \boldsymbol{\Phi}\boldsymbol{\Lambda}\boldsymbol{\Phi}^{\mathrm{T}}$ for orthogonal $\boldsymbol{\Phi}$ and diagonal $\boldsymbol{\Lambda}$, we can introduce $\tilde{\underset{\sim}{X}}_i \equiv \boldsymbol{\Phi}^{\mathrm{T}} \underset{\sim}{X}_i$ to get $\mathbb{E}\left[\tilde{\underset{\sim}{X}}_i \tilde{\underset{\sim}{X}}_i^{\mathrm{T}}\right] = \boldsymbol{\Lambda}$ and $\boldsymbol{S}_n = \boldsymbol{\Phi}\left(\frac{1}{n}\sum_{i=1}^{n} \tilde{\underset{\sim}{X}}_i \tilde{\underset{\sim}{X}}_i^{\mathrm{T}}\right)\boldsymbol{\Phi}^{\mathrm{T}}$. With this transformation, we can characterize the distribution of $\boldsymbol{S}_n$ completely in terms of $\frac{1}{n}\sum_{i=1}^{n} \tilde{\underset{\sim}{X}}_i \tilde{\underset{\sim}{X}}_i^{\mathrm{T}}$.

The next result we present is about the sample principal components. It is motivated by Proposition 2.13 and is originally due to Anderson [1].

**Theorem 2.14.** *For* $n \to \infty$ *and* $p$ *fixed, let* $\boldsymbol{S}_1, \boldsymbol{S}_2, \ldots, \boldsymbol{S}_n$ *be a sequence of random symmetric* $p \times p$ *matrices with* $\sqrt{n}\,\mathrm{vec}\,(\boldsymbol{S}_n - \boldsymbol{\Lambda}) \xrightarrow{d} \mathrm{vec}\,(\boldsymbol{G})\,,$ *for a deterministic* $\boldsymbol{\Lambda} = \mathrm{diag}\,(\lambda_1, \lambda_2, \ldots, \lambda_p)$ *having* $\lambda_1 > \lambda_2 > \cdots > \lambda_p$ *and a random symmetric matrix* $\boldsymbol{G}$*. Let* $\boldsymbol{S}_n = \boldsymbol{U}_n \boldsymbol{L}_n \boldsymbol{U}_n^{\mathrm{T}}$ *be the eigendecomposition of* $\boldsymbol{S}_n$*, with* $\boldsymbol{L}_n = \mathrm{diag}\,(l_{n,1}, l_{n,2}, \ldots, l_{n,p})$ *and* $l_{n,1} \ge l_{n,2} \ge \cdots \ge l_{n,p}$*. If* $\boldsymbol{G} = \mathcal{O}_P(1)$*, and the signs of* $\boldsymbol{U}_n$ *are chosen so that $U_{n,ii} \geq 0$ for $1 \leq i \leq p$, then the elements of $\boldsymbol{U}_n$ and $\boldsymbol{L}_n$ converge jointly as*

$$\sqrt{n}\left(U_{n,ii} - 1\right) \xrightarrow{P} 0 \qquad \text{for } 1 \leq i \leq p, \tag{2.9a}$$

$$\sqrt{n} U_{n,ij} \xrightarrow{d} -\frac{G_{ij}}{\lambda_i - \lambda_j} \qquad \text{for } 1 \leq i, j \leq p \text{ with } i \neq j, \text{ and} \tag{2.9b}$$

$$\sqrt{n}\left(l_{n,i} - \lambda_i\right) \xrightarrow{d} G_{ii} \qquad \text{for } 1 \leq i \leq p. \tag{2.9c}$$

More generally, Anderson treats the case when the $\lambda_i$ are not all unique. The key ingredient to Anderson's proof is a perturbation lemma, which we state and prove below.

**Lemma 2.15.** *For $n \to \infty$ and fixed $p$ let $\boldsymbol{S}_1, \boldsymbol{S}_2, \ldots, \boldsymbol{S}_n \in \mathbb{R}^{p \times p}$ be a sequence of symmetric matrices of the form*

$$\boldsymbol{S}_n = \boldsymbol{\Lambda} + \frac{1}{\sqrt{n}} \boldsymbol{H}_n + o\left(\frac{1}{\sqrt{n}}\right),$$

*where $\boldsymbol{\Lambda} = \operatorname{diag}\left(\lambda_1, \lambda_2, \ldots, \lambda_p\right)$ with $\lambda_1 > \lambda_2 > \cdots > \lambda_p$ and $\boldsymbol{H}_n = \mathcal{O}(1)$. Let $\boldsymbol{S}_n = \boldsymbol{U}_n \boldsymbol{L}_n \boldsymbol{U}_n^{\mathrm{T}}$ be the eigendecomposition of $\boldsymbol{S}_n$, with $\boldsymbol{L}_n = \operatorname{diag}\left(l_{n,1}, l_{n,2}, \ldots, l_{n,p}\right)$. Further suppose that $U_{n,ii} \geq 0$ for $1 \leq i \leq p$ and $l_{n,1} > l_{n,2} > \cdots > l_{n,p}$. Then for all $1 \leq i, j \leq p$ and $i \neq j$ we have*

$$U_{n,ii} = 1 + o\left(\frac{1}{\sqrt{n}}\right), \tag{2.10a}$$

$$U_{n,ij} = -\frac{1}{\sqrt{n}} \frac{H_{n,ij}}{\lambda_i - \lambda_j} + o\left(\frac{1}{\sqrt{n}}\right), \quad \text{and} \tag{2.10b}$$

$$l_{n,i} = \lambda_i + \frac{H_{n,ii}}{\sqrt{n}} + o\left(\frac{1}{\sqrt{n}}\right). \tag{2.10c}$$

*Proof.* Define $p \times p$ matrices $\boldsymbol{E}_n$, $\boldsymbol{F}_n$, and $\boldsymbol{\Delta}_n$ so that

$$\boldsymbol{E}_n = \operatorname{diag}\left(U_{n,11}, U_{n,22}, \ldots, U_{n,pp}\right), \tag{2.11}$$

$$\boldsymbol{F}_n = \sqrt{n}\left(\boldsymbol{U}_n - \boldsymbol{E}_n\right), \tag{2.12}$$

$$\boldsymbol{\Delta}_n = \sqrt{n}\left(\boldsymbol{L}_n - \boldsymbol{\Lambda}\right), \tag{2.13}$$

giving

$$\begin{aligned}
\boldsymbol{U}_n &= \boldsymbol{E}_n + \frac{1}{\sqrt{n}}\boldsymbol{F}_n, \quad \text{and} \\
\boldsymbol{L}_n &= \boldsymbol{\Lambda} + \frac{1}{\sqrt{n}}\boldsymbol{\Delta}_n.
\end{aligned}$$

We have that

$$\begin{aligned}
\boldsymbol{S}_n &= \boldsymbol{\Lambda} + \frac{1}{\sqrt{n}}\boldsymbol{H}_n + o\left(\frac{1}{\sqrt{n}}\right) \\
&= \boldsymbol{U}_n \boldsymbol{L}_n \boldsymbol{U}_n^{\mathrm{T}} \\
&= \boldsymbol{E}_n \boldsymbol{\Lambda} \boldsymbol{E}_n^{\mathrm{T}} + \frac{1}{\sqrt{n}}\left(\boldsymbol{E}_n \boldsymbol{\Delta}_n \boldsymbol{E}_n^{\mathrm{T}} + \boldsymbol{F}_n \boldsymbol{\Lambda} \boldsymbol{E}_n^{\mathrm{T}} + \boldsymbol{E}_n \boldsymbol{\Lambda} \boldsymbol{F}_n^{\mathrm{T}}\right) + \frac{1}{n}\boldsymbol{M}_n
\end{aligned} \tag{2.14}$$

where the elements of $\boldsymbol{M}_n$ are sums of $\mathcal{O}(p)$ terms, with each term a product of elements taken from $\boldsymbol{E}_n$, $\boldsymbol{F}_n$, $\boldsymbol{\Lambda}$, and $\boldsymbol{\Delta}_n$. Also,

$$\begin{aligned}
\boldsymbol{I}_p &= \boldsymbol{U}_n \boldsymbol{U}_n^{\mathrm{T}} \\
&= \boldsymbol{E}_n \boldsymbol{E}_n^{\mathrm{T}} + \frac{1}{\sqrt{n}}\left(\boldsymbol{F}_n \boldsymbol{E}_n^{\mathrm{T}} + \boldsymbol{E}_n \boldsymbol{F}_n^{\mathrm{T}}\right) + \frac{1}{n}\boldsymbol{W}_n,
\end{aligned} \tag{2.15}$$

where $\boldsymbol{W}_n = \boldsymbol{F}_n \boldsymbol{F}_n^{\mathrm{T}}$. From (2.15) we see that for $1 \leq i, j \leq p$ and $i \neq j$ we must have

$$1 = E_{n,ii}^2 + \frac{1}{n} W_{n,ii}, \quad \text{and} \tag{2.16a}$$

$$0 = E_{n,ii} F_{n,ji} + F_{n,ij} E_{n,jj} + \frac{1}{\sqrt{n}} W_{n,ij}. \tag{2.16b}$$

Substituting $E_{n,ii}^2 = 1 - \frac{1}{n} W_{n,ii}$ into equation (2.14), we get

$$H_{n,ii} = E_{n,ii} \Delta_{n,ii} E_{n,ii} + \frac{1}{\sqrt{n}} \left( M_{n,ii} - \lambda_i W_{n,ii} \right) + o\left(1\right), \quad \text{and} \tag{2.17a}$$

$$H_{n,ij} = \lambda_j E_{n,jj} F_{n,ij} + \lambda_i F_{n,ji} E_{n,ii} + \frac{1}{\sqrt{n}} M_{n,ij} + o\left(1\right). \tag{2.17b}$$

Equations (2.16a)–(2.17b) admit the solution

$$E_{n,ii} = 1 + o\left(\frac{1}{\sqrt{n}}\right), \tag{2.18a}$$

$$F_{n,ij} = -\frac{H_{n,ij}}{\lambda_i - \lambda_j} + o\left(1\right), \quad \text{and} \tag{2.18b}$$

$$\Delta_{n,ii} = H_{n,ii} + o\left(1\right). \tag{2.18c}$$

This completes the proof. □

An application of the results of this section is the following theorem, which describes the behavior of principal component analysis for large $n$ and fixed $p$.

**Theorem 2.16.** *Let $\underset{\sim}{X}_1, \underset{\sim}{X}_2, \ldots, \underset{\sim}{X}_n$ be a sequence of iid $\mathcal{N}\left(\underset{\sim}{\mu}, \boldsymbol{\Sigma}\right)$ random vectors in $\mathbb{R}^p$, with sample mean $\bar{\underset{\sim}{X}}_n$ and sample covariance $\boldsymbol{S}_n$. Let $\boldsymbol{\Sigma} = \boldsymbol{\Phi}\boldsymbol{\Lambda}\boldsymbol{\Phi}^{\mathrm{T}}$ be the eigendecomposition of $\boldsymbol{\Sigma}$, with $\boldsymbol{\Lambda} = \operatorname{diag}\left(\lambda_1, \lambda_2, \ldots, \lambda_p\right)$ and $\lambda_1 > \lambda_2 > \cdots > \lambda_p > 0$. Similarly, let $\boldsymbol{S}_n = \boldsymbol{U}_n \boldsymbol{L}_n \boldsymbol{U}_n^{\mathrm{T}}$ be the eigendecomposition of $\boldsymbol{S}_n$, likewise with $\boldsymbol{L}_n = \operatorname{diag}\left(l_{n,1}, l_{n,2}, \ldots, l_{n,p}\right)$, $l_{n,1} > l_{n,2} > \cdots > l_{n,p}$, and signs chosen so that $\left(\boldsymbol{\Phi}^{\mathrm{T}} \boldsymbol{U}_n\right)_{ii} \geq 0$ for $1 \leq i \leq p$. Then*

*(i) $l_{n,i} \overset{a.s.}{\to} \lambda_i$ for $1 \leq i \leq p$, and $\boldsymbol{U}_n \overset{a.s.}{\to} \Phi$ .*

*(ii) After appropriate cenering and scaling, $\{l_{n,i}\}_{i=1}^{p}$ and $\boldsymbol{U}_n$ converge jointly in distribution and their limits are independent. For all $1 \leq i \leq p$*

$$\sqrt{n}\left(l_{n,i} - \lambda_i\right) \overset{d}{\to} \mathcal{N}\left(0,\, 2\lambda_i^2\right),$$

*and*

$$\sqrt{n}\left(\boldsymbol{\Phi}^{\mathrm{T}} \boldsymbol{U}_n - \boldsymbol{I}_p\right) \overset{d}{\to} \boldsymbol{F},$$

*where $\boldsymbol{F}$ is a skew-symmetric matrix with elements above the diagonal independent of each other and distributed as*

$$F_{ij} \sim \mathcal{N}\left(0, \frac{\lambda_i \lambda_j}{(\lambda_i - \lambda_j)^2}\right), \quad \textit{for all } 1 \leq i < j \leq p.$$

*Proof.* Part (i) is a restatement of Proposition 2.11. Part (ii) follows from Proposition 2.13 and Theorem 2.15. □

## 2.3 Modern asymptotics

We now present some results about sample covariance matrices when both the sample size, $n$, and the dimensionality, $p$, go to infinity. Specifically, most of these results suppose that $n \to \infty$, $p \to \infty$, and $\frac{n}{p} \to \gamma$ for a fixed constant $\gamma \in (0, \infty)$. There is no widely-accepted name for $\gamma$, but we will sometimes adopt the terminology of Marčenko and Pastur [56] and call it the *concentration*.

Most of the random matrix theory literature concerning sample covariance matrices is focused on eigenvalues. Given a sequence of sample covariance matrices $\boldsymbol{S}_1, \boldsymbol{S}_2, \ldots, \boldsymbol{S}_n$, with $\boldsymbol{S}_n \in \mathbb{R}^{p \times p}$ and $p = p(n)$ these results generally come in one of two forms. If we label the eigenvalues of $\boldsymbol{S}_n$ as $l_{n,1}, l_{n,2}, \ldots, l_{n,p}$, with $l_{n,1} \geq l_{n,2} \geq \cdots \geq l_{n,p}$, then we can define a random measure

$$F^{\boldsymbol{S}_n} = \frac{1}{p} \sum_{i=1}^{p} \delta_{l_{n,i}}. \tag{2.19}$$

This measure represents a random draw from the set of eigenvalues of $\boldsymbol{S}_n$ that puts equal weight on each eigenvalue. It is called the *spectral measure* of $\boldsymbol{S}_n$. Results about $F^{\boldsymbol{S}_n}$ are generally called results about the "bulk" of the spectrum.

The second major class of results is concerned with the behavior of the extreme eigenvalues $l_{n,1}$ and $l_{n,p}$. Results of this type are called "edge" results.

### 2.3.1 The bulk of the spectrum

To work in a setting where the dimensionality $p$, grows with the sample size, $n$, we introduce a triangular array of sample vectors. The sample covariance matrix $\boldsymbol{S}_n$ is of dimension $p \times p$ and is formed from row $n$ of a triangular array of independent random vectors, $\utilde{X}_{n,1}, \utilde{X}_{n,2}, \ldots, \utilde{X}_{n,n}$. Specifically, $\boldsymbol{S}_n = \frac{1}{n} \sum_{i=1}^{n} \utilde{X}_{n,i} \utilde{X}_{n,i}^{\mathrm{T}}$. We let $\boldsymbol{X}_n$ be the $n \times p$ matrix $\boldsymbol{X}_n = \begin{pmatrix} \utilde{X}_{n,1} & \utilde{X}_{n,2} & \cdots & \utilde{X}_{n,n} \end{pmatrix}^{\mathrm{T}}$, so that $\boldsymbol{S}_n = \frac{1}{n} \boldsymbol{X}_n^{\mathrm{T}} \boldsymbol{X}_n$. Most asymptotic results about sample covariance matrices are expressed in terms of $\boldsymbol{X}_n$ rather than $\boldsymbol{S}_n$. For example, the next theorem about the spectral measure of a large covariance matrix is stated this way.

**Theorem 2.17.** *Let $\boldsymbol{X}_1, \boldsymbol{X}_2, \ldots, \boldsymbol{X}_n$ be a sequence of random matrices of increasing dimension as $n \to \infty$, so that $\boldsymbol{X}_n \in \mathbb{R}^{n \times p}$ and $p = p(n)$. Define $\boldsymbol{S}_n = \frac{1}{n} \boldsymbol{X}_n^{\mathrm{T}} \boldsymbol{X}_n$. If the elements of $\boldsymbol{X}_n$ are iid with $\mathbb{E}|X_{n,11} - \mathbb{E}X_{n,11}|^2 = 1$ and $\frac{n}{p} \to \gamma > 0$, then the empirical spectral measure $F^{\boldsymbol{S}_n}$ almost surely converges in distribution to a deterministic probability measure. This measure, denoted $F_\gamma^{MP}$, is called the Marčenko-Pastur Law. For $\gamma \geq 1$ it has density*

$$f_\gamma^{MP}(x) = \frac{\gamma}{2\pi x} \sqrt{(x - a_\gamma)(b_\gamma - x)}, \quad a_\gamma \leq x \leq b_\gamma, \tag{2.20}$$

*where $a_\gamma = \left(1 - \frac{1}{\sqrt{\gamma}}\right)^2$ and $b_\gamma = \left(1 + \frac{1}{\sqrt{\gamma}}\right)^2$. When $\gamma < 1$, there is an additional point-mass of $(1 - \gamma)$ at the origin.*

Figure 2.1 shows the density $f_\gamma^{\mathrm{MP}}(x)$ for different values of $\gamma$. The reason for choosing the name "concentration" to refer to $\gamma$ becomes apparent in that for larger values of $\gamma$, $F_\gamma^{\mathrm{MP}}$ becomes more and more concentrated around its mean.

The limiting behavior of the empirical spectral measure of a sample covariance matrix was originally studied by Marčenko and Pastur [56] in 1967. Since then, several papers have refined these results, including Grenander and Silverstein [37], Wachter [92], Jonsson [47], Yin and Krishnaiah [98], Yin [96], Silverstein and Bai [82], and Silverstein [81]. These papers either proceed via a combinatorial argument involving the moments of the matrix elements, or else they employ a tool called the Stieltjes transform. Theorem 2.17 is a simplified version of Silverstein and Bai's main result, which more generally considers complex-valued random variables and allows the columns of $\boldsymbol{X}_n$ to have heterogeneous variances.

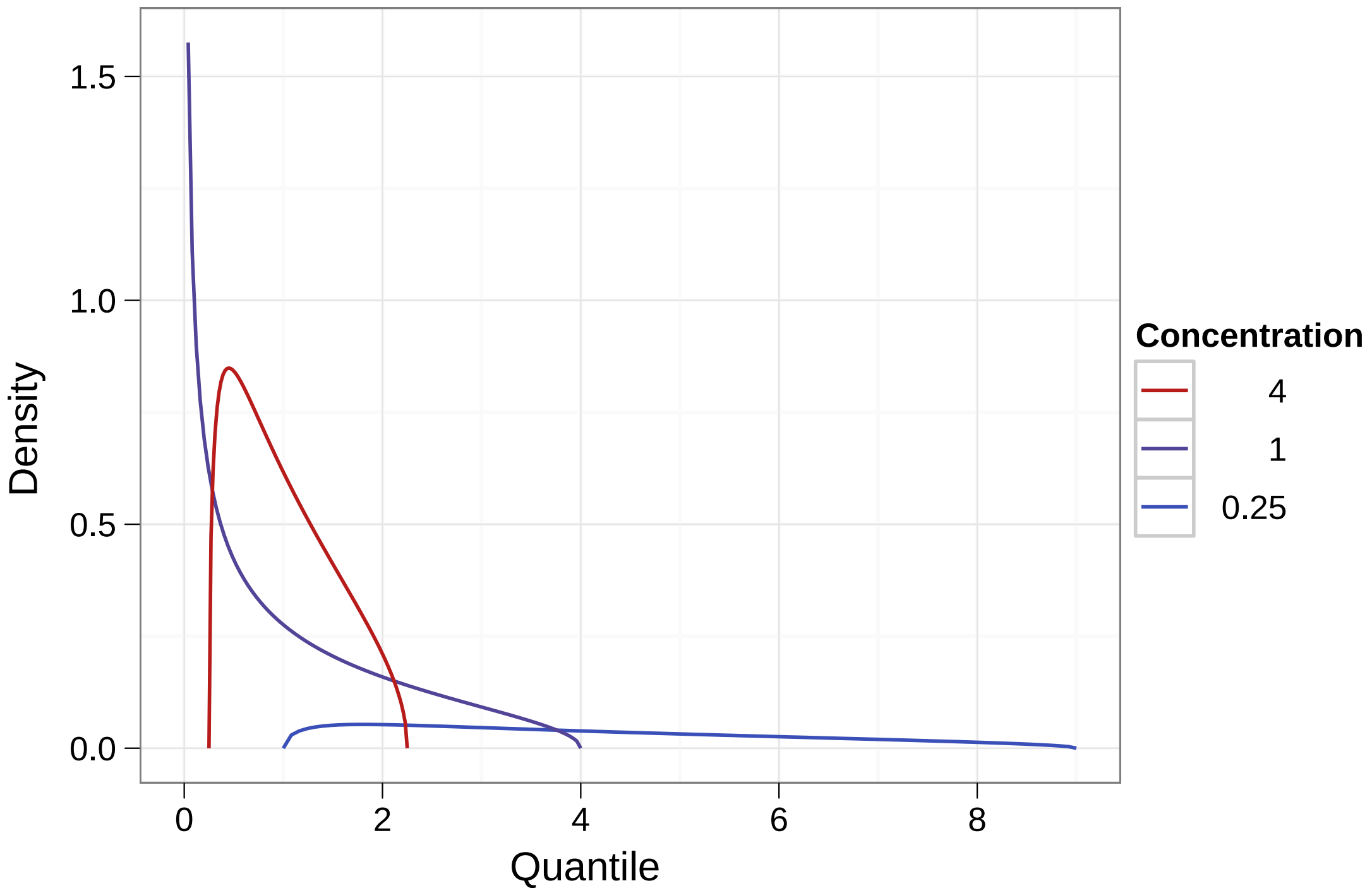

Figure 2.1: THE MARČENKO-PASTUR LAW. Density, $f_\gamma^{\mathrm{MP}}(x)$, plotted against quantile, $x$, for concentration $\gamma = 0.25$, 1, and 4. Concentration is equal to the number of samples per dimension. For $\gamma < 1$, there is an addition point-mass of size $(1 - \gamma)$ at $x = 0$.

The meaning of the phrase "$F^{\boldsymbol{S}_n}$ converges almost surely in distribution to $F_\gamma^{\mathrm{MP}}$" is that for all $x$ which are continuity points of $F_\gamma^{\mathrm{MP}}$,

$$F^{\boldsymbol{S}_n}(x) \overset{a.s.}{\to} F_\gamma^{\mathrm{MP}}(x) \,. \tag{2.21}$$

Equivalently, Theorem 2.17 can be stated as a strong law of large numbers.

**Corollary 2.18** (Wishart LLN)**.** *Let* $\boldsymbol{X}_n$ *and* $\{l_{n,i}\}_{i=1}^p$ *be as in Theorem 2.17. Let*

*$g : \mathbb{R} \to \mathbb{R}$ be any continuous bounded function. Then*

$$\frac{1}{p} \sum_{i=1}^{p} g\left(l_{n,i}\right) \overset{a.s.}{\to} \int g(x)\, dF_{\gamma}^{MP}(x). \tag{2.22}$$

Concerning convergence rates for the quantities in Theorem 2.17, Bai et al. [4] study the total variation distance between $F^{\boldsymbol{S}_n}$ and $F_{\gamma}^{\mathrm{MP}}$. Under suitable conditions on $X_{n,11}$ and $\gamma$, they show that $\|F^{\boldsymbol{S}_n} - F_{\gamma}^{\mathrm{MP}}\|_{\mathrm{TV}} = \mathcal{O}_{\mathrm{P}}\left(n^{-2/5}\right)$ and that with probability one $\|F^{\boldsymbol{S}_n} - F_{\gamma}^{\mathrm{MP}}\|_{\mathrm{TV}} = \mathcal{O}\left(n^{-2/5+\varepsilon}\right)$ for any $\varepsilon > 0$. Guionnet and Zeitouni [39] give concentration of measure results. If $X_{n,11}$ satisfies a Poincaré inequality and $g$ is Lipschitz, they show that for $\delta$ large enough,

$$-\frac{1}{n^2} \log \mathbb{P}\left\{\left|\int g(x)\, dF^{\boldsymbol{S}_n}(x) - \int g(x)\, dF_{\gamma}^{\mathrm{MP}}(x)\right| > \delta\right\} = \mathcal{O}\left(\delta^2\right), \tag{2.23}$$

with an explicit bound on the error. If one is willing to assume that the elements of $\boldsymbol{X}_n$ are Gaussian, then Hiai and Petz [41] give an exact value for the quantity in (2.23). Guionnet [38] gives a survey of other large deviations results.

It is interesting to look at the scaled behavior in equation (2.22) when the quantities are scaled by $p$. Indeed one can prove a Central Limit Theorem (CLT) for functionals of eigenvalues.

**Theorem 2.19** (Wishart CLT). *Let $\boldsymbol{X}_1, \boldsymbol{X}_2, \ldots, \boldsymbol{X}_n$ be a sequence of random $n \times p$ matrices with $p = p(n)$. Assume that $\boldsymbol{X}_n$ has iid elements and that $\mathbb{E}\left[X_{n,11}\right] = 0$, $\mathbb{E}\left[X_{n,11}^2\right] = 1$, and $\mathbb{E}\left[X_{n,11}^4\right] < \infty$. Define $\boldsymbol{S}_n = \frac{1}{n}\boldsymbol{X}_n^{\mathrm{T}}\boldsymbol{X}_n$ and let $F^{\boldsymbol{S}_n}$ be its spectral measure. If $n \to \infty$, $\frac{n}{p} \to \gamma$ and $g_1, g_2, \ldots, g_k$ are real-valued functions analytic on the support of $F_{\gamma}^{MP}$, then the sequence of random vectors*

$$p \cdot \Big( \int g_1(x)\, dF^{\boldsymbol{S}_n}(x) - \int g_1(x)\, dF^{MP}(x), \\ \int g_2(x)\, dF^{\boldsymbol{S}_n}(x) - \int g_2(x)\, dF^{MP}(x), \ldots, \\ \int g_k(x)\, dF^{\boldsymbol{S}_n}(x) - \int g_k(x)\, dF^{MP}(x) \Big)$$

*is tight. Moreover, if* $\mathbb{E}\left[X_{n,11}^4\right] = 3$*, then the sequence converges in distribution to a multivariate normal with mean* $\underline{\mu}$ *and covariance* $\boldsymbol{\Sigma}$*, where*

$$\mu_i = \frac{g_i(a_\gamma) + g_i(b_\gamma)}{4} - \frac{1}{2\pi}\int_{a_\gamma}^{b_\gamma} \frac{g_i(x)}{\sqrt{(x-a_\gamma)(b_\gamma - x)}}\,dx \tag{2.24}$$

*and*

$$\Sigma_{ij} = -\frac{1}{2\pi^2}\iint \frac{g_i(z_1)g_j(z_2)}{\left(m(z_1)-m(z_2)\right)^2}\frac{d}{dz_1}m(z_1)\frac{d}{dz_2}m(z_2)dz_1dz_2. \tag{2.25}$$

*The integrals in* (2.25) *are contour integrals enclosing the support of* $F_\gamma^{MP}$*, and*

$$m(z) = \frac{-(z+1-\gamma^{-1}) + \sqrt{(z-a_\gamma)(z-b_\gamma)}}{2z}, \tag{2.26}$$

*with the square root defined to have positive imaginary part when* $\Im z > 0$*.*

The case when $\mathbb{E}\left[X_{n,11}^4\right] = 3$ is of particular interest because it arises when $X_{n,11} \sim \mathcal{N}(0,1)$.

This theorem was proved by Bai and Silverstein [6], and can be considered a generalization of the work by Johnsson [47]. In computing the variance integral (2.25), it is useful to know that $m$ satisfies the identities

$$m(\bar{z}) = \overline{m(z)},$$

and

$$z = -\frac{1}{m(z)} + \frac{\gamma^{-1}}{m(z)+1}.$$

Bai and Silverstein show how to compute the limiting means and variances for $g(x) = \log x$ and $g(x) = x^r$. They also derive a similar CLT when the elements of $\boldsymbol{X}_n$ are correlated. Pastur and Lytova [66] have recently relaxed some of the assumptions made by Bai and Silverstein.

### 2.3.2 The edges of the spectrum

We now turn our attention to the extreme eigenvalues of a sample covariance matrix. It seems plausible that if $F^{\boldsymbol{S}_n} \xrightarrow{d} F_\gamma^{\mathrm{MP}}$, then the extreme eigenvalues of $\boldsymbol{S}_n$ should converge to the edges of the support of $F_\gamma^{\mathrm{MP}}$. Indeed, under suitable assumptions, this is exactly what happens. For the largest eigenvalue, work on this problem started with Geman [34], and his assumptions were further weakened by Jonsson [48] and Silverstein [76]. Yin et al. [97] prove the result under the weakest possible conditions [7].

**Theorem 2.20.** *Let $\boldsymbol{X}_1, \boldsymbol{X}_2, \ldots, \boldsymbol{X}_n$ be a sequence of random matrices of increasing dimension, with $\boldsymbol{X}_n$ of size $n \times p$, $p = p(n)$, $n \to \infty$, and $\frac{n}{p} \to \gamma \in (0, \infty)$. Let $\boldsymbol{S}_n = \frac{1}{n}\boldsymbol{X}_n^{\mathrm{T}}\boldsymbol{X}_n$ and denote its eigenvalues by $l_{n,1} \geq l_{n,2} \geq \cdots \geq l_{n,p}$. If the elements of $\boldsymbol{X}_n$ are iid with $\mathbb{E}X_{n,11} = 0$, $\mathbb{E}X_{n,11}^2 = 1$, and $\mathbb{E}X_{n,11}^4 < \infty$, then*

$$l_{n,1} \xrightarrow{a.s.} \left(1 + \gamma^{-1/2}\right)^2.$$

For the smallest eigenvalue, the first work was by Silverstein [78], who gives a result when $X_{n,11} \sim \mathcal{N}(0,\, 1)$. Bai and Yin [10] proved a theorem that mirrors Theorem 2.20.

**Theorem 2.21.** *Let $\boldsymbol{X}_n$, $n$, $p$, and $\{l_{n,i}\}_{i=1}^p$ be as in Theorem 2.20. If $\mathbb{E}X_{n,11}^4 < \infty$ and $\gamma \geq 1$, then*

$$l_{n,p} \xrightarrow{a.s.} \left(1 - \gamma^{-1/2}\right)^2.$$

*With the same moment assumption on $X_{n,11}$, if $0 < \gamma < 1$, then*

$$l_{n,p-n+1} \xrightarrow{a.s.} \left(1 - \gamma^{-1/2}\right)^2.$$

For the case when the elements of $\boldsymbol{X}_n$ are correlated, Bai and Silverstein [5] give a general result that subsumes Theorems 2.20 and 2.21.

After appropriate centering and scaling, the largest eigenvalue of a white Wishart matrix converges weakly to a random variable with known distribution. Johansson [44] proved this statement and identified the limiting distribution for complex white Wishart matrices. Johnstone [45] later provided an analogous result for real matrices, which we state below.

**Theorem 2.22.** *Let $\boldsymbol{X}_1, \boldsymbol{X}_2, \ldots, \boldsymbol{X}_n$ be a sequence of random matrices of increasing dimension, with $\boldsymbol{X}_n \in \mathbb{R}^{n\times p}$, $p = p(n)$, and $n \to \infty$ with $\frac{n}{p} \to \gamma \in (0, \infty)$. Define $\boldsymbol{S}_n = \frac{1}{n}\boldsymbol{X}_n^{\mathrm{T}}\boldsymbol{X}_n$ and label its eigenvalues $l_{n,1} \geq l_{n,2} \geq \cdots \geq l_{n,p}$. If the elements of $\boldsymbol{X}_n$ are iid with $X_{n,11} \sim \mathcal{N}(0,\, 1)$, then*

$$\frac{l_{n,1} - \mu_{n,p}}{\sigma_{n,p}} \xrightarrow{d} W_1 \sim F_1^{TW},$$

*where*

$$\mu_{n,p} = \frac{1}{n}\left(\sqrt{n - 1/2} + \sqrt{p - 1/2}\right),$$
$$\sigma_{n,p} = \frac{1}{n}\left(\sqrt{n - 1/2} + \sqrt{p - 1/2}\right)\left(\frac{1}{\sqrt{n-1/2}} + \frac{1}{\sqrt{p-1/2}}\right)^{1/3},$$

*and $F_1^{TW}$ is the Tracy-Widom law of order 1.*

El Karoui [28] extended this result to apply when $\gamma = 0$ or $\gamma = \infty$. With appropriate modifications to $\mu_{n,p}$ and $\sigma_{n,p}$, he later gave a convergence rate of order $(n \wedge p)^{2/3}$ for complex-valued data [29]. Ma [53] gave the analogous result for real-valued data. For correlated complex normals, El Karoui [30] derived a more general version of Theorem 2.22.

The Tracy-Widom distribution, which appears in Theorem 2.22, was established to be the limiting distribution (after appropriate scaling) of the maximum eigenvalue from an $n \times n$ symmetric matrix with independent entries distributed as $\mathcal{N}(0,\, 2)$ along the main diagonal and $\mathcal{N}(0,\, 1)$ otherwise [89] [90]. To describte $F_1^{\mathrm{TW}}$, let $q(x)$ solve the Painlevé II equation

$$q''(x) = xq(x) + 2q^3(x),$$

with boundary condition $q(x) \sim \mathrm{Ai}(x)$ as $x \to \infty$ and $\mathrm{Ai}(x)$ the Airy function. Then it follows that

$$F_1^{\mathrm{TW}}(x) = \exp\left\{-\frac{1}{2}\int_s^\infty q(x) + (x - s)q^2(x)\, dx\right\}.$$

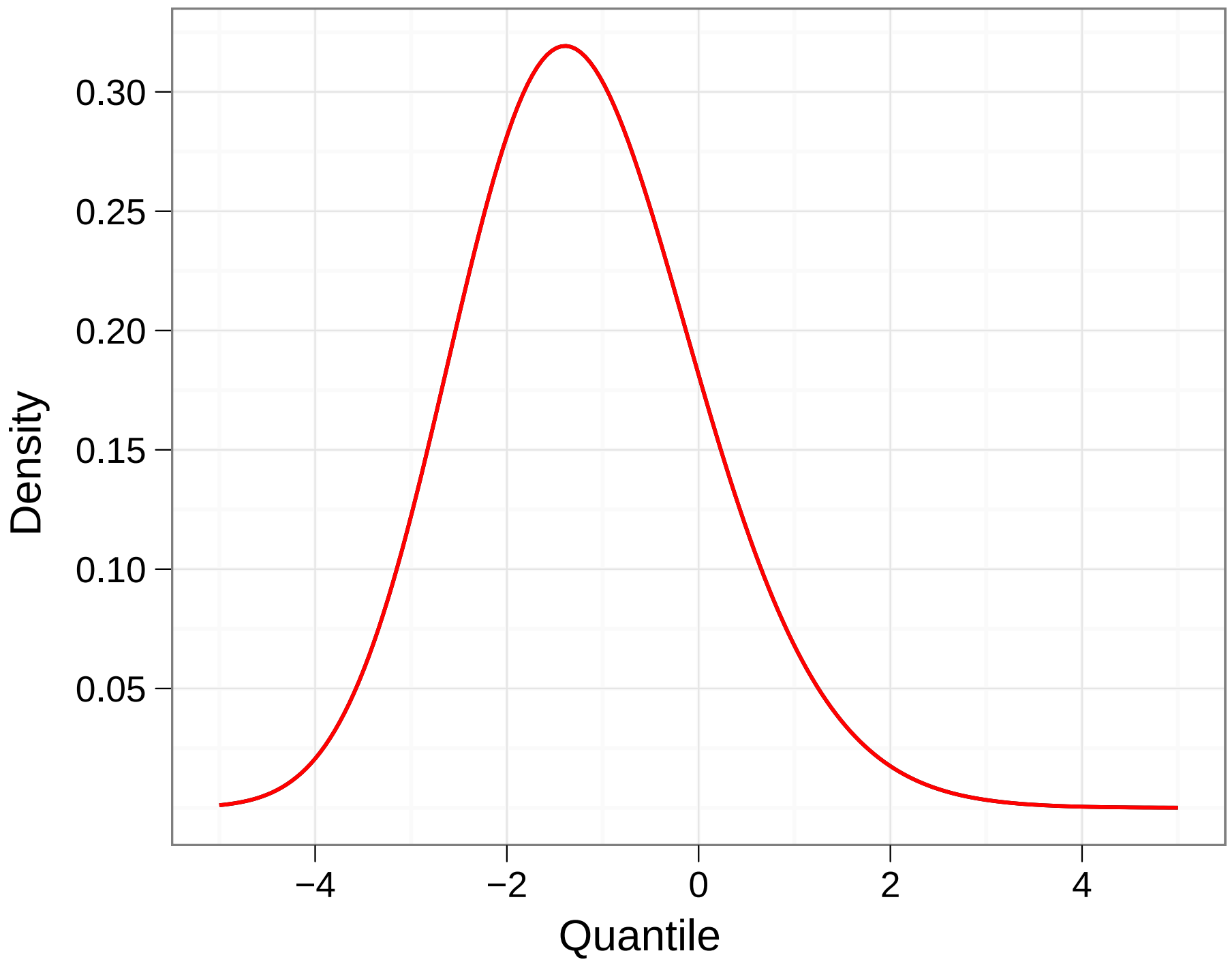

Figure 2.2: THE TRACY-WIDOM LAW. Limiting density of the largest eigenvalue from a white Wishart matrix after appropriate centering and scaling.

Hastings and McLeod [40] study the tail behavior of $q(x)$. Using their analysis, one can show (see, e.g. [70]) that for $s \to -\infty$,

$$F_1^{\text{TW}}(s) \sim \exp\left(-\frac{|s|^3}{24}\right),$$

while for $s \to \infty$,

$$1 - F_1^{\text{TW}}(s) \sim \frac{s^{-3/4}}{4\sqrt{\pi}} \exp\left(-\frac{2}{3}s^{3/2}\right).$$

The density of $F_1^{\text{TW}}$ is shown in Figure 2.2.

A result like Theorem 2.22 holds true for the smallest eigenvalue. We define the *Reflected Tracy-Widom Law* to have distribution function $G_1^{\text{TW}}(s) = 1 - F_1^{\text{TW}}(-s)$. Then we have

**Theorem 2.23.** *With the same assumptions as in Theorem 2.22, if* $\gamma \in (1, \infty)$ *then*

$$\frac{l_{n,p} - \mu^-_{n,p}}{\sigma^-_{n,p}} \xrightarrow{d} W_1 \sim G_1^{TW},$$

*where*

$$\mu^-_{n,p} = \frac{1}{n}\left(\sqrt{n - 1/2} - \sqrt{p - 1/2}\right),$$

$$\sigma^-_{n,p} = \frac{1}{n}\left(\sqrt{n - 1/2} - \sqrt{p - 1/2}\right)\left(\frac{1}{\sqrt{p - 1/2}} - \frac{1}{\sqrt{n - 1/2}}\right)^{1/3}.$$

*If* $\gamma \in (0, 1)$, *then*

$$\frac{l_{n,p-n+1} - \mu^-_{p,n}}{\sigma^-_{p,n}} \xrightarrow{d} W_1 \sim G_1^{TW}.$$

We get the result for $\gamma \in (0, 1)$ by reversing the role of $n$ and $p$. Baker et al [13] proved the result for complex data. Paul [67] extended the result to real data when $\gamma \to \infty$. Ma [53] gives convergence rates. In practice, $\log l_{n,p}$ converges in distribution faster than $l_{n,p}$. Ma recommends appropriate centering and scaling constants for $\log l_{n,p}$ to converge in distribution to a $G^{\mathrm{TW}}$ random variable at rate $(n \wedge p)^{2/3}$.

Theorem 2.23 does not apply when $\frac{n}{p} \to 1$. Edelman [26] derived the limiting distribution of the smallest eigenvalue when $n = p$. It is not known if his result holds more generally when $\frac{n}{p} \to 1$.

**Theorem 2.24.** *Let* $\boldsymbol{X}_n$ *and* $\{l_{n,i}\}_{i=1}^p$ *be as in Theorem 2.22. If* $p(n) = n$, *then for* $t \geq 0$,

$$\mathbb{P}\{n\, l_{n,p} \leq t\} \to \int_0^t \frac{1 + \sqrt{x}}{\sqrt{x}} e^{-(x/2 + \sqrt{x})}\, dx.$$

In addition to the extreme eigenvalues, it is possible to study the joint distribution of top or bottom $k$ sample eigenvalues for fixed $k$ as $n \to \infty$. In light of Theorem 2.17, for fixed $k$ we must have that the top (respectively, bottom) sample eigenvalues converge almost surely to the same limit. Furthermore, Soshnikov [83] showed that applying the centering and scaling from Theorem 2.22 to the top $k$ sample eigenvalues gives a specific limiting distribution.

It is natural to ask if the limiting eigenvalue distributions are specific to Wishart matrices, or if they apply to non-Gaussian data as well. There is compelling evidence that the Tracy-Widom law is universal. Soshnikov [83] extended Theorem 2.22 to more general $\boldsymbol{X}_n$ under the assumption that $X_{n,11}$ is sub-Gaussian and $n - p = \mathcal{O}(p^{1/3})$. Péché [69] later removed the restriction on $n - p$. Tao and Vu [87] showed that Theorem 2.24 applies for general $X_{n,11}$ with $\mathbb{E}X_{n,11} = 0$ and $\mathbb{E}X_{n,11}^2 = 1$.

### 2.3.3 Eigenvectors

Relatively little attention has been focused on the eigenvectors of sample covariance matrices. While many results are known, as of yet there is no complete characterization of the eigenvectors from a general sample covariance matrix. Most of the difficulty in tackling the problem is that it is hard to describe convergence properties of $\boldsymbol{U}_n$, the $p \times p$ matrix of eigenvectors, when $n$ and $p$ go to infinity. The individual $p^2$ elements of $\boldsymbol{U}_n$ do not converge in any meaningful way, so the challenge is to come up with relevant macroscopic characteristics of $\boldsymbol{U}_n$.

Silverstein [74] was perhaps the first to study the eigenvectors of large-dimensional sample covariance matrices. He hypothesized that for sample covariance matrices of increasing dimension, the eigenvector matrix becomes more and more "Haar-like". A random matrix $\boldsymbol{U} \in \mathbb{R}^{p \times p}$ is said to be Haar-distributed over the orthogonal group if for every fixed $p \times p$ orthogonal matrix $\boldsymbol{O}$, the rotated matrix $\boldsymbol{OU}$ has the same distribution as $\boldsymbol{U}$. That is, $\boldsymbol{OU} \overset{d}{=} \boldsymbol{U}$. Silverstein's conjecture was that as $n \to \infty$, $\boldsymbol{U}_n$ behaves more and more like a Haar-distributed matrix. The next theorem displays one aspect of Haar-like behavior.

To state the theorem, we need to define the extension of a scalar function $g : \mathbb{R} \mapsto \mathbb{R}$ to symmetric matrix arguments. If $\boldsymbol{S} = \boldsymbol{ULU}^{\mathrm{T}}$ is the eigendecomposition of the symmetric matrix $\boldsymbol{S} \in \mathbb{R}^{p \times p}$, with $\boldsymbol{L} = \operatorname{diag}(l_1, l_2, \ldots, l_p)$, then we define

$$g(\boldsymbol{S}) = \boldsymbol{U} \Big( \operatorname{diag}\big(g(l_1), g(l_2), \ldots, g(l_p)\big) \Big) \boldsymbol{U}^{\mathrm{T}}.$$

With this notion, we can state Silverstein's result.

**Theorem 2.25.** *Let $\boldsymbol{X}_1, \boldsymbol{X}_2, \ldots, \boldsymbol{X}_n$ be a sequence of random matrices of increasing dimension, with $\boldsymbol{X}_n \in \mathbb{R}^{n \times p}$, $p = p(n)$, and $\boldsymbol{X}_n$ having iid elements with $\mathbb{E}X_{n,11} = 0$, $\mathbb{E}X_{n,11}^2 = 1$, and $\mathbb{E}X_{n,11}^4 < \infty$. Define $\boldsymbol{S}_n = \frac{1}{n}\boldsymbol{X}_n^{\mathrm{T}}\boldsymbol{X}_n$. Let $\underline{a}_1, \underline{a}_2, \ldots, \underline{a}_n$ be a sequence of nonrandom unit vectors with $\underline{a}_n$ in $\mathbb{R}^p$ and let $g : \mathbb{R} \to \mathbb{R}$ be a continuous bounded function. If $n \to \infty$ and $\frac{n}{p} \to \gamma \in (0, \infty)$, then*

$$\underline{a}_n^{\mathrm{T}} g\left(\boldsymbol{S}_n\right) \underline{a}_n \xrightarrow{a.s.} \int g(x)\, dF_\gamma^{MP}(x).$$

Silverstein [74] proves the result for convergence in probability and a specific class of $\boldsymbol{X}_n$. Bai et al. [3] strengthen the result to a larger class of $\boldsymbol{X}_n$ and proves almost-sure convergence. They also consider dependence in $\boldsymbol{X}_n$.

It may not be immediately obvious how Theorem 2.25 is related to eigenvectors. If $\boldsymbol{S}_n = \boldsymbol{U}_n \boldsymbol{L}_n \boldsymbol{U}_n$ is the eigendecomposition of $\boldsymbol{S}_n$, with $\boldsymbol{L}_n = \operatorname{diag}\left(l_{n,1}, l_{n,2}, \ldots, l_{n,p}\right)$, then $g(\boldsymbol{L}_n) = \operatorname{diag}\big(g(l_{n,1}), g(l_{n,2}), \ldots, g(l_{n,p})\big)$ and $g(\boldsymbol{S}_n) = \boldsymbol{U}_n g(\boldsymbol{L}_n) \boldsymbol{U}_n^{\mathrm{T}}$. We let $\underline{b}_n = \boldsymbol{U}_n \underline{a}_n$. Then,

$$\underline{a}_n^{\mathrm{T}} g(\boldsymbol{S}_n) \underline{a}_n = \sum_{i=1}^{p} b_{n,i}^2\, g(l_{n,i}). \tag{2.27}$$

If $\boldsymbol{U}_n$ is Haar-distributed, then $\underline{b}_n$ will be distributed uniformly over the unit sphere in $\mathbb{R}^p$, and the average in (2.27) will put about weight $\frac{1}{p}$ on each eigenvalue. If $\boldsymbol{U}_n$ puts bias in any particular direction then the average will put extra weight on particular eigenvalues.

Silverstein investigated second-order behavior of eigenvectors in [75], [77], [79], and [80]. He demonstrated that certain second-order behavior of $\boldsymbol{U}_n$ depends in a crucial way on the fourth moment of $X_{n,11}$. This greatly restricts the class of $\boldsymbol{X}_n$ for with the eigenvectors of $\boldsymbol{S}_n$ are Haar-like.

**Theorem 2.26.** *Let $\boldsymbol{X}_n$, $\boldsymbol{S}_n$, and $\underline{a}_n$ be as in Theorem 2.25. Suppose also that $\mathbb{E}X_{n,11}^4 = 3$. Let $g_1, g_2, \ldots, g_k$ be real-valued functions analytic on the support of $F_\gamma^{MP}$.*

*Then, the random vector*

$$\sqrt{p} \cdot \Big( \underline{a}_n^{\mathrm{T}} g_1(\boldsymbol{S}_n) \underline{a}_n - \int g_1(x)\, dF_\gamma^{MP}(x),$$

$$\underline{a}_n^{\mathrm{T}} g_2(\boldsymbol{S}_n) \underline{a}_n - \int g_2(x)\, dF_\gamma^{MP}(x), \ldots,$$

$$\underline{a}_n^{\mathrm{T}} g_k(\boldsymbol{S}_n) \underline{a}_n - \int g_k(x)\, dF_\gamma^{MP}(x) \Big)$$

*converges in distribution to a mean-zero multivariate normal with covariance between the $i$th and $j$th components equal to*

$$\int g_i(x) g_j(x)\, dF_\gamma^{MP}(x) - \int g_i(x)\, dF_\gamma^{MP}(x) \cdot \int g_j(x)\, dF_\gamma^{MP}(x).$$

Bai et al. [3] give a similar result for complex-valued and correlated $\boldsymbol{X}_n$. Silverstein [79] showed that if $g_1(x) = x$ and $g_2(x) = x^2$, then the condition $\mathbb{E} X_{n,11}^4 = 3$ is necessary for the random vector in Theorem 2.26 to converge in distribution for all $\underline{a}_n$. However, for the specific choice of $\underline{a}_n = \left( \frac{1}{\sqrt{p}}, \frac{1}{\sqrt{p}}, \ldots, \frac{1}{\sqrt{p}} \right)$, he later showed that the conclusions of Theorem 2.26 hold more generally when $X_{n,11}$ is symmetric and $\mathbb{E} X_{n,11}^4 < \infty$ [80].

# Chapter 3

# Behavior of the SVD in low-rank plus noise models

## 3.1 Introduction

Many modern data sets involve simultaneous measurements of a large number of variables. Some financial applications, such as portfolio selection, involve looking at the market prices of hundreds or thousands of stocks and their evolution over time [57]. Microarray studies involve measuring the expression levels of thousands of genes simultaneously [73]. Text processing involves counting the appearances of thousands of words on thousands of documents [55]. In all of these applications, it is natural to suppose that even though the data are high-dimensional, their dynamics are driven by a relatively small number of latent factors.

Under the hypothesis that one ore more latent factors explain the behavior of a data set, principal component analysis (PCA) [46] is a popular method for estimating these latent factors. When the dimensionality of the data is small relative to the sample size, Anderson's 1963 paper [1] gives a complete treatment of how the procedure behaves. Unfortunately, his results do not apply when the sample size is comparable to the dimensionality.

A further complication with many modern data sets is that it is no longer appropriate to assume the observations are iid. Also, sometimes it is difficult to distinguish between "observation" and "variable". We call such data "transposable". A microarray study involving measurements of $p$ genes for $n$ different patients can be considered transposable: we can either treat each gene as a measurement of the patient, or we can treat each patient as a measurement of the gene. There are correlations both between the genes *and* between the patients, so in fact both interpretations are relevant [27].

One can study latent factor models in a transpose-agnostic way by considering generative models of the form $\boldsymbol{X} = \boldsymbol{U}\boldsymbol{D}\boldsymbol{V}^{\mathrm{T}} + \boldsymbol{E}$. Here, $\boldsymbol{X}$ is the $n \times p$ observed data matrix. The unobserved row and column factors are given by $\boldsymbol{U}$ and $\boldsymbol{V}$, respectively, matrices with $k$ orthonormal columns each, and $k \ll \min(n, p)$. The strengths of the factors are given in the $k \times k$ diagonal matrix $\boldsymbol{D}$, and $\boldsymbol{E}$ is a matrix of noise. A natural estimator for the latent factor term $\boldsymbol{U}\boldsymbol{D}\boldsymbol{V}^{\mathrm{T}}$ can be constructed by truncating the singular value decomposition (SVD) [36] of $\boldsymbol{X}$. The goal of this paper is to study the behavior of the SVD when $n$ and $p$ both tend to infinity, with their ratio tending to a nonzero constant.

In an upcoming paper, Onatski [64] gives a thorough treatment of latent factor models. He assumes that the elements of $\boldsymbol{E}$ are iid Gaussians, that $\sqrt{n}(\boldsymbol{U}^{\mathrm{T}}\boldsymbol{U} - \boldsymbol{I}_k)$ tends to a multivariate Gaussian distribution, and that $\boldsymbol{V}$ and $\boldsymbol{D}$ are both non-random. The contributions of this chapter are twofold. First, we work under a transpose-agnostic generative model that allows randomness in all three of $\boldsymbol{U}$, $\boldsymbol{D}$, and $\boldsymbol{V}$. Second, we give a more complete picture of the almost-sure limits of the sample singular vectors, taking into account the signs of the dot products between the population and sample vectors.

We describe the main results in Section 3.2. Sections 3.3–3.8 are dedicated to proving the two main theorems. Finaly, we discuss related work and extensions in Section 3.9. We owe a substantial debt to Onatski's work. Although most of the details below are different, the general outline and the main points of the argument are the same.

## 3.2 Assumptions, notation, and main results

Here we make explicit what the model and assumptions are, and we present our main results.

### 3.2.1 Assumptions and notation

We will work sequences of matrices indexed by $n$, with

$$\boldsymbol{X}_n = \sqrt{n}\,\boldsymbol{U}_n \boldsymbol{D}_n \boldsymbol{V}_n^{\mathrm{T}} + \boldsymbol{E}_n. \tag{3.1}$$

We denote by $\sqrt{n}\,\boldsymbol{U}_n \boldsymbol{D}_n \boldsymbol{V}_n^{\mathrm{T}}$ the "signal" part of the matrix $\boldsymbol{X}_n$ and $\boldsymbol{E}_n$ the "noise" part. We will often refer to $\boldsymbol{U}_n$ and $\boldsymbol{V}_n$ as the left and right factors of $\boldsymbol{X}_n$, and the matrix $\boldsymbol{D}_n$ will be called the matrix of normalized factor strengths. The first two assumptions concern the signal part:

**Assumption 3.1.** *The factors $\boldsymbol{U}_n$ and $\boldsymbol{V}_n$ are random matrices of dimensions $n \times k$ and $p \times k$, respectively, normalized so that $\boldsymbol{U}_n^{\mathrm{T}} \boldsymbol{U}_n = \boldsymbol{V}_n^{\mathrm{T}} \boldsymbol{V}_n = \boldsymbol{I}_k$. The number of factors, $k$, is a fixed constant. The aspect ratio satisfies $\frac{n}{p} = \gamma + o\left(\frac{1}{\sqrt{n}}\right)$ for a fixed nonzero constant $\gamma \in (0, \infty)$.*

**Assumption 3.2.** *The matrix of factor strengths, $\boldsymbol{D}_n$, is of size $k \times k$ and diagonal, with $\boldsymbol{D}_n = \operatorname{diag}(d_{n,1}, d_{n,2}, \ldots, d_{n,k})$ and $d_{n,1} > d_{n,2} > \cdots > d_{n,k} > 0$. The matrix $\boldsymbol{D}_n$ converges almost surely to a deterministic matrix $\boldsymbol{D} = \operatorname{diag}(\mu_1^{1/2}, \mu_2^{1/2}, \ldots, \mu_k^{1/2})$ with $\mu_1 > \mu_2 > \cdots > \mu_k > 0$. Moreover, the vector $\sqrt{n}(d_{n,1}^2 - \mu_1, d_{n,1}^2 - \mu_2, \ldots, d_{n,K}^2 - \mu_k)$ converges in distribution to a mean-zero multivariate normal with covariance matrix $\boldsymbol{\Sigma}$ having entries $\Sigma_{ij} = \sigma_{ij}$ (possibly degenerate).*

The next assumption concerns the noise part:

**Assumption 3.3.** *The noise matrix $\boldsymbol{E}_n$ is an $n \times p$ matrix with entries $E_{n,ij}$ independent $\mathcal{N}(0, \sigma^2)$ random variables, also independent of $\boldsymbol{U}_n$, $\boldsymbol{D}_n$, and $\boldsymbol{V}_n$.*

For analyzing the SVD of $\boldsymbol{X}_n$, we need to introduce some more notation. We denote the columns of $\boldsymbol{U}_n$ and $\boldsymbol{V}_n$ by $\underline{u}_{n,1}, \underline{u}_{n,2}, \ldots, \underline{u}_{n,k}$ and $\underline{v}_{n,1}, \underline{v}_{n,2}, \ldots, \underline{v}_{n,k}$, respectively. We let $\sqrt{n}\,\hat{\boldsymbol{U}}_n \hat{\boldsymbol{D}}_n \hat{\boldsymbol{V}}_n^{\mathrm{T}}$ be the singular value decomposition of $\boldsymbol{X}_n$ truncated to $k$ terms, where $\hat{\boldsymbol{D}}_n = \operatorname{diag}(\hat{\mu}_{n,1}^{1/2}, \hat{\mu}_{n,2}^{1/2}, \ldots, \hat{\mu}_{n,k}^{1/2})$ and the columns of $\hat{\boldsymbol{U}}$ and $\hat{\boldsymbol{V}}$ are given by $\underline{\hat{u}}_{n,1}, \underline{\hat{u}}_{n,2} \ldots, \underline{\hat{u}}_{n,k}$ and $\underline{\hat{v}}_{n,1}, \underline{\hat{v}}_{n,2}, \ldots, \underline{\hat{v}}_{n,k}$, respectively.

### 3.2.2 Main results

We are now in a position to say what the results are. There are two main theorems, one concerning the sample singular values and the other concerning the sample singular vectors. First we give the result about the singular values.

**Theorem 3.4.** *Under Assumptions 3.1 – 3.3, the vector* $\underline{\hat{\mu}}_n = (\hat{\mu}_{n,1}, \hat{\mu}_{n,2}, \ldots, \hat{\mu}_{n,k})$ *converges almost surely to* $\underline{\bar{\mu}} = (\bar{\mu}_1, \bar{\mu}_2, \ldots, \bar{\mu}_k)$*, where*

$$\bar{\mu}_i = \begin{cases} (\mu_i + \sigma^2)\left(1 + \frac{\sigma^2}{\gamma\mu_i}\right) & \text{when } \mu_i > \frac{\sigma^2}{\sqrt{\gamma}}, \\ \sigma^2\left(1 + \frac{1}{\sqrt{\gamma}}\right)^2 & \text{otherwise.} \end{cases} \tag{3.2}$$

*Moveover,* $\sqrt{n}(\underline{\hat{\mu}} - \underline{\bar{\mu}})$ *converges in distribution to a (possibly degenerate) multivariate normal with covariance matrix* $\bar{\boldsymbol{\Sigma}}$ *whose* $ij$ *element is given by*

$$\bar{\sigma}_{ij} \equiv \begin{cases} \sigma_{ij}\left(1 - \dfrac{\sigma^4}{\gamma\mu_i^2}\right)\left(1 - \dfrac{\sigma^4}{\gamma\mu_j^2}\right) & \\ \quad + \delta_{ij} 2\sigma^2\left(2\mu_i + (1 + \gamma^{-1})\,\sigma^2\right)\left(1 - \dfrac{\sigma^4}{\gamma\mu_i^2}\right) & \text{when } \mu_i,\, \mu_j > \frac{\sigma^2}{\sqrt{\gamma}}, \\ 0 & \text{otherwise.} \end{cases} \tag{3.3}$$

*When* $\sigma_{ii} = 2\mu_i^2$*, and* $\mu_i > \frac{\sigma^2}{\sqrt{\gamma}}$*, the variance of the* $i$*th component simplifies to* $\bar{\sigma}_{ii} = 2(\mu_i + \sigma^2)^2\left(1 - \frac{\sigma^4}{\gamma\mu_i^2}\right)$.

Next, we give the result for the singular vectors:

**Theorem 3.5.** *Suppose Assumptions 3.1 – 3.3 hold. Then the* $k \times k$ *matrix* $\boldsymbol{\Theta}_n \equiv \boldsymbol{V}_n^{\mathrm{T}} \hat{\boldsymbol{V}}_n$ *converges almost surely to a matrix* $\boldsymbol{\Theta} = \operatorname{diag}(\theta_1, \theta_2, \ldots, \theta_k)$*, where*

$$\theta_i^2 = \begin{cases} \left(1 - \frac{\sigma^4}{\gamma\mu_i^2}\right)\left(1 + \frac{\sigma^2}{\gamma\mu_i}\right)^{-1} & \text{when } \mu_i > \frac{\sigma^2}{\sqrt{\gamma}}, \\ 0 & \text{otherwise.} \end{cases} \tag{3.4}$$

*Also, the $k \times k$ matrix $\mathbf{\Phi}_n \equiv \boldsymbol{U}_n^{\mathrm{T}} \hat{\boldsymbol{U}}_n$ converges almost surely to a matrix $\mathbf{\Phi} = \operatorname{diag}(\varphi_1, \varphi_2, \ldots, \varphi_k)$, where*

$$\varphi_i^2 = \begin{cases} \left(1 - \frac{\sigma^4}{\gamma \mu_i^2}\right) \left(1 + \frac{\sigma^2}{\mu_i}\right)^{-1} & \text{when } \mu_i > \frac{\sigma^2}{\sqrt{\gamma}}, \\ 0 & \text{otherwise.} \end{cases} \tag{3.5}$$

*Moreover, $\theta_i$ and $\varphi_i$ almost surely have the same sign.*

### 3.2.3 Notes

Some discussion of the assumptions and the results are in order:

1. Assumptions 3.1 and 3.2 are simpler than the assumptions given in many other papers while still being quite general. For example, Paul's "spiked" covariance model has data of the form

$$\boldsymbol{X} = \boldsymbol{Z} \Xi^{\mathrm{T}} + \boldsymbol{E},$$

   where $\Xi$ is an $p \times k$ matrix of factors and $\boldsymbol{Z}$ is an $n \times k$ matrix of factor loadings whose rows are iid multivariate $\mathcal{N}(0, \boldsymbol{C})$ random variables for covariance matrx $\boldsymbol{C}$ having eigen-decomposition $\boldsymbol{C} = \boldsymbol{Q} \mathbf{\Lambda} \boldsymbol{Q}^{\mathrm{T}}$. Letting $\boldsymbol{Z} = \sqrt{n} \hat{\boldsymbol{P}} \hat{\mathbf{\Lambda}}^{1/2} \hat{\boldsymbol{Q}}^{\mathrm{T}}$ be the SVD of $\boldsymbol{Z}$, Anderson's results [1] give us that $\hat{\mathbf{\Lambda}}$ and $\hat{\boldsymbol{Q}}$ converge almost surely to $\mathbf{\Lambda}$ and $\boldsymbol{Q}$, respectively, and that the diagonal elements of $\sqrt{n}(\hat{\mathbf{\Lambda}} - \mathbf{\Lambda})$ converge to a mean-zero multivariate normal whenever $\boldsymbol{C}$ has no repeated eigenvalues. If we define $\boldsymbol{U} = \hat{\boldsymbol{P}}$, $\boldsymbol{D} = \hat{\mathbf{\Lambda}}^{1/2}$, and $\boldsymbol{V} = \Xi \hat{\boldsymbol{Q}}$, then $\boldsymbol{X} = \sqrt{n} \boldsymbol{U} \boldsymbol{D} \boldsymbol{V}^{\mathrm{T}} + \boldsymbol{E}$, where the factors satisfy Assumptions 3.1 and 3.2. Dropping the normality assumption on the rows of $\boldsymbol{Z}$ poses no problem. Moreover, we can suppose instead of iid that the rows of $\boldsymbol{Z}$ are a martingale difference array with well-behaved low-order moments and still perform a transformation of the variables to get factors of the form we need for Theorems 3.4 and 3.5.

2. There is a sign-indeterminancy in the sample and population singular vectors. We choose them arbitrarily.

3. If instead of almost-sure convergence, $\boldsymbol{D}_n$ converges in probability to $\boldsymbol{D}$, then the theorems still hold with $\underline{\hat{\mu}}_n$, $\boldsymbol{\Theta}_n$, and $\boldsymbol{\Phi}_n$ converging in probability.

4. The assumption that $\sqrt{n}\,(d_1^2 - \mu_1, d_2^2 - \mu_2, \dots d_k^2 - \mu_k)$ converges weakly is only necessary for determining second-order behavior; the first order results still hold without this assumption. If the limiting distribution of the vector of factor strengths $\sqrt{n}\,(d_1^2 - \mu_1, d_2^2 - \mu_2, \dots d_k^2 - \mu_k)$ is non-normal, one can still get at the second-order behavior of the SVD through a small adaptation of the proof.

5. Most of the results in Theorems 3.4 and 3.5 can be gotten from Onatski's results [64]. However, Onatski does not show that $\sqrt{n}(\hat{\mu}_i - \bar{\mu}_i) \xrightarrow{P} 0$ when $\mu_i$ is below the critical threshold. Furthermore, Onatski proves convergence in probability, not almost sure convergence. Lastly, Onatski does not get at the joint behavior between $\boldsymbol{\Theta}_n$ and $\boldsymbol{\Phi}_n$.

## 3.3 Preliminaries

Without loss of generality we will assume that $\sigma^2 = 1$. Until Section 3.8 , we will also assume that $\gamma \geq 1$.

### 3.3.1 Change of basis

A convenient choice of basis will make it easier to study the SVD of $\boldsymbol{X}_n$. Define $\boldsymbol{U}_{n,1} = \boldsymbol{U}_n$, and choose $\boldsymbol{U}_{n,2}$ so that $\begin{pmatrix}\boldsymbol{U}_{n,1} & \boldsymbol{U}_{n,2}\end{pmatrix}$ is an orthogonal matrix. Similarly, put $\boldsymbol{V}_{n,1} = \boldsymbol{V}_n$ and choose $\boldsymbol{V}_{n,2}$ so that $\begin{pmatrix}\boldsymbol{V}_{n,1} & \boldsymbol{V}_{n,2}\end{pmatrix}$ is orthogonal. If we define $\tilde{\boldsymbol{E}}_{n,ij} = \boldsymbol{U}_{n,i}^{\mathrm{T}} \boldsymbol{E}_n \boldsymbol{V}_{n,j}$ and $\boldsymbol{X}_{n,ij} = \boldsymbol{U}_{n,i}^{\mathrm{T}} \boldsymbol{X}_n \boldsymbol{V}_{n,j}$, then in block form,

$$\begin{pmatrix}\boldsymbol{U}_{n,1}^{\mathrm{T}} \\ \boldsymbol{U}_{n,2}^{\mathrm{T}}\end{pmatrix} \boldsymbol{X}_n \begin{pmatrix}\boldsymbol{V}_{n,1} & \boldsymbol{V}_{n,2}\end{pmatrix} = \begin{pmatrix}\sqrt{n}\boldsymbol{D}_n + \tilde{\boldsymbol{E}}_{n,11} & \tilde{\boldsymbol{E}}_{n,12} \\ \tilde{\boldsymbol{E}}_{n,21} & \tilde{\boldsymbol{E}}_{n,22}\end{pmatrix}.$$

Because Gaussian white noise is orthogonally invariant, the matrices $\tilde{\boldsymbol{E}}_{n,ij}$ are all independent with iid $\mathcal{N}(0,\,1)$ elements. Let

$$\tilde{\boldsymbol{E}}_{n,22} = \sqrt{n}\begin{pmatrix}\boldsymbol{O}_{n,1} & \boldsymbol{O}_{n,2}\end{pmatrix}\begin{pmatrix}\boldsymbol{\Lambda}_n^{1/2} \\ 0\end{pmatrix}\boldsymbol{P}_n^{\mathrm{T}} \tag{3.6}$$

be the SVD of $\tilde{\boldsymbol{E}}_{n,22}$, with $\boldsymbol{\Lambda}_n = \operatorname{diag}\left(\lambda_{n,1}, \lambda_{n,2}, \ldots \lambda_{n,p-k}\right)$ . Note that $\tilde{\boldsymbol{E}}_{n,22}^{\mathrm{T}}\tilde{\boldsymbol{E}}_{n,22} \sim \mathcal{W}_{p-k}\left(n-k, \boldsymbol{I}_{p-k}\right)$. Define

$$\begin{aligned}
\tilde{\boldsymbol{X}}_n &= \begin{pmatrix}\boldsymbol{I}_k & 0 \\ 0 & \boldsymbol{O}_{n,1}^{\mathrm{T}} \\ 0 & \boldsymbol{O}_{n,2}^{\mathrm{T}}\end{pmatrix}\begin{pmatrix}\sqrt{n}\boldsymbol{D}_n + \tilde{\boldsymbol{E}}_{n,11} & \tilde{\boldsymbol{E}}_{n,12} \\ \tilde{\boldsymbol{E}}_{n,21} & \tilde{\boldsymbol{E}}_{n,22}\end{pmatrix}\begin{pmatrix}\boldsymbol{I}_k & 0 \\ 0 & \boldsymbol{P}_n\end{pmatrix} \\
&= \begin{pmatrix}\sqrt{n}\boldsymbol{D}_n + \boldsymbol{E}_{n,11} & \boldsymbol{E}_{n,12} \\ \boldsymbol{E}_{n,21} & \sqrt{n}\boldsymbol{\Lambda}_n^{1/2} \\ \boldsymbol{E}_{n,31} & 0\end{pmatrix},
\end{aligned} \tag{3.7}$$

where $\boldsymbol{E}_{n,11} = \tilde{\boldsymbol{E}}_{n,11}$, $\boldsymbol{E}_{n,12} = \tilde{\boldsymbol{E}}_{n,12}\boldsymbol{P}_n$, $\boldsymbol{E}_{n,21} = \boldsymbol{O}_{n,1}^{\mathrm{T}}\tilde{\boldsymbol{E}}_{n,21}$, and $\boldsymbol{E}_{n,31} = \boldsymbol{O}_{n,2}^{\mathrm{T}}\tilde{\boldsymbol{E}}_{n,31}$. Let $\tilde{\boldsymbol{U}}_n\tilde{\boldsymbol{D}}_n\tilde{\boldsymbol{V}}_n$ be the SVD of $\tilde{\boldsymbol{X}}_n$, truncated to $k$ terms. Lastly, put the left and right singular vectors in block form as

$$\tilde{\boldsymbol{U}}_n = \begin{pmatrix}\tilde{\boldsymbol{U}}_{n,1} \\ \tilde{\boldsymbol{U}}_{n,2}\end{pmatrix} \tag{3.8}$$

and

$$\tilde{\boldsymbol{V}}_n = \begin{pmatrix}\tilde{\boldsymbol{V}}_{n,1} \\ \tilde{\boldsymbol{V}}_{n,2}\end{pmatrix}, \tag{3.9}$$

where $\tilde{\boldsymbol{U}}_{n,1}$ and $\tilde{\boldsymbol{V}}_{n,1}$ both $k \times k$ matrices.

We have gotten to $\tilde{\boldsymbol{X}}_n$ via an orthogonal change of basis applied to $\boldsymbol{X}_n$. By carefully choosing this basis, we have assured that:

1. The blocks of $\tilde{\boldsymbol{X}}_n$ are all independent.

2. The elements of the matrices $\boldsymbol{E}_{n,ij}$ are iid $\mathcal{N}\left(0,1\right)$.

3. The elements $\{n\lambda_{n,1}, n\lambda_{n,2}, \ldots, n\lambda_{n,p-k}\}$ are eigenvalues from a white Wishart matrix with $n-k$ degrees of freedom.

4. $\tilde{\boldsymbol{X}}_n$ and $\boldsymbol{X}_n$ have the same singular values. This implies that $\hat{\boldsymbol{D}}_n = \tilde{\boldsymbol{D}}_n$.

5. The left singular vectors of $\boldsymbol{X}_n$ can be recovered from the left singular vectors of $\tilde{\boldsymbol{X}}_n$ via multiplication by an orthogonal matrix. The same holds for the right singular vectors.

6. The dot product matrix $\boldsymbol{U}_n^{\mathrm{T}}\hat{\boldsymbol{U}}_n$ is equal to $\tilde{\boldsymbol{U}}_{n,1}$. Similarly, $\boldsymbol{V}_n^{\mathrm{T}}\hat{\boldsymbol{V}}_n = \tilde{\boldsymbol{V}}_{n,1}$.

This simplified form of the problem makes it much easier to analyze the SVD of $\boldsymbol{X}_n$.

## 3.4 The secular equation

We set $\boldsymbol{S}_n = \frac{1}{n} = \tilde{\boldsymbol{X}}_n^{\mathrm{T}}\tilde{\boldsymbol{X}}_n$. The eigenvalues and eigenvectors of $\boldsymbol{S}_n$ are the squares of the singular values of $\frac{1}{\sqrt{n}}\tilde{\boldsymbol{X}}_n$ and its right singular vectors, respectively. In block form, we have

$$\boldsymbol{S}_n = \begin{pmatrix} \boldsymbol{S}_{n,11} & \boldsymbol{S}_{n,12} \\ \boldsymbol{S}_{n,21} & \boldsymbol{S}_{n,22} \end{pmatrix}, \tag{3.10}$$

where

$$\begin{aligned} \boldsymbol{S}_{n,11} = \boldsymbol{D}_n^2 &+ \frac{1}{\sqrt{n}}\left(\boldsymbol{D}_n\boldsymbol{E}_{n,11} + \boldsymbol{E}_{n,11}^{\mathrm{T}}\boldsymbol{D}_n\right) \\ &+ \frac{1}{n}\left(\boldsymbol{E}_{n,11}^{\mathrm{T}}\boldsymbol{E}_{n,11} + \boldsymbol{E}_{n,21}^{\mathrm{T}}\boldsymbol{E}_{n,21} + \boldsymbol{E}_{n,31}^{\mathrm{T}}\boldsymbol{E}_{n,31}\right), \end{aligned} \tag{3.11a}$$

and

$$\boldsymbol{S}_{n,12} = \frac{1}{\sqrt{n}}\left(\boldsymbol{D}_n\boldsymbol{E}_{n,12} + \boldsymbol{E}_{n,21}^{\mathrm{T}}\boldsymbol{\Lambda}_n^{1/2}\right) + \frac{1}{n}\boldsymbol{E}_{n,11}^{\mathrm{T}}\boldsymbol{E}_{n,12}, \tag{3.11b}$$

$$\boldsymbol{S}_{n,21} = \frac{1}{\sqrt{n}}\left(\boldsymbol{E}_{n,12}^{\mathrm{T}}\boldsymbol{D}_n + \boldsymbol{\Lambda}_n^{1/2}\boldsymbol{E}_{n,21}\right) + \frac{1}{n}\boldsymbol{E}_{n,12}^{\mathrm{T}}\boldsymbol{E}_{n,11}, \tag{3.11c}$$

$$\boldsymbol{S}_{n,22} = \boldsymbol{\Lambda}_n + \frac{1}{n}\boldsymbol{E}_{n,12}^{\mathrm{T}}\boldsymbol{E}_{n,12}. \tag{3.11d}$$

Now we study the eigendecomposition of $\boldsymbol{S}_n$. If $\underline{v} = \begin{pmatrix} \underline{v}_1 \\ \underline{v}_2 \end{pmatrix}$ is an eigenvector of $\boldsymbol{S}_n$ with eigenvalue $\mu$, then

$$\begin{pmatrix} \boldsymbol{S}_{n,11} & \boldsymbol{S}_{n,12} \\ \boldsymbol{S}_{n,21} & \boldsymbol{S}_{n,22} \end{pmatrix} \begin{pmatrix} \underline{v}_1 \\ \underline{v}_2 \end{pmatrix} = \mu \begin{pmatrix} \underline{v}_1 \\ \underline{v}_2 \end{pmatrix}.$$

If $\mu$ is not an eigenvalue of $\boldsymbol{S}_{n,22}$, then we get

$$\underline{v}_2 = -\left(\boldsymbol{S}_{n,22} - \mu \boldsymbol{I}_{p-k}\right)^{-1} \boldsymbol{S}_{n,21}\, \underline{v}_1, \qquad \text{and} \tag{3.12a}$$

$$\left(\boldsymbol{S}_{n,11} - \mu \boldsymbol{I}_k - \boldsymbol{S}_{n,12}\left(\boldsymbol{S}_{n,22} - \mu \boldsymbol{I}_{p-k}\right)^{-1} \boldsymbol{S}_{n,21}\right) \underline{v}_1 = 0. \tag{3.12b}$$

Conversely, if $(\mu, \underline{v}_1)$ is a pair that solves (3.12b) and $\underline{v}_1 \neq 0$, then

$$\underline{v} = \begin{pmatrix} \underline{v}_1 \\ -\left(\boldsymbol{S}_{n,22} - \mu \boldsymbol{I}_{p-k}\right)^{-1} \boldsymbol{S}_{n,21}\, \underline{v}_1 \end{pmatrix} \tag{3.13}$$

is an eigenvector of $\boldsymbol{S}_n$ with eigenvalue $\mu$.

We define

$$\boldsymbol{T}_n(z) = \boldsymbol{S}_{n,11} - z\boldsymbol{I}_k - \boldsymbol{S}_{n,12}\left(\boldsymbol{S}_{n,22} - z\boldsymbol{I}_{p-k}\right)^{-1} \boldsymbol{S}_{n,21} \tag{3.14}$$

$$f_n(z, \underline{x}) = \boldsymbol{T}_n(z)\, \underline{x}, \tag{3.15}$$

and refer to $f_n(z, \underline{x}) = 0$ as the *secular equation.* This terminology comes from numerical linear algebra, where a secular equation is analogous to a characteristic equation; it is an equation whose roots are eigenvalues of a matrix. Typically, secular equations arise in eigenvalue perturbation problems. The name comes from the fact that they originally appeared studying secular perturbations of planetary orbits. A more standard use of the term "secular equation" would involve the equation $\det \boldsymbol{T}_n(z) = 0$. However, for our purposes it is more convenient to work with $f_n$.

### 3.4.1 Outline of the proof

Almost surely $\boldsymbol{S}_n$ and $\boldsymbol{S}_{n,22}$ have no eigenvalues in common, so every eigenvalue-eigenvector pair of $\boldsymbol{S}_n$ is a solution to the secular equation. To study these solutions, we first focus on $\boldsymbol{T}_n(z)$.

It turns out that when $z > (1 + \gamma^{-1/2})^2$, we can find a perturbation expansion for $\boldsymbol{T}_n(z)$. In Section 3.5, we show that for $z$ above this threshold, we can expand

$$\boldsymbol{T}_n(z) = \boldsymbol{T}_0(z) + \frac{1}{\sqrt{n}} \boldsymbol{T}_{n,1}(z),$$

where $\boldsymbol{T}_0(z)$ is deterministic and $\boldsymbol{T}_{n,1}(z)$ converges in distribution to a matrix-valued Gaussian process indexed by $z$. With this expansion, in Section 3.6 we study sequences of solutions to the equation $f_n(z_n, \underset{\sim}{x}_n) = 0$. Using a Taylor series expansion for $z_n > (1 + \gamma^{-1/2})^2$, we write

$$\begin{aligned} z_n &= z_0 + \frac{1}{\sqrt{n}} z_{n,1} + o_{\mathrm{P}}\left(\frac{1}{\sqrt{n}}\right), \\ \underset{\sim}{x}_n &= \underset{\sim}{x}_0 + \frac{1}{\sqrt{n}} \underset{\sim}{x}_{n,1} + o_{\mathrm{P}}\left(\frac{1}{\sqrt{n}}\right), \end{aligned}$$

where $(z^0, \underset{\sim}{x}^0)$ is the limit of $(z_n, \underset{\sim}{x}_n)$ as $n \to \infty$ and $(z_{n,1}, \underset{\sim}{x}_{n,1})$ is the order-$\frac{1}{\sqrt{n}}$ approximation error.

In Section 3.7 we get the singular values and singular vectors of $\frac{1}{\sqrt{n}} \tilde{\boldsymbol{X}}_n$. From every solution pair $(z_n, \underset{\sim}{x}_n)$ satisfying $f_n(z_n, \underset{\sim}{x}_n) = 0$, we can construct an eigenvalue and an eigenvector of $\boldsymbol{S}_n$ using equation (3.13). The eigenvalues are squares of singular values of $\frac{1}{\sqrt{n}} \tilde{\boldsymbol{X}}_n$, and the eigenvectors are right singular vectors. We can get the corresponding left singular vectors by multiplying the right singular vectors by $\tilde{\boldsymbol{X}}_n$ and scaling appropriately. For $z$ values above the critical threshold, we use the perturbation results of the previous two sections. Below the threshold, we use a more direct approach involving the fluctuations of the top eigenvalues of $\boldsymbol{\Lambda}_n$.

Finally, in Section 3.8 we show that the results still hold when $\gamma < 1$. For parts of the proof, we will need some limit theorems for weighted sums from Appendix B.

## 3.5 Analysis of the secular equation

We devote this section to finding a simplified formula for $\boldsymbol{T}_n(z)$ for certain values of $z$. By a bit of algebra and analysis, we find the first- and second-order behavior of the secular equation.

First, we employ the Sherman-Morrison-Woodbury formula [36] to get an expression for $(\boldsymbol{S}_{n,22} - z\boldsymbol{I}_{p-k})^{-1}$. As a reminder, this identity states that for matrices $\boldsymbol{A}$, $\boldsymbol{B}$, and $\boldsymbol{C}$ with compatible dimensions, the following formula for the inverse holds:

$$(\boldsymbol{A} + \boldsymbol{B}\boldsymbol{C})^{-1} = \boldsymbol{A}^{-1} - \boldsymbol{A}^{-1}\boldsymbol{B}\left(\boldsymbol{I} + \boldsymbol{C}\boldsymbol{A}^{-1}\boldsymbol{B}\right)^{-1}\boldsymbol{C}\boldsymbol{A}^{-1}.$$

Using this, we can write

$$\begin{aligned}(\boldsymbol{S}_{n,22} - z\boldsymbol{I}_{p-k})^{-1} &= \left((\boldsymbol{\Lambda}_n - z\boldsymbol{I}_{p-k}) + \frac{1}{n}\boldsymbol{E}_{n,12}^{\mathrm{T}}\boldsymbol{E}_{n12}\right)^{-1} \\ &= (\boldsymbol{\Lambda}_n - z\boldsymbol{I}_{p-k})^{-1} \\ &\quad - \frac{1}{n}(\boldsymbol{\Lambda}_n - z\boldsymbol{I}_{p-k})^{-1} \\ &\qquad \cdot \boldsymbol{E}_{n,12}^{\mathrm{T}}\left(\boldsymbol{I}_k + \frac{1}{n}\boldsymbol{E}_{n,12}(\boldsymbol{\Lambda}_n - z\boldsymbol{I}_{p-k})^{-1}\boldsymbol{E}_{n,12}^{\mathrm{T}}\right)^{-1}\boldsymbol{E}_{n,12} \\ &\qquad \cdot (\boldsymbol{\Lambda}_n - z\boldsymbol{I}_{p-k})^{-1}.\end{aligned} \tag{3.16}$$

Next, we define $\tilde{\boldsymbol{D}}_n = \boldsymbol{D}_n + \frac{1}{\sqrt{n}}\boldsymbol{E}_{n,11}$, so that

$$\begin{aligned}&\boldsymbol{S}_{n,12}\Big(\boldsymbol{S}_{n,22} - z\boldsymbol{I}_{p-k}\Big)^{-1}\boldsymbol{S}_{n,21} \\ &\qquad = \frac{1}{n}\left(\tilde{\boldsymbol{D}}_n^{\mathrm{T}}\boldsymbol{E}_{n,12} + \boldsymbol{E}_{n,21}^{\mathrm{T}}\boldsymbol{\Lambda}_n^{1/2}\right) \\ &\qquad\quad \cdot (\boldsymbol{S}_{n,22} - z\boldsymbol{I}_{p-k})^{-1} \\ &\qquad\quad \cdot \left(\boldsymbol{E}_{n,12}^{\mathrm{T}}\tilde{\boldsymbol{D}}_n + \boldsymbol{\Lambda}_n^{1/2}\boldsymbol{E}_{n,21}\right).\end{aligned} \tag{3.17}$$

There are three important terms coming out of equations (3.16) and (3.17) that involve $\boldsymbol{\Lambda}_n$. These are $\boldsymbol{E}_{n,12}(\boldsymbol{\Lambda}_n - z\boldsymbol{I}_{p-k})^{-1}\boldsymbol{E}_{n,12}^{\mathrm{T}}$, $\boldsymbol{E}_{n,21}^{\mathrm{T}}(\boldsymbol{\Lambda}_n - z\boldsymbol{I}_{p-k})^{-1}\boldsymbol{E}_{n,21}$, and $\boldsymbol{E}_{n,12}\left(\boldsymbol{\Lambda}_n - z\boldsymbol{I}_{p-k}\right)^{-1}\boldsymbol{\Lambda}_n^{1/2}\boldsymbol{E}_{n,21}$. Each term can be written as a weighted sum of outer products.

There is dependence between the weights, but the outer products are iid. For example, with $\boldsymbol{E}_{n,12} = \begin{pmatrix} \underline{E}_{n,12,1} & \underline{E}_{n,12,2} & \cdots & \underline{E}_{n,12,p-k} \end{pmatrix}$, we have

$$\boldsymbol{E}_{n,12}\left(\boldsymbol{\Lambda}_n - z\boldsymbol{I}_{p-k}\right)^{-1}\boldsymbol{E}_{n,12}^{\mathrm{T}}, = \sum_{\alpha=1}^{p-k} \frac{1}{\lambda_{n,\alpha} - z} \cdot \underline{E}_{n,12,\alpha}\,\underline{E}_{n,12,\alpha}^{\mathrm{T}}.$$

From the Central Limit Theorem and the Strong Law of Large Numbers we know that $\frac{1}{p-k}\sum_{\alpha=1}^{p-k}\underline{E}_{n,12,\alpha}\,\underline{E}_{n,12,\alpha}^{\mathrm{T}} \overset{a.s.}{\to} \boldsymbol{I}_k$ and that $\sqrt{p-k}\left(\frac{1}{p-k}\sum_{\alpha=1}^{p-k}\underline{E}_{n,12,\alpha}\,\underline{E}_{n,12,\alpha}^{\mathrm{T}} - \boldsymbol{I}_k\right)$ converges in distribution to mean-zero symmetric matrix whose elements are jointly multivariate normal. In the limiting distribution, the elements have variance 2 along the diagonal and variance 1 off of it; aside from the matrix being symmetric, the unique elements are all uncorrelated. The difficulty in analyzing these terms comes from the dependence between the weights.

When $z$ is in the support of $F_\gamma^{\mathrm{MP}}$, the weights behave erratically, but otherwise they have some nice properties. First of all, $\boldsymbol{\Lambda}_n$ is independent of $\boldsymbol{E}_{n,12}$ and $\boldsymbol{E}_{n,21}$. Secondly, the Wishart LLN (Corollary 2.18) and Theorem 2.20 ensure that for $z > (1+\gamma^{-1/2})^2$, $\frac{1}{p-k}\sum_{\alpha=1}^{p-k}\frac{1}{\lambda_{n,\alpha}-z} \overset{a.s.}{\to} \int \frac{1}{t-z}dF_\gamma^{\mathrm{MP}}(t)$. Moreover, the Wishart CLT (Theorem 2.19) guarantees that the error is of size $\mathcal{O}_{\mathrm{P}}(\frac{1}{n})$. These properties in combination with the limit theorems for weighted sums in Appendix B allow us to get the behavior of $\boldsymbol{E}_{n,12}\left(\boldsymbol{\Lambda}_n - z\boldsymbol{I}_k\right)^{-1}\boldsymbol{E}_{n,12}^{\mathrm{T}}$ and its cousins.

The function

$$\begin{aligned} m(z) &\equiv \int \frac{1}{t-z} dF_\gamma^{\mathrm{MP}}(t) \\ &= \gamma \cdot \frac{-(z-1+\gamma^{-1}) + \sqrt{\left(z-b_\gamma\right)\left(z-a_\gamma\right)}}{2z} \end{aligned} \tag{3.18}$$

is the Stieltjes transform of $F_\gamma^{\mathrm{MP}}$, where $a_\gamma = \left(1-\gamma^{-1/2}\right)^2$ and $b_\gamma = \left(1+\gamma^{-1/2}\right)^2$. When restricted to the complement of the support of $F_\gamma^{\mathrm{MP}}$, $m$ has a well-defined

inverse

$$z(m) = -\frac{1}{m} + \frac{1}{1 + \gamma^{-1} m}. \tag{3.19}$$

Also, $m$ is strictly increasing and convex outside the support of $F_\gamma^{\mathrm{MP}}$. This function appears frequently in the remainder of the chapter.

**Lemma 3.6.** *If $z > b_\gamma$, then*

$$\frac{1}{n} \boldsymbol{E}_{n,12} \left(\boldsymbol{\Lambda}_n - z \boldsymbol{I}_{p-k}\right)^{-1} \boldsymbol{E}_{n,12}^{\mathrm{T}} \overset{a.s.}{\to} \gamma^{-1} m(z) \cdot \boldsymbol{I}_k, \tag{3.20a}$$

$$\frac{1}{n} \boldsymbol{E}_{n,21}^{\mathrm{T}} \left(\boldsymbol{\Lambda}_n - z \boldsymbol{I}_{p-k}\right)^{-1} \boldsymbol{E}_{n,21} \overset{a.s.}{\to} \gamma^{-1} m(z) \cdot \boldsymbol{I}_k, \tag{3.20b}$$

*and*

$$\frac{1}{n} \boldsymbol{E}_{n,12} \left(\boldsymbol{\Lambda}_n - z \boldsymbol{I}_{p-k}\right)^{-1} \boldsymbol{\Lambda}_n^{1/2} \boldsymbol{E}_{n,21} \overset{a.s.}{\to} 0. \tag{3.20c}$$

*Proof.* We prove the result for the first quantity and the other derivations are analogous. For each $1 \leq i, j \leq k$, we have that

$$\left( \frac{1}{n} \boldsymbol{E}_{n,12} \left(\boldsymbol{\Lambda}_n - z \boldsymbol{I}_{p-k}\right)^{-1} \boldsymbol{E}_{n,12}^{\mathrm{T}} \right)_{ij} = \frac{p-k}{n} \cdot \frac{1}{p-k} \sum_{\alpha=1}^{p-k} \frac{\left(\boldsymbol{E}_{n,12}\right)_{i\alpha} \left(\boldsymbol{E}_{n,12}\right)_{j\alpha}}{\lambda_{n,\alpha} - z}.$$

Let $N = p - k$, define weights $W_{N,\alpha} = (\lambda_{n,\alpha} - z)^{-1}$, and let $Y_{N,\alpha} = \left(\boldsymbol{E}_{n,12}\right)_{i\alpha} \left(\boldsymbol{E}_{n,12}\right)_{j\alpha}$. The function

$$g(t) = \begin{cases} \frac{1}{t-z} & \text{if } t \leq b_\gamma, \\ \frac{1}{b_\gamma - z} & \text{otherwise,} \end{cases}$$

is bounded and continuous. Moreover, since $\lambda_{n,1} \overset{a.s.}{\to} b_\gamma$, with probability one $g(\lambda_{n,\alpha})$ is eventually equal to $W_{N,\alpha}$ for all $\alpha$. The Wishart LLN (Corollary 2.18) gives us that $\frac{1}{N} \sum_{\alpha=1}^{N} W_{N,\alpha} \overset{a.s.}{\to} \int \frac{1}{t-z} dF_\gamma^{\mathrm{MP}}(t) = m(z)$. Since $|W_{N,\alpha}| \leq W_{N,1} \overset{\text{a.s.}}{\leq} b_\gamma$, the fourth moments of the weights are uniformly bounded in $N$. The $Y_{N,\alpha}$ are all iid with $\mathbb{E} Y_{N,\alpha} = \delta_{ij}$ and $\mathbb{E} Y_{N,\alpha}^4 < \infty$. Applying these results, the weighed SLLN (Proposition B.2) gives us that $\frac{1}{N} \sum_{\alpha=1}^{N} W_{N,\alpha} Y_{N,\alpha} \overset{a.s.}{\to} m(z) \delta_{ij}$. Since $\frac{p-k}{n} \to \gamma^{-1}$, this completes the proof. □

**Lemma 3.7.** *Considered as functions of $z$, the quantities in Lemma 3.6 and their derivatives converge uniformly over any closed interval $[u, v] \subset \left(b_\gamma, \infty\right)$.*

*Proof.* We show this for the first quantity and the other proofs are similar. Defining

$$e_{n,ij}(z) = \left( \frac{1}{n} \boldsymbol{E}_{n,12} \left( \boldsymbol{\Lambda}_n - z \boldsymbol{I}_{p-k} \right)^{-1} \boldsymbol{E}_{n,12}^{\mathrm{T}} \right)_{ij},$$

we will show that for all $(i,j)$ with $1 \leq i,j \leq k$, $\sup_{z \in [u,v]} |e_{n,ij}(z) - \gamma^{-1} m(z) \delta_{ij}| \overset{a.s.}{\to} 0$.

Let $\varepsilon > 0$ be arbitrary. Note that for $z > b_\gamma$, $m'(z) > 0$ and $m''(z) < 0$. We let $d = m'(u)$ and choose a grid $u = z_1 < z_2 < \cdots < z_{M-1} < z_M = v$, with $|z_l - z_{l+1}| = \frac{\gamma\varepsilon}{2d}$. Then $|\gamma^{-1} m(z_l) - \gamma^{-1} m(z_{l+1})| < \frac{\varepsilon}{2}$. From Lemma 3.6, we can find $N$ large enough such that for $n > N$, $\max_{l \in \{1,\ldots,M\}} |e_{n,ij}(z_l) - \gamma^{-1} m(z_l) \delta_{ij}| < \frac{\varepsilon}{2}$. Also guarantee that $N$ is large enough so that $\lambda_{n,1} < u$ (this is possible since $\lambda_{n,1} \overset{a.s.}{\to} b_\gamma < u$). Let $z \in [u,v]$ be arbitrary and find $l$ such that $z_l \leq z \leq z_{l+1}$. Observe that $e_{n,ij}(z)$ is monotone for $z > \lambda_{n,1}$. Thus, either $e_{n,ij}(z_l) \leq e_{n,ij}(z) \leq e_{n,ij}(z_{l+1})$ or $e_{n,ij}(z_{l+1}) \leq e_{n,ij}(z) \leq e_{n,ij}(z_l)$.

If $i \neq j$, we have for $n > N$, $-\frac{\varepsilon}{2} < e_{n,ij}(z) < \frac{\varepsilon}{2}$. Otherwise, when $i = j$, $e_{n,ij}(z)$ is monotonically increasing and $\gamma^{-1} m(z_l) - \frac{\varepsilon}{2} < e_{n,ij}(z) < \gamma^{-1} m(z_{l+1}) + \frac{\varepsilon}{2}$, so that $\gamma^{-1} m(z) - \varepsilon < e_{n,ij}(z) < \gamma^{-1} m(z) + \varepsilon$. In either case, $|e_{n,ij}(z) - \gamma^{-1} m(z) \delta_{ij}| < \varepsilon$. Since $\frac{d}{dz}\left[\frac{1}{\lambda - z}\right] = -\frac{1}{(\lambda - z)^2}$, which is monotone for $z > \lambda$, the same argument applies to show that the derivatives converge uniformly. □

**Lemma 3.8.** *If $z_1, z_2, \ldots, z_l > b_\gamma$, then jointly for $z \in \{z_1, z_2, \ldots, z_l\}$*

$$\sqrt{n} \left( \frac{1}{n} \boldsymbol{E}_{n,12} \left( \boldsymbol{\Lambda}_n - z \boldsymbol{I}_{p-k} \right)^{-1} \boldsymbol{E}_{n,12}^{\mathrm{T}} - \gamma^{-1} m(z) \boldsymbol{I}_k \right) \equiv \boldsymbol{F}_n(z) \overset{d}{\to} \boldsymbol{F}(z), \tag{3.21a}$$

$$\sqrt{n} \left( \frac{1}{n} \boldsymbol{E}_{n,12} \left( \boldsymbol{\Lambda}_n - z \boldsymbol{I}_{p-k} \right)^{-1} \boldsymbol{\Lambda}_n^{1/2} \boldsymbol{E}_{n,21} \right) \equiv \boldsymbol{G}_n(z) \overset{d}{\to} \boldsymbol{G}(z), \tag{3.21b}$$

$$\sqrt{n} \left( \frac{1}{n} \boldsymbol{E}_{n,21}^{\mathrm{T}} \left( \boldsymbol{\Lambda}_n - z \boldsymbol{I}_{p-k} \right)^{-1} \boldsymbol{E}_{n,21} - \gamma^{-1} m(z) \boldsymbol{I}_k \right) \equiv \boldsymbol{H}_n(z) \overset{d}{\to} \boldsymbol{H}(z), \tag{3.21c}$$

*where the elements of $\boldsymbol{F}(z)$, $\boldsymbol{G}(z)$, and $\boldsymbol{H}(z)$ jointly define a multivariate Gaussian*

*process indexed by $z$, with each matrix independent of the others. The nonzero covariances are defined by*

$$\operatorname{Cov}\Big(F_{ij}(z_1),\, F_{ij}(z_2)\Big) = \gamma^{-1}\,(1+\delta_{ij})\,\frac{m(z_1)-m(z_2)}{z_1-z_2}, \tag{3.22a}$$

$$\operatorname{Cov}\Big(G_{ij}(z_1),\, G_{ij}(z_2)\Big) = \gamma^{-1}\,\frac{z_1\,m(z_1)-z_2\,m(z_2)}{z_1-z_2}, \tag{3.22b}$$

$$\operatorname{Cov}\Big(H_{ij}(z_1),\, H_{ij}(z_2)\Big) = \gamma^{-1}\,(1+\delta_{ij})\,\frac{m(z_1)-m(z_2)}{z_1-z_2}, \tag{3.22c}$$

*with the interpretation when $z_1 = z_2$ that*

$$\frac{m(z_1)-m(z_2)}{z_1-z_2} = m'(z_1), \quad \textit{and} \quad \frac{z_1\,m(z_1)-z_2\,m(z_2)}{z_1-z_2} = m(z_1) + z_1\,m'(z_1).$$

*Proof.* We will need the Wishart CLT (Theorem 2.19) and the strong weighted multivariate CLT (Corollary B.5). To save space we only give the argument for the joint distribution of $\boldsymbol{F}(z_1)$ and $\boldsymbol{F}(z_2)$.

Put $N = p - k$ and consider the $2k^2$-dimensional vector $\underline{Y}_{N,\alpha} = \begin{pmatrix} \tilde{\underline{Y}}_{N,\alpha} \\ \tilde{\underline{Y}}_{N,\alpha} \end{pmatrix}$, where $\tilde{\underline{Y}}_{N,\alpha} = \operatorname{vec}\left(\underline{E}_{n,12,\alpha}\,\underline{E}_{n,12,\alpha}^{\mathrm{T}}\right)$ and $\boldsymbol{E}_{n,12} = \begin{pmatrix} \underline{E}_{n,12,1} & \underline{E}_{n,12,2} & \cdots & \underline{E}_{n,12,N} \end{pmatrix}$. Define the $2k^2$-dimensional weight vector $\underline{W}_{N,\alpha} = \begin{pmatrix} \underline{W}_{N,\alpha,1} \\ \underline{W}_{N,\alpha,2} \end{pmatrix}$, where $\underline{W}_{N,\alpha,i} = \frac{1}{\lambda_\alpha - z_i}\underline{1}$. We have that for $\alpha = 1, 2, \ldots, N$, the $\underline{Y}_{N,\alpha}$ are iid with

$$\mathbb{E}\left[\underline{Y}_{N,1}\right] = \underline{\mu}^Y = \begin{pmatrix} \tilde{\underline{\mu}}^Y \\ \tilde{\underline{\mu}}^Y \end{pmatrix}$$

and $\tilde{\underline{\mu}}^Y = \operatorname{vec}\left(\boldsymbol{I}_k\right)$. Also, we have

$$\mathbb{E}\left[\left(\underline{Y}_{N,1} - \underline{\mu}^Y\right)\left(\underline{Y}_{N,1} - \underline{\mu}^Y\right)^{\mathrm{T}}\right] = \boldsymbol{\Sigma}^Y = \begin{pmatrix} \tilde{\boldsymbol{\Sigma}}^Y & \tilde{\boldsymbol{\Sigma}}^Y \\ \tilde{\boldsymbol{\Sigma}}^Y & \tilde{\boldsymbol{\Sigma}}^Y \end{pmatrix},$$

where $\tilde{\boldsymbol{\Sigma}}^Y = \mathbb{E}\Big[\Big(\operatorname{vec}\left(\underline{E}_{n,12,1}\underline{E}_{n,12,1}^{\mathrm{T}} - \boldsymbol{I}_k\right)\Big)\Big(\operatorname{vec}\left(\underline{E}_{n,12,1}\underline{E}_{n,12,1}^{\mathrm{T}} - \boldsymbol{I}_k\right)\Big)^{\mathrm{T}}\Big]$ is a $k^2 \times k^2$

matrix defined by the relation

$$\mathbb{E}\left[\left(\underline{E}_{n,12,1}\underline{E}_{n,12,1}^{\mathrm{T}} - \boldsymbol{I}_k\right)_{ij}\left(\underline{E}_{n,12,1}\underline{E}_{n,12,1}^{\mathrm{T}} - \boldsymbol{I}_k\right)_{i'j'}\right] = \delta_{(i,j)=(i',j')} + \delta_{(i,j)=(j',i')}.$$

That is, the diagonal elements of $\underline{E}_{n,12,1}\underline{E}_{n,12,1}^{\mathrm{T}} - \boldsymbol{I}_k$ have variance 2 and the off-diagonal elements have variance 1. Aside from the matrix being symmetric, the unique elements are all uncorrelated.

Letting

$$\begin{aligned} \mu_i^W &= \int \frac{dF_\gamma^{\mathrm{MP}}(t)}{t - z_i} = m(z_i), \qquad \text{and} \\ \sigma_{ij}^W &= \int \frac{dF_\gamma^{\mathrm{MP}}(t)}{(t - z_i)(t - z_j)} = \frac{m(z_i) - m(z_j)}{z_i - z_j}, \end{aligned}$$

the Wishart LLN combined with the truncation argument of Lemma 3.6 gives us that $\frac{1}{N}\sum_{\alpha=1}^{N}\frac{1}{\lambda_{n,\alpha}-z_i} \overset{a.s.}{\to} \mu_i^W$ and $\frac{1}{N}\sum_{\alpha=1}^{N}\frac{1}{(\lambda_{n,\alpha}-z_i)(\lambda_{n,\alpha}-z_j)} \overset{a.s.}{\to} \sigma_{ij}^W$. Thus,

$$\begin{aligned} \frac{1}{N}\sum_{\alpha=1}^{N} \underline{W}_{N,\alpha} &\overset{a.s.}{\to} \underline{\mu}^W = \begin{pmatrix} \mu_1^W \underline{1} \\ \mu_2^W \underline{1} \end{pmatrix}, \qquad \text{and} \\ \frac{1}{N}\sum_{\alpha=1}^{N} \underline{W}_{N,\alpha}\,\underline{W}_{N,\alpha}^{\mathrm{T}} &\overset{a.s.}{\to} \boldsymbol{\Sigma}^W = \begin{pmatrix} \sigma_{11}^W \underline{1}\,\underline{1}^{\mathrm{T}} & \sigma_{12}^W \underline{1}\,\underline{1}^{\mathrm{T}} \\ \sigma_{21}^W \underline{1}\,\underline{1}^{\mathrm{T}} & \sigma_{11}^W \underline{1}\,\underline{1}^{\mathrm{T}} \end{pmatrix}. \end{aligned}$$

Moreover, the Wishart CLT tells us that the error in each of the sums is of size $\mathcal{O}_{\mathrm{P}}\left(\frac{1}{N}\right)$.

As in Lemma 3.6, the fourth moments of $\underline{W}_{N,\alpha}$ and $\underline{Y}_{N,\alpha}$ are all well-behaved. Finally, we can invoke the strong weighted CLT (Corollary B.5) to get that the weighted sum $\sqrt{N}\left(\frac{1}{N}\sum_{\alpha=1}^{N}\underline{W}_{N,\alpha} \bullet \underline{Y}_{N,\alpha} - \underline{\mu}^W \bullet \underline{\mu}^T\right)$ converges in distribution to a mean-zero multivariate normal with covariance

$$\boldsymbol{\Sigma}^W \bullet \boldsymbol{\Sigma}^Y = \begin{pmatrix} \sigma_{11}^W \tilde{\boldsymbol{\Sigma}} & \sigma_{12}^W \tilde{\boldsymbol{\Sigma}} \\ \sigma_{21}^W \tilde{\boldsymbol{\Sigma}} & \sigma_{22}^W \tilde{\boldsymbol{\Sigma}} \end{pmatrix}.$$

This completes the proof since

$$\sqrt{n}\,\mathrm{vec}\left(\frac{1}{n}\boldsymbol{E}_{n,12}\left(\boldsymbol{\Lambda}_n - z_i\boldsymbol{I}_{p-k}\right)^{-1}\boldsymbol{E}_{n,12}^{\mathrm{T}} - \gamma^{-1}m(\mu)\boldsymbol{I}_k\right)$$
$$= \gamma^{-1}\sqrt{\frac{n}{p-k}}\cdot\sqrt{N}\left(\frac{1}{N}\sum_{\alpha=1}^{N}\underline{W}_{N,\alpha,i}\bullet\underline{\tilde{Y}}_{N,\alpha} - \underline{\mu}_i^W\bullet\underline{\tilde{\mu}}_i^Y\right)$$

(with $\underline{\mu}_i^W = \mu_i^W\underline{1}$), and $\gamma^{-1}\sqrt{\frac{n}{p-k}} \to \gamma^{-1/2}$. □

**Remark 3.9.** *It is possible to show that the sequences in Lemma 3.8 are tight by an argument similar to the one used in [64]. This implies that the convergence is uniform in $z$. For our purposes, we only need that the finite-dimensional distributions converge.*

With Lemmas 3.6–3.8, we can get a perturbation expansion of $\boldsymbol{T}_n(z)$ for $z > b_\gamma$.

**Lemma 3.10.** *If $[u,v] \subset \big(b_\gamma, \infty\big)$, then $\boldsymbol{T}_n(z) \overset{a.s.}{\to} \boldsymbol{T}_0(z)$ uniformly on $[u,v]$, where*

$$\boldsymbol{T}_0(z) = \frac{1}{1+\gamma^{-1}m(z)}\boldsymbol{D}^2 + \frac{1}{m(z)}\boldsymbol{I}_k. \tag{3.23}$$

**Lemma 3.11.** *If $z > b_\gamma$, then*

$$\boldsymbol{T}_n(z) = \boldsymbol{T}_0(z) + \frac{1}{\sqrt{n}}\boldsymbol{T}_{n,1}(z), \tag{3.24}$$

*where*

$$\begin{aligned}\boldsymbol{T}_{n,1}(z) = &-\big(1+\gamma^{-1}m(z)\big)^{-2}\cdot\boldsymbol{D}\boldsymbol{F}_n(z)\boldsymbol{D}\\ &+\big(1+\gamma^{-1}m(z)\big)^{-1}\\ &\quad\cdot\Big\{\sqrt{n}\left(\boldsymbol{D}_n^2-\boldsymbol{D}^2\right)+\boldsymbol{D}\big(\boldsymbol{E}_{n,11}-\boldsymbol{G}_n(z)\big)+\big(\boldsymbol{E}_{n,11}-\boldsymbol{G}_n(z)\big)^{\mathrm{T}}\boldsymbol{D}\Big\}\\ &-z\,\boldsymbol{H}_n(z)+\sqrt{n}\left(\frac{1}{n}\boldsymbol{E}_{n,31}^{\mathrm{T}}\boldsymbol{E}_{n,31}-(1-\gamma^{-1})\boldsymbol{I}_k\right)+o_P(1)\,.\end{aligned} \tag{3.25}$$

*Proof.* First, we have

$$\boldsymbol{S}_{n,11} = \boldsymbol{D}_n^2 + \boldsymbol{I}_k + \frac{1}{\sqrt{n}}\left(\boldsymbol{D}\boldsymbol{E}_{n,11} + \boldsymbol{E}_{n,11}^{\mathrm{T}}\boldsymbol{D}\right) + \left(\frac{1}{n}\left(\boldsymbol{E}_{n,11}^{\mathrm{T}}\boldsymbol{E}_{n,11} + \boldsymbol{E}_{n,21}^{\mathrm{T}}\boldsymbol{E}_{n,21} + \boldsymbol{E}_{n,31}^{\mathrm{T}}\boldsymbol{E}_{n,31}\right) - \boldsymbol{I}_k\right).$$

Next, we compute

$$\begin{aligned}
&\left(\boldsymbol{I}_k + \frac{1}{n}\boldsymbol{E}_{n,12}\left(\boldsymbol{\Lambda}_n - z\boldsymbol{I}_{p-k}\right)^{-1}\boldsymbol{E}_{n,12}^{\mathrm{T}}\right)^{-1} \\
&\qquad = \left(\boldsymbol{I}_k + \gamma^{-1}m(z)\boldsymbol{I}_k + \frac{1}{\sqrt{n}}\boldsymbol{F}_n(z)\right)^{-1} \\
&\qquad = \left(1 + \gamma^{-1}m(z)\right)^{-1}\boldsymbol{I}_k - \frac{1}{\sqrt{n}}\left(1 + \gamma^{-1}m(z)\right)^{-2}\boldsymbol{F}_n(z) + o_{\mathrm{P}}\left(\frac{1}{\sqrt{n}}\right).
\end{aligned}$$

Using this, after a substantial amount of algebra we get

$$\begin{aligned}
&\boldsymbol{S}_{n,12}\left(\boldsymbol{S}_{n,22} - z\boldsymbol{I}_{p-k}\right)^{-1}\boldsymbol{S}_{n,21} \\
&\quad = z\,m(z)\,\boldsymbol{I}_k + \frac{\gamma^{-1}m(z)}{1 + \gamma^{-1}m(z)}\boldsymbol{D}_n^2 \\
&\quad\quad + \frac{1}{\sqrt{n}}\Bigg\{\frac{1}{1 + \gamma^{-1}m(z)}\left(\boldsymbol{D}\boldsymbol{G}_n(z) + \boldsymbol{G}_n^{\mathrm{T}}(z)\boldsymbol{D}\right) \\
&\quad\quad\quad + \frac{\gamma^{-1}m(z)}{1 + \gamma^{-1}m(z)}\left(\boldsymbol{D}\boldsymbol{E}_{n,11} + \boldsymbol{E}_{n,11}^{\mathrm{T}}\boldsymbol{D}\right) \\
&\quad\quad\quad + \left(\frac{1}{1 + \gamma^{-1}m(z)}\right)^2\boldsymbol{D}\boldsymbol{F}_n(z)\boldsymbol{D} + z\,\boldsymbol{H}_n(z)\Bigg\} \\
&\quad\quad + \frac{1}{n}\boldsymbol{E}_{n,21}^{\mathrm{T}}\boldsymbol{E}_{n,21} + o_{\mathrm{P}}\left(\frac{1}{\sqrt{n}}\right).
\end{aligned}$$

The equations for $\boldsymbol{T}_0$ and $\boldsymbol{T}_{n,1}$ follow. To simplify the form of $\boldsymbol{T}_0$, we use the identity $z \cdot \left(1 + \gamma^{-1}m(z)\right) = -\frac{1}{m(z)} + (1 - \gamma^{-1})$. □

## 3.6 Solutions to the secular equation

We will now study the solutions to $f_n(z, \underline{x}) = 0$, defined in equation (3.15). If $(z, \underline{x})$ is a solution, then so is $(z, \alpha\, \underline{x})$ for any scalar $\alpha$. We restrict our attention to solutions with $\|\underline{x}\|_2 = 1$. We also impose a restriction on the sign of $\underline{x}$, namely we require that the component with the largest magnitude is positive, i.e.

$$\max_i x_i = \max_i |x_i|. \tag{3.26}$$

### 3.6.1 Almost sure limits

First we look at the solutions of $f_0(z, \underline{x}) \equiv \boldsymbol{T}_0(z)\underline{x}$, the limit of $f_n(z, \underline{x})$ for $z > b_\gamma$.

**Lemma 3.12.** *Letting $\bar{k} = \max\{i : \mu_i > \gamma^{-1/2}\}$, if $\mu_1, \mu_2, \ldots, \mu_k$ are all distinct, then there are exactly $\bar{k}$ solutions to the equation $f_0(z, \underline{x}) = 0$. They are given by*

$$(\bar{\mu}_i, \underline{e}_i), \quad i = 1, \ldots, \bar{k},$$

*where $\bar{\mu}_i$ is the unique solution*

$$m(\bar{\mu}_i) = \frac{1}{\mu_i + \gamma^{-1}},$$

*and $\underline{e}_i$ is the $i$th standard basis vector.*

*Proof.* We have that

$$\boldsymbol{T}_0(z) = \frac{1}{1 + \gamma^{-1} m(z)} \boldsymbol{D}^2 + \frac{1}{m(z)} \boldsymbol{I}_k.$$

Since this is diagonal, the equation $f_0(z, \underline{x}) = 0$ holds iff the $i$th diagonal element of $\boldsymbol{T}_0(z)$ is zero and $\underline{x} = \underline{e}_i$. The $i$th diagonal is zero when

$$\frac{\mu_i}{1 + \gamma^{-1} m(z)} + \frac{1}{m(z)} = 0,$$

equivalently

$$m(z) = -\frac{1}{\mu_i + \gamma^{-1}}.$$

Note that $m(z) > -\left(\gamma^{-1/2} + \gamma^{-1}\right)^{-1}$ and $m'(z) > 0$ on $\left(b_\gamma, \infty\right)$. Hence, a unique solution exists exactly when $\mu_i > \gamma^{-1/2}$. □

Given a solution of $f_0(z, \underline{x}) = 0$, it is not hard to believe that there is a sequence of solutions $(z_n, \underline{x}_n)$ such that $f_n(z_n, \underline{x}_n) = 0$, with $z_n$ and $\underline{x}_n$ converging to $z$ and $\underline{x}$, respectively. We dedicate the rest of this section to making this statement precise.

**Lemma 3.13.** *If $\mu > \gamma^{-1/2}$ occurs $l$ times on the diagonal of $\boldsymbol{D}^2$, then with probability one there exist sequences $z_{n,1}, \ldots, z_{n,l}$ such that for $n$ large enough:*

1. $z_{n,i} \neq z_{n,j}$ *for* $i \neq j$
2. $\det \boldsymbol{T}_n(z_{n,i}) = 0$ *for* $i = 1, \ldots, l$.
3. $z_{n,i} \to z_0 = m^{-1}\left(\frac{1}{\mu + \gamma^{-1}}\right)$.

The proof involves a technical lemma, which we state and prove now.

**Lemma 3.14.** *Let $g_n(z)$ be a sequence of continuous real-valued functions that converge uniformly on $(u, v)$ to $g_0(z)$. If $g_0(z)$ is analytic on $(u, v)$, then for $n$ large enough, $g_n(z)$ and $g_0(z)$ have the same number of zeros in $(u, v)$ (counting multiplicity).*

*Proof.* Since $|g_0(z)|$ is bounded away from zero outside the neighborhoods of its zeros, we can assume without loss of generality that $g_0(z)$ has a single zero $z_0 \in (u, v)$ of multiplicity $l$. Define $\tilde{g}_n(z) = \frac{g_n(z)}{|z - z_0|^{l-1}}$. The function $\tilde{g}_0(z)$ is bounded and continuous, and has a simple zero at $z_0$. For $r$ small enough, $\tilde{g}_0(z_0 + r)$ and $\tilde{g}_0(z_0 - r)$ have differing signs. Without loss of generality, say that the first is positive.

The sequence $\tilde{g}_n(z)$ converges uniformly to $\tilde{g}_0(z)$ outside of a neighborhood of $z_0$. Thus, for $n$ large enough, $\tilde{g}_n(z - r) < 0$ and $\tilde{g}_n(z + r) > 0$. Also, either $g_n(z)$ has a zero at $z_0$, or else for a small enough neighborhood around $z_0$, $\operatorname{sgn} \tilde{g}_n(z)$ is constant. Since $\tilde{g}_n(z)$ is continuous outside of a neighborhood of $z_0$, $\tilde{g}_n(z)$ and $g_n(z)$ must have a zero in $(z_0 - r, z_0 + r) \subset (u, v)$. Call this zero $z_{n,1}$. Since $r$ is arbitrary, $z_{n,1} \to z_0$, and $\frac{g_n(z)}{z - z_{n,1}} \to \frac{g_0(z)}{z - z_0}$ uniformly on $(u, v)$. We can now proceed inductively, since $\frac{g_0(z)}{z - z_0}$ has a zero of multiplicity $l - 1$ at $z_0$. □

*Proof of Lemma 3.13.* Let $z_0$ be the unique solution of $m(z_0) = (\mu + \gamma^{-1})^{-1}$. Define $g_n(z) = \det \boldsymbol{T}_n(z)$. Since det is a continuous function, $g_n(z)$ converges uniformly to $g_0(z)$ in any neighborhood of $z_0$. Noting that $g_0(z)$ has a zero of multiplicity $l$ at $z_0$, by Lemma 3.14 we get that for large enough $n$, $g_n(z)$ has $l$ zeros in a neighborhood of $z_0$. By a lemma of Okamoto [63], the zeros of $g_n(z)$ are almost surely simple. □

The last thing we need to do is show that for each sequence $z_{n,i}$ solving the equation $\det \boldsymbol{T}_n(z_{n,i}) = 0$, there is a corresponding sequence of vectors $\underline{x}_{n,i}$ with $f_n(z_{n,i}, \underline{x}_{n,i}) = 0$. Since $\det \boldsymbol{T}_n(z_{n,i}) = 0$, there exists an $\underline{x}_{n,i}$ with $\boldsymbol{T}_n(z_{n,i})\, \underline{x}_{n,i} = 0$. We need to show that the sequence of vectors has a limit.

Every solution pair $(z_{n,i}, \underline{x}_{n,i})$ determines a unique eigenvalue-eigenvector pair through equation (3.13). Since the eigenvalues of $\boldsymbol{S}_n$ are almost surely unique, with the identifiability restriction of (3.26) we must have that $z_{n,i}$ uniquely determines $\underline{x}_{n,i}$. Suppose that $z_{n,i}$ is the $i$th largest solution of $\det \boldsymbol{T}_n(z) = 0$, and that $f_n(z_{n,i}, \underline{x}_{n,i}) = 0$. We will now show that $\underline{x}_{n,i} \xrightarrow{a.s.} \underline{e}_i$.

**Lemma 3.15.** *Suppose that $f_n(z_{n,i}, \underline{x}_{n,i}) = 0$, that $\underline{x}_{n,i}$ satisfies the identifiability restriction* (3.26)*, and that $z_{n,i} \xrightarrow{a.s.} \bar{\mu}_i$. If $\mu_i \neq \mu_j$ for all $j \neq i$, then $\underline{x}_{n,i} \xrightarrow{a.s.} \underline{e}_i$.*

We will use a perturbation lemma, which follows from the $\sin \Theta$ theorem (see Stewart and Sun [85][p. 258]).

**Lemma 3.16.** *Let $(z, \underline{x})$ be an approximate eigenpair of the $k \times k$ matrix $\boldsymbol{A}$ (in the sense that $\boldsymbol{A}\underline{x} \approx z\underline{x}$), with $\|\underline{x}\|_2 = 1$. Let $\underline{r} = \boldsymbol{A}\underline{x} - z\underline{x}$. Suppose that there is a set $\boldsymbol{\mathcal{L}}$ of $k - 1$ eigenvalues of $\boldsymbol{A}$ such that*

$$\delta = \min_{l \in \mathcal{L}} |l - z| > 0.$$

*Then there is an eigenpair $(z_0, \underline{x}_0)$ of $\boldsymbol{A}$ with $\|\underline{x}_0\|_2 = 1$ satisfying*

$$\underline{x}^{\mathrm{T}} \underline{x}_0 \geq \sqrt{1 - \frac{\|\underline{r}\|_2^2}{\delta^2}}$$

*and*

$$|z - z_0| \leq \|\underset{\sim}{r}\|_2.$$

*Proof of Lemma 3.15.* We have that $z_{n,i} \overset{a.s.}{\to} \bar{\mu}_i$ and that $\boldsymbol{T}_n(z_{n,i})\underset{\sim}{x}_{n,i} = 0$. Since $\boldsymbol{T}_n(z_{n,i}) \overset{a.s.}{\to} \boldsymbol{T}_0(\bar{\mu}_i)$, we get

$$\begin{aligned}
\|\boldsymbol{T}_0(\bar{\mu}_i)\underset{\sim}{x}_{n,i}\|_2 &= \|\big(\boldsymbol{T}_n(z_{n,i} - \boldsymbol{T}_0(\bar{\mu}_i)\big)\underset{\sim}{x}_{n,i}\|_2 \\
&\leq \|\boldsymbol{T}_n(z_{n,i} - \boldsymbol{T}_0(\bar{\mu}_i)\|_F \\
&\overset{a.s.}{\to} 0.
\end{aligned}$$

Since $\mu_i$ is distinct, 0 is a simple eigenvalue of $\boldsymbol{T}_0(\bar{\mu}_i)$. Thus, all other eigenvalues have magnitude at least $\delta > 0$ for some $\delta$. Define $\underset{\sim}{r}_n = \boldsymbol{T}_0(\bar{\mu}_i)\underset{\sim}{x}_{n,i}$. By Lemma 3.16, there exists an eigenpair $(z_0, \underset{\sim}{x}_0)$ of $\boldsymbol{T}_0(\bar{\mu}_i)$ with $|z_0| \leq \|\underset{\sim}{r}_n\|_2$ and $\underset{\sim}{x}_{n,i}^{\mathrm{T}}\, \underset{\sim}{x}_0 \geq \sqrt{1 - \frac{\|\underset{\sim}{r}\|_2^2}{\delta^2}}$. Since $\|\underset{\sim}{r}_n\| \overset{a.s.}{\to} 0$, for $n$ large enough we must have $z_0 = 0$ and $\underset{\sim}{x}_0 = \underset{\sim}{e}_i$ or $-\underset{\sim}{e}_i$. Lastly, noting that $\underset{\sim}{x}_0$ and $\underset{\sim}{x}_{n,i}$ are both unit vectors, we get

$$\begin{aligned}
\|\underset{\sim}{x}_{n,i} - \underset{\sim}{x}_0\|_2^2 &= 2 - 2\underset{\sim}{x}_{n,i}^{\mathrm{T}}\, \underset{\sim}{x}_0 \\
&\overset{a.s.}{\to} 0.
\end{aligned}$$

With the identifiability restriction, this forces $\underset{\sim}{x}_{n,i} \overset{a.s.}{\to} \underset{\sim}{e}_i$. □

Finally, we show that eventually the points described in Lemma 3.13 are the only zeros of $f_n(z, \underset{\sim}{x})$ having $z > b_\gamma$. Since $f_n(z, \underset{\sim}{x}) = 0$ implies $\det \boldsymbol{T}_n(z) = 0$, it suffices to show that $\boldsymbol{T}_n(z)$ has no other zeros.

**Lemma 3.17.** *For $n$ large enough, almost surely the equation $\det \boldsymbol{T}_n(z) = 0$ has exactly $\bar{k}$ solutions in $\big(b_\gamma, \infty\big)$ (namely, the $\bar{k}$ points described in Lemma 3.13).*

*Proof.* By Lemma 3.14, for $n$ large enough $\det \boldsymbol{T}_n(z)$ and $\det \boldsymbol{T}_0(z)$ have the same number of solutions in $(u, v) \subset \big(b_\gamma, \infty\big)$. Thus, we only need to show that the solutions of $\det \boldsymbol{T}_n(z)$ are bounded. Since every solution is an eigenvalue of $\boldsymbol{S}_n$, this amounts to showing that the eigenvalues of $\boldsymbol{S}_n$ are bounded. Using the Courant-Fischer min-max characterization of eigenvalues [36], we have

$$\begin{aligned}\|\boldsymbol{S}_n\|_2 &= \frac{1}{\sqrt{n}}\|\boldsymbol{X}_n\|_2^2 \\ &\leq \left(\|\boldsymbol{U}_n\boldsymbol{D}_n\boldsymbol{V}_n^{\mathrm{T}}\|_2 + \frac{1}{\sqrt{n}}\|\boldsymbol{E}_n\|_2\right)^2 \\ &\overset{a.s.}{\to} \left(\sqrt{\mu_1} + \sqrt{b_\gamma}\right)^2.\end{aligned}$$

Thus, the solutions of $\det \boldsymbol{T}_n(z) = 0$ are almost surely bounded. □

### 3.6.2 Second-order behavior

To find the second-order behavior of the solutions to the secular equation, we use a Taylor-series expansion of $f_n(z, \underline{x})$ around the limit points. That is, if $(z_n, \underline{x}_n) \to (z_0, \underline{x}_0)$, we let $Df_n$ be the derivative of $f_n$ and write

$$0 = f_n(z_n, \underline{x}_n) \approx f_n(z_0, \underline{x}_0) + Df_n(z_0, \underline{x}_0)\begin{pmatrix} z_n - z_0 \\ \underline{x}_n - \underline{x}_0 \end{pmatrix}.$$

We now want to solve for $z_n - z_0$ and $\underline{x}_n - \underline{x}_0$. Without the identifiability constraint, there are $k$ equations and $k + 1$ unknowns, but as soon as we impose the condition $\|\underline{x}_n\|_2 = \|\underline{x}_0\|_2 = 1$, the system becomes well-determined.

To make this precise, we first compute

$$Df_n(z, \underline{x}) = \left(\boldsymbol{T}_0'(z)\underline{x} \quad \boldsymbol{T}_0(z)\right) + \mathcal{O}_{\mathrm{P}}\left(\frac{1}{\sqrt{n}}\right), \tag{3.27}$$

with pointwise convergence in $z$. Then, we write

$$f_n(z_n, \underline{x}_n) = f_n(z_0, \underline{x}_0) + Df_n(z_0, \underline{x}_0)\begin{pmatrix} z_n - z_0 \\ \underline{x}_n - \underline{x}_0 \end{pmatrix} + \mathcal{O}_{\mathrm{P}}\left((z_n - z_0)^2 + \|\underline{x}_n - \underline{x}_0\|_2^2\right).$$

If $f_n(z_n, \underline{x}_n) = 0$ and $f_0(z_0, \underline{x}_0) = 0$, then we get

$$0 = \frac{1}{\sqrt{n}} f_{n,1}(z_0, \underline{x}_0) + D f_n(z_0, \underline{x}_0) \begin{pmatrix} z_n - z_0 \\ \underline{x}_n - \underline{x}_0 \end{pmatrix} + \mathcal{O}_{\mathrm{P}}\left( (z_n - z_0)^2 + \|\underline{x}_n - \underline{x}_0\|_2^2 \right).$$

If $(z_n, \underline{x}_n) \overset{a.s.}{\to} (z_0, \underline{x}_0)$, then the differences $z_n - z_0$ and $\underline{x}_n - \underline{x}_0$ must be of size $\mathcal{O}_{\mathrm{P}}\left(\frac{1}{\sqrt{n}}\right)$ and the error term in the Taylor expanson is size $\mathcal{O}_{\mathrm{P}}\left(\frac{1}{n}\right)$. The final simplification we can make is from the length constraint on $\underline{x}_n$ and $\underline{x}_0$. We have

$$\begin{aligned} 1 &= \underline{x}_n^{\mathrm{T}} \underline{x}_n \\ &= \big(\underline{x}_0 + (\underline{x}_n - \underline{x}_0)\big)^{\mathrm{T}} \big(\underline{x}_0 + (\underline{x}_n - \underline{x}_0)\big) \\ &= 1 + 2\underline{x}_0^{\mathrm{T}} (\underline{x}_n - \underline{x}_0) + \|\underline{x}_n - \underline{x}_0\|_2^2, \end{aligned}$$

so that $\underline{x}_0^{\mathrm{T}}(\underline{x}_n - \underline{x}_0) = \mathcal{O}_{\mathrm{P}}\left(\frac{1}{n}\right)$. With a little more effort, we can solve for $z_n - z_0$ and $\underline{x}_n - \underline{x}_0$.

**Lemma 3.18.** *If $(z_n, \underline{x}_n)$ is a sequence converging almost surely to $(\bar{\mu}_i, \underline{e}_i)$, such that $f_n(z_n, \underline{x}_n) = 0$ and $\mu_i \neq \mu_j$ for $i \neq j$, then:*

*(i)*

$$\sqrt{n}(z_n - \bar{\mu}_i) = -\frac{\big(\boldsymbol{T}_{n,1}(\bar{\mu}_i)\big)_{ii}}{\big(\boldsymbol{T}_0'(\bar{\mu}_i)\big)_{ii}} + o_P(1), \tag{3.28}$$

*(ii)*

$$\sqrt{n}\,(x_{n,i} - 1) = o_P(1), \quad \textit{and} \tag{3.29}$$

*(iii)*

$$\sqrt{n}\, x_{n,j} = -\frac{\big(\boldsymbol{T}_{n,1}(\bar{\mu}_i)\big)_{ji}}{\big(\boldsymbol{T}_0(\bar{\mu}_i)\big)_{jj}} + o_P(1) \quad \textit{for } i \neq j. \tag{3.30}$$

*Proof.* We have done most of the work in the exposition above. In particular, we already know that *(ii)* holds. Using the Taylor expansion, we have

$$0 = \frac{1}{\sqrt{n}} \boldsymbol{T}_{n,1}(\bar{\mu}_i)\, \underline{e}_i + \begin{pmatrix} \boldsymbol{T}_0'(\bar{\mu}_i)\, \underline{e}_i & \boldsymbol{T}_0(\bar{\mu}_i) \end{pmatrix} \begin{pmatrix} z_n - \bar{\mu}_i \\ \underline{x}_n - \underline{e}_i \end{pmatrix} + o_{\mathrm{P}}\left(\frac{1}{\sqrt{n}}\right).$$

Since $\boldsymbol{T}_0$ and $\boldsymbol{T}_0'$ are diagonal, using $(ii)$ we get

$$0 = \frac{1}{\sqrt{n}} \big(\boldsymbol{T}_{n,1}(\bar{\mu}_i)\big)_{ii} + \big(\boldsymbol{T}_0'(\bar{\mu}_i)\big)_{ii} (z_n - \bar{\mu}_i) + o_{\mathrm{P}}\left(\frac{1}{\sqrt{n}}\right),$$

$$0 = \frac{1}{\sqrt{n}} \big(\boldsymbol{T}_{n,1}(\bar{\mu}_i)\big)_{ji} + \big(\boldsymbol{T}_0(\bar{\mu}_i)\big)_{jj}\, x_{n,j} + o_{\mathrm{P}}\left(\frac{1}{\sqrt{n}}\right) \qquad \text{for } i \neq j.$$

Therefore, $(i)$ and $(iii)$ follow. □

At this point, it behooves us to do some simplification. For $\mu_i, \mu_j > \gamma^{-1/2}$, we have

$$m(\bar{\mu}_i) = -\frac{1}{\mu_i + \gamma^{-1}}.$$

Using (3.19), this implies

$$\begin{aligned} \bar{\mu}_i &= \mu_i + 1 + \gamma^{-1} + \frac{\gamma^{-1}}{\mu_i} \\ &= \Big(\mu_i + 1\Big)\left(\frac{\mu_i + \gamma^{-1}}{\mu_i}\right). \end{aligned}$$

Note that this agrees with the definition of $\bar{\mu}_i$ in Theorem 3.4. Now,

$$\frac{m(\bar{\mu}_i) - m(\bar{\mu}_j)}{\bar{\mu}_i - \bar{\mu}_j} = \frac{1}{(\mu_i + \gamma^{-1})(\mu_j + \gamma^{-1})} \cdot \frac{\mu_i \mu_j}{\mu_i \mu_j - \gamma^{-1}},$$

so that

$$m'(\bar{\mu}_i) = \frac{1}{(\mu_i + \gamma^{-1})^2} \cdot \frac{\mu_i^2}{\mu_i^2 - \gamma^{-1}}.$$

Also,

$$\frac{\bar{\mu}_i\, m(\bar{\mu}_i) - \bar{\mu}_j\, m(\bar{\mu}_j)}{\bar{\mu}_i - \bar{\mu}_j} = \frac{1}{\mu_i\, \mu_j - \gamma^{-1}}.$$

We can compute

$$\big(\boldsymbol{T}_0(\bar{\mu}_i)\big)_{jj} = (\mu_i - \mu_j)\left(\frac{\mu_i + \gamma^{-1}}{\mu_i}\right) \qquad \text{for } j \neq i,$$

$$\big(\boldsymbol{T}_0'(\bar{\mu}_i)\big)_{ii} = -\left(\frac{\mu_i^2}{\mu_i^2 - \gamma^{-1}}\right)\left(\frac{\mu_i + \gamma^{-1}}{\mu_i}\right).$$

Since $\big(1+\gamma^{-1}m(\bar{\mu}_i)\big)^{-1} = \frac{\mu_i+\gamma^{-1}}{\mu_i}$, we get

$$\begin{aligned}
\boldsymbol{T}_{n,1}(\bar{\mu}_i) = &\left(\frac{1}{\sqrt{n}}\boldsymbol{E}_{n,31}^{\mathrm{T}}\boldsymbol{E}_{n,31} - \sqrt{n}(1-\gamma^{-1})\boldsymbol{I}_k\right) \\
&- \left(\frac{\mu_i+\gamma^{-1}}{\mu_i}\right)^2 \boldsymbol{D}\boldsymbol{F}_n(\bar{\mu}_i)\boldsymbol{D} \\
&+ \left(\frac{\mu_i+\gamma^{-1}}{\mu_i}\right)\Big(\sqrt{n}(\boldsymbol{D}_n^2-\boldsymbol{D}) + \boldsymbol{D}\big(\boldsymbol{E}_{n,11}-\boldsymbol{G}_n(\bar{\mu}_i)\big) \\
&\qquad\qquad + \big(\boldsymbol{E}_{n,11}-\boldsymbol{G}_n(\bar{\mu}_i)\big)^{\mathrm{T}}\boldsymbol{D}\Big) \\
&- (\mu_i+1)\left(\frac{\mu_i+\gamma^{-1}}{\mu_i}\right)\boldsymbol{H}_n(\bar{\mu}_i).
\end{aligned}$$

The first term converges in distribution to a mean-zero multivariate normal with variance $2\,(1-\gamma^{-1})$ along the diagonal and variance $1-\gamma^{-1}$ otherwise; the elements are all uncorrelated except for the obvious symmetry. Also, we have

$$\begin{aligned}
\mathrm{Cov}\big(F_{ij}(\bar{\mu}_1), F_{ij}(\bar{\mu}_2)\big) &= \gamma^{-1}(1+\delta_{ij})\cdot\frac{1}{(\mu_1+\gamma^{-1})(\mu_2+\gamma^{-1})}\cdot\frac{\mu_1\mu_2}{\mu_1\mu_2-\gamma^{-1}}, \\
\mathrm{Cov}\big(G_{ij}(\bar{\mu}_1), G_{ij}(\bar{\mu}_2)\big) &= \gamma^{-1}\cdot\frac{1}{\mu_1\mu_2-\gamma^{-1}}, \\
\mathrm{Cov}\big(H_{ij}(\bar{\mu}_1), H_{ij}(\bar{\mu}_2)\big) &= \gamma^{-1}(1+\delta_{ij})\cdot\frac{1}{(\mu_1+\gamma^{-1})(\mu_2+\gamma^{-1})}\cdot\frac{\mu_1\mu_2}{\mu_1\mu_2-\gamma^{-1}}.
\end{aligned}$$

Therefore, for $j \neq i$ we have variances

$$\begin{aligned}
\mathrm{Var}\Big(\big(\boldsymbol{T}_{n,1}(\bar{\mu}_i)\,\underline{e}_i\big)_i\Big) = \sigma_{ii}\cdot&\left(\frac{\mu_i+\gamma^{-1}}{\mu_i}\right)^2 \\
&+ 2\left(2\mu_i+1+\gamma^{-1}\right)\frac{(\mu_i+\gamma^{-1})^2}{\mu_i^2-\gamma^{-1}} + o(1),
\end{aligned} \tag{3.31a}$$

$$\mathrm{Var}\Big(\big(\boldsymbol{T}_{n,1}(\bar{\mu}_i)\,\underline{e}_i\big)_j\Big) = \left(2\mu_i+1+\gamma^{-1}\right)\frac{(\mu_i+\gamma^{-1})^2}{\mu_i^2-\gamma^{-1}} + o(1), \tag{3.31b}$$

and nontrivial covariances

$$\mathrm{Cov}\Big(\big(\boldsymbol{T}_{n,1}(\bar{\mu}_i)\,\underline{e}_i\big)_i,\ \big(\boldsymbol{T}_{n,1}(\bar{\mu}_j)\,\underline{e}_j\big)_j\Big) = \sigma_{ij}\cdot\frac{\mu_i+\gamma^{-1}}{\mu_i}\cdot\frac{\mu_j+\gamma^{-1}}{\mu_j} + o(1) \tag{3.32a}$$

and

$$\operatorname{Cov}\left(\left(\boldsymbol{T}_{n,1}(\bar{\mu}_i)\,\underset{\sim}{e}_i\right)_j,\ \left(\boldsymbol{T}_{n,1}(\bar{\mu}_j)\,\underset{\sim}{e}_j\right)_i\right) = \left(\mu_i + \mu_j + 1 + \gamma^{-1}\right) \cdot \frac{(\mu_i + \gamma^{-1})(\mu_j + \gamma^{-1})}{\mu_i \mu_j - \gamma^{-1}} + o(1). \tag{3.32b}$$

All other covariances between the elements of $\boldsymbol{T}_{n,1}(\bar{\mu}_i)\,\underset{\sim}{e}_i$ and $\boldsymbol{T}_{n,1}(\bar{\mu}_j)\,\underset{\sim}{e}_j$ are zero.

## 3.7 Singular values and singular vectors

The results about solutions to the secular equation translate directly to results about the singular values and right singular vectors of $\frac{1}{\sqrt{n}}\tilde{\boldsymbol{X}}_n$.

### 3.7.1 Singular values

Every value $z$ with $f_n(z, \underset{\sim}{x}) = 0$ for some $\underset{\sim}{x}$ is the square of a singular value of $\frac{1}{\sqrt{n}}\tilde{\boldsymbol{X}}_n$. Therefore, Section 3.6 describes the behavior of the top $\bar{k}$ singular values. To complete the proof of Theorem 3.4 for $\gamma \geq 1$, we only need to describe what happens to the singular values corresponding to the indices $i$ with $\mu_i \leq \gamma^{-1/2}$.

**Lemma 3.19.** *If $\mu_i \leq \gamma^{-1/2}$ then the $i$th eigenvalue of $\boldsymbol{S}_n$, converges almost surely to $b_\gamma$.*

*Proof.* From Lemma 3.17, we know that for $n$ large enough and $\varepsilon$ small, there are exactly $\bar{k} = \max\{i : \mu_i > \gamma^{-1/2}\}$ eigenvalues of $\boldsymbol{S}_n$ in $\left(b_\gamma + \varepsilon, \infty\right)$. From the eigenvalue interleaving inequalities, we know that the $i$th eigenvalue of $\boldsymbol{S}_n$ is at least as big as the $i$th eigenvalue of $\boldsymbol{S}_{n,22}$.

Denote by $\hat{\mu}_{n,i}$ the $i$th eigenvalue of $\boldsymbol{S}_n$, with $\bar{k} < i \leq k$. Then almost surely,

$$\lim_{n\to\infty} \lambda_{n,i} \leq \varliminf_{n\to\infty} \hat{\mu}_{n,i} \leq \varlimsup_{n\to\infty} \hat{\mu}_{n,i} \leq b_\gamma + \varepsilon.$$

Since $\lambda_{n,i} \overset{a.s.}{\to} b_\gamma$ and $\varepsilon$ is arbitrary, this forces $\hat{\mu}_{n,i} \overset{a.s.}{\to} b_\gamma$. □

**Lemma 3.20.** *If $\mu_i \leq \gamma^{-1/2}$, then $\sqrt{n}(\hat{\mu}_i - \bar{\mu}_i) \xrightarrow{P} 0$, where $\hat{\mu}_i$ is the $i$th eigenvalue of $\boldsymbol{S}_n$.*

*Proof.* We use the same notation as in Lemma 3.19. Recall that in the present situation, $\bar{\mu}_i = b_\gamma$. Since $\lambda_{n,i} = b_\gamma + \mathcal{O}_{\mathrm{P}}(n^{-2/3})$, we have that

$$\sqrt{n}\big(\lambda_{n,i} - b_\gamma\big) \xrightarrow{P} 0.$$

This means that

$$\varliminf_{n\to\infty} \sqrt{n}\big(\hat{\mu}_{n,i} - b_\gamma\big) \geq o_{\mathrm{P}}(1).$$

The upper bound is a little more delicate. By the Courant-Fischer min-max characterization, $\hat{\mu}_i^{1/2}$ is bounded above by $\tilde{\mu}_i^{1/2}$, the $i$th singular value of

$$\frac{1}{\sqrt{n}}\boldsymbol{X}_n + \sqrt{n}\big((\gamma^{-1/2} + \varepsilon)^{1/2} - \mu_i^{1/2}\big)\underline{u}_{n,i}\underline{v}_{n,i}^{\mathrm{T}}$$

for any $\varepsilon > 0$. From the work in Section 3.6, we know that

$$\begin{aligned}
\tilde{\mu}_i &= \big(1 + \gamma^{-1/2} + \varepsilon\big)\left(1 + \frac{1}{\gamma^{1/2} + \gamma\varepsilon}\right) + \mathcal{O}_{\mathrm{P}}\left(\frac{\tilde{\sigma}_i}{\sqrt{n}}\right) \\
&= b_\gamma + \varepsilon^2 \frac{\gamma^{1/2} - 1}{1 + \gamma^{1/2}\varepsilon} + \mathcal{O}_{\mathrm{P}}\left(\frac{\tilde{\sigma}_i}{\sqrt{n}}\right),
\end{aligned}$$

where

$$\begin{aligned}
\tilde{\sigma}_i^2 &= \sigma_{ii}\left(1 - \frac{1}{1 + 2\gamma^{1/2}\varepsilon + \gamma\varepsilon^2}\right)^2 \\
&\quad + 2\big(2(\gamma^{-1/2} + \varepsilon) + 1 + \gamma^{-1}\big)\left(1 - \frac{1}{1 + 2\gamma^{1/2}\varepsilon + \gamma\varepsilon^2}\right) \\
&= \mathcal{O}(\varepsilon).
\end{aligned}$$

Therefore, for all $0 < \varepsilon < 1$, we have

$$\sqrt{n}\left(\hat{\mu}_{n,i} - b_\gamma - \varepsilon^2 \frac{\gamma^{1/2} - 1}{1 + \gamma^{1/2}\varepsilon}\right) \leq \mathcal{O}_{\mathrm{P}}\left(\varepsilon^{1/2}\right).$$

Letting $\varepsilon \to 0$, we get

$$\varlimsup_{n\to\infty} \sqrt{n}\big(\hat{\mu}_{n,i} - b_\gamma\big) \leq o_{\mathrm{P}}(1).$$

Together with the lower bound, this implies $\sqrt{n}\big(\hat{\mu}_{n,i} - b_\gamma\big) \xrightarrow{P} 0$. □

### 3.7.2 Right singular vectors

For $\mu_i > \gamma^{-1/2}$, we can get a right singular vector of $\frac{1}{\sqrt{n}}\tilde{\boldsymbol{X}}_n$ from the sequence of solution pairs $(z_{n,i}, \underline{x}_{n,i})$ satisfiying $f_n(z_{n,i}, \underline{x}_{n,i}) = 0$ and $z_{n,i} \xrightarrow{a.s.} \bar{\mu}_i$. The vector is parallel to

$$\tilde{\underline{x}}_{n,i} = \begin{pmatrix} \underline{x}_n \\ -(\boldsymbol{S}_{n,22} - z_n \boldsymbol{I}_{p-k})^{-1} \boldsymbol{S}_{n,21}\, \underline{x}_n \end{pmatrix}. \tag{3.33}$$

We just need to normalize this vector to have unit length. The length of $\tilde{\underline{x}}_n$ is given by

$$\|\tilde{\underline{x}}_n\|_2^2 = \underline{x}_n^{\mathrm{T}} \left(\boldsymbol{I}_k + \boldsymbol{S}_{n,12}(\boldsymbol{S}_{n,22} - z_n \boldsymbol{I}_{p-k})^{-2}\boldsymbol{S}_{n,21}\right) \underline{x}_n. \tag{3.34}$$

It is straightforward to show that for $z > b_\gamma$,

$$\begin{aligned} \boldsymbol{I}_k + \boldsymbol{S}_{n,12}(\boldsymbol{S}_{n,22} - z\boldsymbol{I}_{p-k})^{-2}\boldsymbol{S}_{n,21} &\xrightarrow{a.s.} -\boldsymbol{T}_0'(z) \\ &= \frac{\gamma^{-1} m'(z)}{\big(1 + \gamma^{-1} m(z)\big)^2} \boldsymbol{D}^2 + \frac{m'(z)}{\big(m(z)\big)^2} \boldsymbol{I}_k \end{aligned}$$

uniformly for $z$ in any compact subset of $\big(b_\gamma, \infty\big)$. It is also not hard to compute for $\mu_i > \gamma^{-1/2}$ that

$$-\boldsymbol{T}_0'(\bar{\mu}_i) = \frac{\gamma^{-1}}{\mu_i^2 - \gamma^{-1}} \boldsymbol{D}^2 + \frac{\mu_i^2}{\mu_i^2 - \gamma^{-1}} \boldsymbol{I}_k.$$

Therefore, if $\mu_i > \gamma^{-1/2}$ and $(z_{n,i}, \underline{x}_{n,i}) \xrightarrow{a.s.} (\bar{\mu}_i, \underline{e}_i)$, then

$$\begin{aligned} \|\tilde{\underline{x}}_{n,i}\|_2^2 &\xrightarrow{a.s.} \frac{\mu_i(\mu_i + \gamma^{-1})}{\mu_i^2 - \gamma^{-1}} \\ &= \frac{1 + \frac{1}{\gamma \mu_i}}{1 - \frac{1}{\gamma \mu^2}}. \end{aligned}$$

The behavior of the right singular vectors when $\mu_i \leq \gamma^{1/2}$ is a little more difficult to get at. We will use a variant of an argument from Paul [68] to show that $\|\tilde{\underline{x}}\|_2 \xrightarrow{a.s.} \infty$, which implies that $\frac{\underline{x}_{n,i}}{\|\tilde{\underline{x}}_{n,i}\|_2} \xrightarrow{a.s.} 0$. We can do this by showing the smallest eigenvalue of $\boldsymbol{S}_{n,12}(\boldsymbol{S}_{n,22} - z_{n,i}\boldsymbol{I}_{p-k})^{-2}\boldsymbol{S}_{n,21}$ goes to $\infty$.

Write

$$\boldsymbol{S}_{n,12}(\boldsymbol{S}_{n,22} - z_{n,i}\boldsymbol{I}_{p-k})^{-2}\boldsymbol{S}_{n,21} = \sum_{\alpha=1}^{p-k} \frac{\underline{s}_{n,\alpha}\,\underline{s}_{n,\alpha}^{\mathrm{T}}}{(\tilde{\lambda}_{n,\alpha} - z_{n,i})^2},$$

where $\underline{s}_{n,1}, \underline{s}_{n,2}, \ldots, \underline{s}_{n,p-k}$ are the eigenvectors of $\boldsymbol{S}_{n,22}$ multiplied by $\boldsymbol{S}_{n,12} = \boldsymbol{S}_{n,21}^{\mathrm{T}}$, and $\tilde{\lambda}_{n,1}, \tilde{\lambda}_{n,2}, \ldots, \tilde{\lambda}_{n,p-k}$ are the eigenvalues of $\boldsymbol{S}_{n,22}$. For $\varepsilon > 0$, define the event $J_n(\varepsilon) = \{z_{n,i} < b_\gamma + \varepsilon\}$. From the interleaving inequality, we have $z_{n,i} \geq \tilde{\lambda}_{n,i}$. With respect to the ordering on positive-definite matrices, we have

$$\begin{aligned}\sum_{\alpha=1}^{p-k} \frac{\underline{s}_{n,\alpha}\,\underline{s}_{n,\alpha}^{\mathrm{T}}}{(\tilde{\lambda}_{n,\alpha} - z_{n,i})^2} &\succeq \sum_{\alpha=i}^{p-k} \frac{\underline{s}_{n,\alpha}\,\underline{s}_{n,\alpha}^{\mathrm{T}}}{(\tilde{\lambda}_{n,\alpha} - z_{n,i})^2} \\ &\succeq \sum_{\alpha=i}^{p-k} \frac{\underline{s}_{n,\alpha}\,\underline{s}_{n,\alpha}^{\mathrm{T}}}{(b_\gamma + \varepsilon - z_{n,i})^2} \qquad \text{on } J_n(\varepsilon).\end{aligned}$$

It is not hard to show that $\left\| \sum_{\alpha=1}^{i-1} \frac{\underline{s}_{n,\alpha}\,\underline{s}_{n,\alpha}^{\mathrm{T}}}{(b_\gamma+\varepsilon-z_{n,i})^2} \right\| \xrightarrow{a.s.} 0$. Therefore,

$$\sum_{\alpha=i}^{p-k} \frac{\underline{s}_{n,\alpha}\,\underline{s}_{n,\alpha}^{\mathrm{T}}}{(b_\gamma + \varepsilon - z_{n,i})^2} \xrightarrow{a.s.} \frac{\gamma^{-1} m'\big(b_\gamma + \varepsilon\big)}{1 + \gamma^{-1} m\big(b_\gamma + \varepsilon\big)} \boldsymbol{D}^2 + \frac{m'\big(b_\gamma + \varepsilon\big)}{\big[m\big(b_\gamma + \varepsilon\big)\big]^2} \boldsymbol{I}_k.$$

As $n \to \infty$, we have $\mathbb{P}\big(J_n(\varepsilon)\big) \to 1$. So, since $m'\big(b_\gamma + \varepsilon\big) \geq \frac{C}{\sqrt{\varepsilon}}$ for some constant $C$, letting $\varepsilon \to 0$ we must have that the smallest eigenvalue of $\boldsymbol{S}_{n,12}(\boldsymbol{S}_{n,22} - z_{n,i}\boldsymbol{I}_{p-k})^{-2}\boldsymbol{S}_{n,21}$ goes to $\infty$.

### 3.7.3 Left singular vectors

We can get the left singular vectors from the right from multiplication by $\frac{1}{\sqrt{n}}\tilde{\boldsymbol{X}}_n$. Specifically, if $\tilde{\underline{v}}_{n,i}$ is a right singular vector of $\frac{1}{\sqrt{n}}\tilde{\boldsymbol{X}}_n$ with singular value $z_{n,i}^{1/2}$, then $\tilde{\underline{u}}_{n,i}$, the corresponding left singular vector, is defined by

$$z_{n,i}^{1/2}\,\tilde{\underline{u}}_{n,i} = \frac{1}{\sqrt{n}} \tilde{\boldsymbol{X}}_n\,\tilde{\underline{v}}_{n,i}.$$

We are only interested in the first $k$ components of $\tilde{\underline{u}}_{n,i}$. If

$$\tilde{\underline{v}}_{n,i} = \frac{1}{\|\tilde{\underline{x}}_{n,i}\|_2} \begin{pmatrix} \underline{x}_{n,i} \\ -(\boldsymbol{S}_{n,22} - z_{n,i}\boldsymbol{I}_{p-k})^{-1}\boldsymbol{S}_{n,21}\,\underline{x}_{n,i} \end{pmatrix},$$

then these are given by $\frac{1}{\|\tilde{\underline{x}}_{n,i}\|_2}\boldsymbol{R}_n(z_{n,i})\,\underline{x}_{n,i}$, where

$$\boldsymbol{R}_n(z) = \boldsymbol{D}_n + \frac{1}{\sqrt{n}}\boldsymbol{E}_{n,11} - \frac{1}{\sqrt{n}}\boldsymbol{E}_{n,12}\left(\boldsymbol{S}_{n,22} - z\boldsymbol{I}_{p-k}\right)^{-1}\boldsymbol{S}_{n,21}.$$

It is not hard to show that $\boldsymbol{R}_n(z) \xrightarrow{a.s.} \boldsymbol{R}_0(z)$, uniformly for $z > b_\gamma$, and $\boldsymbol{R}_n(z) = \boldsymbol{R}_0(z) + \frac{1}{\sqrt{n}}\boldsymbol{R}_{n,1}(z) + o_{\mathrm{P}}\left(\frac{1}{\sqrt{n}}\right)$ pointwise for $z > b_\gamma$. Here,

$$\boldsymbol{R}_0(z) = \frac{1}{1+\gamma^{-1}m(z)}\boldsymbol{D},$$

and

$$\begin{aligned}\boldsymbol{R}_{n,1}(z) = {} & \left(1+\gamma^{-1}m(z)\right)^{-1}\left(\sqrt{n}\left(\boldsymbol{D}_n - \boldsymbol{D}\right) + \boldsymbol{E}_{n,11}\right) \\ & - \left(1+\gamma^{-1}m(z)\right)^{-2}\boldsymbol{F}_n(z)\boldsymbol{D} - \boldsymbol{G}_n(z).\end{aligned}$$

Let $\underline{y}_{n,i} = \frac{1}{\|\tilde{\underline{x}}_{n,i}\|_2}\boldsymbol{R}_n(z_{n,i})\,\underline{x}_{n,i}$. A straightforward calculation shows the following:

**Lemma 3.21.** *If $\mu_i > \gamma^{-1/2}$ and $(z_{n,i}, \underline{x}_{n,i}) \xrightarrow{a.s.} (\bar{\mu}_i, \underline{e}_i)$, then*

$$\underline{y}_{n,i} \xrightarrow{a.s.} \left(\frac{1-\frac{1}{\gamma\mu_i^2}}{1+\frac{1}{\mu_i}}\right)^{1/2}\underline{e}_i.$$

*If $\mu_i \leq \gamma^{-1/2}$ and $z_{n,i} \xrightarrow{a.s.} b_\gamma$, then*

$$\underline{y}_{n,i} \xrightarrow{a.s.} 0.$$

## 3.8 Results for $\gamma \in (0,1)$

Remarkably the formulas for the limiting quantities still hold when $\gamma < 1$. To see this, we can get the behavior for $\gamma < 1$ by taking the transpose of $\boldsymbol{X}_n$ and applying the theorem for $\gamma \geq 1$. We switch the roles of $n$ and $p$, replace $\mu_i$ by $\mu_i' = \gamma\mu_i$, and replace $\gamma$ by $\gamma' = \gamma^{-1}$. Then, for instance, the critical cutoff becomes

$$\begin{aligned} \mu_i' &> \frac{1}{\sqrt{\gamma'}}, \quad \text{i.e.} \\ \gamma\mu_i &> \gamma^{1/2}, \end{aligned}$$

which is the same formula for $\gamma \geq 1$. The almost-sure limit of the square of the $i$th eigenvalue of $\frac{1}{p}\boldsymbol{X}_n\boldsymbol{X}_n^T$ becomes

$$\begin{aligned} \bar{\mu}_i' &= (\gamma\mu_i + 1)\left(1 + \frac{\gamma}{\gamma\mu_i}\right) \\ &= \gamma(\mu_i + 1)\left(1 + \frac{1}{\gamma\mu}\right) \\ &= \gamma\bar{\mu}_i. \end{aligned}$$

The formulas for the other quantities also still remain true.

## 3.9 Related work, extensions, and future work

With the proofs completed, we now discuss some extensions and related work.

### 3.9.1 Related work

When Johnstone [45] worked out the distribution of the largest eigenvalue in the null (no signal) case, he proposed studying "spiked" alternative models. Spiked data consists of vector-valued observations with population covariance of the form

$$\Sigma = \operatorname{diag}(\mu_1, \mu_2, \ldots, \mu_k, 1, \ldots, 1).$$

Work on the spiked model started with Baik et al. [11], Baik & Silverstein [12], and Paul [68]. Baik et al. showed that for complex Gaussian data, a phase transition phenomenon exists, depending on the relative magnitude of the spike. Baik & Silverstein gave the almost-sure limits of the eigenvalues from the sample covariance matrix without assuming Gaussianity. Paul worked with real Gaussian data and gave the limiting distributions when $\gamma > 1$. Paul also gives some results about the eigenvectors.

After this initial work, Bai & Yao [8] [9] derived the almost-sure limits and prove a central limit theorem for eigenvalues for a general class of data that includes colored noise and non-Gaussianity. Chen et al [19] consider another type of spiked model with correlation. Nadler [62] derived the behavior of the first eigenvector in a spiked model with one spike.

The above authors all consider data of the form $\boldsymbol{X} = \boldsymbol{Z}\boldsymbol{\Sigma}^{1/2}$. Onatski [64], like us, examines data of the form $\boldsymbol{X} = \boldsymbol{U}\boldsymbol{D}\boldsymbol{V}^T + \boldsymbol{E}$. With slightly different assumptions than ours, he is able to give the probability limits of the top eigenvalues, along with the marginal distributions of the scaled eigenvalues and singular vectors. However, Onatski does not work in a "transpose-agnostic" framework like we do, so his methods do not allow getting at the joint distribution of the left and right singular vectors.

### 3.9.2 Extensions and future work

We have stopped short of computing the second-order behavior of the singular vectors, but no additional theory is required to get at these quantities. Anyone patient enough can use our results to compute the joint distribution of $\|\tilde{\underline{x}}_{n,i}\|_2$, $\underline{x}_{n,i}$, and $\underline{y}_{n,i}$. This in turn will give the joint behavior of the singular vectors.

Most of the proof remains unchanged for complex Gaussian noise, provided transpose ($^{\mathrm{T}}$) is replaced by conjugate-transpose ($^{\mathrm{H}}$). The variance formulas need a small modification, since the fourth-moment of a real Gaussian is 3 and that of a complex Gaussian is 2.

For colored or non-Gaussian noise, we no longer have orthogonal invariance, so the change of basis in Section 3.3 is a little trickier. It is likely that comparable results can still be found, perhaps using results on the eigenvectors of general sample covariance matrices from Bai et al. [3].

# Chapter 4

# An intrinsic notion of rank for factor models

As Moore's Law progresses, data sets measuring on the order of hundreds or thousands of variables are becoming increasingly more common. Making sense of data of this size is simply not tractable without imposing a simplifying model. One popular simplification is to posit existence of a small number of common factors that drive the dynamics of the data, which are usually estimated by principal component analysis (PCA) or some variation thereof. The $n \times p$ data matrix $\boldsymbol{X}$ is approximated as a low-rank product $\boldsymbol{X} \approx \boldsymbol{U}\boldsymbol{V}^{\mathrm{T}}$, where $\boldsymbol{U}$ and $\boldsymbol{V}$ are $n \times k$ and $p \times k$, respectively, with $k$ much smaller than $n$ and $p$.

The number of algorithms for approximating matrices by low-rank products has exploded in recent years. These algorithms include archetypal analysis [21], the semi-discrete decomposition (SDD) [49], the non-negative matrix factorization (NMF) [52], the plaid model [51], the $CUR$ decomposition [25], and regularized versions thereof. They also include some clustering methods, in particular $k$-means and fuzzy $k$-means [14].

A prevailing question is: How many common factors underlie a data set? Alternately, how should one choose $k$? In general, the answer to this question is application-specific. If we are trying to use $\boldsymbol{X}$ to predict a response, $y$, then the optimal $k$ is the one that gives the best prediction error for $y$. The situation is not always this simple, though. For exploratory analysis, there is no external response, and we want to choose a $k$ that is "intrinsic" to $\boldsymbol{X}$. For other applications, we don't have a single response, $y$, we have *many* responses $y_1, y_2, \ldots, y_m$. We may not even know all of the $y_i$ when we are processing $\boldsymbol{X}$. We want a $k$ that has good average-case or worst-case prediction properties for a large class of responses.

In this chapter, we develop a precise notion of intrinsic rank for "latent factor" matrix data. This choice of $k$ is the one that minimizes average- or worst-case prediction error over all bilinear statistics. We specifically work with the singular value decomposition (SVD), but our definition can be extended to other matrix factorizations as well.

Section 4.1 introduces the latent factor model, the data model from which we base our constructions. Section 4.2 defines intrinsic rank as the minimizer of a loss function. We give a theoretical analysis of the behavior of some natural loss functions in Section 4.3, followed by simulations in Section 4.4. In Section 4.5, we examine the connection between intrinsic rank and the scree plot. Finally, Section 4.6 discusses some extensions and Section 4.7 gives a summary of the chapter.

## 4.1 The latent factor model

We start by describing a model for data generated by a small number of latent factors and additive noise. Suppose that we have $n$ multivariate observations $\underline{x}_1, \underline{x}_2, \ldots, \underline{x}_n \in \mathbb{R}^p$. In a microarray setting, we will have about $n = 50$ arrays measuring the activations of around $p = 5000$ genes (or 50000, or even millions of genes in the case of exon arrays). Alternatively, for financial applications $\underline{x}_i$ will measure the market value of on the order of $p = 1000$ assets on day $i$, and we may be looking at data from the last three years, so $n \approx 1000$. In these situations and others like them, we can often convince ourselves that there aren't really 5000 or 1000 different things going on in the data. Probably, there are a small number, $k$ of unobserved factors driving the dynamics of the data. Typically, in fact, we think $k$ is on the order of around 5 or 10.

To be specific about this intuition, for genomics applications, we don't really think that all $p = 5000$ measured genes are behaving independently. On the contrary, we think that there are a small number of biological processes that determine how much of each protein gets produced. In finance, while it is true that the stock prices of individual companies have a certain degree of independence, often macroscopic effects like industry- and market-wide trends can explain a substantial portion of the value.

### 4.1.1 The spiked model

One way to model latent effects is to assume that the $\underline{x}_i$ are iid and that their covariance is "spiked". We think that $\underline{x}_i$ is a weighted combination of $k$ latent factors corrupted by additive white noise. In this case, $\underline{x}_i$ can be decomposed as

$$\begin{aligned} \underline{x}_i &= \sum_{j=1}^{k} \underline{a}_j s_{i,j} + \underline{\varepsilon}_i \\ &= \boldsymbol{A}\underline{s}_i + \underline{\varepsilon}_i, \end{aligned} \tag{4.1}$$

where $\boldsymbol{A} = \begin{pmatrix} \underline{a}_1 & \underline{a}_2 & \cdots & \underline{a}_k \end{pmatrix}$ is a $p \times k$ matrix of latent factors common to all observations and $\underline{s}_i \in \mathbb{R}^k$ is a vector of loadings for the $i$th observation. We assume that the noise vector $\underline{\varepsilon}_i$ is distributed as $\mathcal{N}(0,\, \sigma^2 \boldsymbol{I}_p)$. If the loadings have mean zero and covariance $\boldsymbol{\Sigma}_\text{S} \in \mathbb{R}^{k \times k}$, and if they are also independent of the noise, then $\underline{x}_i$ has covariance

$$\boldsymbol{\Sigma} \equiv \mathbb{E}\left[\underline{x}_i\, \underline{x}_i^\text{T}\right] = \boldsymbol{A}\boldsymbol{\Sigma}_\text{S}\boldsymbol{A}^\text{T} + \sigma^2 \boldsymbol{I}_p. \tag{4.2}$$

The decomposition in (4.2) can be reparametrized as

$$\boldsymbol{\Sigma} = \boldsymbol{Q}\boldsymbol{\Lambda}\boldsymbol{Q}^\text{T} + \sigma^2 \boldsymbol{I}_p, \tag{4.3}$$

where $\boldsymbol{Q}^\text{T}\boldsymbol{Q} = \boldsymbol{I}_k$ and $\boldsymbol{\Lambda} = \operatorname{diag}(\lambda_1, \lambda_2, \ldots, \lambda_k)$, with $\lambda_1 \geq \lambda_2 \geq \cdots \geq \lambda_k \geq 0$. Equation (4.3) makes it apparent that $\boldsymbol{\Sigma}$ is "spiked", in the sense that most of its eigenvalues are equal, but $k$ eigenvalues stand out above the bulk. The first $k$ eigenvalues are $\lambda_1 + \sigma^2, \lambda_2 + \sigma^2, \ldots, \lambda_k + \sigma^2$, and the remaining $p - k$ eigenvalues are equal to $\sigma^2$.

### 4.1.2 More general matrix models

We can introduce a model more general than the spiked one by specifying a distribution for the $n \times p$ data matrix $\boldsymbol{X} = \begin{pmatrix} \underset{\sim}{x}_1 & \underset{\sim}{x}_2 & \cdots & \underset{\sim}{x}_n \end{pmatrix}^{\mathrm{T}}$ that includes dependence between the rows. In the spiked model, the distribution of $\boldsymbol{X}$ can be described as

$$\boldsymbol{X} \overset{d}{=} \boldsymbol{Z}\boldsymbol{\Lambda}^{1/2}\boldsymbol{Q}^{\mathrm{T}} + \boldsymbol{E}, \tag{4.4}$$

where $\boldsymbol{E} = \begin{pmatrix} \underset{\sim}{\varepsilon}_1 & \underset{\sim}{\varepsilon}_2 & \cdots & \underset{\sim}{\varepsilon}_n \end{pmatrix}^{\mathrm{T}}$ and $\boldsymbol{Z}$ is an $n \times k$ matrix of independent $\mathcal{N}(0,\,1)$ elements. More generally, we can consider data of the form

$$\boldsymbol{X} \overset{d}{=} \sqrt{n}\,\boldsymbol{U}\boldsymbol{D}\boldsymbol{V}^{\mathrm{T}} + \boldsymbol{E}, \tag{4.5}$$

where $\boldsymbol{U}^{\mathrm{T}}\boldsymbol{U} = \boldsymbol{V}^{\mathrm{T}}\boldsymbol{V} = \boldsymbol{I}_k$, and $\boldsymbol{D} = \operatorname{diag}(d_1, d_2, \ldots, d_k)$ with $d_1 \geq d_2 \geq \cdots \geq d_k \geq 0$. We can get (4.5) from (4.4) by letting $\boldsymbol{Z}\boldsymbol{\Lambda}^{1/2} = \sqrt{n}\,\boldsymbol{U}\boldsymbol{D}\boldsymbol{C}^{\mathrm{T}}$ be the SVD of $\boldsymbol{Z}\boldsymbol{\Lambda}^{1/2}$ and defining $\boldsymbol{V} = \boldsymbol{Q}\boldsymbol{C}$. Unlike the spiked model, (4.5) can model dependence between variables as well as dependence between observations.

## 4.2 An intrinsic notion of rank

With Section 4.1's latent factor model in mind, we turn our attention to defining the intrinsic rank of a data set. This definition will be motivated both by the generative model for $\boldsymbol{X}$ and by predictive power considerations. When $\boldsymbol{X} = \sqrt{n}\,\boldsymbol{U}\boldsymbol{D}\boldsymbol{V}^{\mathrm{T}} + \boldsymbol{E}$, we think of $\sqrt{n}\,\boldsymbol{U}\boldsymbol{D}\boldsymbol{V}^{\mathrm{T}}$ as "signal" and $\boldsymbol{E}$ as "noise". We make a distinction between the generative rank and the effective rank.

**Definition 4.1.** *If the $n \times p$ matrix $\boldsymbol{X}$ is distributed as $\boldsymbol{X} = \sqrt{n}\,\boldsymbol{U}\boldsymbol{D}\boldsymbol{V}^{\mathrm{T}} + \boldsymbol{E}$, where $\boldsymbol{U}^{\mathrm{T}}\boldsymbol{U} = \boldsymbol{V}^{\mathrm{T}}\boldsymbol{V} = \boldsymbol{I}_{k_0}$, $\boldsymbol{D}$ is a $k_0 \times k_0$ diagonal matrix with positive diagonal entries, and $\boldsymbol{E}$ is a noise matrix independent of the signal term whose elements are iid $\mathcal{N}(0,\,\sigma^2)$, then we denote by $k_0$ the* generative rank *of $\boldsymbol{X}$.*

Intuitively the generative rank is the rank of the signal part of $\boldsymbol{X}$.

The effective rank is defined in terms of how well the first terms of the SVD of $\boldsymbol{X}$ approximates the signal $\sqrt{n}\,\boldsymbol{U}\boldsymbol{D}\boldsymbol{V}^{\mathrm{T}}$. We let $\boldsymbol{X} = \sqrt{n}\hat{\boldsymbol{U}}\hat{\boldsymbol{D}}\hat{\boldsymbol{V}}^{\mathrm{T}}$ be the full

SVD of $\boldsymbol{X}$, where $\hat{\boldsymbol{U}} = \begin{pmatrix} \hat{\underline{u}}_1 & \hat{\underline{u}}_2 & \cdots & \hat{\underline{u}}_{n\wedge p} \end{pmatrix}$, $\hat{\boldsymbol{V}} = \begin{pmatrix} \hat{\underline{v}}_1 & \hat{\underline{v}}_2 & \cdots & \hat{\underline{v}}_{n\wedge p} \end{pmatrix}$, and $\hat{\boldsymbol{D}} = \operatorname{diag}\left(\hat{d}_1, \hat{d}_2, \ldots, \hat{d}_{n\wedge p}\right)$. If we let $\hat{\boldsymbol{D}}(k) = \operatorname{diag}\left(\hat{d}_1, \hat{d}_2, \ldots, \hat{d}_k, 0, 0, \ldots, 0\right)$, then the SVD truncated to $k$ terms is $\hat{\boldsymbol{X}}(k) = \sqrt{n}\,\hat{\boldsymbol{U}}\hat{\boldsymbol{D}}(k)\hat{\boldsymbol{V}}^{\mathrm{T}}$. We are now in a position to define effective rank

**Definition 4.2.** *Given a loss function $L : \mathbb{R}^{n\times p} \times \mathbb{R}^{n\times p} \to \mathbb{R}$, the effective rank of $\boldsymbol{X}$ with respect to $L$ is equal to*

$$k_L^* \equiv \underset{k}{\operatorname{argmin}}\, L\left(\sqrt{n}\boldsymbol{U}\boldsymbol{D}\boldsymbol{V}^{\mathrm{T}},\, \hat{\boldsymbol{X}}(k)\right). \tag{4.6}$$

The effective rank depends on the choice of loss function. In some settings it is preferable to choose an application-specific loss function. Often, we appeal to simplicity and convenience and choose squared Frobenius loss. Specifically,

$$L_{\mathrm{F}}(\boldsymbol{A},\, \boldsymbol{A}') = \|\boldsymbol{A} - \boldsymbol{A}'\|_{\mathrm{F}}^2. \tag{4.7}$$

One way to motivate this loss function is that it measures average squared error over all bilinear statistics of the form $\underline{\alpha}^{\mathrm{T}}\boldsymbol{A}\,\underline{\beta}$,

$$\frac{1}{np}\,\|\boldsymbol{A} - \boldsymbol{A}'\|_{\mathrm{F}}^2 = \int\limits_{\substack{\|\underline{\alpha}\|_2=1,\\ \|\underline{\beta}\|_2=1}} \left(\underline{\alpha}^{\mathrm{T}}\boldsymbol{A}\,\underline{\beta} - \underline{\alpha}^{\mathrm{T}}\boldsymbol{A}'\underline{\beta}\right)^2 d\underline{\alpha}\, d\underline{\beta}$$

(see Section A.3.2 for details). In the context of the latent factor model, the effective rank with respect to $L_{\mathrm{F}}$ is the rank that gives the best average-case predictions of bilinear statistics of the signal part (with respect to squared-error loss).

A common alternative to Frobenius loss is spectral loss, given by

$$L_2(\boldsymbol{A},\, \boldsymbol{A}') = \|\boldsymbol{A} - \boldsymbol{A}'\|_2^2. \tag{4.8}$$

This can be interpreted as worst-case squared error over the class of all bilinear statistics,

$$\|\boldsymbol{A} - \boldsymbol{A}'\|_2^2 = \sup_{\substack{\|\underline{\alpha}\|_2=1, \\ \|\underline{\beta}\|_2=1}} \left(\underline{\alpha}^{\mathrm{T}} \boldsymbol{A} \underline{\beta} - \underline{\alpha}^{\mathrm{T}} \boldsymbol{A}' \underline{\beta}\right)^2.$$

In the sequel, we denote the optimal ranks with respect to Frobenius and spectral loss as $k_{\mathrm{F}}^*$ and $k_2^*$, respectively.

## 4.3 Loss behavior

In this section we investigate the behavior of the loss functions introduced in Section 4.2. First, we need to be more precise about our working assumptions on the data matrices. The theory is easier if we work in an asymptotic setting, introducing a sequence of data matrices $\boldsymbol{X}_1, \boldsymbol{X}_2, \ldots, \boldsymbol{X}_n$, where $\boldsymbol{X}_n \in \mathbb{R}^{n \times p}$, $p = p(n)$, $n \to \infty$, and $\frac{n}{p} \to \gamma \in (0, \infty)$. We will need three assumptions.

**Assumption 4.3.** *The matrix $\boldsymbol{X}_n \in \mathbb{R}^{n \times p}$ can be decomposed as*

$$\boldsymbol{X}_n = \sqrt{n}\, \boldsymbol{U}_n \boldsymbol{D}_n \boldsymbol{V}_n^{\mathrm{T}} + \boldsymbol{E}_n. \tag{4.9}$$

*Here, $\boldsymbol{U}_n \in \mathbb{R}^{n \times k_0}$, $\boldsymbol{D}_n \in \mathbb{R}^{k_0 \times k_0}$, and $\boldsymbol{V}_n \in \mathbb{R}^{p \times k_0}$. The left and right factors $\boldsymbol{U}_n$ and $\boldsymbol{V}_n$ satisfy $\boldsymbol{U}_n^{\mathrm{T}} \boldsymbol{U}_n = \boldsymbol{V}_n^{\mathrm{T}} \boldsymbol{V}_n = \boldsymbol{I}_{k_0}$. The aspect ratio satisfies $\frac{n}{p} = \gamma + o\left(\frac{1}{\sqrt{n}}\right)$ for a constant $\gamma \in (0, \infty)$. The number of factors $k_0$ is fixed.*

**Assumption 4.4.** *The matrix of normalized factor strengths is diagonal with $\boldsymbol{D}_n = \operatorname{diag}(d_{n,1}, d_{n,2}, \ldots, d_{n,k_0})$. For $1 \leq i \leq k_0$, the diagonal elements satisfy $d_{n,i}^2 \xrightarrow{a.s.} \mu_i$ for deterministic $\mu_i$ satisfiying $\mu_1 > \mu_2 > \cdots > \mu_{k_0} > 0$.*

**Assumption 4.5.** *The noise matrix $\boldsymbol{E}_n$ has iid elements independent of $\boldsymbol{U}_n$, $\boldsymbol{D}_n$, and $\boldsymbol{V}_n$, with $E_{n,11} \sim \mathcal{N}(0, \sigma^2)$.*

These assumptions allow us to apply the results of Chapter 3 to get the first-order behavior of the SVD of $\boldsymbol{X}_n$. As before, we let $\underline{u}_{n,1}, \underline{u}_{n,2}, \ldots, \underline{u}_{n,k_0}$ and $\underline{v}_{n,1}, \underline{v}_{n,2}, \ldots, \underline{v}_{n,k_0}$ denote the columns of $\boldsymbol{U}_n$ and $\boldsymbol{V}_n$, respectively. We set $\boldsymbol{X}_n = \sqrt{n}\, \hat{\boldsymbol{U}}_n \hat{\boldsymbol{D}}_n \hat{\boldsymbol{V}}_n^{\mathrm{T}}$ to be the SVD of $\boldsymbol{X}_n$, where the columns of $\hat{\boldsymbol{U}}_n$ and $\hat{\boldsymbol{V}}_n$ are $\hat{\underline{u}}_{n,1}, \hat{\underline{u}}_{n,2}, \ldots, \hat{\underline{u}}_{n,n \wedge p}$ and $\hat{\underline{v}}_{n,1}, \hat{\underline{v}}_{n,2}, \ldots, \hat{\underline{v}}_{n,n\wedge p}$, respectively. With $\hat{\boldsymbol{D}}_n = \operatorname{diag}\left(\hat{\mu}_{n,1}^{1/2}, \hat{\mu}_{n,2}^{1/2}, \ldots, \hat{\mu}_{n,n\wedge p}^{1/2}\right)$, we set $\hat{\boldsymbol{D}}_n(k) = \operatorname{diag}\left(\hat{\mu}_{n,1}^{1/2}, \hat{\mu}_{n,2}^{1/2}, \ldots, \hat{\mu}_{n,k}^{1/2}, 0, 0, \ldots, 0\right)$ so that $\hat{\boldsymbol{X}}_n(k) = \sqrt{n}\,\hat{\boldsymbol{U}}_n \hat{\boldsymbol{D}}_n(k) \hat{\boldsymbol{V}}_n^{\mathrm{T}}$ is the SVD of $\boldsymbol{X}_n$ truncated to $k$ terms.

We can decompose the columns of $\hat{\boldsymbol{V}}_n$ into sums of two terms, with the first term in the subspace spanned by $\boldsymbol{V}_n$, and the second term orthogonal to it. By setting $\boldsymbol{\Theta}_n = \boldsymbol{V}_n^{\mathrm{T}} \hat{\boldsymbol{V}}_n$ and taking the $QR$ decomposition of $\hat{\boldsymbol{V}}_n - \boldsymbol{V}_n \boldsymbol{\Theta}_n$, the matrix of right factors can be expanded as

$$\hat{\boldsymbol{V}}_n = \boldsymbol{V}_n \boldsymbol{\Theta}_n + \bar{\boldsymbol{V}}_n \bar{\boldsymbol{\Theta}}_n, \tag{4.10}$$

with $\bar{\boldsymbol{V}}_n \in \mathbb{R}^{p \times (p-k_0)}$ satisfying $\bar{\boldsymbol{V}}_n^{\mathrm{T}} \bar{\boldsymbol{V}}_n = \boldsymbol{I}_{p-k_0}$ and $\boldsymbol{V}_n^{\mathrm{T}} \bar{\boldsymbol{V}}_n = 0$. We choose the signs so that $\bar{\boldsymbol{\Theta}}_n$ has non-negative diagonal entries. Note that

$$\boldsymbol{I}_k = \hat{\boldsymbol{V}}_n^{\mathrm{T}} \hat{\boldsymbol{V}}_n = \boldsymbol{\Theta}_n^{\mathrm{T}} \boldsymbol{\Theta}_n + \bar{\boldsymbol{\Theta}}_n^{\mathrm{T}} \bar{\boldsymbol{\Theta}}_n. \tag{4.11}$$

In particular, since $\bar{\boldsymbol{\Theta}}_n$ is upper-triangular, if $\boldsymbol{\Theta}_n$ converges to a diagonal matrix $\boldsymbol{\Theta}$, then $\bar{\boldsymbol{\Theta}}_n$ must also, converge to a diagonal matrix, $\left(\boldsymbol{I}_k - \boldsymbol{\Theta}^2\right)^{1/2}$. This makes the decomposition in equation (4.10) very convenient to work with.

The same trick applies to $\hat{\boldsymbol{U}}_n$. We expand

$$\hat{\boldsymbol{U}}_n = \boldsymbol{U}_n \boldsymbol{\Phi}_n + \bar{\boldsymbol{U}}_n \bar{\boldsymbol{\Phi}}_n, \tag{4.12}$$

with $\bar{\boldsymbol{U}}_n$ and $\bar{\boldsymbol{\Phi}}_n$ defined analogously to $\bar{\boldsymbol{V}}_n$ and $\bar{\boldsymbol{\Theta}}_n$. Again, we have that

$$\boldsymbol{I}_k = \boldsymbol{\Phi}_n^{\mathrm{T}} \boldsymbol{\Phi}_n + \bar{\boldsymbol{\Phi}}_n^{\mathrm{T}} \bar{\boldsymbol{\Phi}}_n. \tag{4.13}$$

Likewise, if $\boldsymbol{\Phi}_n$ converges to a diagonal matrix, then $\bar{\boldsymbol{\Phi}}_n$ must do the same.

We can new get a simplified formula for $\hat{\boldsymbol{X}}_n(k)$. With the decompositions in equations (4.10) and (4.12), we get

$$\hat{\boldsymbol{X}}_n(k) = \sqrt{n} \begin{pmatrix} \boldsymbol{U}_n & \bar{\boldsymbol{U}}_n \end{pmatrix} \begin{pmatrix} \boldsymbol{\Phi}_n \hat{\boldsymbol{D}}_n(k) \boldsymbol{\Theta}_n^{\mathrm{T}} & \boldsymbol{\Phi}_n \hat{\boldsymbol{D}}_n(k) \bar{\boldsymbol{\Theta}}_n^{\mathrm{T}} \\ \bar{\boldsymbol{\Phi}}_n \hat{\boldsymbol{D}}_n(k) \boldsymbol{\Theta}_n^{\mathrm{T}} & \bar{\boldsymbol{\Phi}}_n \hat{\boldsymbol{D}}_n(k) \bar{\boldsymbol{\Theta}}_n^{\mathrm{T}} \end{pmatrix} \begin{pmatrix} \boldsymbol{V}_n^{\mathrm{T}} \\ \bar{\boldsymbol{V}}_n^{\mathrm{T}} \end{pmatrix}. \tag{4.14}$$

We can get the asymptotic limits of the quantities above. For $1 \leq i \leq k_0$, we set

$$\bar{\mu}_i = \begin{cases} (\mu_i + \sigma^2)\left(1 + \frac{\sigma^2}{\gamma\mu_i}\right) & \text{when } \mu_i > \frac{\sigma^2}{\sqrt{\gamma}}, \\ \sigma^2\left(1 + \frac{1}{\sqrt{\gamma}}\right)^2 & \text{otherwise,} \end{cases} \tag{4.15a}$$

$$\theta_i = \begin{cases} \sqrt{\left(1 - \frac{\sigma^4}{\gamma\mu_i^2}\right)\left(1 + \frac{\sigma^2}{\gamma\mu_i}\right)^{-1}} & \text{when } \mu_i > \frac{\sigma^2}{\sqrt{\gamma}}, \\ 0 & \text{otherwise,} \end{cases} \tag{4.15b}$$

$$\varphi_i = \begin{cases} \sqrt{\left(1 - \frac{\sigma^4}{\gamma\mu_i^2}\right)\left(1 + \frac{\sigma^2}{\mu_i}\right)^{-1}} & \text{when } \mu_i > \frac{\sigma^2}{\sqrt{\gamma}}, \\ 0 & \text{otherwise,} \end{cases} \tag{4.15c}$$

while for $i > k_0$, we set $\bar{\mu}_i = \sigma^2\left(1 + \frac{1}{\sqrt{\gamma}}\right)^2$ and $\theta_i = \varphi_i = 0$. For $i \geq 1$, we define

$$\bar{\theta}_i = \sqrt{1 - \theta_i^2}, \tag{4.15d}$$

$$\bar{\varphi}_i = \sqrt{1 - \varphi_i^2}. \tag{4.15e}$$

With

$$\boldsymbol{D}(k) = \operatorname{diag}\left(\bar{\mu}_1^{1/2}, \bar{\mu}_2^{1/2}, \ldots, \bar{\mu}_k^{1/2}, 0, 0, \ldots, 0\right) \in \mathbb{R}^{n\times p} \tag{4.16a}$$

and

$$\boldsymbol{\Theta} = \operatorname{diag}\left(\theta_1, \theta_2, \ldots, \theta_p\right), \tag{4.16b}$$

$$\boldsymbol{\Phi} = \operatorname{diag}\left(\varphi_1, \varphi_2, \ldots, \varphi_n\right), \tag{4.16c}$$

$$\bar{\boldsymbol{\Theta}} = \operatorname{diag}\left(\bar{\theta}_1, \bar{\theta}_2, \ldots, \bar{\theta}_p\right), \tag{4.16d}$$

$$\bar{\boldsymbol{\Phi}} = \operatorname{diag}\left(\bar{\varphi}_1, \bar{\varphi}_2, \ldots, \bar{\varphi}_n\right), \tag{4.16e}$$

Theorems 3.4 and 3.5 give us that for fixed $k$ as $n \to \infty$,

$$\begin{aligned}
\boldsymbol{\Phi}_n \hat{\boldsymbol{D}}_n(k) \boldsymbol{\Theta}_n^{\mathrm{T}} &\xrightarrow{a.s.} \boldsymbol{\Phi} \boldsymbol{D}(k) \boldsymbol{\Theta}^{\mathrm{T}}, \\
\boldsymbol{\Phi}_n \hat{\boldsymbol{D}}_n(k) \bar{\boldsymbol{\Theta}}_n^{\mathrm{T}} &\xrightarrow{a.s.} \boldsymbol{\Phi} \boldsymbol{D}(k) \bar{\boldsymbol{\Theta}}^{\mathrm{T}}, \\
\bar{\boldsymbol{\Phi}}_n \hat{\boldsymbol{D}}_n(k) \boldsymbol{\Theta}_n^{\mathrm{T}} &\xrightarrow{a.s.} \bar{\boldsymbol{\Phi}} \boldsymbol{D}(k) \boldsymbol{\Theta}^{\mathrm{T}}, \\
\bar{\boldsymbol{\Phi}}_n \hat{\boldsymbol{D}}_n(k) \bar{\boldsymbol{\Theta}}_n^{\mathrm{T}} &\xrightarrow{a.s.} \bar{\boldsymbol{\Phi}} \boldsymbol{D}(k) \bar{\boldsymbol{\Theta}}^{\mathrm{T}}.
\end{aligned}$$

This result makes it easy to analyze the loss behavior. Letting $\mu_i = 0$ for $i > k_0$, putting $\bar{\mu}_i(k) = \bar{\mu}_i \, 1\{i \leq k\}$ and

$$\boldsymbol{F}_i(k) = \begin{pmatrix} \mu_i^{1/2} - \varphi_i \, \bar{\mu}_i^{1/2}(k) \, \theta_i & -\varphi_i \, \bar{\mu}_i^{1/2}(k) \, \bar{\theta}_i \\ -\bar{\varphi}_i \, \bar{\mu}_i^{1/2}(k) \, \theta_i & -\bar{\varphi}_i \, \bar{\mu}_i^{1/2}(k) \, \bar{\theta}_i \end{pmatrix} \tag{4.17}$$

we have that for $\| \cdot \|$ being spectral or Frobenius norm, for fixed $k$ as $n \to \infty$,

$$\frac{1}{\sqrt{n}} \left\| \sqrt{n} \, \boldsymbol{U}_n \boldsymbol{D}_n \boldsymbol{V}_n^{\mathrm{T}} - \hat{\boldsymbol{X}}_n(k) \right\| \xrightarrow{a.s.} \left\| \operatorname{diag}\left( \boldsymbol{F}_1(k), \boldsymbol{F}_2(k), \ldots, \boldsymbol{F}_{k \vee k_0}(k) \right) \right\|.$$

Thus,

$$\frac{1}{n} \left\| \sqrt{n} \, \boldsymbol{U}_n \boldsymbol{D}_n \boldsymbol{V}_n^{\mathrm{T}} - \hat{\boldsymbol{X}}_n(k) \right\|_{\mathrm{F}}^2 \xrightarrow{a.s.} \sum_{i=1}^{k \vee k_0} \left\| \boldsymbol{F}_i(k) \right\|_{\mathrm{F}}^2 \tag{4.18}$$

and

$$\frac{1}{n} \left\| \sqrt{n} \, \boldsymbol{U}_n \boldsymbol{D}_n \boldsymbol{V}_n^{\mathrm{T}} - \hat{\boldsymbol{X}}_n(k) \right\|_2^2 \xrightarrow{a.s.} \max_{1 \leq i \leq k \vee k_0} \left\| \boldsymbol{F}_i(k) \right\|_2^2. \tag{4.19}$$

Similar results can be gotten for other orthogonally-invariant norms. A straightforward calculation shows

$$\begin{aligned}
&\boldsymbol{F}_i^{\mathrm{T}}(k) \, \boldsymbol{F}_i(k) \\
&= \begin{pmatrix} \mu_i - 2\varphi_i \mu_i^{1/2} \theta_i \bar{\mu}_i^{1/2}(k) + \theta_i^2 \bar{\mu}_i(k) & -\varphi_i \bar{\theta}_i \mu_i^{1/2} \bar{\mu}_i^{1/2}(k) + \theta_i \bar{\theta}_i \bar{\mu}_i(k) \\ -\varphi_i \bar{\theta}_i \mu_i^{1/2} \bar{\mu}_i^{1/2}(k) + \theta_i \bar{\theta}_i \bar{\mu}_i(k) & \bar{\theta}_i^2 \bar{\mu}_i(k) \end{pmatrix},
\end{aligned}$$

so that

$$\begin{aligned}
\operatorname{tr}\big(\boldsymbol{F}_i^{\mathrm{T}}(k)\,\boldsymbol{F}_i(k)\big) &= \mu_i - 2\varphi_i\theta_i\mu_i^{1/2}\bar{\mu}_i^{1/2}(k) + \bar{\mu}_i(k), \\
\det\big(\boldsymbol{F}_i^{\mathrm{T}}(k)\,\boldsymbol{F}_i(k)\big) &= \bar{\varphi}_i^2\bar{\theta}_i^2\mu_i\bar{\mu}_i(k).
\end{aligned}$$

When $\mu_i > \frac{\sigma^2}{\sqrt{\gamma}}$, we can use the identities $\varphi_i\theta_i\bar{\mu}_i^{1/2}\mu_i^{-1/2} = 1 - \frac{\sigma^4}{\gamma\mu_i^2}$ and $\bar{\varphi}_i^2\bar{\theta}_i^2 = \frac{\sigma^4}{\gamma\mu_i^2}$ to get

$$\operatorname{tr}\big(\boldsymbol{F}_i^{\mathrm{T}}(k)\,\boldsymbol{F}_i(k)\big) = \begin{cases} \frac{\sigma^2}{\gamma\mu_i}\big(3\sigma^2 + (\gamma+1)\mu_i\big) & \text{if } i \leq k, \\ \mu_i & \text{otherwise,} \end{cases} \tag{4.20a}$$

$$\det\big(\boldsymbol{F}_i^{\mathrm{T}}(k)\,\boldsymbol{F}_i(k)\big) = \begin{cases} \left(\frac{\sigma^2}{\gamma\mu_i}\right)^2 (\mu_i + \sigma^2)(\gamma\mu_i + \sigma^2) & \text{if } i \leq k \\ 0 & \text{otherwise.} \end{cases} \tag{4.20b}$$

When $\mu_i \leq \frac{\sigma^2}{\sqrt{\gamma}}$, we have

$$\operatorname{tr}\big(\boldsymbol{F}_i^{\mathrm{T}}(k)\,\boldsymbol{F}_i(k)\big) = \begin{cases} \mu_i + \sigma^2\left(1 + \frac{1}{\sqrt{\gamma}}\right)^2 & \text{if } i \leq k \\ \mu_i & \text{otherwise,} \end{cases} \tag{4.21a}$$

$$\det\big(\boldsymbol{F}_i^{\mathrm{T}}(k)\,\boldsymbol{F}_i(k)\big) = \begin{cases} \mu_i\,\sigma^2\left(1 + \frac{1}{\sqrt{\gamma}}\right)^2 & \text{if } i \leq k \\ 0 & \text{otherwise.} \end{cases} \tag{4.21b}$$

We can use these expressions to compute

$$\|\boldsymbol{F}_i(k)\|_{\mathrm{F}}^2 = \operatorname{tr}\big(\boldsymbol{F}_i^{\mathrm{T}}(k)\,\boldsymbol{F}_i(k)\big), \tag{4.22}$$

$$\begin{aligned}
\|\boldsymbol{F}_i(k)\|_2^2 = \frac{1}{2}\Big\{ &\operatorname{tr}\big(\boldsymbol{F}_i^{\mathrm{T}}(k)\,\boldsymbol{F}_i(k)\big) \\
&+ \sqrt{\big[\operatorname{tr}\big(\boldsymbol{F}_i^{\mathrm{T}}(k)\,\boldsymbol{F}_i(k)\big)\big]^2 - 4\det\big(\boldsymbol{F}_i^{\mathrm{T}}(k)\,\boldsymbol{F}_i(k)\big)}\,\Big\}.
\end{aligned} \tag{4.23}$$

The expression for the limit of $\frac{1}{n}\|\sqrt{n}\,\boldsymbol{U}_n\boldsymbol{D}_n\boldsymbol{V}_n^{\mathrm{T}} - \hat{\boldsymbol{X}}_n(k)\|_2^2$ is fairly complicated. In the Frobenius case, we have

**Proposition 4.6.** *For fixed $k$ as $n \to \infty$, we have*

$$\frac{1}{n}\|\sqrt{n}\boldsymbol{U}_n\boldsymbol{D}_n\boldsymbol{V}_n^{\mathrm{T}} - \hat{\boldsymbol{X}}_n(k)\|_F^2 \\ \overset{a.s.}{\to} \sum_{i=1}^{k} \alpha_i\mu_i + \sum_{i=k+1}^{k_0} \mu_i + \sigma^2\left(1+\frac{1}{\sqrt{\gamma}}\right)^2 \cdot (k-k_0)_+, \quad (4.24)$$

*where*

$$\alpha_i = \begin{cases} \frac{\sigma^2}{\gamma\mu_i^2}\big(3\sigma^2 + (\gamma+1)\mu_i\big) & \text{if } \mu_i > \frac{\sigma^2}{\sqrt{\gamma}}, \\ 1 + \frac{\sigma^2}{\mu_i}\left(1+\frac{1}{\sqrt{\gamma}}\right)^2 & \text{otherwise.} \end{cases} \quad (4.25)$$

Figure 4.1 shows $\alpha_i$ as a function of $\mu_i$. It is beneficial to include the $i$th term when $\alpha_i < 1$, or equivalently $\mu_i > \mu_{\mathrm{F}}^*$, with

$$\mu_{\mathrm{F}}^* \equiv \sigma^2\left(\frac{1+\gamma^{-1}}{2} + \sqrt{\left(\frac{1+\gamma^{-1}}{2}\right)^2 + \frac{3}{\gamma}}\right). \quad (4.26)$$

This gives us the next Corollary.

**Corollary 4.7.** *As $n \to \infty$,*

$$k_F^* \overset{a.s.}{\to} \max\left\{i : \mu_i > \mu_F^*\right\},$$

*provided no $\mu_k$ is exactly equal to $\mu_F^*$.*

The theory in Chapter 3 tells us that when $\mu_i > \frac{\sigma^2}{\sqrt{\gamma}}$, the $i$th signal term is "detectable" in the sense that the $i$th sample singular value and singular vectors are correlated with the population quantities. Proposition 4.6 tells us that when $\mu_i \in \left(\frac{\sigma^2}{\sqrt{\gamma}}, \mu_{\mathrm{F}}^*\right)$, the $i$th signal term is detectable, but it is *not* helpful (in terms of Frobenius norm) to include in the estimate of $\sqrt{n}\boldsymbol{U}_n\boldsymbol{D}_n\boldsymbol{V}_n^{\mathrm{T}}$. Only when $\mu_i$ surpasses the inclusion threshold $\mu_{\mathrm{F}}^*$ is it beneficial to include the $i$th term.

Figure 4.2 shows the detection and inclusion thresholds as functions of $\gamma$.

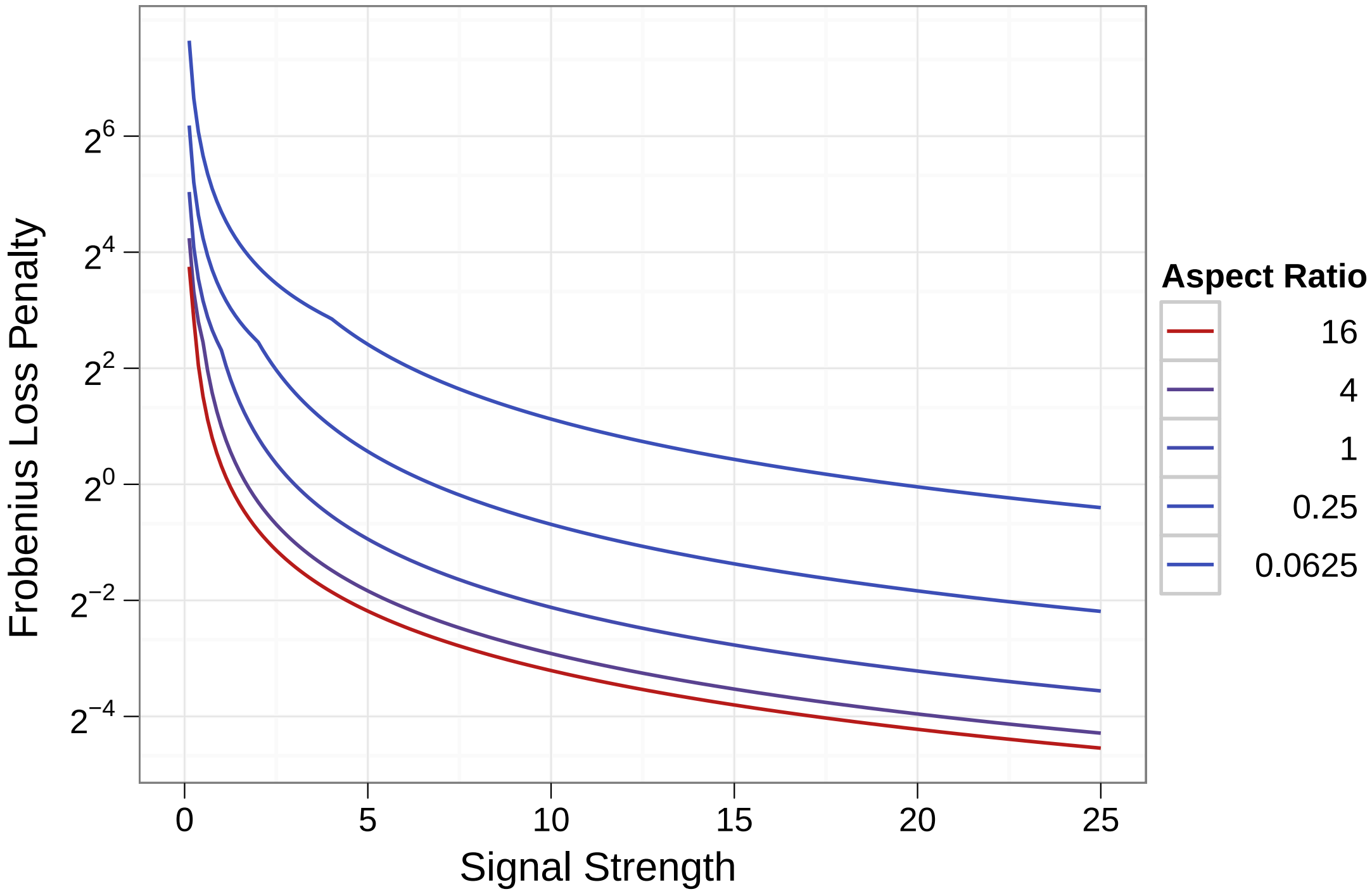

Figure 4.1: Frobenius loss penalty. Relative penalty for including the $i$th factor in the SVD approximation of $\sqrt{n}\boldsymbol{U}_n\boldsymbol{D}_n\boldsymbol{V}_n^{\mathrm{T}}$, with respect to squared Frobenius loss. When the $i$th factor has signal strengh $\mu_i$, the cost for excluding the $i$th term of the SVD is $\mu_i$, and the cost for including it is $\alpha_i \cdot \mu_i$. Here, we plot $\alpha_i$ as a function of $\mu_i$ for various aspect ratios $\gamma = \frac{n}{p}$. The units are chosen so that $\sigma^2 = 1$.

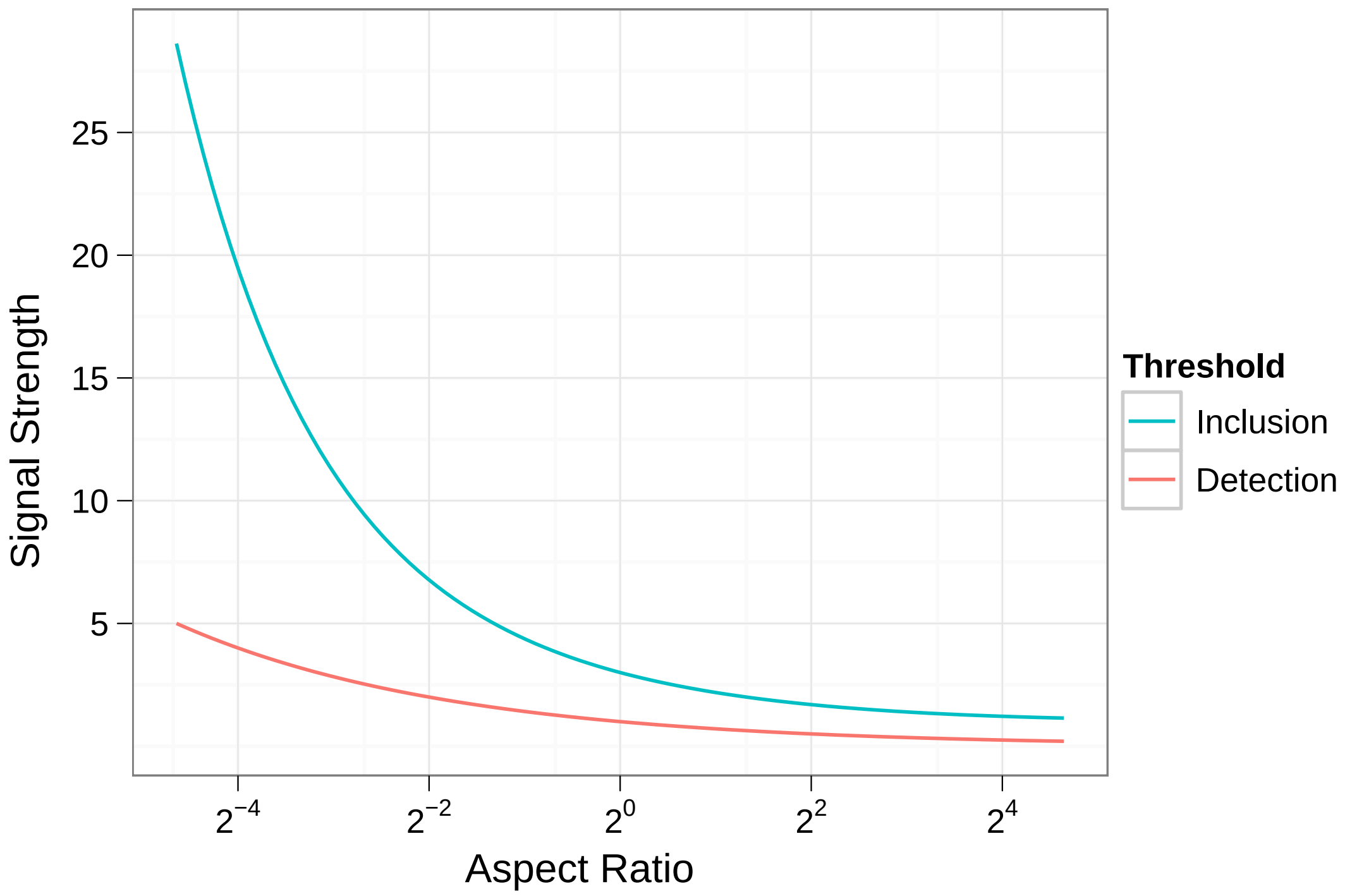

Figure 4.2: SIGNAL STRENGTH THRESHOLDS. Detection threshold $\gamma^{-1/2}$ and inclusion threshold $\mu_{\mathrm{F}}^* = \frac{1+\gamma^{-1}}{2} + \sqrt{\left(\frac{1+\gamma^{-1}}{2}\right)^2 + \frac{3}{\gamma}}$ plotted against the aspect ratio $\gamma = \frac{n}{p}$. When the normalized signal strength $\frac{\mu_i}{\sigma^2}$ is above the detection threshold, the $i$th sample SVD factors are correlated with the population factors. With respect to Frobenius loss, when the normalized signal strength is above the inclusion threshold, it is beneficial to include the $i$th term in the SVD approximation of $\sqrt{n}\,\boldsymbol{U}_n \boldsymbol{D}_n \boldsymbol{V}_n^{\mathrm{T}}$.

## 4.4 Simulations

We confirm the theory of the previous section with a Monte Carlo simulation. We generate a matrix $\boldsymbol{X}$ as follows:

1. The concentration $\gamma$ is one of $\{0.25, 1.0, 4.0\}$.

2. The size of the matrix $s$ is one of $\{144, 400, 1600, 4900\}$. We set the number of rows and columns in the matrix as $n = \sqrt{s\,\gamma}$ and $p = \sqrt{s/\gamma}$, respectively. This ensures that $\gamma = \frac{n}{p}$ and $s = n\,p$.

3. The noise level is set at $\sigma^2 = 1$. The noise matrix $\boldsymbol{E}$ is an $n \times p$ matrix with iid $\mathcal{N}\left(0,\, 1\right)$ elements.

4. The generative rank, $k_0$, is fixed at 5. The normalized factor strengths are set at $(\mu_1,\, \mu_2,\, \mu_3,\, \mu_4,\, \mu_5) = (4\mu_{\mathrm{F}}^*,\, 2\mu_{\mathrm{F}}^*,\, \mu_{\mathrm{F}}^*,\, \frac{1}{2}\mu_{\mathrm{F}}^*,\, \frac{1}{4}\mu_{\mathrm{F}}^*)$, and the factor strength matrix is $\boldsymbol{D} = \operatorname{diag}\left(\mu_1^{1/2}, \mu_2^{1/2}, \ldots, \mu_5^{1/2}\right)$.

5. The left and right factor matrices $\boldsymbol{U}$ and $\boldsymbol{V}$ are of sizes $n \times 5$ and $p \times 5$, respectively. We choose these matrices uniformly at random over the Stiefel manifold according to Algorithm A.1 in Appendix A.

6. We set $\boldsymbol{X} = \sqrt{n}\,\boldsymbol{U}\boldsymbol{D}\boldsymbol{V}^{\mathrm{T}} + \boldsymbol{E}$ and let $\hat{\boldsymbol{X}}(k)$ be the SVD of $\boldsymbol{X}$ truncated to $k$ terms.

After generating $\boldsymbol{X}$, we compute the squared Frobenius loss $\frac{1}{n}\|\sqrt{n}\boldsymbol{U}\boldsymbol{D}\boldsymbol{V}^{\mathrm{T}} - \hat{\boldsymbol{X}}(k)\|_F^2$ and the squared spectral loss $\frac{1}{n}\|\sqrt{n}\boldsymbol{U}\boldsymbol{D}\boldsymbol{V}^{\mathrm{T}} - \hat{\boldsymbol{X}}(k)\|_2^2$ as functions of the rank, $k$. In Figures 4.3 and 4.4, we plot the results over 500 replicates. For the Frobenius case, we should expect the loss to decrease until $k = 2$, stay flat at $k = 3$, and then increase thereafter. This is confirmed by the simulations. The spectral norm behaves similarly to the Frobenius norm.

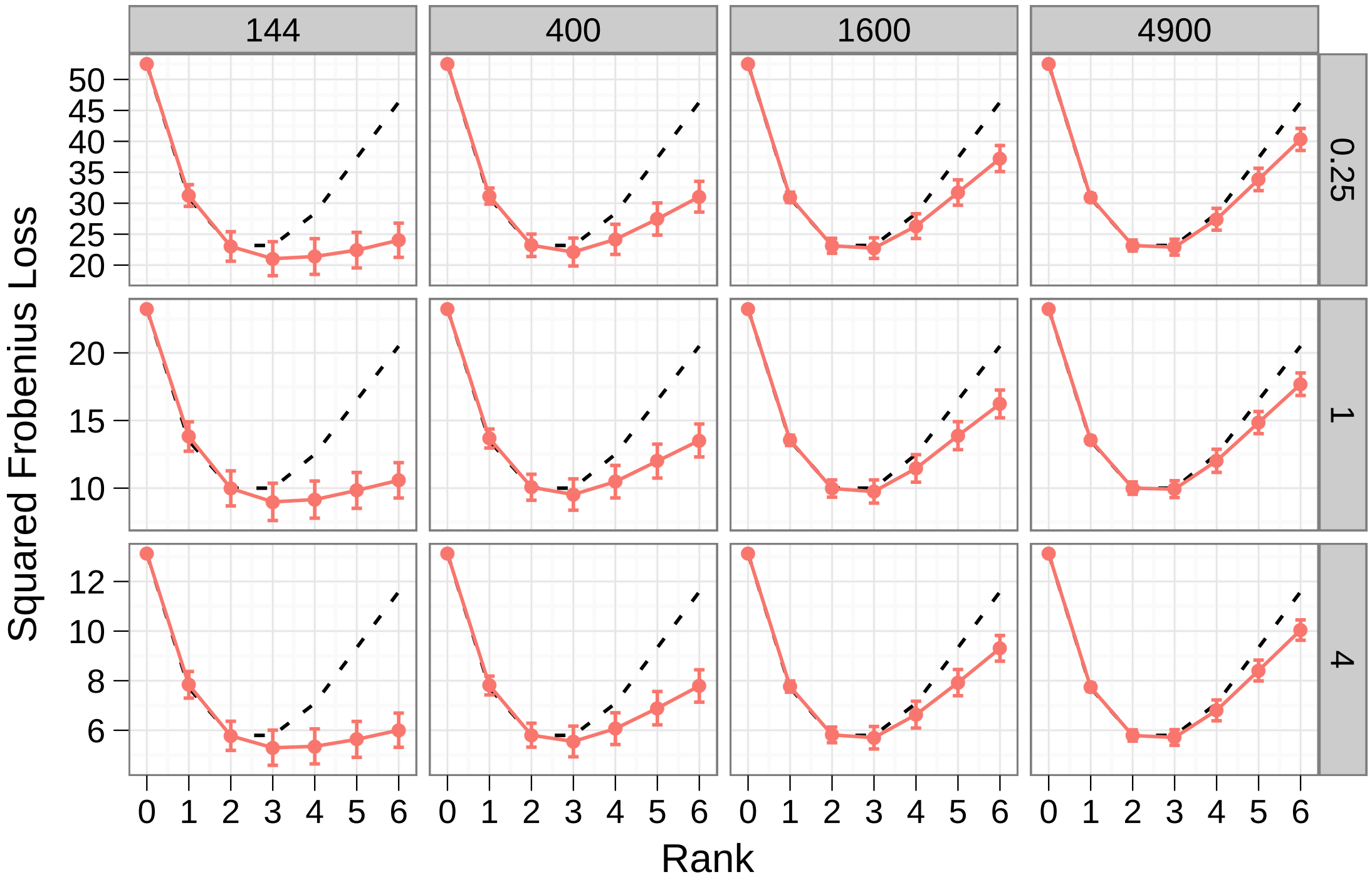

Figure 4.3: SIMULATED FROBENIUS LOSS. Squared Frobenius loss $\frac{1}{n}\|\sqrt{n}\boldsymbol{U}\boldsymbol{D}\boldsymbol{V}^{\mathrm{T}} - \hat{\boldsymbol{X}}(k)\|_F^2$ as a function of the rank, $k$, for $\boldsymbol{X}$ generated according the procedure described in Section 4.4 and $\hat{\boldsymbol{X}}(k)$ equal to the SVD of $\boldsymbol{X}$ truncated to $k$ terms. The concentration $\gamma = \frac{n}{p}$ is one of $\{0.25, 1.0, 4.0\}$ and the size $s = np$ is one of $\{144, 400, 1600, 4900\}$. The solid lines show the means over 500 replicates with the error bars showing one standard deviation. The dashed lines show the predictions from Proposition 4.6. We can see that as the size increases, the simulation agrees more and more with the theory.

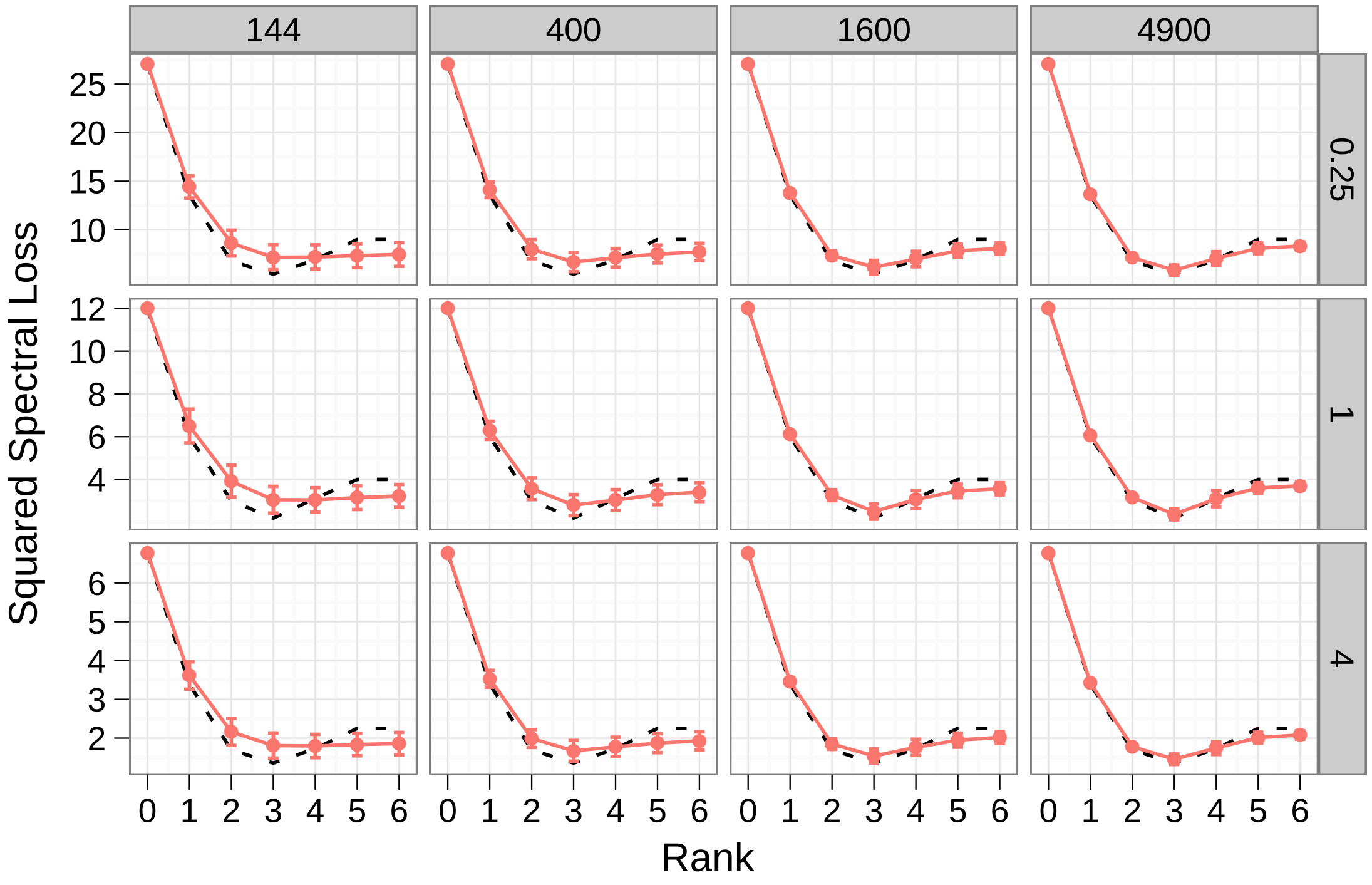

Figure 4.4: SIMULATED SPECTRAL LOSS. Squared spectral loss $\frac{1}{n}\|\sqrt{n}\boldsymbol{U}\boldsymbol{D}\boldsymbol{V}^{\mathrm{T}} - \hat{\boldsymbol{X}}(k)\|_2^2$ as a function of the rank, $k$, with $\boldsymbol{X}$ and $\hat{\boldsymbol{X}}(k)$ as in Figure 4.3. Sample size $s = np$ is shown in the columns and concentration $\gamma = \frac{n}{p}$ is shown in the rows. Solid lines show the means over 500 replicates with the error bars showing one standard deviation; the dashed lines show the predictions from Section 4.3, specifically from equations (4.19), (4.20a–4.21b), and (4.23). The simulations agrees quite well with the theory, especially for large sample sizes.

## 4.5 Relation to the scree plot

Cattell's scree plot [18] is a popular device for choosing the truncation rank in principal component analysis. One plots the square of the singular value, $\hat{d}_k^2$, against the component number, $k$. Typically such a plot exhibits an "elbow", where the slope changes noticeably. This is the point at which Cattell recommends truncating the SVD of the matrix.

In some circumstances, the elbow is close to the Frobenius and spectral loss minimizers, $k_{\mathrm{F}}^*$ and $k_2^*$. In Figure 4.5, we generate a matrix $\boldsymbol{X} = \sqrt{n}\,\boldsymbol{U}\boldsymbol{D}\boldsymbol{V}^{\mathrm{T}} + \boldsymbol{E} \in \mathbb{R}^{n \times p}$, with $n = p = 100$, such that elements of $\boldsymbol{E}$ are iid $\mathcal{N}(0,\,1)$. In the first column, we set $\boldsymbol{D}^2 = \mathrm{diag}(5.0,\,4.75,\,4.5,\,\ldots,\,0.5,\,0.25)$. The figure shows the scree plot along with $\|\sqrt{n}\,\boldsymbol{U}\boldsymbol{D}\boldsymbol{V}^{\mathrm{T}} - \hat{\boldsymbol{X}}(k)\|^2$ for Frobenius and spectral norms, where $\hat{\boldsymbol{X}}(k)$ is the SVD of $\boldsymbol{X}$ truncated to $k$ terms. There is substantial ambiguity in determining the location of the most pronounced elbow. Despite this subtlety, there is indeed an elbow relatively close to $k_{\mathrm{F}}^*$ and $k_2^*$, the minimizers of the two loss functions.

We can easily manipulate the simulation to get a scree plot with a more pronounced elbow in the wrong place. In the second column of Figure 4.5, we augment the factor strength matrix with three additional large values, so that $\boldsymbol{D}^2 = \mathrm{diag}(20.0,\,15.0,\,10.0,\,5.0,\,4.75,\,4.5,\,\ldots,\,0.5,\,0.25)$. With this modification, there is a clear elbow at $k = 4$. However, compared to the optimal values, truncating the SVD at $k = 4$ gives about a 25% worse error with respect to squared Frobenius loss and about 50% worse with respect to squared spectral loss.

In general, we cannot make any assurances about how close the elbow is to $k_{\mathrm{F}}^*$ or $k_2^*$. Through a simulation study, Jackson [42] provides evidence that if the latent factors are strong enough to be easily distinguished from noise, then the elbow is a reasonable estimate of the loss minimizers (he does not actually compute the loss behavior, but this seems likely). However, when there are both well-separated factors and factors near the critical strength level $\mu_{\mathrm{F}}^*$, the second example here illustrates that the elbow might be a poor estimate.

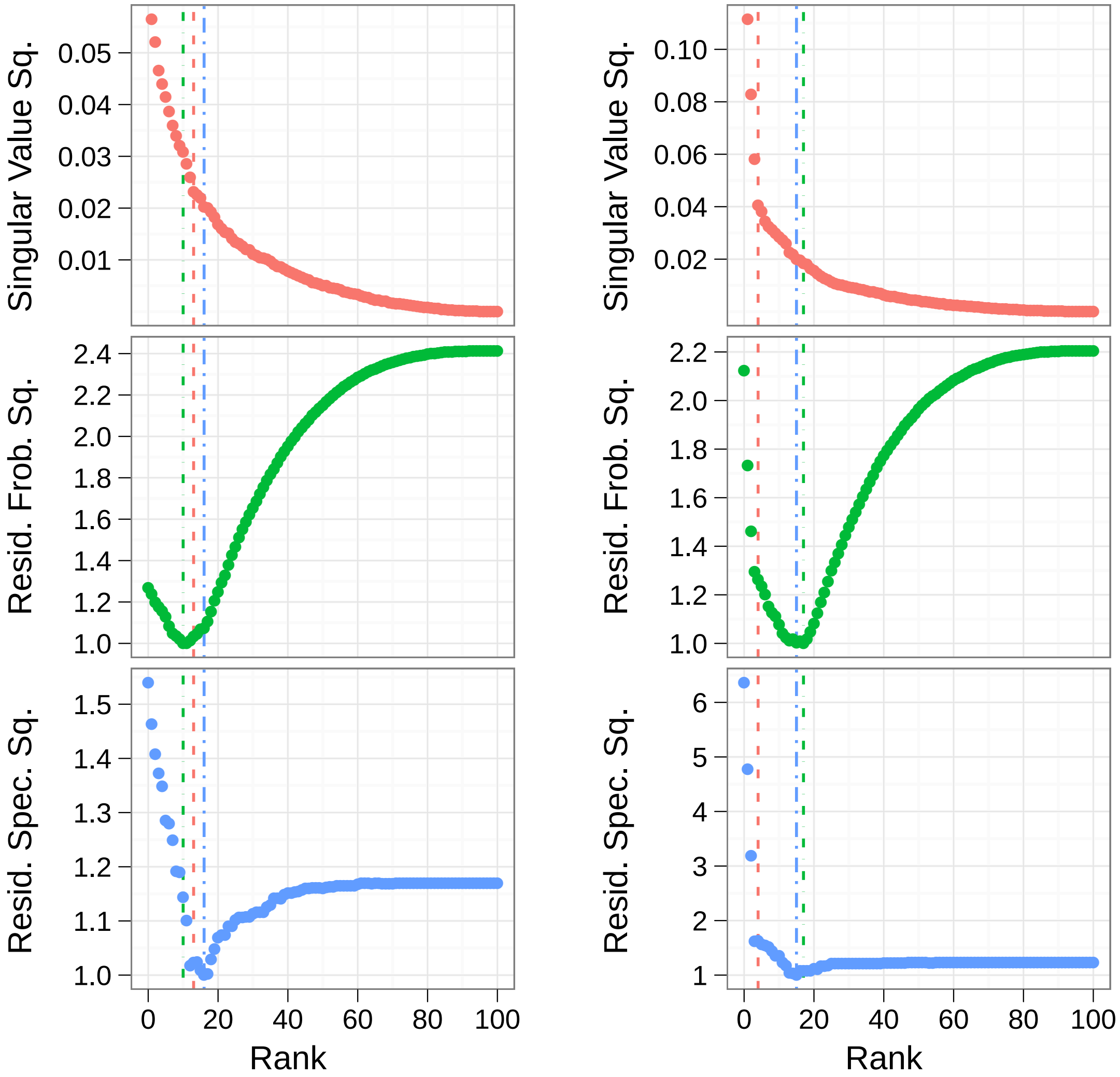

Figure 4.5: Scree plots and loss functions. We generate a matrix as $\boldsymbol{X} = \sqrt{n}\,\boldsymbol{U}\boldsymbol{D}\boldsymbol{V}^{\mathrm{T}} + \boldsymbol{E}$ and set $\hat{\boldsymbol{X}}(k)$ equal to the SVD of $\boldsymbol{X}$ truncated to $k$ terms. The left and right columns show different choices of $\boldsymbol{D}$, described in the text. The top row contains scree plots, where the square of the $k$th singular value of $\boldsymbol{X}$, $\hat{d}_k^2$, is plotted against component number, $k$, with units are chosen so that $\sum_k \hat{d}_k^2 = 1$. The next two rows show $\|\sqrt{n}\,\boldsymbol{U}\boldsymbol{D}\boldsymbol{V}^{\mathrm{T}} - \hat{\boldsymbol{X}}(k)\|_{\mathrm{F}}^2$ and $\|\sqrt{n}\,\boldsymbol{U}\boldsymbol{D}\boldsymbol{V}^{\mathrm{T}} - \hat{\boldsymbol{X}}(k)\|_2^2$, as functions of the rank, $k$ with units chosen so that the minimum value is 1.0. Dashed lines show the elbow of the scree plot and the minimizers of the two loss functions. The elbow of the scree plot (which is fit by eye) gives a reasonable estimate of the loss minimizers in the first column, but not in the second.

## 4.6 Extensions

We have focused specifically on truncating the singular value decomposition because that is the case for which the theory has been most developed. We have also focused specifically on Frobenius and spectral losses. One could easily extend our work to look at the nuclear norm loss (sum of singular values) [32] or Stein's loss [43]. Alternatively, it is not hard to adapt our analysis to study forms like $\|\hat{\boldsymbol{U}}\hat{\boldsymbol{D}}(k)^2\hat{\boldsymbol{U}}^{\mathrm{T}} - \boldsymbol{U}\boldsymbol{D}^2\boldsymbol{U}^{\mathrm{T}}\|$. For matrix decompositions beyond the SVD, extension of our work is more difficult, mainly because very little scholarship has been devoted to their theoretical properties.

We have not examined shrinking the singular values at all, but in some situations this may be beneficial. For example, the Frobenius norm of the $\boldsymbol{F}_i$ matrix from Section 4.3, which is involved in the Frobenius loss $\|\sqrt{n}\boldsymbol{U}\boldsymbol{D}\boldsymbol{V}^{\mathrm{T}} - \hat{\boldsymbol{X}}(k)\|_F^2$, converges to

$$\mu_i - 2\varphi_i\theta_i\mu_i^{1/2}\bar{\mu}_i^{1/2} + \bar{\mu}_i$$

whenever $k \geq i$. Recall that $\bar{\mu}_i$ is the almost-sure limit of the $\hat{d}_i^2$, the square of the $i$th singular value of $\frac{1}{\sqrt{n}}\boldsymbol{X}$. If we shrink the $i$th singular value, replacing $\hat{d}_i$ with $f(\hat{d}_i)$ for continuous $f(\cdot)$, then the Frobenius penalty for including the $i$th term converges to

$$\mu_i - 2\varphi_i\theta_i\mu_i^{1/2} f(\bar{\mu}_i^{1/2}) + f^2(\bar{\mu}_i^{1/2}).$$

This quantity is minimized when $f(\bar{\mu}_i^{1/2}) = \mu_i^{1/2}\varphi_i\theta_i = \bar{\mu}_i^{-1/2}\left(\mu_i - \frac{\sigma^4}{\gamma\mu_i}\right)$. After some algebra, the optimal $f$ takes the form

$$f(\hat{d}_i) = \begin{cases} \left(\hat{d}_i^2 - 2\left(1 + \frac{1}{\gamma}\right)\sigma^2 + \left(1 - \frac{1}{\gamma}\right)^2 \frac{\sigma^4}{\hat{d}_i^2}\right)^{1/2} & \text{when } \hat{d}_i > \sigma\left(1 + \frac{1}{\sqrt{\gamma}}\right), \\ 0 & \text{otherwise.} \end{cases} \tag{4.27}$$

With this shrinkage, it is always beneficial (in terms of Frobenius norm) to include the $i$th term.

## 4.7 Summary and future work

We have described a plausible model for data generated by a small number of latent factors corrupted by additive white noise. With this model in mind, we have motivated two loss functions, the squared Frobenius norm and squared spectral norm, as measuring average- or worst-case quadratic error over the class of all bilinear statistics. These loss functions in turn determine the intrinsic ranks $k^*_{\mathrm{F}}$ and $k^*_2$. We have shown how the losses and the ranks behave for the truncated SVD, both theoretically and through simulation. Finally, we have explored the relation between intrinsic rank and the scree plot, and then discussed some extensions.

We did not describe a way to *estimate* the error from truncating an SVD, nor did we propose a practical method for choosing $k$. For now, our hope is that this work is useful in developing intuition for how the SVD behaves and that it provides a suitable starting point for designing and evaluating such procedures. In later chapters, we explicitly discuss estimation and rank selection.

# Chapter 5

# Cross-validation for unsupervised learning

The problem unsupervised learning (UL) tries to address is this: given some data, describe its distribution. Many estimation problems can be cast as unsupervised learning, including mean and density estimation. However, more commonly unsupervised learning refers to either clustering or manifold learning. A canonical example is principal component analysis (PCA). In PCA, we are given some high-dimensional data, and we look for a lower-dimensional subspace that explains most of the variation in the data. The lower-dimensional subspace describes the distribution of the data. In clustering, the estimated cluster centers give us information about the distribution of the data. The output of every UL method is a summary statistic designed to convey information about the process which generated the data.

Many UL problems involve model selection. For example, in principal component analysis we need to choose how many components to keep. For clustering, we need to choose the number of clusters in the data. Many manifold learning techniques require choosing a bandwidth or a kernel. Often in these contexts, model-selection is done in an ad-hoc manner. Rules of thumb and manual inspection guide most choices for how many components to keep, how many clusters are present, and what is an appropriate kernel. Such informal selection rules can be problematic when different researchers come to different conclusions about what the right model is. Moreover, even when there is an obvious "natural" model to human eyes, it may be hard to pick out in computer-automated analysis. For objectivity and efficiency, it is desirable to have a well-specified automatic model selection procedure.

For concreteness, in this chapter we focus on principal components, though many of the ideas generalize to other methods. We suppose that the data, $\boldsymbol{X}$, is an $n \times p$ matrix generated according to the signal-plus-noise model $\boldsymbol{X} = \sqrt{n}\,\boldsymbol{U}\boldsymbol{D}\boldsymbol{V}^{\mathrm{T}} + \boldsymbol{E}$. We consider the first term to be the "signal" matrix, and the second term to be "noise". Often the signal term is a low-rank product. Our goal is to estimate this term as best as possible by truncating the singular value decomposition (SVD) of $\boldsymbol{X}$. Here "best" means with respect to the metrics introduced in Chapter 4. We are interested in the model selection problem where each model is defined by the number of terms we keep from the SVD of $\boldsymbol{X}$.

We would like our model selection procedure to be non-parametric, if possible. To work in a variety of contexts, the selection procedure cannot assume Gaussianity or independence across samples. We would like the procedure to be driven by the empirical distribution of the data. Cross-validation (CV) is a popular approach to model selection that generally meets these criteria. Therefore, we seek to adapt CV to our purposes.

CV prescribes dividing a data set into a "test" set and a "train" set, fitting a model to the training set, and then evaluating the model's performance on the test set. We repeat the fit/evaluate procedure multiple times over different test/train partitions, and then average over all replicates. Traditionally, the partitions are chosen so that each datum occurs in exactly one test set. As for terminology, for a particular replicate the test set is commonly referred to as the held-out or left-out set, and likewise the train set is often called the held-in or left-in set.

Most often, cross-validation is applied in supervised contexts. In supervised learning (SL) the data consists of a sequence of $(\underline{x}, \underline{y})$ predictor-response pairs. Broadly construed, the goal of supervised learning is to describe the conditional distribution of $\underline{y}$ given $\underline{x}$. This is usually for prediction or classification. In the supervised context, for a particular CV replicate there are four parts of data:

$$\begin{pmatrix} \boldsymbol{X}_{\text{train}} & \boldsymbol{Y}_{\text{train}} \\ \boldsymbol{X}_{\text{test}} & \boldsymbol{Y}_{\text{test}} \end{pmatrix}.$$

Implicit in the description of cross-validation is that the replicates use $\boldsymbol{X}_{\text{test}}$ to predict $\boldsymbol{Y}_{\text{test}}$. So, the held-in data looks like

$$\begin{pmatrix} \boldsymbol{X}_{\text{train}} & \boldsymbol{Y}_{\text{train}} \\ \boldsymbol{X}_{\text{test}} & * \end{pmatrix}.$$

We extrapolate from $\boldsymbol{X}_{\text{test}}$ to predict $\boldsymbol{Y}_{\text{test}}$.

It is not immediately obvious how to apply cross-validation to unsupervised learning. In unsupervised learning there is no $\boldsymbol{Y}$; we instead have the two-way partition

$$\begin{pmatrix} \boldsymbol{X}_{\text{train}} \\ \boldsymbol{X}_{\text{test}} \end{pmatrix}.$$

There is nothing to predict! Renaming $\boldsymbol{X}$ to $\boldsymbol{Y}$ does not make the problem any better, for then the division becomes:

$$\begin{pmatrix} \boldsymbol{Y}_{\text{train}} \\ \boldsymbol{Y}_{\text{test}} \end{pmatrix},$$

with hold-in

$$\begin{pmatrix} \boldsymbol{Y}_{\text{train}} \\ * \end{pmatrix}.$$

Now, there is nothing to extrapolate from to predict $\boldsymbol{Y}_{\text{test}}$! For cross-validation to work in unsupervised learning, we need to consider more general hold-outs.

We look at two different types of hold-outs in this chapter. The first, due to Wold, is "speckled": we leave out random elements of the matrix $\boldsymbol{X}$ and use a missing data algorithm like expectation-maximization (EM) for prediction. The second type of hold-out is "blocked". This type, due to Gabriel, randomly partitions the columns of $\boldsymbol{X}$ into "predictor" and "response" sets and then performs the SL version of cross-validation.

## 5.1 Assumptions, and notation

We will generally assume we have data $\boldsymbol{X} \in \mathbb{R}^{n \times p}$ generated according to the latent factor model

$$\boldsymbol{X} = \sqrt{n}\,\boldsymbol{U}\boldsymbol{D}\boldsymbol{V}^{\mathrm{T}} + \boldsymbol{E}. \tag{5.1}$$

Here, $\boldsymbol{U} \in \mathbb{R}^{n \times k_0}$, $\boldsymbol{V} \in \mathbb{R}^{n \times k_0}$, and $\boldsymbol{D} = \operatorname{diag}(d_1, d_2, \ldots, d_{k_0})$, with $\boldsymbol{U}^{\mathrm{T}}\boldsymbol{U} = \boldsymbol{I}_{k_0}$ $\boldsymbol{V}^{\mathrm{T}}\boldsymbol{V} = \boldsymbol{I}_{k_0}$, and $d_1 \geq d_2 \geq \cdots \geq d_{k_0} > 0$. We call $\sqrt{n}\,\boldsymbol{U}\boldsymbol{D}\boldsymbol{V}^{\mathrm{T}}$ the signal part and $\boldsymbol{E}$ the noise part. In the spirit of data-driven analysis, we avoid putting distributional assumptions on $\boldsymbol{U}$, $\boldsymbol{D}$, $\boldsymbol{V}$, and $\boldsymbol{E}$. This makes the terms unidentifiable. While this indeterminacy can (and should!) bother some readers, for now we will plod on.

We denote the SVD of $\boldsymbol{X}$ by

$$\boldsymbol{X} = \sqrt{n}\,\hat{\boldsymbol{U}}\hat{\boldsymbol{D}}\hat{\boldsymbol{V}}^{\mathrm{T}}, \tag{5.2}$$

with $\hat{\boldsymbol{U}} \in \mathbb{R}^{n \times n \wedge p}$, $\hat{\boldsymbol{V}} \in \mathbb{R}^{p \times n \wedge p}$, and $\hat{\boldsymbol{D}} = \operatorname{diag}(\hat{d}_1, \hat{d}_2, \ldots, \hat{d}_{n \wedge p})$. Here, $\hat{\boldsymbol{U}}^{\mathrm{T}}\hat{\boldsymbol{U}} = \hat{\boldsymbol{V}}^{\mathrm{T}}\hat{\boldsymbol{V}} = \boldsymbol{I}_{n \wedge p}$ and the singular values are ordered $\hat{d}_1 \geq \hat{d}_2 \geq \cdots \geq \hat{d}_{n \wedge p} \geq 0$. We set $\hat{\boldsymbol{D}}(k) = \operatorname{diag}(\hat{d}_1, \hat{d}_2, \ldots, \hat{d}_k, 0, \ldots, 0) \in \mathbb{R}^{n \wedge p \times n \wedge p}$ so that

$$\hat{\boldsymbol{X}}(k) = \sqrt{n}\,\hat{\boldsymbol{U}}\hat{\boldsymbol{D}}(k)\hat{\boldsymbol{V}}^{\mathrm{T}} \tag{5.3}$$

is the SVD of $\boldsymbol{X}$ truncated to $k$ terms. Similarly, we define $\hat{\boldsymbol{U}}(k) \in \mathbb{R}^{n \times k}$ and $\hat{\boldsymbol{V}}(k) \in \mathbb{R}^{p \times k}$ to be the first $k$ rows of $\hat{\boldsymbol{U}}$ and $\hat{\boldsymbol{V}}$, respectively.

We focus on estimating the squared Frobenius model error

$$\mathrm{ME}(k) = \| \sqrt{n}\,\boldsymbol{U}\boldsymbol{D}\boldsymbol{V}^{\mathrm{T}} - \hat{\boldsymbol{X}}(k) \|_{\mathrm{F}}^2 \tag{5.4}$$

or its minimizer,

$$k^*_{\mathrm{ME}} = \operatorname*{argmin}_k \mathrm{ME}(k). \tag{5.5}$$

Here, $\|\cdot\|_{\mathrm{F}}^2$ is the sum of squares of the elements.

Another kind of error relevant to cross-validation is *prediction error*. We let $\boldsymbol{E}'$ be a matrix independent of $\boldsymbol{E}$ but having the same distribution conditionally on $\boldsymbol{U}$, $\boldsymbol{D}$, and $\boldsymbol{V}^{\mathrm{T}}$. We set $\boldsymbol{X}' = \sqrt{n}\,\boldsymbol{U}\boldsymbol{D}\boldsymbol{V}^{\mathrm{T}} + \boldsymbol{E}'$ and define the prediction error

$$\mathrm{PE}(k) = \mathbb{E}\|\boldsymbol{X}' - \hat{\boldsymbol{X}}(k)\|_{\mathrm{F}}^2. \tag{5.6}$$

Likewise, we set

$$k_{\mathrm{PE}}^* = \operatorname*{argmin}_k \mathrm{PE}(k). \tag{5.7}$$

If $\boldsymbol{E}$ is independent of $\boldsymbol{U}$, $\boldsymbol{D}$, and $\boldsymbol{V}$, then

$$\begin{aligned}
\mathrm{PE}(k) &= \mathbb{E}\|\sqrt{n}\,\boldsymbol{U}\boldsymbol{D}\boldsymbol{V}^{\mathrm{T}} - \hat{\boldsymbol{X}}(k) + \boldsymbol{E}'\|_{\mathrm{F}}^2 \\
&= \mathbb{E}\|\sqrt{n}\,\boldsymbol{U}\boldsymbol{D}\boldsymbol{V}^{\mathrm{T}} - \hat{\boldsymbol{X}}(k)\| \\
&\qquad + 2\,\mathbb{E}\left[\mathrm{tr}\left(\left(\sqrt{n}\,\boldsymbol{U}\boldsymbol{D}\boldsymbol{V}^{\mathrm{T}} - \hat{\boldsymbol{X}}(k)\right)^{\mathrm{T}}\boldsymbol{E}'\right)\right] + \mathbb{E}\|\boldsymbol{E}'\|_{\mathrm{F}}^2 \\
&= \mathbb{E}\big[\,\mathrm{ME}(k)\big] + \mathbb{E}\|\boldsymbol{E}\|_{\mathrm{F}}^2.
\end{aligned} \tag{5.8}$$

The prediction error is thus equal to the sum of the expected model error and an irreducible error term.

Finally, we should note that our definitions of prediction error and model error are motivated by the definitions given by Breiman [15] for cross-validating linear regression.

## 5.2 Cross validation strategies

In this section we describe the various hold-out strategies for getting a cross-validation estimate of $\mathrm{PE}(k)$. It is possible to get an estimate of $\mathrm{ME}(k)$ from the estimate of $\mathrm{PE}(k)$ by subtracting an estimate of the irreducible error. For now, though, we choose to focus just on estimating $\mathrm{PE}(k)$.

For performing $K$-fold cross-validation on the matrix $\boldsymbol{X}$, we partition its elements into $K$ hold-out sets. For each of $K$ replicates and for each value of the rank, $k$, we leave out one of the hold-out sets, fit a $k$-term SVD to the left-in set, and evaluate its performance on the left-out set. We thus need to describe the hold-out set, how to fit an SVD to the left-in data, and how to make a prediction of the left-out data. After a brief discussion of why the usual (naive) way of doing hold-outs won't work, we survey both speckled hold-outs (Wold-style), as well as blocked hold-outs (Gabriel-style).

### 5.2.1 Naive hold-outs

The ordinary hold-out strategy will not work for estimating prediction error. Suppose we leave out a subset of the rows of $\boldsymbol{X}$. After permutation, the rows of $\boldsymbol{X}$ are partitioned as $\begin{pmatrix} \boldsymbol{X}_1 \\ \boldsymbol{X}_2 \end{pmatrix}$, where $\boldsymbol{X}_1 \in \mathbb{R}^{n_1 \times p}$ and $\boldsymbol{X}_2 \in \mathbb{R}^{n_2 \times p}$, and $n_1 + n_2 = n$. The only plausible prediction of $\boldsymbol{X}_2$ based on truncating the SVD of $\boldsymbol{X}_1$ is the following:

1. Let $\boldsymbol{X}_1 = \sqrt{n}\,\hat{\boldsymbol{U}}_1 \hat{\boldsymbol{D}}_1 \hat{\boldsymbol{V}}_1^{\mathrm{T}}$ be the SVD of $\boldsymbol{X}_1$, with $\hat{\boldsymbol{D}}_1 = \operatorname{diag}(\hat{d}_1^{(1)}, \hat{d}_2^{(1)}, \ldots, \hat{d}_{n_1 \wedge p}^{(1)})$.
2. Let $\hat{\boldsymbol{D}}_1(k) = \operatorname{diag}(\hat{d}_1^{(1)}, \hat{d}_2^{(1)}, \ldots, \hat{d}_k^{(1)}, 0, \ldots, 0)$ so that $\hat{\boldsymbol{X}}_1(k) \equiv \sqrt{n}\,\hat{\boldsymbol{U}}_1 \hat{\boldsymbol{D}}_1(k) \hat{\boldsymbol{V}}_1^{\mathrm{T}}$ is the SVD of $\boldsymbol{X}_1$ truncated to $k$ terms. Similarly, denote by $\hat{\boldsymbol{V}}_1(k) \in \mathbb{R}^{p \times k}$ the first $k$ columns of $\hat{\boldsymbol{V}}_1$.
3. Let $(^+)$ denote pseudo-inverse and predict the held out rows as $\hat{\boldsymbol{X}}_2(k) = \boldsymbol{X}_2 \hat{\boldsymbol{X}}_1(k)^{\mathrm{T}} \big(\hat{\boldsymbol{X}}_1(k) \hat{\boldsymbol{X}}_1(k)^{\mathrm{T}}\big)^+ \hat{\boldsymbol{X}}_1(k) = \boldsymbol{X}_2 \hat{\boldsymbol{V}}_1(k) \hat{\boldsymbol{V}}_1(k)^{\mathrm{T}}$.

The problem with this procedure is that $\|\boldsymbol{X}_2 - \hat{\boldsymbol{X}}_2(k)\|_{\mathrm{F}}^2$ decreases with $k$ regardless of the true model for $\boldsymbol{X}$. So, it cannot possibly give us a good estimate of the error from truncating the full SVD of $\boldsymbol{X}$.

A similar situation arrises if we leave out only a subset of the columns of $\boldsymbol{X}$. To get a reasonable cross-validation estimate, it is therefore necessary to consider more-general hold-out sets.

### 5.2.2 Wold hold-outs

A Wold-style speckled leave-out is perhaps the most obvious attempt at a more general hold-out. We leave out a subset of the elements of the $\boldsymbol{X}$, then use the left-in elements to predict the rest.

First we need to introduce some more notation. Let $\mathcal{I}$ denote the set of indices of the elements of $\boldsymbol{X}$, so that $\mathcal{I} = \{(i,j) : 1 \le i \le n, 1 \le j \le p\}$. For a subset $I \subset \mathcal{I}$, let $\bar{I}$ denote its complement $\mathcal{I} \setminus I$. We use the symbol $*$ to denote a missing value, and let $\mathbb{R}_* = \mathbb{R} \cup \{*\}$. For $I \subset \mathcal{I}$, we define the matrix $\boldsymbol{X}_I \in \mathbb{R}_*^{n \times p}$ with elements

$$X_{I,ij} = \begin{cases} X_{ij} & \text{if } (i,j) \in I, \\ * & \text{otherwise.} \end{cases} \tag{5.9}$$

Similarly $\boldsymbol{X}_{\bar{I}}$, has elements

$$X_{\bar{I},ij} = \begin{cases} X_{ij} & \text{if } (i,j) \notin I, \\ * & \text{otherwise.} \end{cases} \tag{5.10}$$

Finally, for $\boldsymbol{A} \in \mathbb{R}_*^{n \times p}$, we define

$$\|\boldsymbol{A}\|_{\mathrm{F},I}^2 = \sum_{(i,j) \in I} A_{ij}^2. \tag{5.11}$$

This notation allows us to describe matrices with missing entries.

A Wold-style speckled hold-out is an unstructured random subset $I \subset \mathcal{I}$. The held-in data is the matrix $\boldsymbol{X}_{\bar{I}}$ and the held-out data is the matrix $\boldsymbol{X}_I$. We use a missing value SVD algorithm to fit a $k$-term SVD to $\boldsymbol{X}_{\bar{I}}$. This gives us an approximation $\boldsymbol{U}_k \boldsymbol{D}_k \boldsymbol{V}_k^{\mathrm{T}}$, where $\boldsymbol{U}_k \in \mathbb{R}^{n \times k}$ and $\boldsymbol{V}_k \in \mathbb{R}^{p \times k}$ have orthonormal columns and $\boldsymbol{D} \in \mathbb{R}^{k \times k}$ is diagonal. We evaluate the SVD on the held-out set by $\|\boldsymbol{U}_k \boldsymbol{D}_k \boldsymbol{V}_k^{\mathrm{T}} - \boldsymbol{X}_I\|_{\mathrm{F},I}^2$. Aside from the algorithm for getting $\boldsymbol{U}_k \boldsymbol{D}_k \boldsymbol{V}_k^{\mathrm{T}}$ from $\boldsymbol{X}_{\bar{I}}$, this is a full description of the CV replicate.

When Wold introduced this form of hold-out in 1978 [95], he suggested using an algorithm called nonlinear iterative partial least squares (NIPALS) to fit an SVD to the held-in data. This algorithm, attributed to Fisher and Mackenzie [33] and rediscovered by Wold and Lyttkens [94], never seems to have gained much prominence. We suggest instead using an expectation-maximization (EM) as is consistent with current mainstream practice in Statistics. Either way, there are some subtle issues in taking the SVD of a matrix with missing values. We explore these issues further in Section 5.3, and present a complete algorithm for estimating the factors $\boldsymbol{U}_k \boldsymbol{D}_k \boldsymbol{V}_k^{\mathrm{T}}$.

### 5.2.3 Gabriel hold-outs

A Gabriel-style hold-out works by transforming the unsupervised learning problem into a supervised one. We take a subset of the columns of $\boldsymbol{X}$ and denote them as the *response* columns; the rest are denoted *predictor* columns. Then we take a subset of the rows and consider them as *test* rows; the rest are *train* rows. This partitions the elements of $\boldsymbol{X}$ into four blocks. With permutation matrices $\boldsymbol{P}$ and $\boldsymbol{Q}$, we can write

$$\boldsymbol{P}^{\mathrm{T}} \boldsymbol{X} \boldsymbol{Q} = \begin{pmatrix} \boldsymbol{X}_{11} & \boldsymbol{X}_{12} \\ \boldsymbol{X}_{21} & \boldsymbol{X}_{22} \end{pmatrix}. \tag{5.12}$$

Here, $\boldsymbol{X}_{11}$ consists of the train-predictor block, $\boldsymbol{X}_{12}$ is the train-response block, $\boldsymbol{X}_{21}$ is the test-predictor block, and $\boldsymbol{X}_{22}$ is the test-response block. It is beneficial to think of the blocks as

$$\begin{pmatrix} \boldsymbol{X}_{11} & \boldsymbol{X}_{12} \\ \boldsymbol{X}_{21} & \boldsymbol{X}_{22} \end{pmatrix} = \begin{pmatrix} \boldsymbol{X}_{\text{train}} & \boldsymbol{Y}_{\text{train}} \\ \boldsymbol{X}_{\text{test}} & \boldsymbol{Y}_{\text{test}} \end{pmatrix}.$$

A Gabriel-style replicate has hold-in

$$\begin{pmatrix} \boldsymbol{X}_{11} & \boldsymbol{X}_{12} \\ \boldsymbol{X}_{21} & * \end{pmatrix}$$

and hold-out $\boldsymbol{X}_{22}$.

To fit a model to the hold-in set, we take the SVD of $\boldsymbol{X}_{11}$ and fit a regression function from the predictor columns to the response columns. Normally, the regression is ordinary least squares regression from the principal component loadings to the response columns. Then, to evaluate the function on the hold-out set, we apply the estimated regression function to $\boldsymbol{X}_{21}$ to get a prediction $\hat{\boldsymbol{X}}_{22}$.

In precise terms, suppose that there are $p_1$ predictor columns, $p_2$ response columns, $n_1$ train rows, and $n_2$ test rows. Then $\boldsymbol{X}_{11} \in \mathbb{R}^{n_1 \times p_1}$ and $\boldsymbol{X}_{22} \in \mathbb{R}^{n_2 \times p_2}$ with $p_1 + p_2 = p$ and $n_1 + n_2 = n$. First we fit an SVD to the train-predictor block

$$\boldsymbol{X}_{11} = \sqrt{n}\,\hat{\boldsymbol{U}}_1 \hat{\boldsymbol{D}}_1 \hat{\boldsymbol{V}}_1^{\mathrm{T}}.$$

Then, we truncate the SVD to $k$ terms as $\sqrt{n}\,\hat{\boldsymbol{U}}_1(k)\hat{\boldsymbol{D}}_1(k)\hat{\boldsymbol{V}}_1^{\mathrm{T}}(k)$ in the same way as is discussed in Section 5.1. This defines a projection $\hat{\boldsymbol{V}}_1(k)$ and principal component scores

$$\hat{\boldsymbol{Z}}_1(k) = \boldsymbol{X}_{11}\hat{\boldsymbol{V}}_1(k) = \sqrt{n}\,\hat{\boldsymbol{U}}_1(k)\hat{\boldsymbol{D}}_1(k).$$

Similarly, we can get the principal components scores for the test-predictor block as

$$\hat{\boldsymbol{Z}}_2(k) = \boldsymbol{X}_{21}\hat{\boldsymbol{V}}_1(k).$$

Next, we model the response columns as linear functions of the principal component scores. We fit the model $\boldsymbol{X}_{12} \approx \hat{\boldsymbol{Z}}_1(k)\boldsymbol{B}$ with

$$\hat{\boldsymbol{B}} = \left(\hat{\boldsymbol{Z}}_1(k)^{\mathrm{T}}\hat{\boldsymbol{Z}}_1(k)\right)^{+}\hat{\boldsymbol{Z}}_1(k)^{\mathrm{T}}\boldsymbol{X}_{12} = \frac{1}{\sqrt{n}}\hat{\boldsymbol{D}}_1(k)^{+}\,\hat{\boldsymbol{U}}_1(k)^{\mathrm{T}}\boldsymbol{X}_{12}.$$

Finally, we apply the model the test rows to get a prediction for the test-response block

$$\hat{\boldsymbol{X}}_{22} = \hat{\boldsymbol{Z}}_2(k)\hat{\boldsymbol{B}} = \boldsymbol{X}_{21}\left(\frac{1}{\sqrt{n}}\hat{\boldsymbol{V}}_1(k)\hat{\boldsymbol{D}}_1(k)^{+}\,\hat{\boldsymbol{U}}_1(k)^{\mathrm{T}}\right)\boldsymbol{X}_{12}. \tag{5.13}$$

This gives a complete description of the CV replicate.

**Remark 5.1.** *Typically for Gabriel-style CV, we have two partitions, one for the rows and one for the columns. If the rows are partitioned into $K$ sets and the columns are partitioned into $L$ sets, then we average the estimated prediction error over all $KL$ possible hold-out sets. This corresponds to $\frac{n}{n_2} \approx K$ and $\frac{p}{p_2} \approx L$. In this situation, we say that we are performing $(K, L)$-fold Gabriel-style cross-validation.*

We can get some intuition for why Gabriel-style replicates work by expressing the latent factor decomposition $\boldsymbol{X} = \sqrt{n}\,\boldsymbol{U}\boldsymbol{D}\boldsymbol{V}^{\mathrm{T}} + \boldsymbol{E}$ in block form. We let $\boldsymbol{P} = \begin{pmatrix}\boldsymbol{P}_1 & \boldsymbol{P}_2\end{pmatrix}$ and $\boldsymbol{Q} = \begin{pmatrix}\boldsymbol{Q}_1 & \boldsymbol{Q}_2\end{pmatrix}$ be the block-decompositions of the row and column permutations so that $\boldsymbol{X}_{ij} = \boldsymbol{P}_i^{\mathrm{T}}\boldsymbol{X}\boldsymbol{Q}_j$. We define $\boldsymbol{U}_i = \boldsymbol{P}_i^{\mathrm{T}}\boldsymbol{U}$, $\boldsymbol{V}_j = \boldsymbol{Q}_j^{\mathrm{T}}\boldsymbol{V}$, and

$\boldsymbol{E}_{ij} = \boldsymbol{P}_i^{\mathrm{T}} \boldsymbol{E} \boldsymbol{Q}_j$ so that

$$\begin{aligned}
\boldsymbol{P}^{\mathrm{T}} \boldsymbol{X} \boldsymbol{Q} &= \begin{pmatrix} \boldsymbol{P}_1^{\mathrm{T}} \\ \boldsymbol{P}_2^{\mathrm{T}} \end{pmatrix} \left( \sqrt{n}\, \boldsymbol{U} \boldsymbol{D} \boldsymbol{V}^{\mathrm{T}} + \boldsymbol{E} \right) \begin{pmatrix} \boldsymbol{Q}_1 & \boldsymbol{Q}_2 \end{pmatrix} \\
&= \sqrt{n} \begin{pmatrix} \boldsymbol{U}_1 \boldsymbol{D} \boldsymbol{V}_1^{\mathrm{T}} & \boldsymbol{U}_1 \boldsymbol{D} \boldsymbol{V}_2^{\mathrm{T}} \\ \boldsymbol{U}_2 \boldsymbol{D} \boldsymbol{V}_1^{\mathrm{T}} & \boldsymbol{U}_2 \boldsymbol{D} \boldsymbol{V}_2^{\mathrm{T}} \end{pmatrix} + \begin{pmatrix} \boldsymbol{E}_{11} & \boldsymbol{E}_{12} \\ \boldsymbol{E}_{21} & \boldsymbol{E}_{22} \end{pmatrix}.
\end{aligned}$$

Thus, all four blocks have the same low-rank structure. If the noise is small then $\boldsymbol{X}_{22} \approx \sqrt{n}\, \boldsymbol{U}_2 \boldsymbol{D} \boldsymbol{V}_2^{\mathrm{T}}$ and

$$\hat{\boldsymbol{X}}_{22} \approx \sqrt{n}\, \boldsymbol{U}_2 \boldsymbol{D} \big( \boldsymbol{V}_1^{\mathrm{T}} \hat{\boldsymbol{V}}_1(k) \big)\, \hat{\boldsymbol{D}}(k)^{+} \left( \hat{\boldsymbol{U}}_1(k)^{\mathrm{T}} \boldsymbol{U}_1 \right) \boldsymbol{D} \boldsymbol{V}_2^{\mathrm{T}}$$

If the noise is exactly zero, $k = k_0$, and $\operatorname{rank}(\boldsymbol{X}_{22}) = \operatorname{rank}(\boldsymbol{X})$, then Owen and Perry [65] show that $\hat{\boldsymbol{X}}_{22} = \boldsymbol{X}_{22}$. For other types of noise, the next chapter proves a more general consistency result.

### 5.2.4 Rotated cross-validation

When the underlying signal $\boldsymbol{U}\boldsymbol{D}\boldsymbol{V}^{\mathrm{T}}$ is sparse, it's possible that we will miss it in the training set. For example, if $\boldsymbol{U} = \begin{pmatrix} 1 & 0 & 0 & 0 \end{pmatrix}^{\mathrm{T}}$ and $\boldsymbol{V} = \begin{pmatrix} 1 & 0 & 0 & 0 & 0 \end{pmatrix}^{\mathrm{T}}$, then

$$\boldsymbol{U}\boldsymbol{D}\boldsymbol{V}^{\mathrm{T}} = \begin{pmatrix} d_1 & 0 & 0 & 0 & 0 \\ 0 & 0 & 0 & 0 & 0 \\ 0 & 0 & 0 & 0 & 0 \\ 0 & 0 & 0 & 0 & 0 \end{pmatrix}.$$

Either the test set will observe this factor, or the train set, but not both. Only the set that contains $X_{11}$ will be affected by the factor.

One way to adjust for sparse factors is to randomly rotate the rows and columns of $\boldsymbol{X}$ before performing the cross-validation. We generate $\boldsymbol{P} \in \mathbb{R}^{n \times n}$ and $\boldsymbol{Q} \in \mathbb{R}^{p \times p}$, uniformly random orthogonal matrices, by employing Algorithm A.1. Then, we set $\tilde{\boldsymbol{X}} = \boldsymbol{P}\boldsymbol{X}\boldsymbol{Q}^{\mathrm{T}}$ and perform ordinary cross-validation on $\tilde{\boldsymbol{X}}$. Regardless of the factor structure in $\boldsymbol{X}$, the signal part of $\tilde{\boldsymbol{X}}$ will be uniformly spread across all elements of the matrix. We call this procedure *rotated cross-validation*, abbreviated RCV.

RCV is similar in spirit to the generalized cross-validation (GCV) described in Craven & Wahba [20] and Golub et al. [35]. GCV is an orthogonally-invariant way of performing leave-one-out cross validation for regression. The difference is that GCV rotates to a specific non-random configuration of the data that gives equal weight to all observations in the rotated space, while RCV rotates to a random configuration that gives equal weight in expectation.

## 5.3 Missing-value SVDs

To perform a Wold-style cross-validation we need to be able to compute the SVD of a matrix with missing entries. This is a difficult problem. First of all, the problem is not very well defined. If $\boldsymbol{A}$ is a matrix with missing entries, there are potentially many different ways of factoring $\boldsymbol{A}$ as an SVD-like product. Often, one attempts to force uniqueness by finding the complete matrix $\boldsymbol{A}' \in \mathbb{R}^{n,p}$ of minimum rank such that $A'_{ij} = A_{ij}$ for all non-missing elements of $\boldsymbol{A}$. Even then, $\boldsymbol{A}'$ may not be unique. Take

$$\boldsymbol{A} = \begin{pmatrix} 1 & * \\ * & * \end{pmatrix}.$$

Then

$$\begin{pmatrix} 1 & 0 \\ 0 & 0 \end{pmatrix}, \quad \begin{pmatrix} 1 & 1 \\ 0 & 0 \end{pmatrix}, \quad \begin{pmatrix} 1 & 0 \\ 1 & 0 \end{pmatrix}, \quad \text{and} \quad \begin{pmatrix} 1 & 1 \\ 1 & 1 \end{pmatrix}$$

are all rank-1 matrices that agree with $\boldsymbol{A}$ on its non-missing entries. We might discriminate between these by picking the matrix with minimum Frobenius norm. In this case, the matrix

$$\begin{pmatrix} 1 & * \\ * & 1 \end{pmatrix}$$

presents an interesting problem. We can either complete it as

$$\begin{pmatrix} 1 & 1 \\ 1 & 1 \end{pmatrix} \quad \text{or} \quad \begin{pmatrix} 1 & 0 \\ 0 & 1 \end{pmatrix}.$$

The first option has rank 1 and Frobenius norm 2. The second option has higher rank, 2, but lower Frobenius norm, $\sqrt{2}$. Which criterion is more important? It it not clear what the "right" SVD of $\boldsymbol{A}$ is.

We can alleviate the problem by considering a sequence of SVDs rather than a single one. For each $k = 0, 1, 2, \ldots$, define $\boldsymbol{A}'_k$ to the rank-$k$ matrix of minimum Frobenius norm that agrees with $\boldsymbol{A}$ on its non-missing elements. If no such matrix exists, let $I \subset \mathcal{I}$ be indices of the non-missing elements of $\boldsymbol{A}$ and define the candidate sets

$$\begin{aligned} \mathcal{A}_k &= \{\boldsymbol{A}_k \in \mathbb{R}^{n \times p} : \operatorname{rank}(\boldsymbol{A}_k) = k\} \\ \mathcal{C}_k &= \{\boldsymbol{A}_k \in \mathcal{A}_k : \|\boldsymbol{A} - \boldsymbol{A}_k\|_{\mathrm{F},I} = \min_{\boldsymbol{B}_k \in \mathcal{A}_k} \|\boldsymbol{A} - \boldsymbol{B}_k\|_{\mathrm{F},I}\}. \end{aligned}$$

Lastly, put

$$\boldsymbol{A}'_k = \operatorname*{argmin}_{\boldsymbol{A}_k \in \mathcal{C}_k} \|\boldsymbol{A}_k\|_{\mathrm{F},I}.$$

We then define the rank-$k$ SVD of $\boldsymbol{A}$ to be the equal to the SVD of $\boldsymbol{A}'_k$.

There are still some problems with these SVDs. First of all, in general there may be no relationship between $\boldsymbol{A}'_k$ and $\boldsymbol{A}'_{k+1}$; the two matrices may be completely different and have completely different SVDs. We lose the nesting property of ordinary SVDs, where the rank-$k$ SVD is contained in the rank-$(k+1)$ SVD. Secondly, although it seems plausible, we do not have any guarantees that $\boldsymbol{A}'_k$ is unique. Situations may arise where two different rank-$k$ matrices have the same norms and the same residual norms on the non-missing elements of $\boldsymbol{A}$. Finally, finding $\boldsymbol{A}'_k$ is a non-convex problem. The function $\|\boldsymbol{A} - \boldsymbol{A}_k\|_{\mathrm{F},I}$ can have more than one local maximum. We need to be aware of these deficiencies.

Rather than get too deep into the missing-value SVD rabbit-hole, we choose instead to live with an approximation. We acknowledge that computing the best rank-$k$ approximation as defined above is computationally infeasible for large $n$ and $p$. Instead of proving theorems about the global optimum, we focus on an algorithm for computing a local solution.

We use an EM-algorithm to estimate $\boldsymbol{A}'_k$. The inputs to the algorithm are $k$, a non-negative integer rank, and $\boldsymbol{A} \in \mathbb{R}_*^{n \times p}$, a matrix with missing values. The output is $\boldsymbol{A}'_k$, a rank-$k$ approximation of $\boldsymbol{A}$. The algorithm proceeds by iteratively estimating the missing values of $\boldsymbol{A}$ by the values from the first $k$ terms of the SVD of the completed matrix. We give detailed pseudocode for the procedure as Algorithm 5.1. This is essentially the same algorithm as the SVDimpute algorithm given in Troyanksaya et al. [91], except that we use a different convergence criterion. SVDimpute stops when successive estimates of the missing values of $\boldsymbol{A}$ differ by less than "the empirically determined threshold of 0.01." We instead stop when relative difference of the residual sum of squares (RSS) between the non-missing entries and the rank-$k$ SVD is small (usually $1.0 \times 10^{-4}$ or less). Our reason for using a different convergence rule is that the analysis of the EM algorithm in Dempster et al. [22] shows that the RSS decreases with each iteration, but makes no assurances about the missing values converging. Regardless of which convergence criterion is used, the algorithm is very easy to implement.

**Algorithm 5.1** Rank-$k$ SVD approximation with missing values

1. Let $I = \{(i, j) : A_{ij} \neq *\}$.

2. For $1 \leq j \leq p$ let $\mu_j$ be the mean of the non-missing values in column $j$ of $\boldsymbol{A}$, or 0 if all of the entries in column $j$ are missing.

3. Define $\boldsymbol{A}^{(0)} \in \mathbb{R}^{n \times p}$ by

$$A_{ij}^{(0)} = \begin{cases} A_{ij} & \text{if } (i, j) \in I, \\ \mu_j & \text{otherwise.} \end{cases}$$

4. Initialize the iteration count $N \leftarrow 0$.

5. (M-Step) Let

$$\boldsymbol{A}^{(N)} = \sum_{i=1}^{n \wedge p} d_i^{(N)} \underline{u}_i^{(N)} \underline{v}_i^{(N)\mathrm{T}}$$

   be the SVD of $\boldsymbol{A}^{(N)}$ and let $\boldsymbol{A}_k'^{(N)}$ be the SVD truncated to $k$ terms, so that

$$\boldsymbol{A}_k'^{(N)} = \sum_{i=1}^{k} d_i^{(N)} \underline{u}_i^{(N)} \underline{v}_i^{(N)\mathrm{T}}.$$

6. (E-Step) Define $\boldsymbol{A}^{(N+1)} \in \mathbb{R}^{n \times p}$ as

$$A_{ij}^{(N+1)} = \begin{cases} A_{ij} & \text{if } (i, j) \in I, \\ A_{k,ij}'^{(N)} & \text{otherwise.} \end{cases}$$

7. Set

$$\mathrm{RSS}^{(N)} = \| \boldsymbol{A} - \boldsymbol{A}_k'^{(N)} \|_{\mathrm{F},I}^2.$$

   If $|\mathrm{RSS}^{(N)} - \mathrm{RSS}^{(N-1)}|$ is small, declare convergence and output $\boldsymbol{A}_k'^{(N)}$ as $\boldsymbol{A}_k'$. Otherwise, increment $N \leftarrow N + 1$ and go to Step 5.

## 5.4 Simulations

We performed two sets of simulations to gauge the performance of Gabriel- and Wold-style cross-validation. In the first set of simulations, we compare estimated prediction error with true prediction error. In the second set, we evaluate the methods' abilities to estimate $k^*_{\mathrm{PE}}$, the optimal rank.

Each simulation generates a data matrix $\boldsymbol{X}$ as $\boldsymbol{X} = \sqrt{n}\,\boldsymbol{U}\boldsymbol{D}\boldsymbol{V}^{\mathrm{T}} + \boldsymbol{E}$. We fix the dimensions and number of generating signals as $n = 100$, $p = 50$, and $k_0 = 6$. For "weak" factors we set $\boldsymbol{D} = \boldsymbol{D}_{\mathrm{weak}} = \mathrm{diag}\,(10, 9, 8, 7, 6, 5)$ and for "strong" factors, we set $\boldsymbol{D} = \boldsymbol{D}_{\mathrm{strong}} = \sqrt{n}\boldsymbol{D}_{\mathrm{weak}}$. We generate $\boldsymbol{U}$, $\boldsymbol{V}$, and $\boldsymbol{E}$ independently of each other. To avoid ambiguity in what constitutes "signal" and what constitutes "noise", we ensure that the elements of $\boldsymbol{E}$ are uncorrelated with each other.

We consider two types of factors. For "Gaussian" factors, we put the elements of $\boldsymbol{U}$ distributed iid with $U_{11} \sim \mathcal{N}\big(0,\, \frac{1}{n}\big)$ and the elements of $\boldsymbol{V}$ iid with $V_{11} \sim \mathcal{N}\big(0,\, \frac{1}{p}\big)$, also independent of $\boldsymbol{U}$. For "Sparse" factors we use sparsity parameter $s = 10\%$ and set

$$\begin{aligned} \mathbb{P}\{U_{11} = 0\} &= 1 - s, \\ \mathbb{P}\Big\{U_{11} = -\frac{1}{\sqrt{s\,n}}\Big\} = \mathbb{P}\Big\{U_{11} = +\frac{1}{\sqrt{s\,n}}\Big\} &= \frac{s}{2}. \end{aligned}$$

Similarly, we put

$$\begin{aligned} \mathbb{P}\{V_{11} = 0\} &= 1 - s, \\ \mathbb{P}\Big\{V_{11} = -\frac{1}{\sqrt{s\,p}}\Big\} = \mathbb{P}\Big\{V_{11} = +\frac{1}{\sqrt{s\,p}}\Big\} &= \frac{s}{2}. \end{aligned}$$

The scalings in both cases are chosen so that $\mathbb{E}[\boldsymbol{U}^{\mathrm{T}}\boldsymbol{U}] = \mathbb{E}[\boldsymbol{V}^{\mathrm{T}}\boldsymbol{V}] = \boldsymbol{I}_{k_0}$. Gaussian factors are uniformly spread out in the observations and variables, while sparse factors are only observable in a small percentage of the matrix entries (about 1%).

We use three types of noise. For "white" noise, we generate the elements of $\boldsymbol{E}$ iid with $E_{11} \sim \mathcal{N}\,(0,\, 1)$. For "heavy" noise, we use iid elements with $E_{11} \sim \sigma_\nu^{-1}\, t_\nu$, and $\nu = 3$. Here $t_\nu$ is a $t$ random variable with $\nu$ degrees of freedom, and $\sigma_\nu = \sqrt{\nu/(\nu - 2)}$ is chosen so that $\mathbb{E}[E_{11}^2] = 1$. Heavy noise is so-called because it has a heavy tail. Lastly, for "colored" noise, we first generate $\sigma_1^2, \ldots, \sigma_n^2 \sim \text{Inverse-}\chi^2(\nu_1)$ and $\tau_1^2, \ldots, \tau_p^2 \sim \text{Inverse-}\chi^2(\nu_2)$ independently, with $\nu_1 = \nu_2 = 3$. Then we generate the elements of $\boldsymbol{E}$ independently as $E_{ij} \sim c_{\nu_1,\nu_2}^{-1} \cdot \mathcal{N}(0, \sigma_i^2 + \tau_j^2)$, where $c_{\nu_1,\nu_2} = \sqrt{1/(\nu_1 - 2) + 1/(\nu_2 - 2)}$. Again, $c_{\nu_1,\nu_2}$ is chosen so that $\mathbb{E}[E_{ij}^2] = 1$. Colored noise simulates heteroscedasticity. All three types of noise are plausible for real-world data.

Obviously, this set of simulations comes with a number of caveats. We are only looking at two choices for the signal strengths, one choice of $n$ and $p$, and a single hold-out size. Moreover, this example uses relatively small $n$ and $p$, potentially too small for consistency asymptotics to kick in. Despite these deficiencies, the simulations still convey substantial information about the behavior of the procedures under consideration.

### 5.4.1 Prediction error estimation

Our goal with the first simulation was to get intuition for the behavior of Gabriel- and Wold-style cross validation as prediction error estimators. We generated random data of the form $\boldsymbol{X} = \sqrt{n}\boldsymbol{U}\boldsymbol{D}\boldsymbol{V}^{\mathrm{T}} + \boldsymbol{E}$, described above. With $\hat{\boldsymbol{X}}(k)$ being the SVD of $\boldsymbol{X}$ truncated to $k$ terms, the (normalized) true prediction error is given by

$$\mathrm{PE}(k) = \|\sqrt{n}\boldsymbol{U}\boldsymbol{D}\boldsymbol{V}^{\mathrm{T}} - \hat{\boldsymbol{X}}(k)\|_{\mathrm{F}}^2 + 1.$$

Cross-validation gives us an estimate $\widehat{\mathrm{PE}}(k)$ of the prediction error curve. We wanted to see how $\widehat{\mathrm{PE}}(k)$ compares to $\mathrm{PE}(k)$.

Figures 5.1 and 5.2 show the true and estimated prediction error curves from $(2, 2)$-fold Gabriel CV and 5-fold Wold CV, along with their RCV variants. The plots only show one set of curves for each factor and noise instance, but other replicates showed similar behavior.

For most of the simulations, the cross-validation estimates of prediction error are generally conservative. The only exception is with weak factors and colored noise, perhaps because of the ambiguity in what constitutes "signal" and what constitutes "noise" in this simulation. This global upward bias agrees with previous studies of cross-validation (e.g. [15] and [17]). The RCV versions of the methods generally brought down the bias. Some authors have observed a downward bias at the minimizer of $\widehat{\mathrm{PE}}(k)$ for ordinary cross-validation ([16], [88]). This bias does not appear to be present here.

A striking difference between Wold- and Gabriel-style CV is their behavior for $k$ greater than $k^*_{\mathrm{PE}}$. Gabriel-style CV does a better job at estimating the true $\mathrm{PE}(k)$, which is relatively flat. Wold-style CV, on the other hand, increases steeply for $k$ past the minimizer. In some situations, the Wold-style behavior is more desirable, but as the heavy-noise examples in Figure 5.2 illustrate, the steep increase is not always in the right place. Gabriel-style cross-validation is better at conveying ambiguity when the underlying dimensionality of the data is unclear.

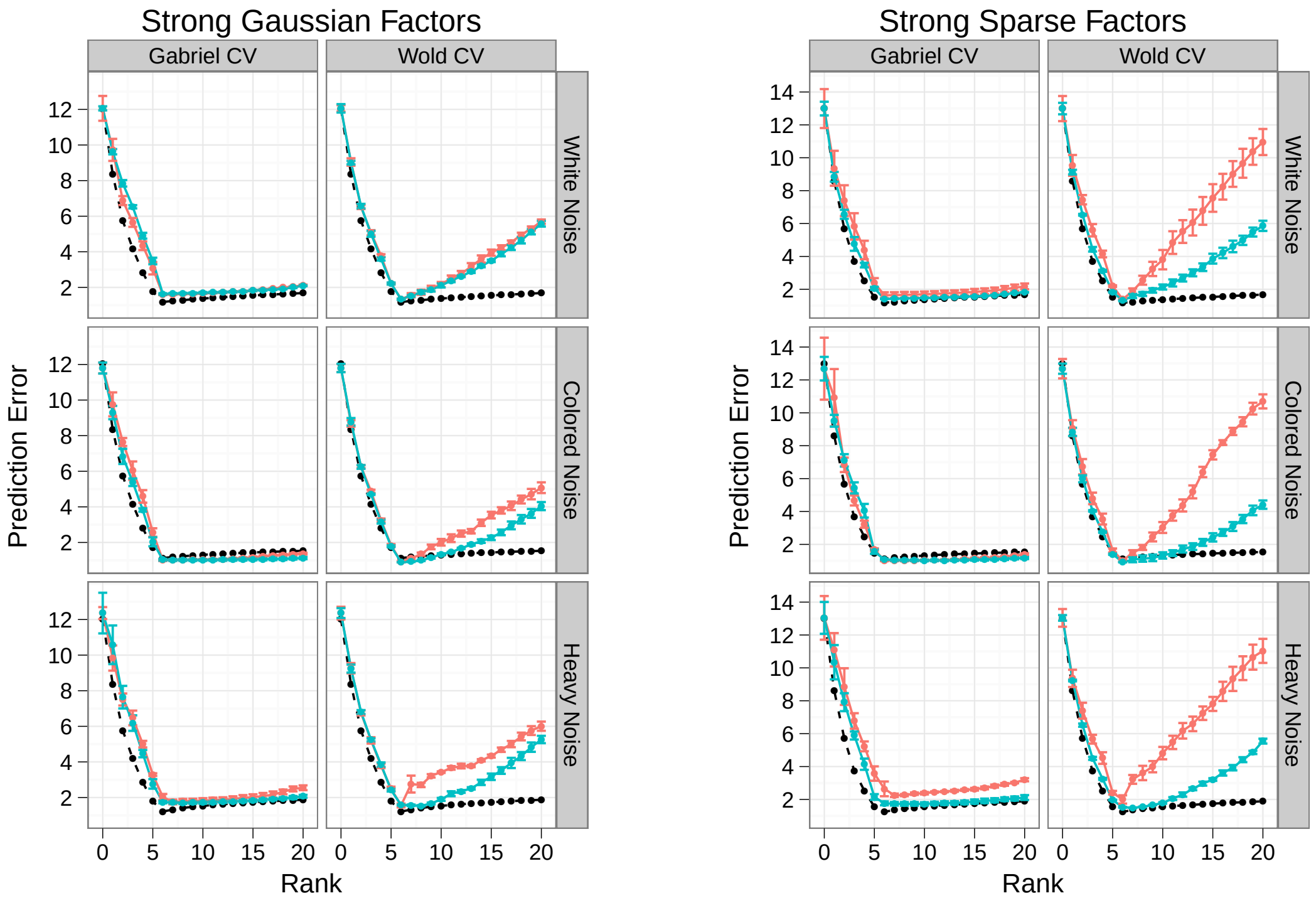

Figure 5.1: Cross-validation with strong factors. Estimated prediction error curves for Gabriel- and Wold-style cross-validation, both original and rotated (RCV) versions, with strong factors in the data. The true prediction error is shown in black, the CV curves are red, and the RCV curves are blue. Error bars are computed from the CV replicates. Despite their upward bias, the methods do well at estimating $\mathrm{PE}(k)$ for $k \leq k^*_{\mathrm{PE}}$.

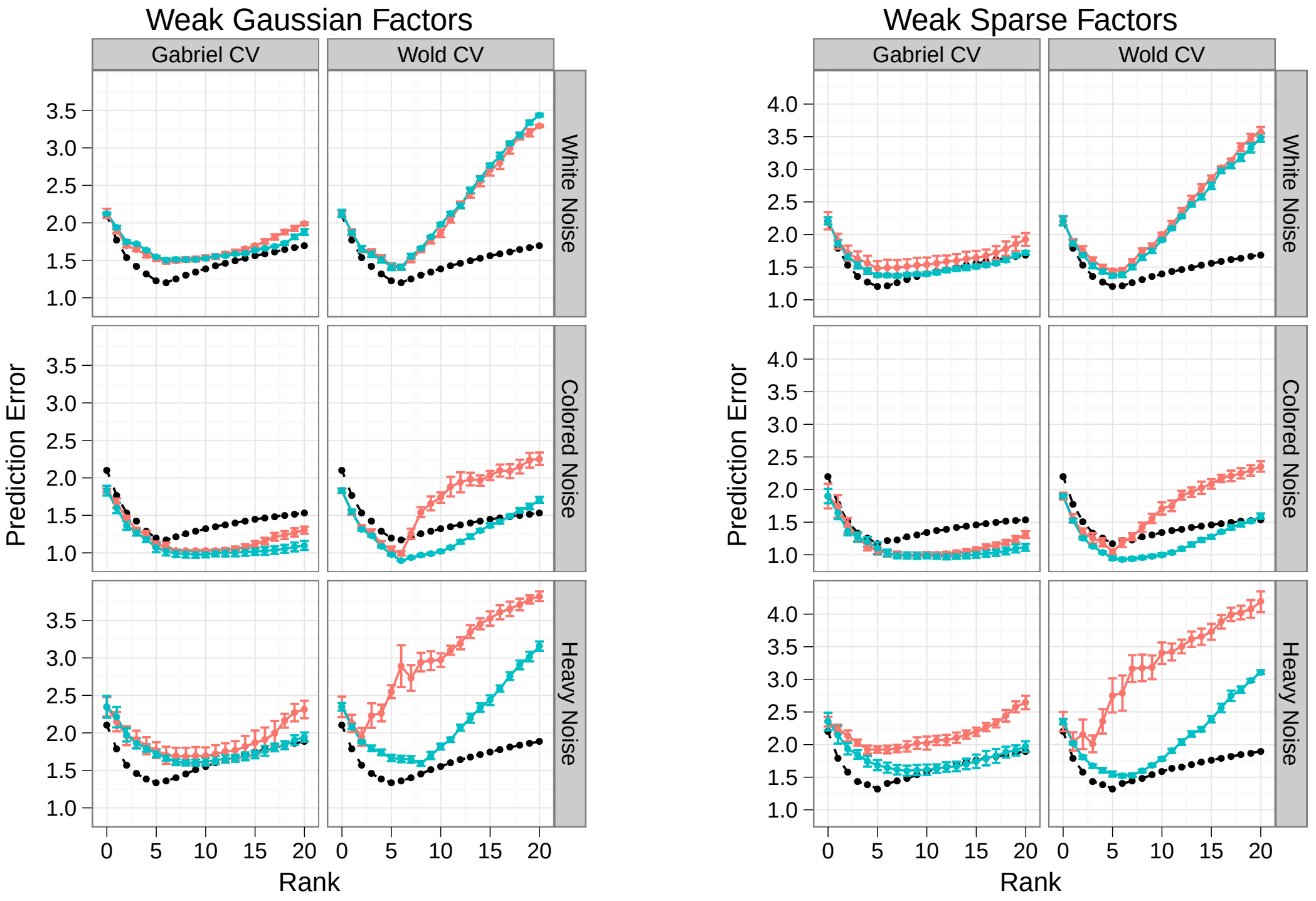

Figure 5.2: Cross-validation with weak factors. Estimated prediction error curves for Gabriel- and Wold-style cross-validation, both original and rotated (RCV) versions, with weak factors in the data. As in Figure 5.1 the true prediction error is shown in black, the CV curves are red, and the RCV curves are blue. Error bars give are computed from the CV replicates. The methods have a harder time estimating PE($k$) and its minimizer.

### 5.4.2 Rank estimation

In the next simulation, we see how well cross-validation works at estimating the optimal rank, especially as compared to other rank-selection methods. We generate data in the same manner as before, and record how far off the minimizer of $\widehat{\mathrm{PE}}(k)$ is from the minimizer of $\mathrm{PE}(k)$.

We compared the four CV-based rank estimation methods with seven other methods. They are as follows:

- AIC: Rao & Edelman's AIC-based estimator [71].
- $\mathrm{BIC}_1$, $\mathrm{BIC}_2$, and $\mathrm{BIC}_3$: Bai & Ng's BIC-based estimators [2].
- $F$: Faber & Kowalski's modification of Malinowski's $F$-test, with a significance level of 0.05 [31, 54].
- MDL: Wax & Kailaith's estimator based on the minimum description length principle [93] .
- UIP: Kritchman & Nadler's estimator based on Roy's union-intersection principle and their background noise estimator, with a significance level of 0.001 [50, 72].

Kritchman and Nadler [50] give concise descriptions of four of the estimators. The other estimators, Bai and Ng's BICs, are defined as the minimizers of

$$\mathrm{BIC}_1(k) = \log \|\boldsymbol{X} - \hat{\boldsymbol{X}}(k)\|_\mathrm{F}^2 + k\,\frac{n+p}{n\,p}\log\frac{n\,p}{n+p}, \tag{5.14a}$$

$$\mathrm{BIC}_2(k) = \log \|\boldsymbol{X} - \hat{\boldsymbol{X}}(k)\|_\mathrm{F}^2 + k\,\frac{n+p}{n\,p}\log C_{n,p}, \tag{5.14b}$$

$$\mathrm{BIC}_3(k) = \log \|\boldsymbol{X} - \hat{\boldsymbol{X}}(k)\|_\mathrm{F}^2 + k\,\frac{\log C_{n,p}^2}{C_{n,p}}, \tag{5.14c}$$

where $C_{n,p} = \min(\sqrt{n}, \sqrt{p})$.

Tables 5.1–5.4 summarize the results of 100 replicates. For the strong factors in white noise, almost all of the methods correctly estimate the true PE-minimizing rank. When the noise is non-white, Wold-style CV seems to be the clear winner.

| | Estimated Rank | | | | | | | | | | | | | | | |
|---|---|---|---|---|---|---|---|---|---|---|---|---|---|---|---|---|
| Method | −7 | −6 | −5 | −4 | −3 | −2 | −1 | 0 | +1 | +2 | +3 | +4 | +5 | +6 | +7 | > 7 |
| *White Noise* | | | | | | | | | | | | | | | | |
| CV-Gabriel | | | | | | | | 99 | 1 | | | | | | | |
| RCV-Gabriel | | | | | | | | 97 | 2 | 1 | | | | | | |
| CV-Wold | | | | | | | | 100 | | | | | | | | |
| RCV-Wold | | | | | | | | 100 | | | | | | | | |
| AIC | | | | | | | | 99 | 1 | | | | | | | |
| $BIC_1$ | | | | | | | | 100 | | | | | | | | |
| $BIC_2$ | | | | | | | | 100 | | | | | | | | |
| $BIC_3$ | | | | | | | | 100 | | | | | | | | |
| $F$ | | | | | | | | 100 | | | | | | | | |
| MDL | | | | | | | | 100 | | | | | | | | |
| UIP | | | | | | | | 100 | | | | | | | | |
| *Colored Noise* | | | | | | | | | | | | | | | | |
| CV-Gabriel | | | | | | | | 32 | 42 | 18 | 5 | 2 | 1 | | | |
| RCV-Gabriel | | | | | | | | 6 | 7 | 18 | 13 | 15 | 26 | 9 | 4 | 2 |
| CV-Wold | | | | | | | | 97 | 3 | | | | | | | |
| RCV-Wold | | | | | | | | 22 | 39 | 29 | 7 | 2 | 1 | | | |
| AIC | | | | | | | | | | | | | 2 | 9 | 19 | 70 |
| $BIC_1$ | | | | | | | | 14 | 26 | 31 | 20 | 4 | 4 | 1 | | |
| $BIC_2$ | | | | | | | | 23 | 41 | 24 | 9 | 1 | 2 | | | |
| $BIC_3$ | | | | | | | | | 1 | 1 | 3 | 4 | 5 | 3 | 4 | 79 |
| $F$ | | | | | | | | | 4 | 11 | 29 | 27 | 15 | 12 | 1 | 1 |
| MDL | | | | | | | | 13 | 24 | 36 | 20 | 3 | 3 | 1 | | |
| UIP | | | | | | | | | 1 | 5 | 14 | 28 | 23 | 19 | 9 | 1 |
| *Heavy Noise* | | | | | | | | | | | | | | | | |
| CV-Gabriel | | | | | | | | 61 | 35 | 4 | | | | | | |
| RCV-Gabriel | | | | | | | | 42 | 27 | 12 | 14 | 5 | | | | |
| CV-Wold | | | | | | | | 99 | 1 | | | | | | | |
| RCV-Wold | | | | | | | | 68 | 26 | 5 | 1 | | | | | |
| AIC | | | | | | | | 3 | 20 | 22 | 32 | 18 | 4 | 1 | | |
| $BIC_1$ | | | | | | | | 65 | 27 | 7 | 1 | | | | | |
| $BIC_2$ | | | | | | | | 70 | 26 | 3 | 1 | | | | | |
| $BIC_3$ | | | | | | | | 32 | 27 | 20 | 13 | 4 | 4 | | | |
| $F$ | | | | | | | | 38 | 29 | 27 | 6 | | | | | |
| MDL | | | | | | | | 64 | 28 | 7 | 1 | | | | | |
| UIP | | | | | | | | 18 | 35 | 27 | 18 | 2 | | | | |

Table 5.1: RANK ESTIMATION WITH STRONG GAUSSIAN FACTORS. Difference between the estimated rank and the true minimizer of $\mathrm{PE}(k)$ for 100 replicates of strong Gaussian factors with various types of noise.

However, as Figure 5.1 demonstrates, there is not much Frobenius loss penalty for slightly overestimating the rank. Therefore, Tables 5.1 and 5.2 may be exaggerating the advantage of Wold-style CV.

For the weak factors in white noise, the AIC, $F$ and UIP methods fare well, and the performance of the cross-validation-based methods is mediocre. For non-white noise and weak factors, none of the methods perform very well. This is probably due to the inherent ambiguity between what constitutes “signal” and what constitutes “noise” in these simulations.

| Method | Estimated Rank | | | | | | | | | | | | | | | |
|---|---|---|---|---|---|---|---|---|---|---|---|---|---|---|---|---|
| | −7 | −6 | −5 | −4 | −3 | −2 | −1 | 0 | +1 | +2 | +3 | +4 | +5 | +6 | +7 | > 7 |
| *White Noise* | | | | | | | | | | | | | | | | |
| CV-Gabriel | | | | | | | 8 | 73 | 11 | 7 | 1 | | | | | |
| RCV-Gabriel | | | | | | | | 98 | 2 | | | | | | | |
| CV-Wold | | | | | | 1 | 8 | 91 | | | | | | | | |
| RCV-Wold | | | | | | | 1 | 99 | | | | | | | | |
| AIC | | | | | | | | 100 | | | | | | | | |
| $BIC_1$ | | | | | | | 1 | 99 | | | | | | | | |
| $BIC_2$ | | | | | | | 1 | 99 | | | | | | | | |
| $BIC_3$ | | | | | | | | 100 | | | | | | | | |
| $F$ | | | | | | | | 100 | | | | | | | | |
| MDL | | | | | | | | 100 | | | | | | | | |
| UIP | | | | | | | | 100 | | | | | | | | |
| *Colored Noise* | | | | | | | | | | | | | | | | |
| CV-Gabriel | | | | | | | 4 | 34 | 19 | 24 | 8 | 6 | 5 | | | |
| RCV-Gabriel | | | | | | | | 1 | 5 | 13 | 13 | 22 | 20 | 10 | 8 | 8 |
| CV-Wold | | | | | 1 | 1 | 8 | 83 | 4 | 3 | | | | | | |
| RCV-Wold | | | | | | | | 26 | 32 | 25 | 12 | 2 | 2 | | | 1 |
| AIC | | | | | | | | | | | | | 2 | 10 | 19 | 69 |
| $BIC_1$ | | | | | | | | 17 | 23 | 29 | 20 | 3 | 4 | 2 | | 2 |
| $BIC_2$ | | | | | | | | 24 | 36 | 27 | 9 | 2 | 1 | 1 | | |
| $BIC_3$ | | | | | | | | | | 3 | 2 | 2 | 4 | 7 | 2 | 80 |
| $F$ | | | | | | | | | 4 | 9 | 32 | 28 | 14 | 11 | | 2 |
| MDL | | | | | | | | 16 | 24 | 31 | 19 | 6 | 2 | 1 | | 1 |
| UIP | | | | | | | | | | 4 | 17 | 27 | 23 | 18 | 9 | 2 |
| *Heavy Noise* | | | | | | | | | | | | | | | | |
| CV-Gabriel | | | | | | | 5 | 52 | 29 | 11 | 2 | 1 | | | | |
| RCV-Gabriel | | | | | | | | 39 | 24 | 19 | 11 | 6 | 1 | | | |
| CV-Wold | | | | | 1 | 1 | 11 | 83 | 4 | | | | | | | |
| RCV-Wold | | | | | | | | 66 | 28 | 5 | 1 | | | | | |
| AIC | | | | | | | | 2 | 21 | 24 | 33 | 11 | 6 | 3 | | |
| $BIC_1$ | | | | | | | 1 | 63 | 27 | 6 | 3 | | | | | |
| $BIC_2$ | | | | | | | 1 | 71 | 22 | 5 | 1 | | | | | |
| $BIC_3$ | | | | | | | | 28 | 31 | 19 | 14 | 7 | 1 | | | |
| $F$ | | | | | | | | 32 | 41 | 14 | 12 | 1 | | | | |
| MDL | | | | | | | | 63 | 28 | 6 | 3 | | | | | |
| UIP | | | | | | | | 20 | 32 | 27 | 15 | 6 | | | | |

Table 5.2: Rank estimation with strong sparse factors. Difference between the estimated rank and the true minimizer of $\mathrm{PE}(k)$ for 100 replicates of strong sparse factors with various types of noise.

| | Estimated Rank | | | | | | | | | | | | | | | |
|---|---|---|---|---|---|---|---|---|---|---|---|---|---|---|---|---|
| Method | −7 | −6 | −5 | −4 | −3 | −2 | −1 | 0 | +1 | +2 | +3 | +4 | +5 | +6 | +7 | > 7 |
| *White Noise* | | | | | | | | | | | | | | | | |
| CV-Gabriel | | | | | | | 3 | 56 | 32 | 9 | | | | | | |
| RCV-Gabriel | | | | | | | 5 | 66 | 20 | 8 | 1 | | | | | |
| CV-Wold | | | | | | 4 | 31 | 65 | | | | | | | | |
| RCV-Wold | | | | | | 3 | 32 | 65 | | | | | | | | |
| AIC | | | | | | | 3 | 95 | 2 | | | | | | | |
| $BIC_1$ | | | | | 1 | 13 | 46 | 40 | | | | | | | | |
| $BIC_2$ | | | | 1 | 8 | 29 | 47 | 15 | | | | | | | | |
| $BIC_3$ | | | | | | | | 99 | 1 | | | | | | | |
| $F$ | | | | | | | | 99 | 1 | | | | | | | |
| MDL | | | | | | 6 | 44 | 50 | | | | | | | | |
| UIP | | | | | | | | 99 | 1 | | | | | | | |
| *Colored Noise* | | | | | | | | | | | | | | | | |
| CV-Gabriel | | | | | | | 2 | 9 | 22 | 21 | 25 | 10 | 3 | 5 | 2 | 1 |
| RCV-Gabriel | | | | | | | | 2 | 4 | 11 | 16 | 14 | 12 | 19 | 9 | 13 |
| CV-Wold | 1 | 4 | 4 | 7 | 9 | 17 | 14 | 34 | 4 | | | | 2 | | 1 | 1 |
| RCV-Wold | | | | | | | 2 | 29 | 24 | 17 | 8 | 6 | 7 | 2 | 4 | 1 |
| AIC | | | | | | | | | | | | | 4 | 14 | 14 | 68 |
| $BIC_1$ | | | | | | | 2 | 26 | 22 | 20 | 9 | 8 | 3 | 2 | 4 | 4 |
| $BIC_2$ | | | | | | | 4 | 39 | 28 | 12 | 5 | 4 | 1 | 3 | 3 | 1 |
| $BIC_3$ | | | | | | | | | 1 | 1 | 5 | 11 | 1 | 3 | 4 | 74 |
| $F$ | | | | | | | | 1 | 7 | 16 | 24 | 17 | 13 | 10 | 1 | 11 |
| MDL | | | | | | | 1 | 25 | 23 | 21 | 8 | 9 | 4 | 1 | 4 | 4 |
| UIP | | | | | | | | | 2 | 11 | 12 | 22 | 19 | 15 | 4 | 15 |
| *Heavy Noise* | | | | | | | | | | | | | | | | |
| CV-Gabriel | | | | 1 | | 1 | 2 | 10 | 51 | 25 | 8 | 1 | 1 | | | |
| RCV-Gabriel | | | | | | | | 13 | 37 | 24 | 13 | 9 | 2 | 1 | | 1 |
| CV-Wold | 1 | 4 | 6 | 4 | 13 | 17 | 21 | 32 | 1 | | | | | | 1 | |
| RCV-Wold | | | | | 1 | 2 | 13 | 59 | 16 | 4 | 2 | 1 | 1 | | | 1 |
| AIC | | | | | | | | 7 | 15 | 28 | 25 | 14 | 7 | 1 | 2 | 1 |
| $BIC_1$ | | | | | 1 | 2 | 16 | 58 | 16 | 3 | 2 | | 1 | | | 1 |
| $BIC_2$ | | | | | 4 | 15 | 23 | 46 | 8 | 1 | 1 | 1 | | 1 | | |
| $BIC_3$ | | | | | | | | 38 | 21 | 23 | 8 | 4 | 4 | 1 | | 1 |
| $F$ | | | | | | | | 45 | 22 | 19 | 11 | 1 | 1 | | | 1 |
| MDL | | | | | | 2 | 14 | 61 | 15 | 3 | 3 | | 1 | | | 1 |
| UIP | | | | | | | | 27 | 29 | 27 | 9 | 4 | 3 | | | 1 |

Table 5.3: RANK ESTIMATION WITH WEAK GAUSSIAN FACTORS. Difference between the estimated rank and the true minimizer of $\mathrm{PE}(k)$ for 100 replicates of weak Gaussian factors with various types of noise.

| Method | Estimated Rank | | | | | | | | | | | | | | | |
|---|---|---|---|---|---|---|---|---|---|---|---|---|---|---|---|---|
| | −7 | −6 | −5 | −4 | −3 | −2 | −1 | 0 | +1 | +2 | +3 | +4 | +5 | +6 | +7 | > 7 |
| *White Noise* | | | | | | | | | | | | | | | | |
| CV-Gabriel | | | | | | 8 | 16 | 39 | 27 | 8 | 2 | | | | | |
| RCV-Gabriel | | | | | | 1 | 8 | 53 | 27 | 9 | 1 | 1 | | | | |
| CV-Wold | | | 2 | 2 | 11 | 16 | 29 | 40 | | | | | | | | |
| RCV-Wold | | | | | 2 | 15 | 29 | 54 | | | | | | | | |
| AIC | | | | | | | 6 | 77 | 16 | 1 | | | | | | |
| $BIC_1$ | | | | 1 | 4 | 21 | 45 | 29 | | | | | | | | |
| $BIC_2$ | | | 1 | 4 | 14 | 27 | 37 | 17 | | | | | | | | |
| $BIC_3$ | | | | | | 1 | 2 | 86 | 11 | | | | | | | |
| $F$ | | | | | | | | 85 | 15 | | | | | | | |
| MDL | | | | | 2 | 19 | 42 | 37 | | | | | | | | |
| UIP | | | | | | | | 73 | 26 | 1 | | | | | | |
| *Colored Noise* | | | | | | | | | | | | | | | | |
| CV-Gabriel | | | | | | 2 | 5 | 16 | 19 | 19 | 12 | 9 | 3 | 10 | 3 | 2 |
| RCV-Gabriel | | | | | | | | 1 | 2 | 7 | 12 | 15 | 19 | 11 | 8 | 25 |
| CV-Wold | 1 | 3 | 6 | 11 | 10 | 17 | 21 | 25 | 2 | 2 | | | 1 | | | 1 |
| RCV-Wold | | | | | | | 2 | 22 | 21 | 19 | 12 | 11 | 3 | 5 | 3 | 2 |
| AIC | | | | | | | | | | | | | 2 | 8 | 9 | 81 |
| $BIC_1$ | | | | | | | 1 | 22 | 23 | 20 | 10 | 9 | 4 | 2 | 3 | 6 |
| $BIC_2$ | | | | | 1 | | 7 | 37 | 17 | 13 | 9 | 7 | 1 | 3 | 4 | 1 |
| $BIC_3$ | | | | | | | | | | 2 | 2 | 2 | 4 | 8 | 7 | 75 |
| $F$ | | | | | | | | | 4 | 6 | 21 | 22 | 14 | 13 | 6 | 14 |
| MDL | | | | | | | | 22 | 19 | 22 | 12 | 10 | 5 | 2 | 3 | 5 |
| UIP | | | | | | | | | | 5 | 9 | 22 | 19 | 17 | 10 | 18 |
| *Heavy Noise* | | | | | | | | | | | | | | | | |
| CV-Gabriel | | | | | | 2 | 4 | 22 | 25 | 24 | 20 | 3 | | | | |
| RCV-Gabriel | | | | | | 1 | 4 | 9 | 18 | 20 | 18 | 12 | 13 | 2 | 1 | 2 |
| CV-Wold | 1 | 1 | 4 | 10 | 16 | 19 | 29 | 18 | | 1 | | | | | 1 | |
| RCV-Wold | | | | | | 2 | 19 | 51 | 7 | 16 | 3 | 1 | | | 1 | |
| AIC | | | | | | | | 2 | 17 | 16 | 19 | 22 | 8 | 4 | 10 | 2 |
| $BIC_1$ | | | | | 1 | 4 | 21 | 51 | 5 | 13 | 3 | 1 | | | 1 | |
| $BIC_2$ | | | | 2 | 4 | 16 | 32 | 31 | 4 | 9 | | 1 | | | 1 | |
| $BIC_3$ | | | | | | | 3 | 25 | 23 | 17 | 12 | 10 | 4 | 4 | 1 | 1 |
| $F$ | | | | | | | 2 | 29 | 21 | 23 | 10 | 7 | 5 | 2 | | 1 |
| MDL | | | | | | 3 | 19 | 52 | 6 | 15 | 3 | 1 | | | 1 | |
| UIP | | | | | | | | 20 | 23 | 22 | 13 | 12 | 3 | 5 | 1 | 1 |

Table 5.4: RANK ESTIMATION WITH WEAK SPARSE FACTORS. Difference between the estimated rank and the true minimizer of PE($k$) for 100 replicates of weak sparse factors with various types of noise.

## 5.5 Real data example

We conclude this chapter with a neuroscience application. The Neural Prosthetic Systems Laboratory at Stanford University (NPSL) is interested in studying the motor cortex region of the brain. Essentially, they want to know what the correspondence is between neural activity in that part of the brain and motor activity (movement). Research is still at a very fundamental level, and the basic question of how many things are being represented is still unanswered. It is thought that desired position, speed, and velocity get expressed as neural activity, but conjectures about the dimensionality of the neural responses vary from 7 to 20 or more.

NPSL has designed and carried out an experiment meant to measure the dimensionality of neural response for a two-dimensional motion task. The experiment involves measuring the activity in 49 neurons as a monkey performs 27 different movement tasks (conditions).

For a particular neuron and condition, a simplified explanation of the experiment is as follows:

1. At time $t = 0\,\text{ms}$, start recording neural activity in the monkey.

2. At time $t = 400\,\text{ms}$ (TARGET-ON), show the monkey a target. The monkey is not allowed to move at this point.

3. At a random time between time $t = 400\,\text{ms}$ and time $t = 1560\,\text{ms}$, allow the monkey to move.

4. At time $t = 1560\,\text{ms}$ (MOVEMENT), the monkey starts to move and point at the target.

5. Record activity up to but not including time $t = 2110\,\text{ms}$.

Each condition includes a target position and a configuration of obstacles. The same monkey is used for every trial. Measurements are taken at $5\,\text{ms}$ intervals, so that there are 422 total time points. It is necessary to do some registration, scaling, and interpolation before doing more serious data analysis, but the details of those processes are not important for our purposes. Figure 5.3 shows the preprocessed responses for each neuron and condition.

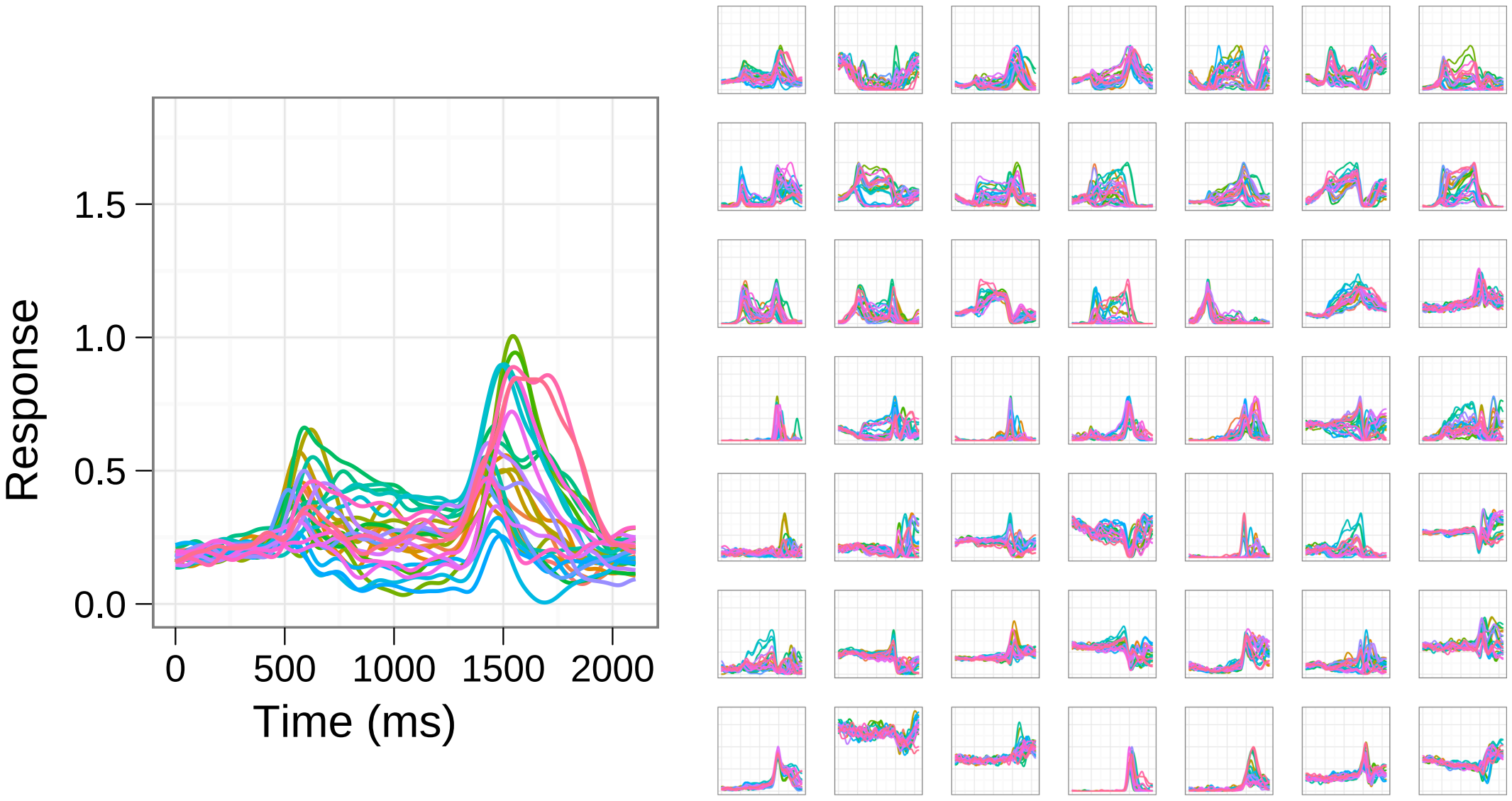

Figure 5.3: MOTOR CORTEX DATA. Response rates in 47 neurons for 27 movement tasks. The subplots show the normalized response rates in a single neuron as functions of time. Each color corresponds to a different movement task. The plot on the left is a zoomed-in view of the data for the first neuron.

The data from the NPSL motor cortex experiment can be put into a matrix where each neuron is thought of as a variable, and the timepoints of each condition are thought of as observations. This gives us a matrix $\boldsymbol{X}$ with $p = 49$ variables and $n = 27 \cdot 422 = 11394$ observations. Of course, the rows of $\boldsymbol{X}$ are nothing like iid, and the noise in $\boldsymbol{X}$ is not white. Parametric methods are not likely to give very reliable estimates of the dimensionality of $\boldsymbol{X}$, but cross-validation stands a reasonable chance.

After centering the columns of $\boldsymbol{X}$, we performed Wold- and Gabriel-style cross-validation to estimate the dimensionality of the signal part of $\boldsymbol{X}$. For Gabriel-style CV, we first tried both $(2, 2)$-fold and $(2, 49)$-fold; both resulting $\widehat{PE}(k)$ curves had their minima at the maximum $k$. For 5-fold Wold-style CV, there is a minimum at $k = 13$. The BIC, $F$, MDL, and UIP estimators all chose $k = 48$ as the dimensionality, while the AIC estimator chose $k = 47$. We show the cross-validation estimated prediction curves in Figure 5.4. It is likely that the true dimensionality of $\boldsymbol{X}$ is high.

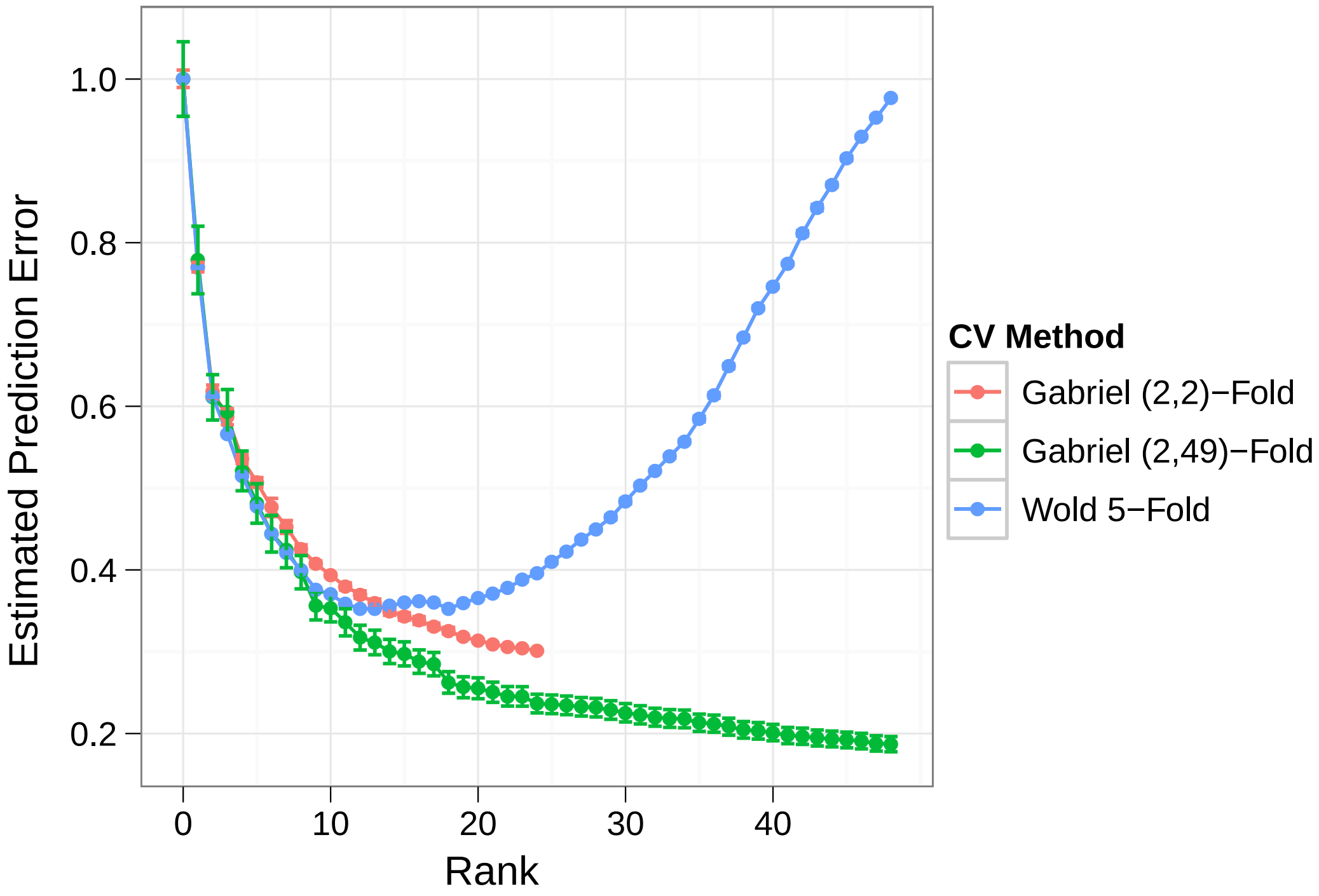

Figure 5.4: Motor cortex estimated prediction error. Prediction error as a function of rank, estimated by three cross-validation methods. The units are normalized so that the maximum prediction error is 1.0. Error bars show one standard error, estimated from the folds. The prediction error estimate from Wold-style CV shows a minimum at $k = 13$, but the Gabriel-style CV estimates always decrease with $k$. It is likely that the true dimensionality of the data is high.

## 5.6 Summary and future work

We have described two different forms of cross-validation appropriate for model selection in unsupervised learning. Wold-style CV uses a "speckled" leave-out, and Gabriel-style CV uses a "blocked" leave-out. We have defined two forms of error associated with SVD/PCA-like models, the prediction error and the model error. Through simulations, we have shown that both forms of CV can be considered to give estimates of prediction error. Both methods perform well, but Wold-style CV seems to be more robust to badly-behaved noise. We have applied these cross-validation methods to a data analysis problem from a neuroscience experiment.

We have focused on latent factor models and the singular value decomposition. However, it is relatively easy to translate the two styles of cross-validation presented here to other unsupervised learning methods. For Wold-style hold-outs, an EM-like algorithm can usually be applied to the models in many unsupervised learning contexts. The Gabriel-style philosophy of "treat some of the variables as response and the others as predictors" can also be applied more broadly. We are in the process of investigating both cross-validation strategies for clustering and kernel-based manifold learning.

This chapter leaves a number of open questions. Mainly, we have not provided any theoretical results here, only simulations. A quote from Downton's discussion of Stone's 1974 paper on cross-validation equally applies to our work:

> A current nine-day wonder in the press concerns the exploits of a Mr. Uri Geller who appears to be able to bend metal objects without touching them; [The author] seems to be attempting to bend statistics without touching them. My attitude to both of these phenomena is one of open-minded scepticism; I do not believe in either of these prestigious activities, on the other hand they both deserve serious scientific examination [86].

Despite Dowton's skepticism, cross-validation has proven to be an invaluable tool for supervised learning. It is our hope that with some additional work, CV can be just as valuable for unsupervised learning. In the next chapter, we provide some theoretical justification for Gabriel-style cross-validation, but analysis of Wold-style cross-validation is still an open problem.

# Chapter 6

# A theoretical analysis of bi-cross-validation

In this chapter we will determine the optimal leave out-size for Gabriel-style cross-validation of an SVD, also known as bi-cross-validation (BCV), along with proving a weak form of consistency. In Chapter 4, we rigorously defined the rank estimation problem, and in Chapter 5 we introduced Gabriel-style cross validation. Here, we provide theoretic justification for Gabriel-style CV.

First, a quick review of the problem. We are given $\boldsymbol{X}$, an $n \times p$ matrix generating by a "signal-plus-noise" process,

$$\boldsymbol{X} = \sqrt{n}\,\boldsymbol{U}\boldsymbol{D}\boldsymbol{V}^{\mathrm{T}} + \boldsymbol{E}.$$

Here, $\boldsymbol{U} \in \mathbb{R}^{n \times k_0}$, $\boldsymbol{D} \in \mathbb{R}^{k_0 \times k_0}$, $\boldsymbol{V} \in \mathbb{R}^{p \times k_0}$, and $\boldsymbol{E} \in \mathbb{R}^{n \times p}$. The first term, $\sqrt{n}\,\boldsymbol{U}\boldsymbol{D}\boldsymbol{V}^{\mathrm{T}}$, is the low-rank "signal" part. We call $\boldsymbol{U}$ and $\boldsymbol{V}$ the matrices of left and right factors, respectively. They are normalized so that $\boldsymbol{U}^{\mathrm{T}}\boldsymbol{U} = \boldsymbol{V}^{\mathrm{T}}\boldsymbol{V} = \boldsymbol{I}_{k_0}$. The factor "strengths" are given in $\boldsymbol{D}$, a diagonal matrix of the form $\boldsymbol{D} = \mathrm{diag}(d_1, d_2, \ldots, d_{k_0})$, with $d_1 \geq d_2 \geq \cdots \geq d_{k_0} \geq 0$. Also, typically $k_0$ is much smaller than $n$ and $p$. Lastly, $\boldsymbol{E}$ consists of "noise". Although more general types of noise are possible, for simplicity we will assume that $\boldsymbol{E}$ is independent of $\boldsymbol{U}$, $\boldsymbol{D}$, and $\boldsymbol{V}$. We think of the signal part as the important part of $\boldsymbol{X}$, and the noise part is inherently uninteresting.

The rank estimation problem is find the optimal number of terms of the SVD to keep to estimate the signal part. We let $\boldsymbol{X} = \sqrt{n}\hat{\boldsymbol{U}}\hat{\boldsymbol{D}}\hat{\boldsymbol{V}}^{\mathrm{T}}$ be the SVD of $\boldsymbol{X}$, where $\hat{\boldsymbol{U}} \in \mathbb{R}^{n \times n\wedge p}$ and $\hat{\boldsymbol{V}} \in \mathbb{R}^{p \times n\wedge p}$ have orthonormal columns, and $\hat{\boldsymbol{D}} = \operatorname{diag}(\hat{d}_1, \hat{d}_2, \ldots, \hat{d}_{n\wedge p})$ with $\hat{d}_1 \geq \hat{d}_2 \geq \cdots \geq \hat{d}_{n\wedge p}$. For $0 \leq k \leq n \wedge p$, we define $\hat{\boldsymbol{D}}(k) = \operatorname{diag}(\hat{d}_1, \hat{d}_2, \ldots, \hat{d}_k, 0, 0, \ldots, 0)$ so that $\hat{\boldsymbol{X}}(k) \equiv \sqrt{n}\hat{\boldsymbol{U}}\hat{\boldsymbol{D}}(k)\hat{\boldsymbol{V}}^{\mathrm{T}}$ is the SVD of $\boldsymbol{X}$ truncated to $k$ terms. The model error with respect to Frobenius loss is given by

$$\mathrm{ME}(k) = \frac{1}{n\,p}\|\sqrt{n}\boldsymbol{U}\boldsymbol{D}\boldsymbol{V}^{\mathrm{T}} - \hat{\boldsymbol{X}}(k)\|_{\mathrm{F}}^2$$

The optimal rank is defined with respect to this criterion is

$$k^* = \operatorname*{argmin}_k \mathrm{ME}(k).$$

The problem we consider is how to estimate $\mathrm{ME}(k)$ or $k^*$.

Closely related to model error is the *prediction error*. For prediction error, we conjure up a noise matrix $\boldsymbol{E}'$ with the same distribution as $\boldsymbol{E}$ and let $\boldsymbol{X}' = \sqrt{n}\boldsymbol{U}\boldsymbol{D}\boldsymbol{V}^{\mathrm{T}} + \boldsymbol{E}'$. The prediction error is defined as

$$\mathrm{PE}(k) \equiv \frac{1}{n\,p}\mathbb{E}\|\boldsymbol{X}' - \hat{\boldsymbol{X}}(k)\|_{\mathrm{F}}^2,$$

which can be expressed as

$$\mathrm{PE}(k) = \mathbb{E}[\mathrm{ME}(k)] + \frac{1}{n\,p}\mathbb{E}\|\boldsymbol{E}\|_{\mathrm{F}}^2.$$

The minimizer of PE is the same as the minimizer of $\mathbb{E}[\mathrm{ME}(k)]$, and one can get an estimate of ME from an estimate of PE by subtracting an estimate of the noise level.

The previous chapter suggests using Gabriel-style cross-validation for estimating the optimal rank. Owen & Perry [65] call this procedure bi-cross-validation (BCV). For fold $(i, j)$ of BCV, we permute the rows of $\boldsymbol{X}$ with matrices $\boldsymbol{P}^{(i)}$ and $\boldsymbol{Q}^{(j)}$, then partition the result into four blocks as

$$\boldsymbol{P}^{(i)\mathrm{T}} \boldsymbol{X} \boldsymbol{Q}^{(j)} = \begin{pmatrix} \boldsymbol{X}_{11} & \boldsymbol{X}_{12} \\ \boldsymbol{X}_{21} & \boldsymbol{X}_{22} \end{pmatrix}.$$

We take the SVD of the upper-left block and evaluate its predictive performance on the lower-right block. If $\boldsymbol{X}_{11} = \sqrt{n}\,\hat{\boldsymbol{U}}_1 \hat{\boldsymbol{D}}_1 \hat{\boldsymbol{V}}^{\mathrm{T}}$ is the SVD of $\boldsymbol{X}_{11}$ and $\hat{\boldsymbol{X}}_{11}(k) = \sqrt{n}\,\hat{\boldsymbol{U}}_1 \hat{\boldsymbol{D}}_1(k) \hat{\boldsymbol{V}}^{\mathrm{T}}$ is its truncation to $k$ terms (with $\hat{\boldsymbol{D}}_1(k)$ defined analogously to $\hat{\boldsymbol{D}}(k)$), then the BCV estimate of prediction error from this fold is given by

$$\widehat{\mathrm{PE}}\big(k; i, j\big) = \frac{1}{n_2\, p_2} \| \boldsymbol{X}_{22} - \boldsymbol{X}_{21} \hat{\boldsymbol{X}}_{11}(k)^{+} \boldsymbol{X}_{12} \|_{\mathrm{F}}^2.$$

Here, $^{+}$ denotes pseudo-inverse and $\boldsymbol{X}_{22}$ has dimensions $n_2 \times p_2$. For $(K, L)$-fold BCV, the final estimate is the average over all folds:

$$\widehat{\mathrm{PE}}(k) = \frac{1}{KL} \sum_{i=1}^{K} \sum_{j=1}^{L} \widehat{\mathrm{PE}}\big(k; i, j\big).$$

From $\widehat{\mathrm{PE}}(k)$ we can get an estimate of the optimal rank as $\hat{k} = \operatorname{argmin}_k \widehat{\mathrm{PE}}(k)$.

In this chapter, we give a theoretical analysis of $\widehat{\mathrm{PE}}(k)$. This allows us to determine the bias inherent in $\widehat{\mathrm{PE}}(k)$ and its consistency properties for estimating $k^*$, along with guidance for choosing the number of folds ($K$ and $L$). Section 6.1 sets out our assumptions and notation. Section 6.2 gives our main results. Then, Sections 6.3 and 6.4 are devoted to proofs, followed by a discussion in Section 6.5.

## 6.1 Assumptions and notation

The theory becomes easier if we work in an asymptotic framework. For that, we introduce a sequence of data matrices indexed by $n$:

$$\boldsymbol{X}_n = \sqrt{n}\, \boldsymbol{U}_n \boldsymbol{D}_n \boldsymbol{V}_n^{\mathrm{T}} + \boldsymbol{E}_n. \tag{6.1}$$

Here, $\boldsymbol{X}_n \in \mathbb{R}^{n \times p}$ with $p = p(n)$ and $\frac{n}{p} \to \gamma \in (0, \infty)$. Even though the dimensions of $\boldsymbol{X}_n$ grow, we assume that the number of factors is fixed at $k_0$. The first set of assumptions is as follows:

**Assumption 6.1.** *We have a sequence of random matrices $\boldsymbol{X}_n \in \mathbb{R}^{n \times p}$ with $n \to \infty$ and $p = p(n)$ also going to infinity. Their ratio converges to a fixed constant $\gamma \in (0, \infty)$ as $\frac{n}{p} = \gamma + o\left(\frac{1}{\sqrt{n}}\right)$.*

**Assumption 6.2.** *The matrix $\boldsymbol{X}_n$ is generated as $\boldsymbol{X}_n = \sqrt{n}\,\boldsymbol{U}_n \boldsymbol{D}_n \boldsymbol{V}_n^{\mathrm{T}} + \boldsymbol{E}_n$. Here, $\boldsymbol{U}_n \in \mathbb{R}^{n \times k_0}$, $\boldsymbol{D}_n \in \mathbb{R}^{k_0 \times k_0}$, $\boldsymbol{V}_n \in \mathbb{R}^{p \times k_0}$, and $\boldsymbol{E}_n \in \mathbb{R}^{n \times p}$. The number of factors, $k_0$, is fixed.*

**Assumption 6.3.** *The matrices of left and right factors, $\boldsymbol{U}_n$ and $\boldsymbol{V}_n$, have orthonormal columns, i.e. $\boldsymbol{U}_n^{\mathrm{T}} \boldsymbol{U}_n = \boldsymbol{V}_n^{\mathrm{T}} \boldsymbol{V}_n = \boldsymbol{I}_{k_0}$. Their columns are denoted by $\underline{u}_{n,1}, \underline{u}_{n,2}, \ldots, \underline{u}_{n,k_0}$ and $\underline{v}_{n,1}, \underline{u}_{n,2}, \ldots, \underline{v}_{n,k_0}$, respectively.*

**Assumption 6.4.** *The matrix of factor strengths is diagonal:*

$$\boldsymbol{D}_n = \operatorname{diag}(d_{n,1}, d_{n,2}, \ldots, d_{n,k_0}). \tag{6.2}$$

*The strengths converge as $d_{n,i}^2 \xrightarrow{a.s.} \mu_i$ and $d_{n,i}^2 - \mu_i = \mathcal{O}_P\left(\frac{1}{\sqrt{n}}\right)$, strictly ordered as $\mu_1 > \mu_2 > \cdots > \mu_{k_0} > 0$.*

**Assumption 6.5.** *The noise matrix $\boldsymbol{E}_n$ is independent of $\boldsymbol{U}_n$, $\boldsymbol{D}_n$, and $\boldsymbol{V}_n$. Its elements are iid with $E_{n,11} \sim \mathcal{N}(0,\, \sigma^2)$.*

These assumptions are standard for latent factor models.

We can apply the work of Chapter 4 to get the behavior of the model error. We let $\boldsymbol{X}_n = \sqrt{n}\,\hat{\boldsymbol{U}}_n \hat{\boldsymbol{D}}_n \hat{\boldsymbol{V}}_n^{\mathrm{T}}$ be the SVD of $\boldsymbol{X}_n$ and let $\hat{\boldsymbol{X}}_n(k) = \sqrt{n}\,\hat{\boldsymbol{U}}_n \hat{\boldsymbol{D}}_n(k) \hat{\boldsymbol{V}}_n^{\mathrm{T}}$ be its truncation to $k$ terms. Then the model error is

$$\mathrm{ME}_n(k) = \frac{1}{n\,p} \left\| \sqrt{n}\,\boldsymbol{U}_n \boldsymbol{D}_n \boldsymbol{V}_n^{\mathrm{T}} - \hat{\boldsymbol{X}}_n(k) \right\|_{\mathrm{F}}^2. \tag{6.3}$$

With Assumtions 6.1–6.5, we can apply Proposition 4.6 to get that for fixed $k$ as

$n \to \infty$,

$$p \cdot \mathrm{ME}_n(k) \xrightarrow{a.s.} \sum_{i=1}^{k} \alpha_i \mu_i + \sum_{i=k+1}^{k_0} \mu_i + \sigma^2 \left(1 + \frac{1}{\sqrt{\gamma}}\right)^2 \cdot (k - k_0)_+, \tag{6.4}$$

where

$$\alpha_i = \begin{cases} \frac{\sigma^2}{\gamma \mu_i^2}\left(3\sigma^2 + (\gamma + 1)\mu_i\right) & \text{if } \mu_i > \frac{\sigma^2}{\sqrt{\gamma}}, \\ 1 + \frac{\sigma^2}{\mu_i}\left(1 + \frac{1}{\sqrt{\gamma}}\right)^2 & \text{otherwise.} \end{cases} \tag{6.5}$$

Defining $k_n^*$ as the minimizer of $\mathrm{ME}_n(k)$, we also get that

$$k_n^* \xrightarrow{a.s.} \max\left\{i : \mu_i > \mu_{\mathrm{crit}}\right\}, \tag{6.6}$$

where

$$\mu_{\mathrm{crit}} \equiv \sigma^2 \left( \frac{1 + \gamma^{-1}}{2} + \sqrt{\left(\frac{1 + \gamma^{-1}}{2}\right)^2 + \frac{3}{\gamma}} \right), \tag{6.7}$$

provided no $\mu_i$ is exactly eqaul to $\mu_{\mathrm{crit}}$. We therefore know how $\mathrm{ME}_n(k)$ and its minimizer behave.

To study the bi-cross-validation estimate of prediction error, we need to introduce some more assumptions and notation. As we are only analyzing first-order behavior, we can restrict our analysis to the prediction error estimate from a single fold. We let $\boldsymbol{P}_n \in \mathbb{R}^{n \times n}$ and $\boldsymbol{Q}_n \in \mathbb{R}^{p \times p}$ be permutation matrices for the fold, partitioned as $\boldsymbol{P}_n = \begin{pmatrix} \boldsymbol{P}_{n,1} & \boldsymbol{P}_{n,2} \end{pmatrix}$ and $\boldsymbol{Q}_n = \begin{pmatrix} \boldsymbol{Q}_{n,1} & \boldsymbol{Q}_{n,2} \end{pmatrix}$, with $\boldsymbol{P}_{n,1} \in \mathbb{R}^{n \times n_1}$, $\boldsymbol{P}_{n,2} \in \mathbb{R}^{n \times n_2}$, $\boldsymbol{Q}_{n,1} \in \mathbb{R}^{p \times p_1}$, and $\boldsymbol{Q}_{n,2} \in \mathbb{R}^{p \times p_2}$. Note that $n = n_1 + n_2$ and that $p = p_1 + p_2$. We define $\boldsymbol{X}_{n,ij} = \boldsymbol{P}_{n,i}^{\mathrm{T}} \boldsymbol{X} \boldsymbol{Q}_{n,j}$, $\boldsymbol{E}_{n,ij} = \boldsymbol{P}_{n,i}^{\mathrm{T}} \boldsymbol{E} \boldsymbol{Q}_{n,j}$, $\boldsymbol{U}_{n,i} = \boldsymbol{P}_{n,i}^{\mathrm{T}} \boldsymbol{U}_n$, and $\boldsymbol{V}_{n,j} = \boldsymbol{Q}_{n,j}^{\mathrm{T}} \boldsymbol{V}_n$. Then in block form,

$$\begin{aligned} \boldsymbol{P}_n^{\mathrm{T}} \boldsymbol{X}_n \boldsymbol{Q}_n &= \begin{pmatrix} \boldsymbol{X}_{n,11} & \boldsymbol{X}_{n,12} \\ \boldsymbol{X}_{n,21} & \boldsymbol{X}_{n,22} \end{pmatrix} \\ &= \sqrt{n} \begin{pmatrix} \boldsymbol{U}_{n,1} \boldsymbol{D}_n \boldsymbol{V}_{n,1}^{\mathrm{T}} & \boldsymbol{U}_{n,1} \boldsymbol{D}_n \boldsymbol{V}_{n,2}^{\mathrm{T}} \\ \boldsymbol{U}_{n,2} \boldsymbol{D}_n \boldsymbol{V}_{n,1}^{\mathrm{T}} & \boldsymbol{U}_{n,2} \boldsymbol{D}_n \boldsymbol{V}_{n,2}^{\mathrm{T}} \end{pmatrix} + \begin{pmatrix} \boldsymbol{E}_{n,11} & \boldsymbol{E}_{n,12} \\ \boldsymbol{E}_{n,21} & \boldsymbol{E}_{n,22} \end{pmatrix} \end{aligned} \tag{6.8}$$

This is the starting point of our analysis.

Now we look at the estimate of prediction error. We let $\boldsymbol{X}_{n,11} = \sqrt{n}\,\hat{\boldsymbol{U}}_{n,1}\hat{\boldsymbol{D}}_{n,1}\hat{\boldsymbol{V}}_{n,1}^{\mathrm{T}}$ be the SVD of $\boldsymbol{X}_{n,11}$. Here,

$$\hat{\boldsymbol{U}}_{n,1} = \begin{pmatrix} \underline{\hat{u}}_{n,1}^{(1)} & \underline{\hat{u}}_{n,2}^{(1)} & \cdots & \underline{\hat{u}}_{n,n_1\wedge p_1}^{(1)} \end{pmatrix}, \tag{6.9a}$$

$$\hat{\boldsymbol{V}}_{n,1} = \begin{pmatrix} \underline{\hat{v}}_{n,1}^{(1)} & \underline{\hat{v}}_{n,2}^{(1)} & \cdots & \underline{\hat{v}}_{n,n_1\wedge p_1}^{(1)} \end{pmatrix}, \tag{6.9b}$$

and

$$\hat{\boldsymbol{D}}_{n,1} = \operatorname{diag}\left(\hat{d}_{n,1}^{(1)}, \hat{d}_{n,2}^{(1)}, \ldots, \hat{d}_{n,n_1\wedge p_1}^{(1)}\right). \tag{6.9c}$$

For convenience, we define $\hat{\mu}_{n,i}^{(1)} = \left(\hat{d}_{n,i}^{(1)}\right)^2$. For $0 \leq k \leq n_1 \wedge p_1$, we let

$$\hat{\boldsymbol{D}}_{n,1}(k) = \operatorname{diag}\left(\hat{d}_{n,1}^{(1)}, \hat{d}_{n,2}^{(1)}, \ldots, \hat{d}_{n,k}^{(1)}, 0, 0, \ldots, 0\right) \tag{6.10}$$

so that $\hat{\boldsymbol{X}}_{n,11}(k) \equiv \sqrt{n}\,\hat{\boldsymbol{U}}_{n,1}\hat{\boldsymbol{D}}_{n,1}(k)\hat{\boldsymbol{V}}_{n,1}^{\mathrm{T}}$ is the SVD of $\boldsymbol{X}_{n,11}$ truncated to $k$ terms. This matrix has pseudo-inverse $\hat{\boldsymbol{X}}_{n,11}(k)^{+} = \frac{1}{\sqrt{n}}\hat{\boldsymbol{V}}_{n,1}\hat{\boldsymbol{D}}_{n,1}(k)^{+}\hat{\boldsymbol{U}}_{n,1}^{\mathrm{T}}$. Therefore, the BCV rank-$k$ prediction of $\boldsymbol{X}_{22}$ is

$$\begin{aligned}
\hat{\boldsymbol{X}}_{n,22}(k) &= \boldsymbol{X}_{n,21}\hat{\boldsymbol{X}}_{n,11}^{+}(k)\boldsymbol{X}_{n,12} \\
&= \left(\sqrt{n}\,\boldsymbol{U}_{n,2}\boldsymbol{D}_n\boldsymbol{V}_{n,1}^{\mathrm{T}} + \boldsymbol{E}_{n,21}\right)\left(\hat{\boldsymbol{X}}_{n,11}(k)^{+}\right)\left(\sqrt{n}\,\boldsymbol{U}_{n,1}\boldsymbol{D}_n\boldsymbol{V}_{n,2}^{\mathrm{T}} + \boldsymbol{E}_{n,12}\right) \\
&= \sqrt{n}\,\boldsymbol{U}_{n,2}\boldsymbol{D}_n\boldsymbol{V}_{n,1}^{\mathrm{T}}\hat{\boldsymbol{V}}_{n,1}\hat{\boldsymbol{D}}_{n,1}(k)^{+}\hat{\boldsymbol{U}}_{n,1}^{\mathrm{T}}\boldsymbol{U}_{n,1}\boldsymbol{D}_n\boldsymbol{V}_{n,2}^{\mathrm{T}} \\
&\quad + \boldsymbol{U}_{n,2}\boldsymbol{D}_n\boldsymbol{V}_{n,1}^{\mathrm{T}}\hat{\boldsymbol{V}}_{n,1}\hat{\boldsymbol{D}}_{n,1}(k)^{+}\hat{\boldsymbol{U}}_{n,1}^{\mathrm{T}}\boldsymbol{E}_{n,12} \\
&\quad + \boldsymbol{E}_{n,21}\hat{\boldsymbol{V}}_{n,1}\hat{\boldsymbol{D}}_{n,1}(k)^{+}\hat{\boldsymbol{U}}_{n,1}^{\mathrm{T}}\boldsymbol{U}_{n,1}\boldsymbol{D}_n\boldsymbol{V}_{n,2}^{\mathrm{T}} \\
&\quad + \frac{1}{\sqrt{n}}\boldsymbol{E}_{n,21}\hat{\boldsymbol{V}}_{n,1}\hat{\boldsymbol{D}}_{n,1}(k)^{+}\hat{\boldsymbol{U}}_{n,1}^{\mathrm{T}}\boldsymbol{E}_{n,12} \\
&= \sqrt{n}\,\boldsymbol{U}_{n,2}\boldsymbol{D}_n\boldsymbol{\Theta}_{n,1}\hat{\boldsymbol{D}}_{n,1}(k)^{+}\boldsymbol{\Phi}_{n,1}^{\mathrm{T}}\boldsymbol{D}_n\boldsymbol{V}_{n,2}^{\mathrm{T}} \\
&\quad + \boldsymbol{U}_{n,2}\boldsymbol{D}_n\boldsymbol{\Theta}_{n,1}\hat{\boldsymbol{D}}_{n,1}(k)^{+}\tilde{\boldsymbol{E}}_{n,12} \\
&\quad + \tilde{\boldsymbol{E}}_{n,21}\hat{\boldsymbol{D}}_{n,1}(k)^{+}\boldsymbol{\Phi}_{n,1}^{\mathrm{T}}\boldsymbol{D}_n\boldsymbol{V}_{n,2}^{\mathrm{T}} \\
&\quad + \frac{1}{\sqrt{n}}\tilde{\boldsymbol{E}}_{n,21}\hat{\boldsymbol{D}}_{n,1}(k)^{+}\tilde{\boldsymbol{E}}_{n,12},
\end{aligned} \tag{6.11}$$

where $\boldsymbol{\Theta}_{n,1} = \boldsymbol{V}_{n,1}^{\mathrm{T}} \hat{\boldsymbol{V}}_{n,1}$, $\boldsymbol{\Phi}_{n,1} = \boldsymbol{U}_{n,1}^{\mathrm{T}} \hat{\boldsymbol{U}}_{n,1}$, $\tilde{\boldsymbol{E}}_{n,12} = \hat{\boldsymbol{U}}_{n,1}^{\mathrm{T}} \boldsymbol{E}_{n,12}$, and $\tilde{\boldsymbol{E}}_{n,21} = \boldsymbol{E}_{n,21} \hat{\boldsymbol{V}}_{n,1}$. Note that $\tilde{\boldsymbol{E}}_{n,12}$ and $\tilde{\boldsymbol{E}}_{n,21}$ have iid $\mathcal{N}(0,\, \sigma^2)$ entries, also independent of the other terms that make up $\hat{\boldsymbol{X}}_{n,22}(k)$ and $\boldsymbol{X}_{n,22}$. By conditioning on $\tilde{\boldsymbol{E}}_{n,12}$ and $\tilde{\boldsymbol{E}}_{n,21}$, we can see that $\hat{\boldsymbol{X}}_{22}(k)$ is in general a biased estimate of $\sqrt{n}\, \boldsymbol{U}_{n,2} \boldsymbol{D}_n \boldsymbol{V}_{n,2}^{\mathrm{T}}$.

To analyze $\hat{\boldsymbol{X}}_{22}(k)$, we need to impose some additional assumptions. The first assumption is fairly banal and involves the leave-out sizes.

**Assumption 6.6.** *There exist fixed $K, L \in (0, \infty)$ (not necessarily integers), such that $\frac{n}{n_2} = K + o\left(\frac{1}{\sqrt{n}}\right)$ and $\frac{p}{p_2} = L + o\left(\frac{1}{\sqrt{n}}\right)$.*

The next assumption is not quite so innocent and involves the distribution of the factors.

**Assumption 6.7.** *The departure from orthogonality for the held-in factors $\boldsymbol{U}_{n,1}$ and $\boldsymbol{V}_{n,1}$ is of order $\frac{1}{\sqrt{n}}$. Specifically,*

$$\sup_n \mathbb{E} \left\| \sqrt{n_1} \Big( \boldsymbol{U}_{n,1}^{\mathrm{T}} \boldsymbol{U}_{n,1} - \frac{n_1}{n} \boldsymbol{I}_{k_0} \Big) \right\|_F^2 < \infty, \quad \textit{and}$$
$$\sup_n \mathbb{E} \left\| \sqrt{p_1} \Big( \boldsymbol{V}_{n,1}^{\mathrm{T}} \boldsymbol{V}_{n,1} - \frac{p_1}{p} \boldsymbol{I}_{k_0} \Big) \right\|_F^2 < \infty.$$

This assumption is there so that we can apply the theory in Chapter 3 to get at the behavior of the SVD of $\boldsymbol{X}_{n,11}$. It is satisfied, for example, if we are performing rotated cross-validation (see subsection 5.2.4) or if the factors are generated by certain stationary processes. Proposition A.5 in Appendix A is helpful for verifying Assumption 6.7. It is likely that the theory presented below holds under a weaker condition, but a detailed analysis is beyond the scope of this chapter.

## 6.2 Main results

The BCV estimate of prediction error from a single replicate is given by

$$\widehat{\mathrm{PE}}_n(k) = \frac{1}{n_2\, p_2} \| \boldsymbol{X}_{n,22} - \hat{\boldsymbol{X}}_{n,22}(k) \|_{\mathrm{F}}^2. \tag{6.12}$$

It turns out that $\mathbb{E}\left[\widehat{\mathrm{PE}}_n(k)\right]$ is dominated by the irreducible error. However, if we have a $\sqrt{np}$-consistent estimate of $\sigma^2$, then we can get an expression for the scaled limit of the estimated model error. This expression is the main result of the chapter.

**Theorem 6.8.** *Suppose that $\hat{\sigma}_n^2$ is a sequence of $\sqrt{np}$-consistent estimators of $\sigma^2$ satisfying $\mathbb{E}\left[\sqrt{np}\,(\hat{\sigma}_n^2 - \sigma^2)\right] \to 0$. Define the BCV estimate of model error from a single replicate as*

$$\widehat{\mathrm{ME}}_n(k) = \widehat{\mathrm{PE}}_n(k) - \hat{\sigma}_n^2. \tag{6.13}$$

*Then, for fixed $k$ as $n \to \infty$,*

$$\mathbb{E}\left[p \cdot \widehat{\mathrm{ME}}_n(k)\right] \to \sum_{i=1}^{k \wedge k_0} \beta_i\, \mu_i + \sum_{i=k+1}^{k_0} \mu_i + \eta \cdot (k - k_0)_+, \tag{6.14}$$

*where*

$$\beta_i = \begin{cases} \dfrac{\frac{\sigma^2}{\gamma \mu_i^2}\left(3\sigma^2 + \left(\gamma \frac{K-1}{K} + \frac{L-1}{L}\right)\mu_i\right)}{\rho + \left(\gamma \frac{K-1}{K} + \frac{L-1}{L}\right)\frac{\sigma^2}{\gamma \mu_i} + \frac{\sigma^4}{\gamma \mu_i^2}} & \text{when } \mu_i > \frac{\sigma^2}{\sqrt{\rho\gamma}}, \\ 1 + \frac{\eta}{\mu_i} & \text{otherwise,} \end{cases} \tag{6.15a}$$

$$\rho = \frac{K-1}{K} \cdot \frac{L-1}{L}, \tag{6.15b}$$

*and*

$$\eta = \frac{\sigma^2}{\left(\sqrt{\gamma} + \sqrt{\frac{K}{K-1}} \cdot \sqrt{\frac{L-1}{L}}\right)^2}. \tag{6.15c}$$

It is interesting to compare $\beta_i$ with the expression for $\alpha_i$ in equation (6.5), which appears in the scaled limit of the true model error, $p \cdot \mathrm{ME}_n(k)$. Although the BCV estimator of model error is biased, this bias is small for large $\mu_i$, $K$, and $L$.

A corollary gives the behavior of the minimizer of the expected model error estimate. We let $\hat{k}_n$ be the rank that minimizes $\mathbb{E}\left[\widehat{\mathrm{ME}}_n(k)\right]$. Note that $\hat{k}_n$ is a deterministic quantity. The next two results follow from Theorem 6.8.

**Corollary 6.9.** *As* $n \to \infty$*,*

$$\hat{k}_n \to \max\left\{ i : \mu_i > \frac{\sigma^2}{\sqrt{\gamma}} \cdot \sqrt{\frac{2}{\rho}} \right\}, \tag{6.16}$$

*(provided no* $\mu_i$ *is exactly equal to* $\frac{\sigma^2}{\sqrt{\gamma}} \cdot \sqrt{\frac{2}{\rho}}$*, in which case the limit is ambiguous).*

We can use Corollary 6.9 to guide our choice of $K$ and $L$. If we choose them carefully, then $\hat{k}_n$ and $k_n^*$ will converge to the same value.

**Corollary 6.10.** *If*

$$\sqrt{\rho} = \frac{\sqrt{2}}{\sqrt{\bar{\gamma}} + \sqrt{\bar{\gamma} + 3}}, \tag{6.17}$$

*where*

$$\bar{\gamma} = \left( \frac{\gamma^{1/2} + \gamma^{-1/2}}{2} \right)^2, \tag{6.18}$$

*then* $\hat{k}_n$ *and* $k_n^*$ *converge to the same value (provided no* $\mu_i$ *is exactly equal to* $\frac{\sigma^2}{\sqrt{\gamma}} \cdot \sqrt{\frac{2}{\rho}}$*).*

Interestingly, the first-order-optimal choices of $K$ and $L$ do not depend on the aspect ratio of the original matrix. All that matters is the ratio of the number of elements in $\boldsymbol{X}_{n,11}$ to the number of elements in $\boldsymbol{X}_n$.

For a square matrix, $\gamma = \bar{\gamma} = 1$, and the optimal $\rho$ is $\frac{2}{9}$. If we choose $K = L$, then this requires

$$\frac{K-1}{K} = \frac{\sqrt{2}}{3},$$

so that

$$K = \left( 1 - \frac{\sqrt{2}}{3} \right)^{-1} \approx 1.89.$$

For general aspect ratios ($\gamma \neq 1$), this requires

$$K = \frac{3}{3 - 2\left(\sqrt{\bar{\gamma} + 3} - \sqrt{\bar{\gamma}}\right)}.$$

For very large or very small aspect ratios ($\gamma \to 0$ or $\gamma \to \infty$), $K \to 1$. In these situations one should leave out almost all of the matrix when performing bi-cross-validation.

The remainder of the chapter is devoted to proving Theorem 6.8.

## 6.3 The SVD of the held-in block

The first step in analyzing the BCV estimate of prediction error is to see how the SVD of $\boldsymbol{X}_{n,11}$ behaves. With Assumptions 6.1–6.7, we can start this analysis. Our strategy is to use Assumption 6.7 to apply a matrix perturbation argument in combination with Theorems 3.4 and 3.5.

We show that we can apply Theorem 3.5 to $\sqrt{n}\,\boldsymbol{U}_{n,1}\boldsymbol{D}_n\boldsymbol{V}_{n,1}^{\mathrm{T}} + \boldsymbol{E}_{n,11}$ even though $\boldsymbol{U}_{n,1}$ and $\boldsymbol{V}_{n,1}$ do not have orthogonal columns. First we set

$$\gamma_1 = \frac{K-1}{K} \cdot \frac{L}{L-1} \cdot \gamma \tag{6.19}$$

and note that $\frac{n_1}{p_1} = \frac{n_1}{n} \cdot \frac{p}{p_1} \cdot \frac{n}{p} = \gamma_1 + o\left(\frac{1}{\sqrt{n_1}}\right)$. Next, for $1 \leq i \leq k_0$, we define

$$\mu_{1,i} = \frac{L-1}{L} \cdot \mu_i, \tag{6.20a}$$

$$\bar{\mu}_{1,i} = \begin{cases} \frac{K-1}{K} \cdot (\mu_{1,i} + \sigma^2)\left(1 + \frac{\sigma^2}{\gamma_1 \mu_{1,i}}\right) & \text{when } \mu_{1,i} > \frac{\sigma^2}{\sqrt{\gamma_1}}, \\ \frac{K-1}{K} \cdot \sigma^2 \left(1 + \frac{1}{\sqrt{\gamma_1}}\right)^2 & \text{otherwise.} \end{cases} \tag{6.20b}$$

$$\theta_{1,i} = \begin{cases} \sqrt{\frac{L-1}{L}} \cdot \sqrt{\left(1 - \frac{\sigma^4}{\gamma_1 \mu_{1,i}^2}\right)\left(1 + \frac{\sigma^2}{\gamma_1 \mu_{1,i}}\right)^{-1}} & \text{when } \mu_{1,i} > \frac{\sigma^2}{\sqrt{\gamma_1}}, \\ 0 & \text{otherwise,} \end{cases} \tag{6.20c}$$

$$\varphi_{1,i} = \begin{cases} \sqrt{\frac{K-1}{K}} \cdot \sqrt{\left(1 - \frac{\sigma^4}{\gamma_1 \mu_{1,i}^2}\right)\left(1 + \frac{\sigma^2}{\mu_{1,i}}\right)^{-1}} & \text{when } \mu_{1,i} > \frac{\sigma^2}{\sqrt{\gamma_1}}, \\ 0 & \text{otherwise.} \end{cases} \tag{6.20d}$$

For $i > k_0$, we put $\bar{\mu}_{1,i} = \frac{K-1}{K} \cdot \sigma^2 \left(1 + \frac{1}{\sqrt{\gamma_1}}\right)^2$. Now, for $k \geq 1$ we let $\boldsymbol{\Theta}_1(k)$ and $\boldsymbol{\Phi}_1(k)$ be $k_0 \times k$ matrices with entries

$$\theta_{1,ij}^{(k)} = \begin{cases} \theta_{1,i} & \text{if } i = j, \\ 0 & \text{otherwise,} \end{cases} \quad \text{and} \quad \varphi_{1,ij}^{(k)} = \begin{cases} \varphi_{1,i} & \text{if } i = j, \\ 0 & \text{otherwise,} \end{cases} \tag{6.21}$$

respectively. In this section, we prove the following:

**Proposition 6.11.** *For fixed $k$ as $n \to \infty$, the first $k$ columns of $\mathbf{\Theta}_{1,n}$ and $\mathbf{\Phi}_{1,n}$ converge in probability to $\mathbf{\Theta}_1(k)$ and $\mathbf{\Phi}_1(k)$, respectively. Likewise, for $1 \leq i \leq k$, $\hat{d}^{(1)}_{n,i} \xrightarrow{P} \bar{\mu}^{1/2}_{1,i}$.*

We prove the proposition by leveraging the work of Chapter 3. We have that $\boldsymbol{X}_{n,11} = \sqrt{n}\,\boldsymbol{U}_{n,1}\boldsymbol{D}_n\boldsymbol{V}^{\mathrm{T}}_{n,1} + \boldsymbol{E}_{n,11}$. The first term does not satisfy the conditions of Theorems 3.4 and 3.5 since $\boldsymbol{U}_{n,1}$ and $\boldsymbol{V}_{n,1}$ do not have orthonormal columns. Moreover, the scaling is $\sqrt{n}$ instead of $\sqrt{n_1}$. We introduce scaling constants and group the terms as

$$\boldsymbol{X}_{n,11} = \sqrt{n_1}\Big(\sqrt{\frac{n}{n_1}}\boldsymbol{U}_{n,1}\Big)\Big(\sqrt{\frac{p_1}{p}}\boldsymbol{D}_n\Big)\Big(\sqrt{\frac{p}{p_1}}\boldsymbol{V}_{n,1}\Big)^{\mathrm{T}} + \boldsymbol{E}_{n,11}. \tag{6.22}$$

With this scaling,

$$\boldsymbol{E}\left[\Big(\sqrt{\frac{n}{n_1}}\boldsymbol{U}_{n,1}\Big)^{\mathrm{T}}\Big(\sqrt{\frac{n}{n_1}}\boldsymbol{U}_{n,1}\Big)\right] = \boldsymbol{E}\left[\Big(\sqrt{\frac{p}{p_1}}\boldsymbol{V}_{n,1}\Big)^{\mathrm{T}}\Big(\sqrt{\frac{p}{p_1}}\boldsymbol{V}_{n,1}\Big)\right] = \boldsymbol{I}_{k_0},$$

and the diagonal elements of $\left(\sqrt{\frac{p_1}{p}}\boldsymbol{D}_n\right)$ converge to $\mu^{1/2}_{1,1}, \mu^{1/2}_{1,2}, \ldots, \mu^{1/2}_{1,k_0}$. So, Proposition 6.11 should at least be plausible.

We prove the result by showing that $\left(\sqrt{\frac{n}{n_1}}\boldsymbol{U}_{n,1}\right)\left(\sqrt{\frac{p_1}{p}}\boldsymbol{D}_n\right)\left(\sqrt{\frac{p}{p_1}}\boldsymbol{V}_{n,1}\right)^{\mathrm{T}}$ is *almost* an SVD. We denote its $k_0$-term SVD by $\tilde{\boldsymbol{U}}_{n,1}\tilde{\boldsymbol{D}}_{n,1}\tilde{\boldsymbol{V}}^{\mathrm{T}}_{n,1}$, and demonstrate the following:

**Lemma 6.12.** *Three properties hold:*

*(1) For $1 \leq i \leq k_0$, $|\frac{p_1}{p}d^2_{n,i} - \tilde{d}^2_{n,i}| = \mathcal{O}_P\left(\frac{1}{\sqrt{n}}\right)$.*

*(2) For any sequence of vectors $\underline{x}_1, \underline{x}_2, \ldots, \underline{x}_n$ with $\underline{x}_n \in \mathbb{R}^n$,*

$$\left\|\Big(\sqrt{\frac{n}{n_1}}\boldsymbol{U}_{n,1}\Big)^{\mathrm{T}}\underline{x}_n - \tilde{\boldsymbol{U}}^{\mathrm{T}}_{n,1}\,\underline{x}_n\right\|_2 = \mathcal{O}_P\left(\frac{\|\underline{x}_n\|_2}{\sqrt{n}}\right).$$

*(3) For any sequence of vectors $\underline{y}_1, \underline{y}_2, \ldots, \underline{y}_n$ with $\underline{y}_n \in \mathbb{R}^p$,*

$$\left\|\Big(\sqrt{\frac{p}{p_1}}\boldsymbol{V}_{n,1}\Big)^{\mathrm{T}}\underline{y}_n - \tilde{\boldsymbol{V}}^{\mathrm{T}}_{n,1}\,\underline{y}_n\right\|_2 = \mathcal{O}_P\left(\frac{\|\underline{y}_n\|_2}{\sqrt{n}}\right).$$

*Above, $\tilde{d}_{n,i}$ is the ith diagonal entry of $\tilde{\boldsymbol{D}}_n$, which are assumed to be sorted in descending order.*

Proposition 6.11 is then a direct consequence of Lemma 6.12 combined with Theorems 3.4 and 3.5.

*Proof of Lemma 6.12.* First, let $\sqrt{\frac{n}{n_1}}\boldsymbol{U}_{n,1} = \bar{\boldsymbol{U}}_{n,1}\boldsymbol{R}_n$ be a $QR$-decomposition so that $\bar{\boldsymbol{U}}_{n,1} \in \mathbb{R}^{n_1 \times k_0}$ has orthonormal columns and $\boldsymbol{R}_n$ is an upper-triangular matrix. Define $\boldsymbol{R}_{n,1} = \sqrt{n}(\boldsymbol{R}_{n,1} - \boldsymbol{I}_{k_0})$ so that $\boldsymbol{R}_n = \boldsymbol{I}_{k_0} + \frac{1}{\sqrt{n}}\boldsymbol{R}_{n,1}$. We can write

$$\frac{n}{n_1}\boldsymbol{U}_{n,1}^{\mathrm{T}}\boldsymbol{U}_{n,1} = \boldsymbol{I}_{k_0} + \frac{1}{\sqrt{n}}\left(\boldsymbol{R}_{n,1} + \boldsymbol{R}_{n,1}^{\mathrm{T}}\right) + \frac{1}{n}\boldsymbol{R}_{n,1}^{\mathrm{T}}\boldsymbol{R}_{n,1}.$$

By Assumption 6.7, $\frac{n}{n_1}\boldsymbol{U}_{n,1}^{\mathrm{T}}\boldsymbol{U}_{n,1} - I_{k_0} = \mathcal{O}_{\mathrm{P}}\left(\frac{1}{\sqrt{n}}\right)$. Therefore, $\boldsymbol{R}_{n,1} = \mathcal{O}_{\mathrm{P}}(1)$.

The same argument applies to show that there exits a $\bar{\boldsymbol{V}}_{n,1} \in \mathbb{R}^{p_1 \times k_0}$ with orthonormal columns such that

$$\sqrt{\frac{p}{p_1}}\boldsymbol{V}_{n,1} = \tilde{\boldsymbol{V}}_{n,1}\left(\boldsymbol{I}_{k_0} + \frac{1}{\sqrt{n}}\boldsymbol{S}_{n,1}\right),$$

and $\boldsymbol{S}_{n,1}$ is upper-triangular with $\boldsymbol{S}_{n,1} = \mathcal{O}_{\mathrm{P}}(1)$.

We now look at

$$\begin{aligned}\left(\boldsymbol{I}_{k_0} + \frac{1}{\sqrt{n}}\boldsymbol{R}_{n,1}\right)\left(\boldsymbol{D}_n\right)\left(\boldsymbol{I}_{k_0} + \frac{1}{\sqrt{n}}\boldsymbol{S}_{n,1}\right)^{\mathrm{T}}& \\ &= \boldsymbol{D}_n + \frac{1}{\sqrt{n}}\left(\boldsymbol{R}_{n,1}\boldsymbol{D}_n + \boldsymbol{D}_n\boldsymbol{S}_{n,1}^{\mathrm{T}}\right) + \frac{1}{n}\boldsymbol{R}_{n,1}\boldsymbol{D}_n\boldsymbol{S}_{n,1}^{\mathrm{T}}.\end{aligned}$$

Since the diagonal elements of $\boldsymbol{D}_n$ are distinct, we can apply Lemma 2.15 twice to get that there exist $k_0 \times k_0$ matrices $\bar{\boldsymbol{R}}_{n,1}$, $\bar{\boldsymbol{S}}_{n,1}$ and $\boldsymbol{\Delta}_n$ of size $\mathcal{O}_{\mathrm{P}}(1)$ such that

$$\begin{aligned}\left(\boldsymbol{I}_{k_0} + \frac{1}{\sqrt{n}}\boldsymbol{R}_{n,1}\right)\left(\boldsymbol{D}_n\right)\left(\boldsymbol{I}_{k_0} + \frac{1}{\sqrt{n}}\boldsymbol{S}_{n,1}\right)^{\mathrm{T}}& \\ &= \left(\boldsymbol{I}_{k_0} + \frac{1}{\sqrt{n}}\bar{\boldsymbol{R}}_{n,1}\right)\left(\boldsymbol{D}_n + \frac{1}{\sqrt{n}}\boldsymbol{\Delta}_n\right)\left(\boldsymbol{I}_{k_0} + \frac{1}{\sqrt{n}}\bar{\boldsymbol{S}}_{n,1}\right)^{\mathrm{T}},\end{aligned}$$

with $\boldsymbol{I}_{k_0} + \frac{1}{\sqrt{n}}\bar{\boldsymbol{R}}_{n,1}$ and $\boldsymbol{I}_{k_0} + \frac{1}{\sqrt{n}}\bar{\boldsymbol{S}}_{n,1}$ both orthogonal matrices, and $\boldsymbol{\Delta}_n$ diagonal.

We define $\tilde{\boldsymbol{U}}_{n,1} = \bar{\boldsymbol{U}}_{n,1}\Big(\boldsymbol{I}_{k_0} + \frac{1}{\sqrt{n}}\bar{\boldsymbol{R}}_{n,1}\Big)$, $\tilde{\boldsymbol{V}}_{n,1} = \bar{\boldsymbol{V}}_{n,1}\Big(\boldsymbol{I}_{k_0} + \frac{1}{\sqrt{n}}\bar{\boldsymbol{S}}_{n,1}\Big)$, and $\tilde{\boldsymbol{D}}_n = \sqrt{\frac{p_1}{p}}\Big(\boldsymbol{D}_n + \frac{1}{\sqrt{n}}\boldsymbol{\Delta}\Big)$. Now, as promised, $\tilde{\boldsymbol{U}}_{n,1}\tilde{\boldsymbol{D}}_n\tilde{\boldsymbol{V}}_{n,1}^{\mathrm{T}}$ is the SVD of

$$\Big(\sqrt{\frac{n}{n_1}}\boldsymbol{U}_{n,1}\Big)\Big(\sqrt{\frac{p_1}{p}}\boldsymbol{D}_n\Big)\Big(\sqrt{\frac{p}{p_1}}\boldsymbol{V}_{n,1}\Big)^{\mathrm{T}}.$$

We can see immediately the property (1) holds. For property (2), note that

$$\begin{aligned}
\tilde{\boldsymbol{U}}_{n,1} &= \bar{\boldsymbol{U}}_{n,1}\Big(\boldsymbol{I}_{k_0} + \frac{1}{\sqrt{n}}\bar{\boldsymbol{R}}_{n,1}\Big) \\
&= \sqrt{\frac{n}{n_1}}\boldsymbol{U}_{n,1}\Big(\boldsymbol{I}_{k_0} + \frac{1}{\sqrt{n}}\boldsymbol{R}_{n,1}\Big)^{-1}\Big(\boldsymbol{I}_{k_0} + \frac{1}{\sqrt{n}}\bar{\boldsymbol{R}}_{n,1}\Big) \\
&= \sqrt{\frac{n}{n_1}}\boldsymbol{U}_{n,1}\Big(\boldsymbol{I}_{k_0} + \frac{1}{\sqrt{n}}\tilde{\boldsymbol{R}}_{n,1}\Big)
\end{aligned}$$

for some $\tilde{\boldsymbol{R}}_{n,1} = \mathcal{O}_{\mathrm{P}}(1)$. Therefore,

$$\tilde{\boldsymbol{U}}_{n,1}^{\mathrm{T}}\underline{x}_n - \boldsymbol{U}_{n,1}^{\mathrm{T}}\underline{x}_n = \frac{1}{\sqrt{n}}\tilde{\boldsymbol{R}}_{n,1}^{\mathrm{T}}\boldsymbol{U}_{n,1}^{\mathrm{T}}\underline{x}_n$$

so that

$$\begin{aligned}
\|\tilde{\boldsymbol{U}}_{n,1}^{\mathrm{T}}\underline{x}_n - \boldsymbol{U}_{n,1}^{\mathrm{T}}\underline{x}_n\|_2 &\leq \frac{1}{\sqrt{n}}\|\tilde{\boldsymbol{R}}_{n,1}\|_{\mathrm{F}} \cdot \|\boldsymbol{U}_{n,1}\|_{\mathrm{F}} \cdot \|\underline{x}_n\|_2 \\
&= \mathcal{O}_{\mathrm{P}}\left(\frac{\|\underline{x}_n\|_2}{\sqrt{n}}\right).
\end{aligned}$$

A similar argument applies to show that property (3) holds. □

## 6.4 The prediction error estimate

In this section, we study the estimate of prediction error

$$\widehat{\mathrm{PE}}_n(k) = \frac{1}{n_2\,p_2}\|\boldsymbol{X}_{n,22} - \hat{\boldsymbol{X}}_{n,22}(k)\|_{\mathrm{F}}^2$$

We can expand this as

$$
\begin{aligned}
\widehat{\mathrm{PE}}_n(k) &= \frac{1}{n_2\, p_2} \operatorname{tr}\Big( \big(\boldsymbol{X}_{n,22} - \hat{\boldsymbol{X}}_{n,22}(k)\big)\big(\boldsymbol{X}_{n,22} - \hat{\boldsymbol{X}}_{n,22}(k)\big)^{\mathrm{T}} \Big) \\
&= \frac{1}{n_2\, p_2} \operatorname{tr}\Big( \big(\sqrt{n}\,\boldsymbol{U}_{n,2}\boldsymbol{D}_n\boldsymbol{V}_{n,2}^{\mathrm{T}} + \boldsymbol{E}_{n,22} - \hat{\boldsymbol{X}}_{n,22}(k)\big) \\
&\qquad\qquad \cdot \big(\sqrt{n}\,\boldsymbol{U}_{n,2}\boldsymbol{D}_n\boldsymbol{V}_{n,2}^{\mathrm{T}} + \boldsymbol{E}_{n,22} - \hat{\boldsymbol{X}}_{n,22}(k)\big)^{\mathrm{T}} \Big) \\
&= \frac{1}{n_2\, p_2} \Big( \big\|\sqrt{n}\,\boldsymbol{U}_{n,2}\boldsymbol{D}_n\boldsymbol{V}_{n,2}^{\mathrm{T}} - \hat{\boldsymbol{X}}_{n,22}(k)\big\|_{\mathrm{F}}^2 \\
&\qquad + 2 \operatorname{tr}\Big( \boldsymbol{E}_{n,22}\big(\sqrt{n}\,\boldsymbol{U}_{n,2}\boldsymbol{D}_n\boldsymbol{V}_{n,2}^{\mathrm{T}} - \hat{\boldsymbol{X}}_{n,22}(k)\big)^{\mathrm{T}} \Big) \\
&\qquad + \|\boldsymbol{E}_{n,22}\|_{\mathrm{F}}^2 \Big).
\end{aligned} \tag{6.23}
$$

It has expectation

$$
\mathbb{E}\Big[\widehat{\mathrm{PE}}_n(k)\Big] = \mathbb{E}\Big[\frac{1}{n_2\, p_2}\big\|\sqrt{n}\,\boldsymbol{U}_{n,2}\boldsymbol{D}_n\boldsymbol{V}_{n,2}^{\mathrm{T}} - \hat{\boldsymbol{X}}_{n,22}(k)\big\|\Big] + \sigma^2. \tag{6.24}
$$

The first term is the expected model approximation error and the second term is the irreducible error.

The expected model approximation error expands into four terms. We have

$$
\begin{aligned}
&\mathbb{E}\big\|\sqrt{n}\,\boldsymbol{U}_{n,2}\boldsymbol{D}_n\boldsymbol{V}_{n,2}^{\mathrm{T}} - \hat{\boldsymbol{X}}_{n,22}(k)\big\|_{\mathrm{F}}^2 \\
&\quad = \mathbb{E}\Big[ \operatorname{tr}\Big( \big(\sqrt{n}\,\boldsymbol{U}_{n,2}(\boldsymbol{D}_n - \boldsymbol{D}_n\boldsymbol{\Theta}_{n,1}\hat{\boldsymbol{D}}_{n,1}(k)^{+}\boldsymbol{\Phi}_{n,1}\boldsymbol{D}_n)\boldsymbol{V}_{n,2}^{\mathrm{T}} \\
&\qquad\qquad - \boldsymbol{U}_{n,2}\boldsymbol{D}_n\boldsymbol{\Theta}_{n,1}\hat{\boldsymbol{D}}_{n,1}(k)^{+}\tilde{\boldsymbol{E}}_{n,12} \\
&\qquad\qquad - \tilde{\boldsymbol{E}}_{n,21}\hat{\boldsymbol{D}}_{n,1}(k)^{+}\boldsymbol{\Phi}_{n,1}^{\mathrm{T}}\boldsymbol{D}_n\boldsymbol{V}_{n,2}^{\mathrm{T}} \\
&\qquad\qquad - \frac{1}{\sqrt{n}}\tilde{\boldsymbol{E}}_{n,21}\hat{\boldsymbol{D}}_{n,1}(k)^{+}\tilde{\boldsymbol{E}}_{n,12}\big) \\
&\qquad \cdot \big(\sqrt{n}\,\boldsymbol{U}_{n,2}(\boldsymbol{D}_n - \boldsymbol{D}_n\boldsymbol{\Theta}_{n,1}\hat{\boldsymbol{D}}_{n,1}(k)^{+}\boldsymbol{\Phi}_{n,1}\boldsymbol{D}_n)\boldsymbol{V}_{n,2}^{\mathrm{T}} \\
&\qquad\qquad - \boldsymbol{U}_{n,2}\boldsymbol{D}_n\boldsymbol{\Theta}_{n,1}\hat{\boldsymbol{D}}_{n,1}(k)^{+}\tilde{\boldsymbol{E}}_{n,12} \\
&\qquad\qquad - \tilde{\boldsymbol{E}}_{n,21}\hat{\boldsymbol{D}}_{n,1}(k)^{+}\boldsymbol{\Phi}_{n,1}^{\mathrm{T}}\boldsymbol{D}_n\boldsymbol{V}_{n,2}^{\mathrm{T}} \\
&\qquad\qquad\qquad - \frac{1}{\sqrt{n}}\tilde{\boldsymbol{E}}_{n,21}\hat{\boldsymbol{D}}_{n,1}(k)^{+}\tilde{\boldsymbol{E}}_{n,12}\big)^{\mathrm{T}}\Big)\Big].
\end{aligned} \tag{6.25}
$$

By first conditioning on everything but $\tilde{\boldsymbol{E}}_{n,12}$ and $\tilde{\boldsymbol{E}}_{n,21}$, the cross-terms cancel and we get

$$\begin{aligned}
&\mathbb{E}\big\|\sqrt{n}\,\boldsymbol{U}_{n,2}\boldsymbol{D}_n\boldsymbol{V}_{n,2}^{\mathrm{T}} - \hat{\boldsymbol{X}}_{n,22}(k)\big\|_{\mathrm{F}}^2 \\
&\qquad = \mathbb{E}\big\|\sqrt{n}\,\boldsymbol{U}_{n,2}(\boldsymbol{D}_n - \boldsymbol{D}_n\boldsymbol{\Theta}_{n,1}\hat{\boldsymbol{D}}_{n,1}(k)^{+}\boldsymbol{\Phi}_{n,1}\boldsymbol{D}_n)\boldsymbol{V}_{n,2}^{\mathrm{T}}\big\|_{\mathrm{F}}^2 \\
&\qquad\quad + \mathbb{E}\big\|\boldsymbol{U}_{n,2}\boldsymbol{D}_n\boldsymbol{\Theta}_{n,1}\hat{\boldsymbol{D}}_{n,1}(k)^{+}\tilde{\boldsymbol{E}}_{n,12}\big\|_{\mathrm{F}}^2 \\
&\qquad\quad + \mathbb{E}\big\|\tilde{\boldsymbol{E}}_{n,21}\hat{\boldsymbol{D}}_{n,1}(k)^{+}\boldsymbol{\Phi}_{n,1}^{\mathrm{T}}\boldsymbol{D}_n\boldsymbol{V}_{n,2}^{\mathrm{T}}\big\|_{\mathrm{F}}^2 \\
&\qquad\quad + \mathbb{E}\Big\|\frac{1}{\sqrt{n}}\tilde{\boldsymbol{E}}_{n,21}\hat{\boldsymbol{D}}_{n,1}(k)^{+}\tilde{\boldsymbol{E}}_{n,12}\Big\|_{\mathrm{F}}^2. \qquad (6.26)
\end{aligned}$$

Since $\tilde{\boldsymbol{E}}_{n,12}$ and $\tilde{\boldsymbol{E}}_{n,21}$ are made of iid $\mathcal{N}(0,\,\sigma^2)$ random variables, the last three terms are fairly easy to analyze. We can use the following lemma:

**Lemma 6.13.** *Let $\boldsymbol{Z} \in \mathbb{R}^{m\times n}$ be a random matrix with uncorrelated elements, all having mean 0 and variance 1 (but not necessarily from the same distribution). If $\boldsymbol{A} \in \mathbb{R}^{n\times p}$ is independent of $\boldsymbol{Z}$, then*

$$\mathbb{E}\,\|\boldsymbol{Z}\boldsymbol{A}\|_F^2 = m \cdot \boldsymbol{E}\,\|\boldsymbol{A}\|_F^2 .$$

*Proof.* The square of the $ij$ element of the product is given by

$$\begin{aligned}
(\boldsymbol{Z}\boldsymbol{A})_{ij}^2 &= \Big(\sum_{\alpha=1}^{n} Z_{i\alpha}A_{\alpha j}\Big)^2 \\
&= \sum_{\alpha=1}^{n} Z_{i\alpha}^2 A_{\alpha j}^2 + \sum_{\alpha\neq\beta} Z_{i\alpha}A_{\alpha j}Z_{i\beta}A_{\beta j}
\end{aligned}$$

This has expectation

$$\mathbb{E}\left[(\boldsymbol{Z}\boldsymbol{A})_{ij}^2\right] = \sum_{\alpha=1}^{n} \mathbb{E}\left[A_{\alpha j}^2\right],$$

so that

$$\begin{aligned}
\mathbb{E}\left\|\boldsymbol{Z}\boldsymbol{A}\right\|_{\mathrm{F}}^{2} &= \sum_{i=1}^{m}\sum_{j=1}^{p}\mathbb{E}\left[(\boldsymbol{Z}\boldsymbol{A})_{ij}^{2}\right] \\
&= \sum_{i=1}^{m}\sum_{j=1}^{p}\sum_{\alpha=1}^{n}\mathbb{E}\left[A_{\alpha j}^{2}\right] \\
&= m\cdot\mathbb{E}\left\|\boldsymbol{A}\right\|_{\mathrm{F}}^{2}. \qquad \square
\end{aligned}$$

We also need a technical result to ensure that after appropriate scaling, the expectations are finite.

**Lemma 6.14.** *For fixed $k$ as $n\to\infty$, $\hat{d}_{n,k}^{(1)}$ is almost surely bounded away from zero.*

*Proof.* We have that $\hat{d}_{n,k}^{(1)}$ is the $k$th singular value of $\frac{1}{\sqrt{n}}\boldsymbol{X}_{n,11} = \boldsymbol{U}_{n,1}\boldsymbol{D}_n\boldsymbol{V}_{n,1}^{\mathrm{T}} + \frac{1}{\sqrt{n}}\boldsymbol{E}_{n,11}$. This is the same as the $k$th eigenvalue of

$$\begin{aligned}
\frac{1}{n}\boldsymbol{X}_{n,11}\boldsymbol{X}_{n,11}^{\mathrm{T}} = \boldsymbol{U}_{n,1}\boldsymbol{D}_n\boldsymbol{V}_{n,1}^{\mathrm{T}}\boldsymbol{V}_{n,1}\boldsymbol{D}_n\boldsymbol{U}_{n,1}^{\mathrm{T}} &+ \frac{1}{\sqrt{n}}\boldsymbol{U}_{n,1}\boldsymbol{D}_n\boldsymbol{V}_{n,1}^{\mathrm{T}}\boldsymbol{E}_{n,11}^{\mathrm{T}} \\
&+ \frac{1}{\sqrt{n}}\boldsymbol{E}_{n,11}\boldsymbol{V}_{n,1}\boldsymbol{D}_n\boldsymbol{U}_{n,1}^{\mathrm{T}} + \frac{1}{n}\boldsymbol{E}_{n,11}\boldsymbol{E}_{n,11}^{\mathrm{T}}.
\end{aligned}$$

For each $n$, we choose $\boldsymbol{O}_n = \begin{pmatrix}\boldsymbol{O}_{n,1} & \boldsymbol{O}_{n,2}\end{pmatrix} \in \mathbb{R}^{n\times n}$ to be an orthogonal matrix with $\boldsymbol{O}_{n,2}\in\mathbb{R}^{n\times n-k_0}$ and $\boldsymbol{O}_{n,2}^{\mathrm{T}}\boldsymbol{U}_{n,1} = 0$. Then, with $\lambda_k(\cdot)$ denoting the $k$th eigenvalue, we have

$$\begin{aligned}
\hat{d}_{n,k}^{(1)} &= \lambda_k\left(\frac{1}{n}\boldsymbol{X}_{n,11}\boldsymbol{X}_{n,11}^{\mathrm{T}}\right) \\
&= \lambda_k\left(\boldsymbol{O}_n^{\mathrm{T}}\Big(\frac{1}{n}\boldsymbol{X}_{n,11}\boldsymbol{X}_{n,11}^{\mathrm{T}}\Big)\boldsymbol{O}_n\right) \\
&= \lambda_k\left(\begin{pmatrix}\boldsymbol{A}_{n,11} & \boldsymbol{A}_{n,12} \\ \boldsymbol{A}_{n,21} & \boldsymbol{A}_{n,22}\end{pmatrix}\right),
\end{aligned}$$

for $\boldsymbol{A}_{n,ij} = \frac{1}{n}\boldsymbol{O}_{n,i}^{\mathrm{T}}\boldsymbol{X}_{n,11}^{\mathrm{T}}\boldsymbol{X}_{n,11}\boldsymbol{O}_{n,j}$. Note that $\boldsymbol{A}_{n,22} = \frac{1}{n}\boldsymbol{O}_{n,2}^{\mathrm{T}}\boldsymbol{E}_{n,11}\boldsymbol{E}_{n,11}^{\mathrm{T}}\boldsymbol{O}_{n,2}$ is an $(n-k)\times(n-k)$ matrix with iid $\mathcal{N}(0,\,\sigma^2)$ entries.

Define $G_n = \lambda_k(\boldsymbol{A}_{n,22})$ By the eigenvalue interlacing inequality [36], we have that $\hat{d}_{n,k}^{(1)} \geq G_n$. Moreover, Theorems 2.17 and 2.20 give us that $G_n \stackrel{a.s.}{\to} \sigma^2 \left(1 + \frac{1}{\sqrt{\gamma_1}}\right)^2$. Hence, almost surely for $n$ large enough $\hat{d}_{n,k}^{(1)} \geq G_n \geq \sigma^2$. □

With these lemmas, we can prove the following:

**Lemma 6.15.** *The terms in the expected model approximation error converge as:*

$$\mathbb{E}\left[\frac{p}{n_2\, p_2}\left\|\sqrt{n}\,\boldsymbol{U}_{n,2}(\boldsymbol{D}_n - \boldsymbol{D}_n\boldsymbol{\Theta}_{n,1}\hat{\boldsymbol{D}}_{n,1}(k)^{+}\boldsymbol{\Phi}_{n,1}\boldsymbol{D}_n)\boldsymbol{V}_{n,2}^{\mathrm{T}}\right\|_F^2\right] \to \sum_{i=1}^{k\wedge k_0} \mu_i \left(1 - \sqrt{\frac{\mu_i}{\bar{\mu}_{1,i}}}\theta_{1,i}\varphi_{1,i}\right)^2 + \sum_{i=k+1}^{k_0} \mu_i, \tag{6.27a}$$

$$\mathbb{E}\left[\frac{p}{n_2\, p_2}\left\|\boldsymbol{U}_{n,2}\boldsymbol{D}_n\boldsymbol{\Theta}_{n,1}\hat{\boldsymbol{D}}_{n,1}(k)^{+}\tilde{\boldsymbol{E}}_{n,12}\right\|_F^2\right] \to \frac{\sigma^2}{\gamma}\cdot\sum_{i=1}^{k\wedge k_0}\frac{\mu_i}{\bar{\mu}_{1,i}}\theta_{1,i}^2, \tag{6.27b}$$

$$\mathbb{E}\left[\frac{p}{n_2\, p_2}\left\|\tilde{\boldsymbol{E}}_{n,21}\hat{\boldsymbol{D}}_{n,1}(k)^{+}\boldsymbol{\Phi}_{n,1}^{\mathrm{T}}\boldsymbol{D}_n\boldsymbol{V}_{n,2}^{\mathrm{T}}\right\|_F^2\right] \to \sigma^2\cdot\sum_{i=1}^{k\wedge k_0}\frac{\mu_i}{\bar{\mu}_{1,i}}\varphi_{1,i}^2, \tag{6.27c}$$

*and*

$$\mathbb{E}\left[\frac{p}{n_2\, p_2}\left\|\frac{1}{\sqrt{n}}\tilde{\boldsymbol{E}}_{n,21}\hat{\boldsymbol{D}}_{n,1}(k)^{+}\tilde{\boldsymbol{E}}_{n,12}\right\|_F^2\right] \to \frac{\sigma^2}{\gamma}\cdot\sum_{i=1}^{k}\frac{\sigma^2}{\bar{\mu}_{1,i}}. \tag{6.27d}$$

*Proof.* The squared Frobenius norm in the first term is equal to

$$\begin{aligned}
&\operatorname{tr}\Big(\big(\sqrt{n}\,\boldsymbol{U}_{n,2}(\boldsymbol{D}_n - \boldsymbol{D}_n\boldsymbol{\Theta}_{n,1}\hat{\boldsymbol{D}}_{n,1}(k)^{+}\boldsymbol{\Phi}_{n,1}\boldsymbol{D}_n)\boldsymbol{V}_{n,2}^{\mathrm{T}}\big)\\
&\qquad\cdot\big(\sqrt{n}\,\boldsymbol{U}_{n,2}(\boldsymbol{D}_n - \boldsymbol{D}_n\boldsymbol{\Theta}_{n,1}\hat{\boldsymbol{D}}_{n,1}(k)^{+}\boldsymbol{\Phi}_{n,1}\boldsymbol{D}_n)\boldsymbol{V}_{n,2}^{\mathrm{T}}\big)^{\mathrm{T}}\Big)\\
&= n\cdot\operatorname{tr}\Big(\big(\boldsymbol{D}_n - \boldsymbol{D}_n\boldsymbol{\Theta}_{n,1}\hat{\boldsymbol{D}}_{n,1}(k)^{+}\boldsymbol{\Phi}_{n,1}\boldsymbol{D}_n\big)\cdot\big(\boldsymbol{V}_{n,2}^{\mathrm{T}}\boldsymbol{V}_{n,2}\big)\\
&\qquad\cdot\big(\boldsymbol{D}_n - \boldsymbol{D}_n\boldsymbol{\Theta}_{n,1}\hat{\boldsymbol{D}}_{n,1}(k)^{+}\boldsymbol{\Phi}_{n,1}\boldsymbol{D}_n\big)\cdot\big(\boldsymbol{U}_{n,2}^{\mathrm{T}}\boldsymbol{U}_{n,2}\big)\Big).
\end{aligned}$$

Now, $\boldsymbol{U}_{n,2}^{\mathrm{T}}\boldsymbol{U}_{n,2} \xrightarrow{P} \frac{1}{K}\boldsymbol{I}_{k_0}$, $\boldsymbol{V}_{n,2}^{\mathrm{T}}\boldsymbol{V}_{n,2} \xrightarrow{P} \frac{1}{L}\boldsymbol{I}_{k_0}$, and

$$\begin{aligned}
&\boldsymbol{D}_n - \boldsymbol{D}_n\boldsymbol{\Theta}_{n,1}\hat{\boldsymbol{D}}_{n,1}(k)^{+}\boldsymbol{\Phi}_{n,1}\boldsymbol{D}_n \\
&\quad\xrightarrow{P} \operatorname{diag}\Big(\mu_1^{1/2} - \mu_1^{1/2}\theta_{1,1}\bar{\mu}_{1,1}^{-1/2}(k)\,\varphi_{1,1}\mu_1^{1/2}, \quad \mu_2^{1/2} - \mu_2^{1/2}\theta_{1,2}\bar{\mu}_{1,2}^{-1/2}(k)\,\varphi_{1,2}\mu_2^{1/2}, \quad \ldots, \\
&\qquad\qquad \mu_{k_0}^{1/2} - \mu_{k_0}^{1/2}\theta_{1,k_0}\bar{\mu}_{1,k_0}^{-1/2}(k)\,\varphi_{1,k_0}\mu_{k_0}^{1/2}\Big),
\end{aligned}$$

where

$$\bar{\mu}_{1,i}(k) = \begin{cases} \bar{\mu}_{1,i} & \text{if } i \leq k, \\ 0 & \text{otherwise.} \end{cases}$$

We can apply the Bounded Convergence Theorem to get the result for the first term since the elements of $\boldsymbol{U}_{n,2}$, $\boldsymbol{V}_{n,2}$, $\boldsymbol{\Theta}_{n,1}$ and $\boldsymbol{\Phi}_{n,1}$ are bounded by 1 and since Lemma 6.14 ensures that the elements of $\hat{\boldsymbol{D}}_{n,1}(k)^{+}$ are bounded as well.

The last three terms can be gotten similarly by applying Lemmas 6.13 and 6.14.

□

We can now get an expression for the limit of the estimated model approximation error. Specifically,

$$\begin{aligned}
&\mathbb{E}\left[\frac{p}{n_2\,p_2}\left\|\sqrt{n}\,\boldsymbol{U}_{n,2}\boldsymbol{D}_n\boldsymbol{V}_{n,2}^{\mathrm{T}} - \hat{\boldsymbol{X}}_{n,22}(k)\right\|_{\mathrm{F}}^{2}\right] \\
&\qquad= \sum_{i=1}^{k\wedge k_0}\left\{\mu_i\left(1 - \sqrt{\frac{\mu_i}{\bar{\mu}_{1,i}}}\theta_{1,i}\varphi_{1,i}\right)^2 + \sigma^2\frac{\mu_i}{\bar{\mu}_i}\left(\gamma^{-1}\theta_{1,i}^2 + \varphi_{1,i}^2\right) + \frac{\sigma^4}{\gamma\bar{\mu}_{1,i}}\right\} \\
&\qquad\qquad + \sum_{i=k+1}^{k_0}\mu_i + \frac{\sigma^2}{\gamma}\cdot\sum_{i=k_0+1}^{k}\frac{\sigma^2}{\bar{\mu}_{1,i}} \\
&\qquad= \sum_{i=1}^{k\wedge k_0}\beta_i\,\mu_i + \sum_{i=k+1}^{k_0}\mu_i + \frac{\sigma^2}{\gamma}\left(1 + \frac{1}{\sqrt{\gamma_1}}\right)^{-2}\cdot(k - k_0)_{+},
\end{aligned} \tag{6.28}$$

where

$$\beta_i = \left(1 - \sqrt{\frac{\mu_i}{\bar{\mu}_{1,i}}}\theta_{1,i}\varphi_{1,i}\right)^2 + \frac{\sigma^2}{\bar{\mu}_{1,i}}\left(\gamma^{-1}\theta_{1,i}^2 + \varphi_{1,i}^2\right) + \frac{1}{\mu_i}\frac{\sigma^4}{\gamma\bar{\mu}_{1,i}}$$
$$= 1 - 2\sqrt{\frac{\mu_i}{\bar{\mu}_{1,i}}}\theta_{1,i}\varphi_{1,i} + \frac{\mu_i}{\bar{\mu}_{1,i}}\theta_{1,i}^2\varphi_{1,i}^2 + \frac{\sigma^2}{\bar{\mu}_{1,i}}\left(\gamma^{-1}\theta_{1,i}^2 + \varphi_{1,i}^2\right) + \frac{1}{\mu_i}\frac{\sigma^4}{\gamma\bar{\mu}_{1,i}}. \tag{6.29}$$

If $\mu_{1,i} \le \frac{\sigma^2}{\sqrt{\gamma_1}}$, then $\theta_{1,i} = \varphi_{1,i} = 0$ and $\bar{\mu}_{1,i} = \frac{K-1}{K} \cdot \sigma^2 \left(1 + \frac{1}{\sqrt{\gamma_1}}\right)^2$, so that

$$\beta_i = 1 + \frac{1}{\gamma} \cdot \frac{\sigma^2}{\mu_i} \cdot \left(1 + \frac{1}{\sqrt{\gamma_1}}\right)^{-2}.$$

In the opposite situation ($\mu_{1,i} > \frac{\sigma^2}{\sqrt{\gamma_1}}$), we define

$$\rho = \frac{L-1}{L} \cdot \frac{K-1}{K}$$

and get the simplifications

$$\frac{\bar{\mu}_{1,i}}{\mu_i} = \rho \cdot \left(1 + (1 + \gamma_1^{-1})\frac{\sigma^2}{\mu_{1,i}} + \frac{\sigma^4}{\gamma_1 \mu_i^2}\right),$$
$$\theta_{1,i}\varphi_{1,i} = \rho \cdot \sqrt{\frac{\mu_i}{\bar{\mu}_{1,i}}} \cdot \left(1 - \frac{\sigma^4}{\gamma_1 \mu_{1,i}^2}\right),$$

so that

$$- 2\sqrt{\frac{\mu_i}{\bar{\mu}_{1,i}}}\theta_{1,i}\varphi_{1,i} + \frac{\mu_i}{\bar{\mu}_i}\theta_{1,i}^2\varphi_{1,i}^2$$
$$= -\rho^2 \left(\frac{\mu_i}{\bar{\mu}_{1,i}}\right)^2 \left(1 - \frac{\sigma^4}{\gamma_1 \mu_{1,i}^2}\right)\left(1 + 2(1 + \gamma_1^{-1})\frac{\sigma^2}{\mu_{1,i}} + 3\frac{\sigma^4}{\gamma_1 \mu_{1,i}^2}\right),$$

$$\frac{\sigma^2}{\bar{\mu}_{1,i}}\left(\gamma^{-1}\theta_{1,i}^2 + \varphi_{1,i}^2\right) = \rho^2 \left(\frac{\mu_i}{\bar{\mu}_{1,i}}\right)^2 \left(1 - \frac{\sigma^4}{\gamma_1 \mu_{1,i}^2}\right)\left((1 + \gamma_1^{-1})\frac{\sigma^2}{\mu_{1,i}} + 2\frac{\sigma^4}{\gamma_1 \mu_{1,i}^2}\right),$$

and

$$\frac{1}{\mu_i}\frac{\sigma^4}{\gamma\bar{\mu}_{1,i}} = \rho^2 \left(\frac{\mu_i}{\bar{\mu}_{1,i}}\right)^2 \left(\frac{\sigma^4}{\gamma_1 \mu_{1,i}^2}\right)\left(1 + (1 + \gamma_1^{-1})\frac{\sigma^2}{\mu_{1,i}} + \frac{\sigma^4}{\gamma_1 \mu_{1,i}^2}\right).$$

Putting it all together, we get

$$\begin{aligned}
\beta_i &= 1 + \rho^2 \left(\frac{\mu_i}{\bar{\mu}_{1,i}}\right)^2 \cdot \Bigg\{ \left(1 - \frac{\sigma^4}{\gamma_1 \mu_{1,i}^2}\right)\left(-1 - (1+\gamma_1^{-1})\frac{\sigma^2}{\mu_{1,i}} - \frac{\sigma^4}{\gamma_1 \mu_{1,i}^2}\right) \\
&\qquad + \left(\frac{\sigma^4}{\gamma_1 \mu_{1,i}^2}\right)\left(1 + (1+\gamma_1^{-1})\frac{\sigma^2}{\mu_{1,i}} + \frac{\sigma^4}{\gamma_1 \mu_{1,i}^2}\right)\Bigg\} \\
&= 1 + \rho^2 \left(\frac{\mu_i}{\bar{\mu}_{1,i}}\right)^2 \left(2\frac{\sigma^4}{\gamma_1 \mu_{1,i}^2} - 1\right)\left(1 + (1+\gamma_1^{-1})\frac{\sigma^2}{\mu_{1,i}} + \frac{\sigma^4}{\gamma_1 \mu_{1,i}^2}\right) \\
&= 1 + \frac{2\frac{\sigma^4}{\gamma_1 \mu_{1,i}^2} - 1}{1 + (1+\gamma_1^{-1})\frac{\sigma^2}{\mu_{1,i}} + \frac{\sigma^4}{\gamma_1 \mu_{1,i}^2}} \\
&= \frac{3\frac{\sigma^4}{\gamma_1 \mu_{1,i}^2} + (1+\gamma_1^{-1})\frac{\sigma^2}{\mu_{1,i}}}{1 + (1+\gamma_1^{-1})\frac{\sigma^2}{\mu_{1,i}} + \frac{\sigma^4}{\gamma_1 \mu_{1,i}^2}} \\
&= \frac{\frac{\sigma^2}{\gamma_1 \mu_{1,i}^2}\left(3\sigma^2 + (\gamma_1 + 1)\mu_{1,i}\right)}{1 + (1+\gamma_1^{-1})\frac{\sigma^2}{\mu_{1,i}} + \frac{\sigma^4}{\gamma_1 \mu_{1,i}^2}}.
\end{aligned} \tag{6.30}$$

In particular, note that $\beta_i < 1$ when $2\frac{\sigma^4}{\gamma_1 \mu_{1,i}^2} - 1 < 0$, or equivalently $\mu_i > \frac{\sigma^2}{\sqrt{\gamma}} \cdot \sqrt{\frac{2}{\rho}}$. Getting the final expressions for $\beta_i$ and $\eta$ in Theorem 6.8 is a matter of routine algebra.

## 6.5 Summary and future work

We have provided an analysis of the first-order behavior of bi-cross-validation. This analysis has shown that BCV gives a biased estimate of prediction error, with an explicit expression for the bias. Fortunately, the bias is not too bad when the signal strength is large and the leave-out sizes are small. Importantly, our analysis gives guidance as to how the leave-out sizes should be chosen. Our theoretical analysis agrees with the simulations done by Owen & Perry [65], who observed that despite bias in prediction error estimates, larger hold-out sizes tend to perform better at estimating $k_{\mathrm{F}}^*$.

The form of consistency we give is rather weak since we did not analyze the variance of the BCV prediction error estimate. As a follow-up, it would be worthwhile to address this limitation. For now, though, we can get some comfort from empirical observations in [65] that the variance of the estimator is not too large. Indeed, based on these simulations, it is entirely possible that the variance becomes negligible for large $n$ and $p$.

# Appendix A

# Properties of random projections

We use this section to present some results about random projection matrices. We call a symmetric $p \times p$ matrix $\boldsymbol{P}$ a projection if its eigenvalues are in the set $\{0, 1\}$. Any projection matrix of rank $k \leq p$ can be decomposed as $\boldsymbol{P} = \boldsymbol{V}\boldsymbol{V}^{\mathrm{T}}$ for some $\boldsymbol{V}$ satisfying $\boldsymbol{V}^{\mathrm{T}}\boldsymbol{V} = \boldsymbol{I}_k$. We call the set

$$\mathcal{V}_k(\mathbb{R}^p) = \left\{\boldsymbol{V} \in \mathbb{R}^{p \times k} : \boldsymbol{V}^{\mathrm{T}}\boldsymbol{V} = \boldsymbol{I}_k\right\} \subseteq \mathbb{R}^{p \times k} \tag{A.1}$$

the *rank-k Stiefel manifold of* $\mathbb{R}^p$. It is the set of orthonormal $k$-frames in $\mathbb{R}^p$. Similarly, we call the set

$$\mathcal{G}_k(\mathbb{R}^p) = \left\{\boldsymbol{V}\boldsymbol{V}^{\mathrm{T}} : \boldsymbol{V} \in \mathcal{V}_k(\mathbb{R}^p)\right\} \subseteq \mathbb{R}^{p \times p} \tag{A.2}$$

the *rank-k Grassmannian of* $\mathbb{R}^p$; this is the set of rank $k$ projection matrices. If $\begin{pmatrix}\boldsymbol{V} & \bar{\boldsymbol{V}}\end{pmatrix}$ is an $p \times p$ Haar-distributed orthogonal matrix and $\boldsymbol{V}$ is $p \times k$, then we say that $\boldsymbol{V}$ is uniformly distributed over $\mathcal{V}_k(\mathbb{R}^p)$ and that $\boldsymbol{V}\boldsymbol{V}^{\mathrm{T}}$ is uniformly distributed over $\mathcal{G}_k(\mathbb{R}^p)$.

## A.1 Uniformly distributed orthonormal $k$-frames

We first present some results about a matrix $\boldsymbol{V}$ distributed uniformly over $\mathcal{V}_k(\mathbb{R}^p)$. We denote this distribution by $\boldsymbol{V} \sim \mathrm{Unif}\big(\mathcal{V}_k(\mathbb{R}^p)\big)$. In the special case of $k = 1$, the distribution is equivalent to drawing a random vector uniformly from the unit sphere in $\mathbb{R}^p$. We denote this distribution by $\underline{V} \sim \text{Unif}\big(\mathcal{S}^{p-1}\big)$.

### A.1.1 Generating random elements

The easiest way to generate a random element of $\mathcal{V}_k(\mathbb{R}^p)$ is to let $\boldsymbol{Z}$ be a $p \times k$ matrix of iid $\mathcal{N}(0,\,1)$ random variables, take the $QR$ decomposition $\boldsymbol{Z} = \boldsymbol{QR}$, and let $\boldsymbol{V}$ be equal to the first $k$ columns of $\boldsymbol{Q}$. In practice, there is a bias in the way standard $QR$ implementations choose the signs of the columns of $\boldsymbol{Q}$. To get around this, we recommend using Algorithm A.1, below.

**Algorithm A.1** Generate a random orthonormal $k$-frame

1. Draw $\boldsymbol{Z}$, a random $p \times k$ matrix whose elements are iid $\mathcal{N}(0,\,1)$ random variables.
2. Compute $\boldsymbol{Z} = \boldsymbol{QR}$, the $QR$-decomposition of $\boldsymbol{Z}$. Set $\boldsymbol{Q}_1$ to be the $p \times k$ matrix containing the first $k$ columns of $\boldsymbol{Q}$.
3. Draw $\boldsymbol{S}$, a random $k \times k$ diagonal matrix with iid diagonal entries such that $\mathbb{P}\{S_{11} = -1\} = \mathbb{P}\{S_{11} = +1\} = \frac{1}{2}$.
4. Return $\boldsymbol{V} = \boldsymbol{Q}_1 \boldsymbol{S}$.

This algorithm has a time complexity of $\mathcal{O}(pk^2)$. Diaconis and Shahshahani [23] present an alternative approach called the subgroup algorithm which can be used to generate $\boldsymbol{V}$ as a product of $k$ Householder reflections. Their algorithm has time complexity $\mathcal{O}(pk)$. Mezzadri [60] gives a simple description of the subgroup algorithm.

### A.1.2 Mixed moments

Since we can flip the sign of any row or column of $\boldsymbol{V}$ and not change its distribution, the mixed moments of the elements of $\boldsymbol{V}$ vanish unless the number of elements from any row or column is even (counting multiplicity). For example, $\mathbb{E}[V_{11}^2 V_{21}] = 0$ since we can flip the sign of the second row of $\boldsymbol{V}$ to get

$$V_{11}^2 V_{21} \overset{d}{=} V_{11}^2(-V_{21}) = -V_{11}^2 V_{21}.$$

This argument does not apply to $V_{11}V_{12}V_{22}V_{21}$ or $V_{11}^2 V_{22}^2$.

In the special case when $k = 1$, the vector $\left(V_{11}^2, V_{21}^2, \ldots, V_{p1}^2\right)$ is distributed as Dirichlet $\left(\frac{1}{2}, \frac{1}{2}, \ldots, \frac{1}{2}\right)$. This follows from the fact that if $Y_1, Y_2, \ldots, Y_p$ are independently distributed Gamma random variables and $Y_i$ has shape $a_i$ and scale $s$, then with $S = \sum_{i=1}^p Y_i$, the vector $\left(\frac{Y_1}{S}, \frac{Y_2}{S}, \ldots, \frac{Y_p}{S}\right)$ is distributed Dirichlet $(a_1, a_2, \ldots, a_p)$. Using the standard formulas for Dirichlet variances and covariances, we get the mixed moments up to fourth order. They are summarized in the next lemma.

**Lemma A.1.** *If* $\boldsymbol{V} \sim \operatorname{Unif}\big(\mathcal{V}_1(\mathbb{R}^p)\big)$, *then*

$$\mathbb{E}\left[V_{11}^2\right] = \frac{1}{p}, \tag{A.3a}$$

$$\mathbb{E}\left[V_{11}V_{21}\right] = 0, \tag{A.3b}$$

$$\mathbb{E}\left[V_{11}^4\right] = \frac{3}{p\,(p+2)}, \tag{A.3c}$$

$$\mathbb{E}\left[V_{11}^2 V_{21}^2\right] = \frac{1}{p\,(p+2)}. \tag{A.3d}$$

*The odd mixed moments are all equal to zero.*

Using Theorem 4 from Diaconis and Shahshahani [24], which gives the moments of the traces of Haar-distributed orthogonal matrices, we can compute the mixed moments of $\boldsymbol{V}$ for more general $k$. Meckes [59] gives an alternative derivation of these results.

**Lemma A.2.** *If* $\boldsymbol{V} \sim \operatorname{Unif}\big(\mathcal{V}_k(\mathbb{R}^p)\big)$ *and* $k > 1$, *then the nonzero mixed moments of* *the elements* $\boldsymbol{V}$ *up to fourth order are defined by*

$$\mathbb{E}\left[V_{11}^2\right] = \frac{1}{p}, \tag{A.4a}$$

$$\mathbb{E}\left[V_{11}^4\right] = \frac{3}{p\,(p+2)}, \tag{A.4b}$$

$$\mathbb{E}\left[V_{11}^2 V_{21}^2\right] = \frac{1}{p\,(p+2)}, \tag{A.4c}$$

$$\mathbb{E}\left[V_{11}^2 V_{22}^2\right] = \frac{p+1}{p\,(p-1)(p+2)}, \tag{A.4d}$$

$$\mathbb{E}\left[V_{11} V_{12} V_{22} V_{21}\right] = \frac{-1}{p\,(p-1)(p+2)}. \tag{A.4e}$$

*Proof.* The first three equations follow directly from the previous lemma. We can get the other moments from the moments of $\boldsymbol{O}$, a Haar-distibuted $p \times p$ orthogonal matrix. For the fourth equation, we use that

$$\mathbb{E}\left[\mathrm{tr}(\boldsymbol{O})\right]^4 = \sum_{r,s,t,u} \mathbb{E}\left[O_{rr} O_{ss} O_{tt} O_{uu}\right].$$

Only the terms with even powers of $O_{ii}$ are nonzero. Thus, we have

$$\begin{aligned}\mathbb{E}\left[\mathrm{tr}(\boldsymbol{O})\right]^4 &= \binom{p}{1}\mathbb{E}\left[O_{11}^4\right] + \binom{p}{2}\binom{4}{2}\mathbb{E}\left[O_{11}^2 O_{22}^2\right] \\ &= p\,\mathbb{E}\left[O_{11}^4\right] + 3p\,(p-1)\,\mathbb{E}\left[O_{11}^2 O_{22}^2\right].\end{aligned}$$

Theorem 4 of Diaconis and Shahshahani [24] gives that $\mathbb{E}\left[\mathrm{tr}(\boldsymbol{O})\right]^4 = 3$. Combined with Lemma A.1, we get that

$$\begin{aligned}\mathbb{E}\left[O_{11}^2 O_{22}^2\right] &= \frac{1}{3p\,(p-1)}\left\{\mathbb{E}\left[\mathrm{tr}(\boldsymbol{O})\right]^4 - p\,\mathbb{E}\left[O_{11}^4\right]\right\} \\ &= \frac{1}{3p\,(p-1)}\left\{3 - p\cdot\frac{3}{p\,(p+1)}\right\} \\ &= \frac{p+1}{p\,(p-1)(p+2)}.\end{aligned}$$

For the last equation, we use $\mathbb{E}\left[\mathrm{tr}(\boldsymbol{O}^4)\right]$. We have that

$$(\boldsymbol{O}^4)_{ij} = \sum_{r,s,t} O_{ir}O_{rs}O_{st}O_{tj}.$$

We would like to compute $\mathbb{E}\left[(\boldsymbol{O}^4)_{11}\right]$. Note that unless $r = s = t = 1$, there are only three situations when $\mathbb{E}\left[O_{1r}O_{rs}O_{st}O_{t1}\right] \neq 0$. Two of them are demonstrated visually by the configurations

$$\underset{r=1,\, s=t,\, s\neq 1}{\begin{array}{c c} & \begin{array}{ccccc} 1 & & s & & \end{array} \\ \begin{array}{c} 1 \\ \\ s \\ \\ \end{array} & \begin{pmatrix} O_{1r} & \cdots & O_{rs} & \cdots \\ \vdots & \ddots & \vdots & \\ O_{t1} & \cdots & O_{st} & \cdots \\ \vdots & & \vdots & \end{pmatrix} \end{array}} \quad \text{and} \quad \underset{t=1,\, r=s,\, r\neq 1}{\begin{array}{c c} & \begin{array}{ccccc} 1 & & r & & \end{array} \\ \begin{array}{c} 1 \\ \\ r \\ \\ \end{array} & \begin{pmatrix} O_{t1} & \cdots & O_{1r} & \cdots \\ \vdots & \ddots & \vdots & \\ O_{st} & \cdots & O_{rs} & \cdots \\ \vdots & & \vdots & \end{pmatrix} \end{array}}.$$

The other nonzero term is when $s = 1$, $r = t$, and $r \neq 1$, so that $O_{1r}O_{rs}O_{st}O_{t1} = O_{1r}^2O_{r1}^2$. In all other configurations there is a row or a column that only contains one of $\{O_{1r}, O_{rs}, O_{st}, O_{t1}\}$. Since we can multiply a row or a column of $\boldsymbol{O}$ by $-1$ and not change the distribution of $\boldsymbol{O}$, for this choice of $r$, $s$, and $t$ we must have $O_{1r}O_{rs}O_{st}O_{t1} \overset{d}{=} -O_{1r}O_{rs}O_{st}O_{t1}$. This in turn implies that $\mathbb{E}\left[O_{1r}O_{rs}O_{st}O_{t1}\right] = 0$. With these combinatorics in mind, we have that

$$\begin{aligned} \mathbb{E}\left[(\boldsymbol{O}^4)_{11}\right] &= \sum_{s\neq 1}\mathbb{E}\left[O_{11}O_{1s}O_{ss}O_{s1}\right] + \sum_{r\neq 1}\mathbb{E}\left[O_{1r}O_{rr}O_{r1}O_{11}\right] + \sum_{r\neq 1}\mathbb{E}\left[O_{1r}^2O_{r1}^2\right] + \mathbb{E}\left[O_{11}^4\right] \\ &= 2(p-1)\mathbb{E}\left[O_{11}O_{12}O_{22}O_{21}\right] + (p-1)\mathbb{E}\left[O_{12}^2O_{21}^2\right] + \mathbb{E}\left[O_{11}^4\right] \\ &= 2(p-1)\mathbb{E}\left[O_{11}O_{12}O_{22}O_{21}\right] + (p-1)\mathbb{E}\left[O_{11}^2O_{22}^2\right] + \mathbb{E}\left[O_{11}^4\right] \end{aligned}$$

Again applying Theorem 4 of [24], we have that $\mathbb{E}\left[\mathrm{tr}(\boldsymbol{O}^4)\right] = 1$. Combined with

Lemma A.1, we have

$$\begin{aligned}\mathbb{E}\left[O_{11}O_{12}O_{22}O_{21}\right] &= \frac{1}{2(p-1)}\left\{\frac{1}{p}\mathbb{E}\left[\operatorname{tr}(\boldsymbol{O}^4)\right] - \mathbb{E}\left[O_{11}^4\right] - (p-1)\mathbb{E}\left[O_{11}^2O_{22}^2\right]\right\} \\ &= \frac{1}{2(p-1)}\left\{\frac{1}{p} - \frac{3}{p\,(p+2)} - (p-1)\frac{p+1}{p\,(p-1)(p+2)}\right\} \\ &= \frac{-1}{p\,(p-1)(p+2)}.\end{aligned}$$

□

## A.2 Uniformly distributed projections

When $\boldsymbol{P} \in \mathbb{R}^{p\times p}$ is chosen uniformly over the set of rank-$k$ $p \times p$ projection matrices, we say $\boldsymbol{P} \sim \operatorname{Unif}\big(\mathcal{G}_k(\mathbb{R}^p)\big)$. With the results of the previous section, we can derive the moments of random projection matrices.

**Lemma A.3.** *If $\boldsymbol{P} \sim \operatorname{Unif}\big(\mathcal{G}_k(\mathbb{R}^p)\big)$, then*

$$\mathbb{E}\left[\boldsymbol{P}\right] = \frac{k}{p}\boldsymbol{I}_p. \tag{A.5}$$

*Proof.* Write $\boldsymbol{P} = \boldsymbol{V}\boldsymbol{V}^{\mathrm{T}}$, where $\boldsymbol{V} \sim \operatorname{Unif}\big(\mathcal{V}_k(\mathbb{R}^p)\big)$. For $1 \leq i, j \leq p$ we have

$$P_{ij} = \sum_{r=1}^{k} V_{ir}V_{jr},$$

so

$$\mathbb{E}\left[P_{ij}\right] = \begin{cases} k\,\mathbb{E}\left[V_{11}^2\right], & \text{when } i = j, \\ k\,\mathbb{E}\left[V_{11}V_{21}\right], & \text{otherwise.}\end{cases}$$

The result now follows from Lemma A.1. □

**Lemma A.4.** *Let $\boldsymbol{P} \sim \operatorname{Unif}\big(\mathcal{G}_k(\mathbb{R}^p)\big)$. If $1 \leq i, j, i', j' \leq p$, then*

$$\begin{aligned}\operatorname{Cov}\left[P_{ij}, P_{i'j'}\right] = \frac{1}{(p-1)(p+2)}\left(\frac{k}{p}\right)\left(1 - \frac{k}{p}\right) \\ \cdot\Big(p\,\delta_{(i,j)=(i',j')} + p\,\delta_{(i,j)=(j',i')} - 2\,\delta_{(i,i')=(j,j')}\Big).\end{aligned} \tag{A.6}$$

*This gives us that aside from the obvious symmetry ($P_{ij} = P_{ji}$), the off-diagonal elements of $\boldsymbol{P}$ are uncorrelated with each other and with the diagonal elements.*

*Proof.* We need to perform six computations. As before, we use the representation $\boldsymbol{P} = \boldsymbol{V}\boldsymbol{V}^{\mathrm{T}}$, where $\boldsymbol{V} \sim \mathrm{Unif}\big(\mathcal{V}_k(\mathbb{R}^p)\big)$. We have

$$\begin{aligned}
\mathbb{E}\left[P_{11}^2\right] &= \mathbb{E}\Big[\Big(\sum_{i=1}^{k} V_{1i}^2\Big)^2\Big] \\
&= \mathbb{E}\Big[\sum_{i=1}^{k} V_{1i}^4 + \sum_{i \neq j} V_{1i}^2 V_{1j}^2\Big] \\
&= k \cdot \frac{3}{p\,(p+2)} + k\,(k-1) \cdot \frac{1}{n\,(p+2)} \\
&= \frac{k\,(k+2)}{p\,(p+2)} \\
&= \frac{2}{p+2}\left(\frac{k}{p}\right)\left(1 - \frac{k}{p}\right) + \left(\frac{k}{p}\right)^2,
\end{aligned}$$

which gives us that

$$\mathrm{Var}\left[P_{11}\right] = \frac{2}{p+2}\left(\frac{k}{p}\right)\left(1 - \frac{k}{p}\right).$$

Next,

$$\begin{aligned}
\mathbb{E}\left[P_{12}^2\right] &= \mathbb{E}\Big[\Big(\sum_{i=1}^{k} V_{1i} V_{2i}\Big)^2\Big] \\
&= \mathbb{E}\Big[\sum_{i=1}^{k} V_{1i}^2 V_{2i}^2 + \sum_{i \neq j} V_{1i} V_{2i} V_{1j} V_{2j}\Big] \\
&= k \cdot \frac{1}{p\,(p+2)} + k\,(k-1) \cdot \frac{-1}{p\,(p-1)(p+2)} \\
&= \frac{p}{(p-1)(p+2)}\left(\frac{k}{p}\right)\left(1 - \frac{k}{p}\right),
\end{aligned}$$

so that

$$\mathrm{Var}\left[P_{12}\right] = \frac{p}{(p-1)(p+2)}\left(\frac{k}{p}\right)\left(1 - \frac{k}{p}\right).$$

Also,

$$
\begin{aligned}
\mathbb{E}\left[P_{11}P_{22}\right] &= \mathbb{E}\Big[\Big(\sum_{i=1}^{k} V_{1i}^2\Big)\Big(\sum_{j=1}^{k} V_{2j}^2\Big)\Big] \\
&= \mathbb{E}\Big[\sum_{i=1}^{k} V_{1i}^2 V_{2i}^2 + \sum_{i=1}^{k}\sum_{j\neq i} V_{1i}^2 V_{2j}^2\Big] \\
&= k \cdot \frac{1}{p\,(p+2)} + k\,(k-1) \cdot \frac{p+1}{p\,(p-1)(p+2)} \\
&= \frac{k}{p\,(p+2)} \left(1 + (p+1)\frac{k-1}{p-1}\right) \\
&= \frac{-2}{(p-1)(p+2)} \left(\frac{k}{p}\right)\left(1 - \frac{k}{p}\right) + \left(\frac{k}{p}\right)^2,
\end{aligned}
$$

so that

$$
\operatorname{Cov}\left[P_{11}, P_{22}\right] = \frac{-2}{(p-1)(p+2)} \left(\frac{k}{p}\right)\left(1 - \frac{k}{p}\right).
$$

Since $\boldsymbol{P}$ is symmetric, we have

$$
\operatorname{Cov}\left[P_{12}, P_{21}\right] = \operatorname{Var}\left[P_{12}\right].
$$

The other covariances are all zero. This is because

$$
\begin{aligned}
\mathbb{E}\left[P_{11}P_{12}\right] &= \mathbb{E}\Big[\sum_{i,j} V_{1i}^2\, V_{1j} V_{2j}\Big], \\
\mathbb{E}\left[P_{12}P_{23}\right] &= \mathbb{E}\Big[\sum_{i,j} V_{1i} V_{2i} V_{2j} V_{3j}\Big],
\end{aligned}
$$

and

$$
\mathbb{E}\left[P_{12}P_{34}\right] = \mathbb{E}\Big[\sum_{i,j} V_{1i} V_{2i} V_{3j} V_{4j}\Big].
$$

Each term in these sums has an element that appears only once in a row. Thus, the expectations are all 0. □

## A.3 Applications

We now present two applications of the results in this section.

### A.3.1 Projections of orthonormal $k$-frames

Suppose we have $\boldsymbol{U} \in \mathbb{R}^{p\times k}$, an orthonormal $k$-frame, and we randomly project the columns of $\boldsymbol{U}$ into $\mathbb{R}^q$, with $q < p$. If we denote the projection matrix by $\boldsymbol{V}^{\mathrm{T}}$ and set $\tilde{\boldsymbol{U}} = \boldsymbol{V}^{\mathrm{T}}\boldsymbol{U}$, it is natural to ask how close $\tilde{\boldsymbol{U}}$ is to being an orthonormal $k$-frame. We can prove the following:

**Proposition A.5.** *Suppose $\boldsymbol{U} \in \mathbb{R}^{p\times k}$ satisfies $\boldsymbol{U}^{\mathrm{T}}\boldsymbol{U} = \boldsymbol{I}_k$. Let $\boldsymbol{V} \sim \mathrm{Unif}\big(\mathcal{V}_q(\mathbb{R}^p)\big)$ with $k \leq q \leq p$ and set $\tilde{\boldsymbol{U}} = \sqrt{p/q}\boldsymbol{V}^{\mathrm{T}}\boldsymbol{U}$. Then there exists a decomposition $\tilde{\boldsymbol{U}} = \tilde{\boldsymbol{U}}_0 + \frac{1}{\sqrt{q}}\tilde{\boldsymbol{U}}_1$ such that $\tilde{\boldsymbol{U}}_0^{\mathrm{T}}\tilde{\boldsymbol{U}}_0 = \boldsymbol{I}_k$ and*

$$\mathbb{E}\|\tilde{\boldsymbol{U}}_1\|_F^2 \leq \frac{1}{2}\,k\,(k+1)\left(\frac{q}{p}\right)^2\left(1-\frac{q}{p}\right). \tag{A.7}$$

*In particular, this implies that $\mathbb{E}\|\tilde{\boldsymbol{U}}_1\|_F^2 \leq \frac{2}{27}\,k\,(k+1)$.*

The main ingredients of the proof are a perturbation lemma and a result about $\tilde{\boldsymbol{U}}^{\mathrm{T}}\tilde{\boldsymbol{U}}$, stated below.

**Lemma A.6.** *Suppose $\boldsymbol{U} \in \mathbb{R}^{p\times k}$ satisfies $\boldsymbol{U}^{\mathrm{T}}\boldsymbol{U} = \boldsymbol{I}_k$. Let $\boldsymbol{V} \sim \mathrm{Unif}\big(\mathcal{V}_q(\mathbb{R}^p)\big)$ with $k \leq q \leq p$ and set $\tilde{\boldsymbol{U}} = \sqrt{p/q}\boldsymbol{V}^{\mathrm{T}}\boldsymbol{U}$. Then $\mathbb{E}\left[\tilde{\boldsymbol{U}}^{\mathrm{T}}\tilde{\boldsymbol{U}}\right] = \boldsymbol{I}_k$ and*

$$\boldsymbol{E}\left\|\sqrt{q}\,\big(\tilde{\boldsymbol{U}}^{\mathrm{T}}\tilde{\boldsymbol{U}} - \boldsymbol{I}_k\big)\right\|_F^2 \leq k\,(k+1)\left(\frac{q}{p}\right)^2\left(1-\frac{q}{p}\right), \tag{A.8a}$$

*with a matching lower bound of*

$$\boldsymbol{E}\left\|\sqrt{q}\,\big(\tilde{\boldsymbol{U}}^{\mathrm{T}}\tilde{\boldsymbol{U}} - \boldsymbol{I}_k\big)\right\|_F^2 \geq k\,(k+1)\left(\frac{q}{p}\right)^2\left(1-\frac{q}{p}\right)\left(\frac{p}{p+2}\right). \tag{A.8b}$$

*Proof.* Set $\boldsymbol{P} = \boldsymbol{V}\boldsymbol{V}^{\mathrm{T}}$ so that $\tilde{\boldsymbol{U}}^{\mathrm{T}}\tilde{\boldsymbol{U}} = \frac{p}{q}\boldsymbol{U}^{\mathrm{T}}\boldsymbol{P}\boldsymbol{U}$. Now,

$$\boldsymbol{E}\left[\tilde{\boldsymbol{U}}^{\mathrm{T}}\tilde{\boldsymbol{U}}\right] = \frac{p}{q}\boldsymbol{U}^{\mathrm{T}}\mathbb{E}\left[\boldsymbol{P}\right]\boldsymbol{U} = \boldsymbol{I}_k.$$

Also, since $\boldsymbol{O}^{\mathrm{T}}\boldsymbol{P}\boldsymbol{O} \overset{d}{=} \boldsymbol{P}$ for any $p \times p$ orthogonal matrix $\boldsymbol{O}$, we must that $\boldsymbol{U}^{\mathrm{T}}\boldsymbol{P}\boldsymbol{U}$ has the same distribution as the upper $k \times k$ submatrix of $\boldsymbol{P}$. This implies that

$$\begin{aligned}
\boldsymbol{E}\left\|\tilde{\boldsymbol{U}}^{\mathrm{T}}\tilde{\boldsymbol{U}} - \boldsymbol{I}_k\right\|_{\mathrm{F}}^2 &= \sum_{i=1}^{k}\sum_{j=1}^{k}\mathrm{Var}\left[\boldsymbol{P}_{ij}\right] \\
&= \frac{q}{p}\left(1 - \frac{q}{p}\right)\left\{k \cdot \frac{2}{p+2} + k\,(k-1)\cdot\frac{p}{(p-1)(p+2)}\right\} \\
&= \frac{p\,k\,(k+1)}{(p-1)(p+2)}\left(\frac{q}{p}\right)\left(1 - \frac{q}{p}\right)\left(1 - \frac{2}{p\,(k+1)}\right).
\end{aligned}$$

The lower and upper bounds follow. □

The next ingredient is a perturbation theorem due to Mirsky, which we take from Stewart [84] and state as a lemma.

**Lemma A.7** (Mirsky). *If $\boldsymbol{A}$ and $\boldsymbol{A} + \boldsymbol{E}$ are in $\mathbb{R}^{n\times p}$, then*

$$\sum_{i=1}^{n\wedge p}\big(\sigma_i(\boldsymbol{A}+\boldsymbol{E}) - \sigma_i(\boldsymbol{A})\big)^2 \leq \|\boldsymbol{E}\|_F^2, \tag{A.9}$$

*where $\sigma_i(\cdot)$ denotes the $i$th singular value.*

We can now proceed to the rest of the proof.

*Proof of Proposition A.5.* We have that $\tilde{\boldsymbol{U}}^{\mathrm{T}}\tilde{\boldsymbol{U}} = \boldsymbol{I}_k + \boldsymbol{E}$, where $\mathbb{E}\|\sqrt{q}\boldsymbol{E}\|_{\mathrm{F}}^2 \leq C$ and $C$ is given in Lemma A.6. We can apply Lemma A.7 to get

$$\sum_{i=1}^{k}\big(\sigma_i(\tilde{\boldsymbol{U}}^{\mathrm{T}}\tilde{\boldsymbol{U}}) - 1\big)^2 \leq \|\boldsymbol{E}\|_{\mathrm{F}}^2.$$

Setting $\varepsilon_i = \sigma_i(\tilde{\boldsymbol{U}}^{\mathrm{T}}\tilde{\boldsymbol{U}}) - 1$, we have $\sigma_i(\tilde{\boldsymbol{U}}) = \sqrt{1+\varepsilon_i}$. Note that $\mathbb{E}\left[q\sum_i \varepsilon_i^2\right] \leq \mathbb{E}\|\sqrt{q}\,\boldsymbol{E}\|_{\mathrm{F}}^2 \leq C$. Let $\boldsymbol{R}$ and $\boldsymbol{S}$ be $p \times k$ and $k \times k$ matrices with orthonormal columns

such that

$$\tilde{\boldsymbol{U}} = \boldsymbol{R}(\boldsymbol{I}_k + \boldsymbol{\Delta})\boldsymbol{S}^{\mathrm{T}}$$

is the SVD of $\tilde{\boldsymbol{U}}$, where $\boldsymbol{\Delta} = \operatorname{diag}(\Delta_1, \Delta_2, \dots, \Delta_k)$, and $\Delta_i = \sqrt{1+\varepsilon_i} - 1$. Set $\tilde{\boldsymbol{U}}_0 = \boldsymbol{R}\boldsymbol{S}^{\mathrm{T}}$ and $\tilde{\boldsymbol{U}}_1 = \sqrt{q}\,\boldsymbol{R}\boldsymbol{\Delta}\boldsymbol{S}^{\mathrm{T}}$. Then,

$$\tilde{\boldsymbol{U}}_0^{\mathrm{T}}\tilde{\boldsymbol{U}}_0 = \boldsymbol{S}\boldsymbol{R}^{\mathrm{T}}\boldsymbol{R}\boldsymbol{S}^{\mathrm{T}} = \boldsymbol{S}\boldsymbol{S}^{\mathrm{T}} = \boldsymbol{I}_k$$

and

$$\mathbb{E}\|\tilde{\boldsymbol{U}}_1\|_{\mathrm{F}}^2 = \sum_{i=1}^{k} \mathbb{E}\left[q\,\Delta_i^2\right] \leq \frac{1}{2}\sum_{i=1}^{k} \mathbb{E}\left[q\,\varepsilon_i^2\right] \leq \frac{C}{2},$$

where we have used that $\sqrt{1+\varepsilon_i} \geq 1 + \frac{1}{2}\varepsilon_i - \frac{1}{2}\varepsilon_i^2$. □

### A.3.2 A probabilistic interpretation of the Frobenius norm

As another application of the results in this section, we give a probabilistic representation of the Frobenius norm of a matrix. It is commonly known that for any $n \times p$ matrix $\boldsymbol{A}$,

$$\sup_{\substack{\|\underline{x}\|_2=1,\\ \|\underline{y}\|_2=1}} \left(\underline{x}^{\mathrm{T}}\boldsymbol{A}\,\underline{y}\right)^2 = \|\boldsymbol{A}\|_2^2, \tag{A.10}$$

where $\|\cdot\|_2$ is the spectral norm, equal largest singular value (see, e.g. [36]). The function $f(\underline{x},\underline{y}) = \underline{x}^{\mathrm{T}}\boldsymbol{A}\,\underline{y}$ is a general bilinear form on $\mathbb{R}^n \times \mathbb{R}^p$. The square of the spectral norm of $\boldsymbol{A}$ gives the maximum value of $\left(f(\underline{x},\underline{y})\right)^2$ when $\underline{x}$ and $\underline{y}$ are both unit vectors. It turns out that the Frobenius norm of $\boldsymbol{A}$ is related to the *average* value of $\left(f(\underline{X},\underline{Y})\right)^2$ when $\underline{X}$ and $\underline{Y}$ are random unit vectors.

**Proposition A.8.** *If $\boldsymbol{A} \in \mathbb{R}^{n\times p}$, then*

$$\int\limits_{\substack{\|\underline{x}\|_2=1,\\ \|\underline{y}\|_2=1}} \left(\underline{x}^{\mathrm{T}}\boldsymbol{A}\,\underline{y}\right)^2 d\underline{x}\,d\underline{y} = \frac{1}{np}\|\boldsymbol{A}\|_F^2. \tag{A.11}$$

*Proof.* There are two steps to the proof. First, we show that if $\underline{X} \sim \operatorname{Unif}\left(\mathcal{S}^{n-1}\right)$ and

$\underline{a}$ is arbitrary, then

$$\mathbb{E}\left[\underline{X}^{\mathrm{T}}\underline{a}\right]^2 = \frac{1}{n}\|\underline{a}\|_2^2.$$

Next, we show that if $\underline{Y} \sim \mathrm{Unif}\big(\mathcal{S}^{p-1}\big)$, then

$$\mathbb{E}\|\boldsymbol{A}\,\underline{Y}\|_2^2 = \frac{1}{p}\|\boldsymbol{A}\|_{\mathrm{F}}^2.$$

The result follows from these two facts. To see the first part, since $\underline{X}$ is orthogonally invariant we have

$$\underline{X}^{\mathrm{T}}\underline{a} \overset{d}{=} \underline{X}^{\mathrm{T}}\left(\|\underline{a}\|_2\,\underline{e}_1\right) = \|\underline{a}\|_2\,\underline{X}_1.$$

To see the second part, let $\boldsymbol{A} = \boldsymbol{U}\boldsymbol{\Sigma}\boldsymbol{V}^{\mathrm{T}}$ be the SVD of $\boldsymbol{A}$ and write

$$\|\boldsymbol{A}\,\underline{Y}\|_2^2 = \|\boldsymbol{\Sigma}\boldsymbol{V}^{\mathrm{T}}\underline{Y}\|_2^2 \overset{d}{=} \|\boldsymbol{\Sigma}\underline{Y}\|_2^2 = \sum_{i=1}^{n\wedge p}\sigma_i^2(\boldsymbol{A})\,Y_i^2. \qquad \square$$

# Appendix B

# Limit theorems for weighted sums

In this appendix we state and prove some limit theorems for weighted sums of iid random variables. First, we give a weak law of large numbers (WLLN):

**Proposition B.1** (Weighted WLLN)**.** *Let $X_{n,1}, X_{n,2}, \ldots, X_{n,n}$ be a triangular array of random variables, iid across each row, with $\mathbb{E}X_{n,1} = \mu$ and $\mathbb{E}X_{n,1}^2$ uniformly bounded in $n$. Also, let $W_{n,1}, W_{n,2}, \ldots, W_{n,n}$ be another triangular array of random variables independent of the $X_{n,i}$ (but not necessarily of each other). Define $\bar{W}_n = \frac{1}{n}\sum_{i=1}^n W_{n,i}$. If $\bar{W}_n \xrightarrow{P} \bar{W}$ and $\mathbb{E}\left[\frac{1}{n}\sum_{i=1}^n W_{n,i}^2\right]$ is uniformly bounded in $n$, then*

$$\frac{1}{n}\sum_{i=1}^n W_{n,i}\, X_{n,i} \xrightarrow{P} \bar{W}\mu. \tag{B.1}$$

We do not give a proof of Proposition B.1, but instead derive a strong law of large numbers (SLLN) below. The proof of the weak law is similar. Here is the strong law:

**Proposition B.2** (Weighted SLLN)**.** *Let $X_{n,1}, X_{n,2}, \ldots, X_{n,n}$ be a triangular array of random variables, iid across each row, with $\mathbb{E}X_{n,1} = \mu$ and $\mathbb{E}X_{n,1}^4$ uniformly bounded in $n$. Also, let $W_{n,1}, W_{n,2}, \ldots, W_{n,n}$ be another triangular array of random variables independent of the $X_{n,i}$ (but not necessarily of each other). Define $\bar{W}_n = \frac{1}{n}\sum_{i=1}^n W_{n,i}$. If $\bar{W}_n \xrightarrow{a.s.} \bar{W}$ and $\mathbb{E}\left[\frac{1}{n}\sum_{i=1}^n W_{n,i}^4\right]$ is uniformly bounded in $n$, then*

$$\frac{1}{n}\sum_{i=1}^n W_{n,i}\, X_{n,i} \xrightarrow{a.s.} \bar{W}\mu. \tag{B.2}$$

*Proof.* Define $S_n = \sum_{i=1}^n W_{n,i}\, X_{n,i}$ and let $\mathcal{F}_n^W = \sigma(W_{n,1}, W_{n,2}, \ldots, W_{n,n})$. We have that

$$\frac{1}{n} S_n - \bar{W}_n\, \mu = \frac{1}{n} \sum_{i=1}^n W_{n,i}\, (X_{n,i} - \mu)$$

Note that

$$\begin{aligned} &\mathbb{E}\left[ \frac{1}{n^2} \left( \sum_{i=1}^n W_{n,i}\, (X_{n,i} - \mu) \right)^4 \,\middle|\, \mathcal{F}_n^W \right] \\ &\quad = \frac{1}{n^2} \left( \mathbb{E}\, [X_{n,1} - \mu]^4 \sum_{i=1}^n W_{n,i}^4 + 3\, \mathbb{E}\left[ (X_{n,1} - \mu)^2 (X_{n,2} - \mu)^2 \right] \sum_{i \neq j} W_{n,i}^2 W_{n,j}^2 \right) \\ &\quad \leq C_1 \left[ \frac{1}{n} \sum_{i=1}^n W_{n,i}^2 \right]^2 \end{aligned}$$

for some constant $C_1$ bounding $\mathbb{E}\, [X_{n,1} - \mu]^4$ and $3\, \mathbb{E}\, [(X_{n,1} - \mu)^2 (X_{n,2} - \mu)^2]$. Therefore, the full expectation is bounded by some other constant $C_2$. We have just shown that

$$\mathbb{E} \left[ \frac{1}{n} S_n - \bar{W}_n\, \mu \right]^4 \leq \frac{C}{n^2}$$

for some constant $C$. Applying Chebyschev's inequality, we get

$$P \left\{ \left| \frac{1}{n} S_n - \bar{W}_n\, \mu \right| > \varepsilon \right\} \leq \frac{C}{n^2 \varepsilon^4}.$$

Invoking the first Borel-Cantelli Lemma, see the sum converges almost surely. ☐

Next, we derive a central limit theorem (CLT). To prove it we will need a CLT for dependent variables, which we take from McLeish [58]:

**Lemma B.3.** *Let $X_{n,i}, \mathcal{F}_{n,i}, i = 1, \ldots, n$ be a martingale difference array. If the Lindeberg condition $\sum_{i=1}^n \mathbb{E}\left[ X_{n,i}^2; |X_{n,i}| > \varepsilon \right] \to 0$ is satisfied and $\sum_{i=1}^n X_{n,i}^2 \xrightarrow{P} \sigma^2$ then $\sum_{i=1}^n X_{n,i} \xrightarrow{d} \mathcal{N}(0,\, \sigma^2)$.*

With this Lemma, we can prove a CLT for weighted sums.

**Proposition B.4** (Weighted CLT)**.** *Let $\underline{X}_{n,i}, i = 1, \ldots, n$ be a triangular array of random vectors in $\mathbb{R}^p$ with $\underline{X}_{n,i}$ iid such that $\mathbb{E}\underline{X}_{n,1} = \underline{\mu}^X$,* $\mathrm{Cov}[\underline{X}_{n,1}] = \mathbf{\Sigma}^X$, *and all* *mixed fourth moments of the elements of* $\underline{X}_{n,1}$ *are uniformly bounded in* $n$*. Let* $\underline{W}_{n,i}$*,* $i = 1, \ldots, n$ *be another triangular array of random vectors in* $\mathbb{R}^p$*, independent of the* $\underline{X}_{n,i}$ *but not necessarily of each other. Assume that* $\frac{1}{n}\sum_{i=1}^n \underline{W}_{n,i}\underline{W}_{n,i}^{\mathrm{T}} \xrightarrow{P} \mathbf{\Sigma}^W$*, and that for all sets of indices* $j_1, j_2, j_3, j_4$*, with each index between* 1 *and* $p$*, we have that* $\mathbb{E}\left[\frac{1}{n}\sum_{i=1}^n W_{n,ij_1}W_{n,ij_2}W_{n,ij_3}W_{n,ij_4}\right]$ *is uniformly bounded in* $n$*. Then*

$$\sqrt{n}\left[\frac{1}{n}\sum_{i=1}^n \underline{W}_{n,i} \bullet \underline{X}_{n,i} - \frac{1}{n}\sum_{i=1}^n \underline{W}_{n,i} \bullet \underline{\mu}^X\right] \xrightarrow{d} \mathcal{N}\left(0,\, \mathbf{\Sigma}^W \bullet \mathbf{\Sigma}^X\right), \tag{B.3}$$

*where* $\bullet$ *denotes Hadamard (elementwise) product.*

*Proof.* Let

$$\underline{S}_n = \sqrt{n}\left[\frac{1}{n}\sum_{i=1}^n \underline{W}_{n,i} \bullet \underline{X}_{n,i} - \frac{1}{n}\sum_{i=1}^n \underline{W}_{n,i} \bullet \underline{\mu}^X\right].$$

We will use the Cramér-Wold device. Let $\underline{\theta} \in \mathbb{R}^p$ be arbitrary. Define

$$Y_{n,i} = \frac{1}{\sqrt{n}}\sum_{j=1}^p \theta_j W_{n,ij}\,(X_{n,ij} - \mu_j^X),$$

so that $\underline{\theta}^{\mathrm{T}}\underline{S}_n = \sum_{i=1}^n Y_{n,i}$. Also, with $\mathcal{F}_{n,i} = \sigma(Y_{n,1}, Y_{n,2}, \ldots, Y_{n,i-1})$, the collection $\{Y_{n,i}, \mathcal{F}_{n,i}\}$ is a martingale difference array. We will use Lemma B.3 to prove the result. First, we compute the variance as

$$\begin{aligned}
\sum_{i=1}^n Y_{n,i}^2 &= \frac{1}{n}\sum_{i=1}^n \left[\sum_{j=1}^p \theta_j W_{n,ij}\,(X_{n,ij} - \mu_j^X)\right]^2 \\
&= \frac{1}{n}\sum_{i=1}^n \sum_{j=1}^n \sum_{k=1}^p \theta_j\theta_k W_{n,ij}W_{n,ik}(X_{n,ij} - \mu_j^X)(X_{n,ik} - \mu_k^X) \\
&\xrightarrow{P} \sum_{j=1}^p \sum_{k=1}^p \theta_j\theta_k \Sigma_{jk}^W \Sigma_{jk}^X \\
&= \underline{\theta}^{\mathrm{T}}\left(\mathbf{\Sigma}^W \bullet \mathbf{\Sigma}^X\right)\underline{\theta},
\end{aligned}$$

where we have used the fourth moment assumptions and Proposition B.1 to get the convergence. Lastly we check the Lindeberg condition. We have that

$$\begin{aligned}
\sum_{i=1}^{n}\mathbb{E}\left[Y_{n,i}^2; |Y_{n,i}| > \varepsilon\right]
&\leq \frac{1}{\varepsilon^2}\sum_{i=1}^{n}\mathbb{E}\left[Y_{n,i}^4\right] \\
&= \frac{1}{\varepsilon^2 n^2}\sum_{i=1}^{n}\sum_{j_1,\ldots,j_4}\Big\{\theta_{j_1}\theta_{j_2}\theta_{j_3}\theta_{j_4}\cdot\mathbb{E}\left[W_{n,ij_1}W_{n,ij_2}W_{n,ij_3}W_{n,ij_4}\right] \\
&\qquad \cdot\mathbb{E}\Big[(X_{n,ij_1}-\mu_{j_1}^X)(X_{n,ij_2}-\mu_{j_2}^X)(X_{n,ij_3}-\mu_{j_3}^X)(X_{n,ij_4}-\mu_{j_4}^X)\Big]\Big\} \\
&\leq \frac{C_1}{\varepsilon^2 n}\sum_{j_1,\ldots,j_4}\mathbb{E}\left[\frac{1}{n}\sum_{i=1}^{n}W_{n,ij_1}W_{n,ij_2}W_{n,ij_3}W_{n,ij_4}\right] \\
&\leq \frac{C_1C_2p^4}{\varepsilon^2 n} \\
&\to 0
\end{aligned}$$

where $C_1$ is a constant bounding $|\theta_j|^4$ and the centered fourth moments of $X_{n,ij}$, and $C_2$ bounds the fourth moments of the weights. □

For some classes of weights, we can get a stronger result.

**Corollary B.5.** *With the same assumptions as in Proposition B.4, if the mean of the weights converges sufficiently fast as* $\sqrt{n}\left[\frac{1}{n}\sum_{i=1}^{n}\underline{W}_{n,i}-\underline{\mu}^W\right]\xrightarrow{P}0$, *then*

$$\sqrt{n}\left[\frac{1}{n}\sum_{i=1}^{n}\underline{W}_{n,i}\bullet\underline{X}_{n,i}-\underline{\mu}^W\bullet\underline{\mu}^X\right]\xrightarrow{d}\mathcal{N}\left(0,\,\mathbf{\Sigma}^W\bullet\mathbf{\Sigma}^X\right). \tag{B.4}$$